\documentclass{article}
\usepackage{amssymb,hyperref}
\usepackage[english]{babel}
\usepackage{amsmath, graphics}
\usepackage{geometry} 
\usepackage{color}

\newcommand{\mathsym}[1]{{}}
\newcommand{\unicode}[1]{{}}

\def\noi{\noindent}
\def\nq{\hspace*{-1em}}
\def\nqq{\hspace*{-2em}}
\def\nn{\nonumber\\ {}}
\def\nll{\nonumber \nl\hspace{2.2em} }

\def\lal{&&\nqq {}}
\def\nnn{\nonumber\\ \lal }
\def\nl{\\ \lal }
\def\nll{\nonumber \nl\hspace{2.2em} }

\def\noi{\noindent}
\def\nqq{\hspace{-2em}}

\def\barr{\left(\begin{array}}
\def\earr{\end{array}\right)}
\def\beq#1{\begin{equation}\label{#1}}
\def\eeq{\end{equation}}
\def\ber#1{\begin{eqnarray}\label{#1} &&\nqq}
\def\eer{\end{eqnarray}}

\newcommand{\bear}[1]{\begin{eqnarray}\label{#1}}
\newcommand{\ear}{\end{eqnarray}}
\newcommand{\bearr}[1]{\begin{eqnarray}\lal \label{#1}}
\newcommand{\earrr}{\end{eqnarray}}

\renewcommand{\theequation}{\arabic{section}.\arabic{equation}}
\catcode`\@=11 \@addtoreset{equation}{section}\catcode`\@=12
\newcommand{\N}{ {\mathbb N} }
\newcommand{\R}{ {\mathbb R} }

\newcommand{\fnm}{\footnotemark}
\newcommand{\fnt}{\footnotetext}

\begin{document}

 \vspace{15pt}

 \begin{center}
 \large\bf

  Fluxbrane polynomials and Melvin-like solutions for
   simple Lie algebras

 \vspace{15pt}

\vspace{15pt}

 \normalsize\bf
      S. V. Bolokhov\fnm[1]\fnt[1]{bolokhov-sv@rudn.ru}$^{, a}$
      and  V. D. Ivashchuk\fnm[2]\fnt[2]{ivas@vniims.ru}$^{, a, b}$

 \vspace{7pt}

 \it 
 
 (a) \  Institute of Gravitation and Cosmology,
  RUDN University,
  6 Miklukho-Maklaya Str.,  Moscow 117198, Russia \\
 
 (b) \ \ \ Center for Gravitation and Fundamental
 Metrology,  VNIIMS, 46 Ozyornaya Str., Moscow 119361, Russia  \\

 \end{center}
 \vspace{15pt}

 \small\noi

 \begin{abstract}
  This review dealt with generalized Melvin  solutions for  
  simple finite-dimensional Lie algebras. 
  Each solution appears in a model which includes a metric and $n$ scalar fields coupled to $n$ Abelian 2-forms with
  dilatonic coupling vectors determined by simple Lie algebra  of rank $n$.
  The set of $n$ moduli functions $H_s(z)$ comply with $n$ non-linear (ordinary) differential equations (of second order) with certain boundary conditions
  set. Earlier, it was hypothesized that these moduli functions should
  be polynomials in $z$  (so-called ``fluxbrane'' polynomials) 
   depending upon certain parameters $p_s > 0$, $s = 1,\dots,n$.
  Here,  we presented  explicit relations for the polynomials 
  corresponding to  Lie algebras of ranks  $n = 1,2,3,4,5$  and 
  exceptional algebra $E_6$. 
   Certain relations for the polynomials (e.g., symmetry and duality ones) were outlined.
  In a general case where polynomial conjecture holds,  2-form flux integrals  are finite. 
   The use of fluxbrane polynomials to dilatonic black hole solutions was also explored.

 \end{abstract}


\large 

 \section{Introduction}
  In this review article, we dealt with a certain generalization of the  Melvin solution~\cite{Melv},
   which was studied  earlier in Ref.~\cite{GI-09}. It occurs in the  model 
  which contains metric, $n$ Abelian 2-forms $ F^s =    dA^s$ and $l$ scalar fields 
  $\varphi^{\alpha}$ ($l \geq n$ ).  
  This  solution is governed by a certain non-degenerate  matrix 
  $(A_{s l})$ (``quasi-Cartan'' matrix), $s, l = 1, \dots, n$.  
  It is a particular case of   the so-called generalized fluxbrane  
  solutions presented earlier in  Ref.~\cite{Iflux}.  
  
  The original $4d$ Melvin's solution, which describes the gravitational field of a magnetic flux tube,  
  has numerous multidimensional analogs, supported by  certain configurations of form fields. These
  analogs are usually referred to as fluxbranes. The fluxbrane 
    solutions originally appear in superstring/brane models. For generalized  Melvin and  fluxbrane solutions, see ~\cite{BronShikin,GW,GM,DGKT,DGGidH,DGGH1,DGGH2,GR,CGS,CosG,Saf1,GutSt,CosHC,Emp,Saf2,PapF,BSaf,RTs3,CGSaf,FFS,EmpGut,GIM,KKR} and references therein. 
    
  It is important to note that the solutions from Ref.~\cite{Iflux} are governed by 
  a set of moduli functions   $H_s(z) > 0$ defined on the interval $(0, +\infty)$. Here, 
  $z = \rho^2$ and $\rho$ is a radial variable. These functions 
  obey a set of  $n$ non-linear ordinary differential equations of second \linebreak order---so-called 
  master equations,  governed by the matrix $(A_{s s'})$ (in fact, they are equivalent to 
    Toda-like equations).  The moduli functions should also obey the  boundary 
      conditions:  $H_{s}(+ 0) = 1$,   $s = 1,...,n$.

 Here,  we assume that the matrix $(A_{s s'})$ is just a Cartan matrix of a 
 simple finite-dimensional Lie algebra $\cal G$ of rank $n$. 
 (Obviously, $A_{ss} = 2$ for all $s$).
 Due to ``polynomial conjecture'' from  Ref.~\cite{Iflux}, the
 solutions to master equations with the  boundary conditions imposed 
 have a  polynomial structure: 

  \beq{1.3}
  H_{s}(z) = 1 + \sum_{k = 1}^{n_s} P_s^{(k)} z^k,
  \eeq
  where $P_s^{(k)}$ are constants (
 $P_s^{(n_s)} \neq 0$)  and 
 \beq{1.4}
 n_s = 2 \sum_{s' =1}^{n} A^{s s'}.
 \eeq 
 
 Here,  we denote $(A^{s s'}) = (A_{s s'})^{-1}$.
  The parameters $n_s$ are integers, which  are called components  of a twice-dual
 Weyl vector in the basis of simple roots~\cite{FS}.

In Refs.~\cite{GI-09,GI}, a program  (in Maple)  for the calculation of these (fluxbrane) polynomials for
a classical series of Lie  algebras  ($A_n$, $B_n$, $C_n$, $D_n$) was presented.

The  fluxbrane polynomials $H_s$ define  
special solutions to  open Toda chain equations~\cite{T,B,K,OP}, corresponding 
to  simple finite-dimensional Lie algebra $\cal G$,

 \beq{1.T}
 \frac{d^2 y^s}{du^2} = -  4 P_s \exp( \sum_{l = 1}^{n} A_{s l} y^{l} ),
 \eeq
  where
 \beq{1.h}
  H_s = \exp(- y^s(u) -  n_s u),
 \eeq
 
  $P_s > 0$, $s = 1, ..., n$, and $z = e^{-2u}$.
  These special solutions obey
   \beq{1.hh}
     y^s(u) = -  n_s u + o(1),
   \eeq
  as $u \to + \infty$.

In Section 2, we describe the generalized  Melvin   solution related to a 
simple  finite- \\ dimensional Lie algebra $\cal G$ ~\cite{GI-09}.  
In Section 3 and in Appendicies A and B,
 we present  fluxbrane polynomials  
for Lie algebras of ranks $n =1,2,3,4,5$ and also for $E_6$. 
Here,  we also outline 
the so-called symmetry and duality identities for these polynomials and
consider certain relations between them. 
In Section 4,  we present calculations of  2-form flux integrals $\Phi^s = \int_{M_{*}} F^s$ 
over a certain $2d$ submanifold  $M_{*}$. 
It is amazing that these integrals (fluxes) are  finite for all parameters of fluxbrane polynomials.
In Section 5, we outline  possible applications of fluxbrane polynomials to dilatonic black hole solutions.
 
 It should be noted that definitions of fluxbrane polynomials can be easily extended to (finite dimensional) 
 semisimple Lie algebras.  

\section{The solutions}

We consider a  model governed by the action
 \beq{2.1}
 S=\int d^Dx \sqrt{|g|} \biggl \{R[g]-
 h_{\alpha\beta}g^{MN}\partial_M\varphi^{\alpha}\partial_N\varphi^{\beta}-\frac{1}{2}
 \sum_{s =1}^{n}\exp[2\lambda_s(\varphi)](F^s)^2 \biggr \},
 \eeq
 where $g=g_{MN}(x)dx^M\otimes dx^N$ is a metric,
 $\varphi=(\varphi^{\alpha})\in\R^l$ is a set of scalar fields,
 $(h_{\alpha\beta})$ is a  constant symmetric non-degenerate
 $l\times l$ matrix $(l\in \N)$,    $ F^s =    dA^s
          =  \frac{1}{2} F^s_{M N}  dx^{M} \wedge  dx^{N}$
 is a $2$-form,  $\lambda_s$ is a 1-form on $\R^l$:
 $\lambda_s(\varphi)=\lambda_{s \alpha}\varphi^\alpha$,
 $s = 1,..., n$; $\alpha=1,\dots,l$.
 Here $(\lambda_{s \alpha})$, $s =1,\dots, n$, are dilatonic 
 coupling vectors.
 In (\ref{2.1})
 we denote $|g| =   |\det (g_{MN})|$, $(F^s)^2  =
 F^s_{M_1 M_{2}} F^s_{N_1 N_{2}}  g^{M_1 N_1} g^{M_{2} N_{2}}$, 
 $s = 1,\dots, n$.

 Here we deal with a family of exact
solutions to field equations which correspond to the action
(\ref{2.1}) and depend on one variable $\rho$. These solutions
are defined on the manifold
 \beq{2.2}
  M = (0, + \infty)  \times M_1 \times M_2,
 \eeq
 where $M_1$ is a one-dimensional manifold (say $S^1$ or $\R$) and
 $M_2$ is a ($D-2$)-dimensional Ricci-flat manifold. The solution 
 (from the family under consideration) reads \cite{GI-09}
 \bear{2.30}
  g= \Bigl(\prod_{s = 1}^{n} H_s^{2 h_s /(D-2)} \Bigr)
  \biggl\{ w d\rho \otimes d \rho  +
  \Bigl(\prod_{s = 1}^{n} H_s^{-2 h_s} \Bigr) \rho^2 d\phi \otimes d\phi +
    g^2  \biggr\},
 \\  \label{2.31}
  \exp(\varphi^\alpha)=
  \prod_{s = 1}^{n} H_s^{h_s  \lambda_{s}^\alpha},
 \\  \label{2.32a}
  F^s=  q_s \left( \prod_{l = 1}^{n}  H_{l}^{- A_{s l}} \right) \rho d\rho \wedge d \phi,
  \ear
 $s = 1,\dots, n$; $\alpha = 1,\dots, l$, where $w = \pm 1$, $g^1 = d\phi \otimes d\phi$ is a
  metric on $M_1$ and $g^2$ is a  Ricci-flat metric on
 $M_{2}$.  Here $q_s \neq 0$ are integration constants  
 ($q_s = - Q_s$ in notations of Ref. \cite{GI-09}), $s = 1,\dots, n$.

 The moduli functions $H_s(z) > 0$, $z = \rho^2$, obey the master equations
\beq{1.1}
  \frac{d}{dz} \left( \frac{ z}{H_s} \frac{d}{dz} H_s \right) =
   P_s \prod_{l = 1}^{n}  H_{l}^{- A_{s l}},
  \eeq
 with  the following boundary conditions
 \beq{1.2}
   H_{s}(+ 0) = 1,
 \eeq
 where
 \beq{2.21}
  P_s =  \frac{1}{4} K_s q_s^2,
 \eeq
 $s = 1,\dots,n$.   For $w = +1$ the boundary conditions (\ref{1.2}) are necessary  
 to avoid a conic singularity (for our metric  (\ref{2.30})) in the limit $\rho =  +0$.

 The parameters  $h_s$  obey the relations
  \beq{2.16}
  h_s = K_s^{-1}, \qquad  K_s = B_{s s} > 0,
  \eeq
 where
 \beq{2.17}
  B_{ss'} \equiv
  1 +\frac{1}{2-D}+  \lambda_{s \alpha} \lambda_{s' \beta}   h^{\alpha\beta},
  \eeq
 $s, s' = 1,..., n$, with $(h^{\alpha\beta})=(h_{\alpha\beta})^{-1}$.
 In relations above we denote 
 $\lambda_{s}^{\alpha} = h^{\alpha\beta}  \lambda_{s \beta}$
 and
 \beq{2.18}
  (A_{ss'}) = \left( 2 B_{s s'}/B_{s' s'} \right).
 \eeq
 This is the so-called quasi-Cartan matrix. 
 
   The constants   $B_{s s'}$ and $K_s = B_{s s}$
  are related to scalar products of so-called ``brane vectors'' $U^s$,   belonging  to a certain linear space (in our case it has dimension $l+2$). We have  $B_{s s'} =(U^s,U^{s'})$ and  $K_s = (U^s,U^{s})$, with certain scalar product $(.,.)$ defined in Refs.  \cite{IMJ,IM-top}. Such scalar  products  appear in  various solutions with branes
   (black branes, fluxbranes, $S$-branes etc), e.g. in calculations of certain physical parameters
   (Hawking temperature, black hole/brane entropy, PPN parameters etc), see Ref. \cite{IM-top}.

  By product relation (\ref{2.18}) defines generalized intersection rules for 
  branes \cite{IMJ}, while numbers $K_s$ are invariant under dimensional reductions
    with typical value $K_s = 2$  for brane $U$-vectors which appear 
   in numerous supergravity models,  e.g. for $D = 10,11$   \cite{Ivas-Symmetry-17}.
   
 It can be readily shown that if the matrix $(h_{\alpha\beta})$
 is of Euclidean signature,   $l \geq n$, 
 and  $(A_{ss'})$ is a Cartan matrix 
 of certain simple Lie algebra of rank $n$, then
 there exist co-vectors 
 $\lambda_1, \dots, \lambda_n$  obeying (\ref{2.18}).

Our solution  is nothing more than  a special case of the
fluxbrane (for $w = +1$,  $M_1 = S^1$) and $S$-brane
($w = -1$) solutions from Refs. \cite{Iflux} and \cite{GIM}, respectively.

If we put $w = +1$ and choose Ricci-flat metric $g^2$ of
pseudo-Euclidean signature on manifold $M_2$
of  dimension $d_2 > 2$ , we obtain a higher dimensional 
generalization of the  Melvin's solution \cite{Melv}. 
 
The Melvin's solution does not contain scalar fields. It corresponds (in our notations) to $D = 4$,
$n = 1$, $w = +1$, $l =0$, $M_1 = S^1$ ($0 < \phi <  2 \pi$),  $M_2 = \R^2$,
$g^2 = -  dt \otimes dt + d x \otimes d x$
and Lie algebra ${\cal G} = A_1 = {\rm sl}(2)$. 
  
For the case of $w = -1$ and $g^2$ of Euclidean signature one can obtain a cosmological solution
with a horizon (as $\rho = + 0$) if $M_1 = \R$ ($ - \infty < \phi < + \infty$).

 \section{Examples of solutions for certain Lie algebras }
 
  Here we deal with the generalized Melvin-like solution for $n = l$, $w = + 1$ and $M_1 = S^1$,  
  which corresponds to  simple (finite dimensional) Lie algebra  of certain rank $n$ with the 
  Cartan matrix $A =  (A_{sl})$.
   
   We put here $h_{\alpha\beta} = \delta_{\alpha \beta}$ and denote  
   $(\lambda_{s \alpha }) =  (\lambda_{s}^{\alpha}) = \vec{\lambda}_{s}$, $s = 1, \dots, n$.
   
 Due to (\ref{2.16})-(\ref{2.18}) we get
       \begin{equation}
              \label{2.3.17}
       K_s =   \frac{D - 3}{D -2} +  \vec{\lambda}_{s}^2,
         \end{equation} 
       $h_s =  K_s^{-1}$,  and  
         \beq{2.3.18}
           \vec{\lambda}_{s} \vec{\lambda}_{l} = 
               \frac{1}{2} K_l A_{sl}  - \frac{D - 3}{D -2} \equiv G_{sl},
         \eeq    
        $s,l = 1, \dots, n$;  (\ref{2.3.17}) just follows from  (\ref{2.3.18}). 
 
     {\bf Remark 1.} 
          {\em For large enough $K_s$ in (\ref{2.3.18}) (or large enough $\vec{\lambda}_{s}^2$ ) 
          there exist vectors   $\vec{\lambda}_s$  obeying   (\ref{2.3.18}) (and hence (\ref{2.3.17})).
          Indeed, the matrix $G = (G_{sl})$ is positive definite
          if $K_s > K_{*}$,  where $K_{*}$ is some positive number. Hence there exists
          a matrix $\Lambda$, such that $\Lambda^{T} \Lambda = G$. We put
          $(\Lambda_{as}) = (\lambda_{s}^a)$ and get a set of vectors obeying
           (\ref{2.3.18}).}

  Here as in \cite{I-14}  we use another parameters $p_s$   instead of $P_s$:
    \beq{4.g}
     p_s =  P_s/n_s, 
    \eeq
 $s = 1, ..., n$.
 This is done to avoid big denominators for $P_s^{(k)}$ in relation (\ref{1.3}).
 
 {\bf Remark 2.} 
 {\em The parameters  $p_s$ give us the coefficients $P_s^{(1)}$ in  (\ref{1.3}):
  \beq{4.g1}
    P_s^{(1)}  =  P_s = p_s n_s , 
     \eeq
  $s = 1, ..., n$. These relations can be readily obtained by putting $z = +0$ into master equations
  (\ref{1.1}) and using the boundary conditions (\ref{1.2}). 
   Moreover, for given Lie algebra one can deduce  recurrent relations
   for higher coefficients   $P_s^{(k+1)}$ in (\ref{1.3}) as functions 
   of $P_s^{(k)}, \dots, P_s^{(1)} = P_s $   and solve the chain of recurrent relations,  
   justifying that  $P_s^{(n_s + 1)} = 0$, for all $s = 1, ..., n$. 
   }   

We note also that for a special choice  of $p_s$ parameters: $p_s = p > 0 $, the polynomials
 have the following simple form \cite{Iflux}
 \beq{1.s}
 H_{s}(z) = (1 +  p z)^{n_s},
 \eeq
 $s = 1, ..., n$. This relation may be considered as nice
 tool for a verification of general solutions for polynomials.

  \subsection{Rank-$1$ case}
  
   {\bf $A_1$-case.} We start with the simplest example which occurs for 
  the Lie algebra $A_1 = sl(2)$. 
   We obtain  \cite{Iflux}
   \begin{equation}
     \label{1.A.1}
    H_{1}  = 1 + p_1 z.
   \end{equation}
   Here  $n_1 = 1$.
   
      In this case (due to (\ref{2.16})-(\ref{2.18})) 
   we have
  \begin{equation}
               \label{1.3.17}
        K_1 =   \frac{D - 3}{D -2} +  \vec{\lambda}_{1}^2,
          \end{equation} 
        $h_1 =  K_1^{-1}$.
  
         Due to  (\ref{1.A.1}) we get the following asymptotical behaviour 
         \begin{equation}
           \label{1.A.1as}
            H_1 = H_1(z, p_1 )  \sim p_1 z \equiv  H_1^{as}(z, p_1),
          \end{equation}
         as $z \to \infty$.

          Relations (\ref{1.A.1}) and (\ref{1.A.1as})  imply 
           the following  identity.
           
          { \bf Duality relation.}             
          
          {\bf Proposition 1.} {\em  The fluxbrane polynomial
           corresponding to Lie algebra $A_1$
            obeys    for all $p_1 \neq 0$ and $z \neq 0$
            the identity
           \begin{equation}
             \label{1.3.12}
              H_{1}(z, p_1) = H_1^{as}(z, p_1) H_1(z^{-1}, p_1^{-1}).
            \end{equation}
             }

   \subsection{Rank-$2$ case}
  
  Now we proceed with the solutions which correspond to  simple  Lie algebras ${\cal G}$ of rank $2$, i.e. 
  the matrix   $A =  (A_{sl})$ is just a  Cartan matrix 
  \beq{2.A.5}
       \left(A_{ss'}\right)=
     \left( \begin{array}{*{6}{c}}
        2 & -1\\
        -k& 2\\ 
       \end{array} 
    \right) ,\quad \eeq
  where $k = 1,2,3$ for ${\cal G} = A_2,  C_2 \cong B_2,  G_2$, respectively \cite{BolIvas_R2-17}. 
  
  The matrix $A$ is described graphically by the Dynkin diagrams presented in Figure 1 
  (for any of these three Lie algebras).
   \vspace{25pt}
    
    \setlength{\unitlength}{1mm}
   \begin{figure}[h]
   \centering
   \begin{picture}(40, 0)
   \put(2,5){\circle*{2}}
   \put(12,5){\circle*{2}}
   \put(1,5){\line(1,0){10}}
   \put(0,-0.5){$1$}
   \put(10,-0.5){$2$}
    
   \put(22,5){\circle*{2}}
   \put(32,5){\circle*{2}}
   \put(22,5.5){\line(1,0){10}}
   \put(22,4.5){\line(1,0){10}}
   \put(20,-0.5){$1$}
   \put(30,-0.5){$2$}
   \put(25,3.9){\large $<$}
   
   \put(42,5){\circle*{2}}
   \put(52,5){\circle*{2}}
   \put(42,5.7){\line(1,0){10}}
   \put(42,5){\line(1,0){10}}
   \put(42,4.3){\line(1,0){10}}
   \put(40,-0.5){$1$}
   \put(50,-0.5){$2$}
   \put(45,3.9){\large $<$}
   \end{picture}
   \caption{The Dynkin diagrams for the Lie algebras $A_2$, $C_2$, $G_2$, respectively.}
   \end{figure}
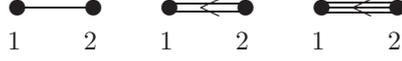

            It follows from  (\ref{2.16})-(\ref{2.18}) that 
           \beq{2.3.26}
            \frac{h_1}{h_2} = \frac{K_{2}}{K_{1}} =  \frac{A_{21}}{A_{12}} = k,
             \eeq
          where $k = 1,2,3$ for ${\cal G} = A_2,  C_2,  G_2$, respectively.

  {\bf Polynomials.}
  
   {\bf $A_2$-case.} For the Lie algebra $A_2 = sl(3)$ 
            we have \cite{Iflux,GIM,I-14} 
             \bear{2.A.6}
               H_{1} = 1 + 2 p_1 z +  p_1 p_2 z^{2}, \\
              \label{2.A.7}
               H_{2} = 1 + 2 p_2 z +  p_1 p_2 z^{2}.
              \ear

            {\bf $C_2$-case.}
             In case of  Lie algebra $C_2 = so(5)$ we get 
             the following polynomials \cite{GIM,I-14}
             \bear{2.C.2}
               H_1 = 1+ 3 p_1 z+ 3 p_1 p_2 z^2 + p_1^2 p_2 z^3,
                           \\ \label{2.C.3}
               H_2 = 1+ 4 p_2 z+ 6 p_1 p_2 z^2 + 4 p_1^2 p_2 z^3
                    +  p_1^2 p_2^2 z^4.
             \ear
                        
             { \bf $G_2$-case. }
             For the Lie algebra $G_2$ 
             the fluxbrane polynomials read \cite{GIM,I-14}
             \vspace{15pt}
            \bear{2.G.2}
              H_{1} = 1+ 6 p_1 z+ 15 p_1 p_2 z^2 + 20 p_1^2 p_2 z^3 + \\ \nonumber
                15 p_1^3 p_2 z^4 + 6 p_1^3 p_2^2 z^5 + p_1^4 p_2^2 z^6 ,
            \\ \label{2.G.3}
               H_2 =  1+  10 p_2 z + 45 p_1 p_2 z^2  +  120 p_1^2 p_2 z^3
              +  p_1^2 p_2( 135 p_1 + 75 p_2) z^4 \\ \nonumber
                            + 252 p_1^3 p_2^2 z^5
               + p_1^3 p_2^2 \biggl(75 p_1 + 135 p_2 \biggr)z^6   +  120 p_1^4 p_2^3  z^7
               \\ \nonumber
               + 45 p_1^5 p_2^3 z^8 +  10 p_1^6 p_2^3 z^9
                  + p_1^{6} p_2^{4}  z^{10}.
              \ear

 Let us denote
       \begin{equation}
         \label{2.3e.5}
          H_s = H_s(z) = H_s(z, (p_i) ),
        \end{equation}
       $s = 1, 2$; where $(p_i) = (p_1,p_2)$.
       
        We have the following asymptotical  relations for  polynomials 
       \begin{equation}
         \label{2.3e.6}
          H_s = H_s(z, (p_i) )  \sim \left( \prod_{l=1}^{2} (p_l)^{\nu^{sl}} \right) z^{n_s} \equiv 
          H_s^{as}(z, (p_i)),
        \end{equation}
       $s = 1, 2$, as $z \to \infty$. 
       
       Here $\nu = (\nu^{sl})$ is the integer valued matrix 
        \begin{equation}
          \label{2.3e.7}
          \nu =  \left(
          \begin{array}{cccccc}
           1 & 1  \\
           1 & 1  
          \end{array}
          \right),
          \qquad  
          \left(
          \begin{array}{cccccc}
                          2 & 1  \\
                          2 & 2  
                         \end{array}
                         \right),
          \qquad 
            \left(  
            \begin{array}{cccccc}
                          4 & 2  \\
                          6 & 4  
                         \end{array}
                       \right),
           \end{equation}
       for Lie algebras $A_2,  C_2,  G_2$, respectively.

       For last two cases ($C_2$ and $G_2$) we have $\nu = 2 A^{-1}$ 
        ($A^{-1}$ is inverse Cartan matrix).
       For the  $A_2$-case  the matrix $\nu$ reads 
       \begin{equation}
         \label{2.3e.8}
         \nu = A^{-1} (I + P),
        \end{equation}
        where $I$ is $2 \times 2$ identity matrix and
       \begin{equation}
          \label{2.3e.9}
          P  = \left(
          \begin{array}{cccccc}
           0 & 1  \\
           1 & 0  
           \end{array}
          \right).
          \end{equation}
       is permutation matrix. 
       It corresponds to the permutation $\sigma \in S_2$ ($S_2$ is symmetric group)
       \begin{equation}
         \label{2.3.10}
          \sigma: (1,2) \mapsto (2,1),
        \end{equation}
        by the following relation  $P = (P^i_j) = (\delta^i_{\sigma(j)})$. 
        Here $\sigma$   is the generator of the group 
        $S_2 = \{ \sigma, id \}$ - the group of symmetry of the Dynkin diagram (for $A_2$) 
        which is isomorphic to the group $Z_2$.
        
         Here  in all cases we get 
         \begin{equation}
           \label{2.3e.10a}
            \sum_{l= 1}^2 \nu^{sl} = n_s,
          \end{equation}
          $s = 1, 2$.
       
        Now we denote $\hat{p}_i = p_{\sigma(i)} $ for the $A_{2}$-case
        and $\hat{p}_i = p_{i}$ for $C_2$ and $G_2$ cases, $i= 1,2$.         
        We call the ordered set $(\hat{p}_i)$ as dual one to the ordered set $(p_i)$.
        Using the relations for  polynomials   we obtain  the following identities 
        (which can be redily verified just ``by hands'').
        
           {\bf Symmetry relations.}
                      \\
           {\bf Proposition 2.1.} {\em  
            The fluxbrane polynomials for $A_2$ obey for all $p_i$ and $z$
           the identities:
                  \beq{2.3.11}
                   H_{\sigma(s)}(z, (p_i) )\, = H_s(z, (\hat{p}_i)),
                  \eeq
                       $s= 1, 2$. }
                      \\[1em]

        {\bf Duality relations.}
               
        {\bf Proposition 2.2.} {\em  The fluxbranes polynomials
         corresponding to Lie algebras $A_2$,
         $C_2$ and $G_2$ obey
          for all $p_i > 0$ and $z > 0$
          the identities
         \begin{equation}
           \label{2.3.12}
            H_{s}(z, (p_i) ) = H_s^{as}(z, (p_i)) H_s(z^{-1}, (\hat{p}_i^{-1})),
          \end{equation}
          $s = 1, 2$. }
         
        We call  relations (\ref{2.3.12})  as duality ones.
 
 \subsection{Rank-$3$ algebras}
 
  Now we deal with polynomials which correspond to  simple  Lie algebras ${\cal G}$ of rank $3$, when 
  the matrix   $A =  (A_{sl})$ is coinciding with one of the Cartan matrix 
  \beq{3.A.5}
       \left(A_{ss'}\right)=
       \begin{pmatrix}
        2 & -1 & 0 \\
        -1 & 2 & -1 \\
        0 & -1 & 2
       \end{pmatrix}\!,
    \quad
     \begin{pmatrix}
     2 & -1 & 0 \\
     -1 & 2 & -2 \\
     0 & -1 & 2
    \end{pmatrix}\!,
    \quad
    \begin{pmatrix}
       2 & -1 & 0 \\
       -1 & 2 & -1 \\
       0 & -2 & 2
      \end{pmatrix}
       \eeq
  for ${\cal G} = A_3,  B_3,  C_3$, respectively \cite{BolIvas_R3-18}. 
  
  Any of these matricies is described graphically by a Dynkin diagram 
  pictured in Figure 2.
   \vspace{25pt}

   \setlength{\unitlength}{1mm}
   \begin{figure}[ph]
   \centering
   \begin{picture}(80, 0)
   \put(2,5){\circle*{2}}
   \put(12,5){\circle*{2}}
   \put(22,5){\circle*{2}}
   \put(1,5){\line(1,0){20}}
   \put(0,-0.5){$1$}
   \put(10,-0.5){$2$}
   \put(20,-0.5){$3$}
   
   \put(32,5){\circle*{2}}
   \put(42,5){\circle*{2}}
   \put(52,5){\circle*{2}}
   \put(32,5){\line(1,0){10}}
   \put(42,5.5){\line(1,0){10}}
   \put(42,4.5){\line(1,0){10}}
   \put(30,-0.5){$1$}
   \put(40,-0.5){$2$}
   \put(50,-0.5){$3$}
   \put(45,3.9){\large $>$}
   
   \put(62,5){\circle*{2}}
    \put(72,5){\circle*{2}}
    \put(82,5){\circle*{2}}
    \put(62,5){\line(1,0){10}}
    \put(72,5.5){\line(1,0){10}}
    \put(72,4.5){\line(1,0){10}}
    \put(60,-0.5){$1$}
    \put(70,-0.5){$2$}
    \put(80,-0.5){$3$}
    \put(75,3.9){\large $<$}
   \end{picture}
   \caption{The Dynkin diagrams for the Lie algebras $A_3$, $B_3$, $C_3$, respectively.\label{fig1}}
   \end{figure}
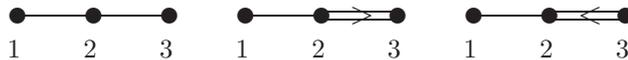
       
         It follows from  (\ref{2.16}), (\ref{2.18}) that 
         \beq{3.3.26}
         \frac{h_s}{h_l} = \frac{K_{l}}{K_{s}} = \frac{B_{ll}}{B_{ss}} = 
            \frac{B_{ls}}{B_{ss}} \frac{B_{ll}}{B_{sl}} = \frac{A_{ls}}{A_{sl}}
          \eeq
         for any $s \neq l$ obeying $A_{sl}, A_{ls} \neq 0$. This
         implies
         \beq{3.3.26K}
          K_1 =K_2 = K, \qquad K_3 = K, \ \frac{1}{2} K, \ 2K  
          \eeq
          or 
          \beq{3.3.26h}
          h_1 = h_2 = h, \qquad h_3 = h, \ 2h, \ \frac{1}{2} h ,   
          \eeq
          ($h = K^{-1}$) for ${\cal G} =  A_3,  B_3,  C_3$, respectively.

       {\bf Polynomials.}
       The set of moduli functions   $(H_1(z),H_2(z), H_3(z))$, obeying Eqs. (\ref{1.1}) and (\ref{1.2})
       with the matrix $A =  (A_{sl})$ from  (\ref{3.A.5}) 
       are polynomials with powers $(n_1,n_2,n_3) = (3,4,3), (6,10,6), (5,8,9)$ for
       ${\cal G} = A_3,  B_3,  C_3$, respectively.
       We get the following polynomials \cite{BolIvas_R3-18}.

            {\bf $A_3$-case.} For the Lie algebra $A_3 \cong sl(4)$ 
            we have \cite{GI,I-14}
             \bear{3.A.6}
               &&H_1=1+3 p_1 z +3 p_1 p_2 z^2 +p_1 p_2 p_3 z^3,\\
              \label{A.7}
               &&H_2=1+4 p_2 z +(3 p_1 p_2+3 p_2 p_3)z^2 + 4 p_1 p_2 p_3 z^3 + p_1 p_2^2 p_3 z^4, \\
               &&H_3=1+3 p_3 z +3 p_2 p_3 z^2 + p_1 p_2 p_3 z^3.
              \ear

            {\bf $B_3$-case.}
             In case of Lie algebra $B_3 \cong so(7)$  the fluxbrane polynomials read  \cite{GI}
  \begin{eqnarray}\label{3.B.3}
  &&\nq H_1=1+6 p_1 z +15 p_1 p_2 z^2 +20 p_1 p_2 p_3 z^3+15 p_1 p_2 p_3^2 z^4 +6 p_1 p_2^2 p_3^2 z^5+p_1^2 p_2^2 p_3^2 z^6,  \quad 
   \\
  &&\nq H_2=1+10 p_2 z+ \left(15 p_1 p_2+30 p_2 p_3\right)z^2+\left(80 p_1 p_2 p_3+40 p_2 p_3^2\right)z^3 \nn
  &&\;\; + \left(50 p_1 p_2^2 p_3+135 p_1 p_2 p_3^2+25 p_2^2 p_3^2\right)z^4 + 252 p_1 p_2^2 p_3^2 z^5\nn
  &&\;\; +\left(25 p_1^2 p_2^2 p_3^2+135 p_1 p_2^3 p_3^2+50 p_1 p_2^2 p_3^3\right)z^6 +\left(40 p_1^2 p_2^3 p_3^2+80 p_1 p_2^3  p_3^3\right)z^7 \nn
  &&\;\;+\left(30 p_1^2 p_2^3 p_3^3+15 p_1 p_2^3 p_3^4\right)z^8 +10  p_1^2 p_2^3 p_3^4z^9+ p_1^2 p_2^4 p_3^4 z^{10},
  \\
  &&\nq H_3=1+6 p_3z+15 p_2 p_3 z^2 +\left(10 p_1 p_2 p_3+10 p_2 p_3^2\right)z^3+15 p_1 p_2 p_3^2 z^4 \nn
  &&\;\;+6 p_1 p_2^2 p_3^2 z^5 +p_1 p_2^2 p_3^3 z^6 . 
  \end{eqnarray}
                           
             { \bf $C_3$-case. }
             For the Lie algebra $C_3 \cong sp(3)$ we obtain  (with the use of MATHEMATICA)  the following polynomials
             \vspace{10pt}
  \begin{eqnarray}\label{3.C.3}
  &&\nq H_1=1+5 p_1 z +10  p_1 p_2 z^2 +10 p_1 p_2 p_3 z^3 +5 p_1 p_2^2 p_3 z^4 + p_1^2 p_2^2 p_3 z^5 ,
  \\
  &&\nq H_2=1+8 p_2 z + \left(10 p_1 p_2+18 p_2 p_3\right)z^2+ \left(40 p_1 p_2 p_3+16 p_2^2 p_3\right) z^3 +70 p_1 p_2^2 p_3 z^4 \nn
  &&\;\;+\left(16 p_1^2 p_2^2 p_3+40 p_1 p_2^3 p_3\right)z^5+\left(18 p_1^2 p_2^3 p_3+10 p_1 p_2^3 p_3^2\right)z^6 \nn
  &&\;\;+8 p_1^2 p_2^3 p_3^2z^7 +p_1^2 p_2^4 p_3^2 z^8, 
  \\
  &&\nq H_3=1+9 p_3 z+36 p_2 p_3 z^2 +\left(20 p_1 p_2 p_3+64 p_2^2 p_3\right)z^3+\left(90 p_1 p_2^2 p_3+36 p_2^2 p_3^2\right)z^4 \nn
  &&\;\; +\left(36 p_1^2 p_2^2 p_3+90 p_1 p_2^2 p_3^2\right)z^5 + \left(64 p_1^2 p_2^2 p_3^2+20 p_1 p_2^3 p_3^2\right)z^6 \nn
  &&\;\;+ 36 p_1^2 p_2^3 p_3^2 z^7 +9 p_1^2 p_2^4 p_3^2 z^8 + p_1^2 p_2^4 p_3^3 z^9 . 
  \end{eqnarray}
            
        We denote
        \begin{equation}
          \label{3.3e.5}
           H_s = H_s(z) = H_s(z, (p_i) ),
         \end{equation}
         where $(p_i) = (p_1,p_2,p_3)$.
        
        Due to obtained relations for  polynomials we get the asymptotical behaviour 
        \begin{equation}
          \label{3.3e.6}
           H_s = H_s(z, (p_i) )  \sim \left( \prod_{l=1}^{3} (p_l)^{\nu^{sl}} \right) z^{n_s} \equiv 
           H_s^{as}(z, (p_i)),
         \end{equation}
         as $z \to \infty$. 
        
        Here $\nu = (\nu^{sl})$ is the integer valued matrix 
         \begin{equation}
           \label{3.3e.7}
           \nu =  \left(
           \begin{array}{cccccc}
            1 & 1 & 1 \\
            1 & 2 & 1 \\
            1 & 1 & 1
           \end{array}
           \right)\!,
           \quad  
           \left(
           \begin{array}{cccccc}
                           2 & 2 & 2  \\
                           2 & 4 & 4 \\
                           1 & 2 & 3  
                          \end{array}
                          \right)\!,
           \quad 
             \left(  
             \begin{array}{cccccc}
                           2 & 2 & 1  \\
                           2 & 4 & 2 \\
                           2 & 4 & 3 
                          \end{array}
                        \right)\!,
            \end{equation}
        for Lie algebras $A_3,  B_3,  C_3$, respectively.

       For Lie algebras  $B_3$ and $C_3$  we have 
        \begin{equation}
          \label{3.3e.8a}
        \nu = 2 A^{-1}, 
        \end{equation}
        where $A^{-1}$ is inverse Cartan matrix.    For the  $A_3$-case   
        the matrix $\nu$ reads  as follows
        \begin{equation}
          \label{3.3e.8b}
          \nu = A^{-1} (I + P),
         \end{equation}
         where $I$ is $3 \times 3$ identity matrix and
        \begin{equation}
           \label{3.3e.9}
           P  = \left(
           \begin{array}{cccccc}
            0 & 0 & 1  \\
            0 & 1 & 0  \\
            1 & 0 & 0
            \end{array}
           \right)
           \end{equation}
        is permutation matrix, which corresponds to the permutation $\sigma \in S_3$ ($S_3$ is symmetric group)
        \begin{equation}
          \label{3.3.10}
           \sigma: (1,2,3) \mapsto (3,2,1),
         \end{equation}
         by the following formula  $P = (P^i_j) = (\delta^i_{\sigma(j)})$. 
         Here $\sigma$   is the generator of the group 
         $G = \{ \sigma, {\rm id} \}$ - the group of symmetry of the Dynkin diagram (for $A_3$), which 
         is isomorphic to the group $\mathbb{Z}_2$.
         
         Here in all three cases we have 
          \begin{equation}
            \label{3.3e.10a}
             \sum_{l= 1}^3 \nu^{sl} = n_s,
           \end{equation}
           $s = 1, 2, 3$.

         Now we introduce notations: $\hat{p}_i = p_{\sigma(i)} $ for the $A_3$ 
         and $\hat{p}_i = p_{i}$ for $B_3$ and $C_3$ algebras, $i= 1,2,3$.         
         Using  relations for rank-3 polynomials   we obtain (with a help of MATHEMATICA)
          the following identities. 
                   
               {\bf Symmetry relations.}
              \\
                  {\bf Proposition 3.1.} {\em  
                    The fluxbrane polynomials for $A_3$ algebra  obey for all $p_i$ and $z$
                            the identities:
                   \beq{3.3.11}
                   H_{\sigma(s)}(z, (p_i) )\, = H_s(z, (\hat{p}_i)),
                   \eeq
                     $s= 1, 2,3$.  }
                      \\[1em]

          {\bf Duality relations.}
         
         \vskip 1em 
         \noindent {\bf Proposition 3.2.} {\em  The fluxbranes polynomials
          corresponding to Lie algebras $A_3$,
          $B_3$ and $C_3$ obey for all $p_i > 0$ and $z > 0$
          the identities
          \begin{equation}
            \label{3.3.12}
             H_{s}(z, (p_i) ) = H_s^{as}(z, (p_i)) H_s(z^{-1}, (\hat{p}_i^{-1})),
           \end{equation}
           $s = 1, 2, 3$. }
          
        
 \subsection{Rank-$4$ algebras}
 
 In this subsection we deal with the solutions related to Lie algebras ${\cal G}$ of rank $4$, 
 i.e. the matrix  $A = (A_{sl})$ coincides with one of the Cartan matrices 
 \begin{align} 
      \left(A_{ss'}\right) &=
      \begin{pmatrix}
      2 & -1 & 0 & 0 \\
      -1 & 2 & -1 & 0 \\
      0 & -1 & 2 & -1 \\
      0 & 0 & -1 & 2 \\
      \end{pmatrix},\!
   \quad
    \begin{pmatrix}
     2 & -1 & 0 & 0 \\
     -1 & 2 & -1 & 0 \\
     0 & -1 & 2 & -2 \\
     0 & 0 & -1 & 2 \\
   \end{pmatrix},\! 
   \quad
   \begin{pmatrix}
      2 & -1 & 0 & 0 \\
      -1 & 2 & -1 & 0 \\
      0 & -1 & 2 & -1 \\
      0 & 0 & -2 & 2 \\
     \end{pmatrix}, 
    \quad \nonumber \\
 &\hspace{6em}   \begin{pmatrix}
         2 & -1 & 0 & 0 \\
         -1 & 2 & -1 & -1 \\
         0 & -1 & 2 & 0 \\
         0 & -1 & 0 & 2 \\
     \end{pmatrix},
      \begin{pmatrix}
             2 & -1 & 0 & 0 \\
             -1 & 2 & -2 & 0 \\
             0 & -1 & 2 & -1 \\
             0 &0 & -1 & 2 \\
         \end{pmatrix}
   \label{4.A.5}
      \end{align}
 for ${\cal G} = A_4, \, B_4, \, C_4, \, D_4, \, F_4$, respectively. 
 
 These matrices are graphically described by using the Dynkin diagrams pictured in Figure 3.
  \vspace{25pt}
   
   \setlength{\unitlength}{1mm}
  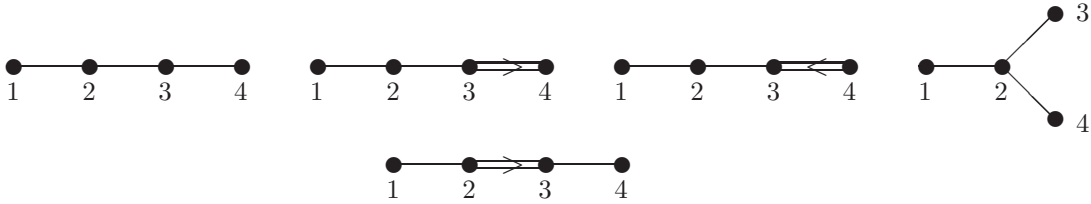
\begin{figure}[h]
  \centering
  \begin{picture}(140, 25)
  \put(2,25){\circle*{2}}
  \put(12,25){\circle*{2}}
  \put(22,25){\circle*{2}}
  \put(32,25){\circle*{2}}
  \put(1,25){\line(1,0){30}}
  \put(0.3,20.5){$1$}
  \put(10.3,20.5){$2$}
  \put(20.3,20.5){$3$}
  \put(30.3,20.5){$4$}
   
  \put(42,25){\circle*{2}}
  \put(52,25){\circle*{2}}
  \put(62,25){\circle*{2}}
  \put(72,25){\circle*{2}}
  \put(42,25){\line(1,0){20}}
  \put(62,25.5){\line(1,0){10}}
  \put(62,24.5){\line(1,0){10}}
  \put(40.3,20.5){$1$}
  \put(50.3,20.5){$2$}
  \put(60.3,20.5){$3$}
  \put(70.3,20.5){$4$}
  \put(65.3,23.9){\large $>$}
  
  \put(82,25){\circle*{2}}
  \put(92,25){\circle*{2}}
  \put(102,25){\circle*{2}}
  \put(112,25){\circle*{2}}
  \put(82,25){\line(1,0){20}}
  \put(102,25.5){\line(1,0){10}}
  \put(102,24.5){\line(1,0){10}}
  \put(80.3,20.5){$1$}
  \put(90.3,20.5){$2$}
  \put(100.3,20.5){$3$}
  \put(110.3,20.5){$4$}
  \put(105.3,23.9){\large $<$}
   
  \put(122,25){\circle*{2}}
  \put(132,25){\circle*{2}}
  \put(139,32){\circle*{2}}
  \put(139,18){\circle*{2}}
  \put(121,25){\line(1,0){10}}
  \put(131.3,24.5){\line(1,1){7.5}}
  \put(132,25){\line(1,-1){7.5}}
  \put(120.3,20.5){$1$}
  \put(130.3,20.5){$2$}
  \put(141,31){$3$}
  \put(141,16.3){$4$}

  \put(52,12){\circle*{2}}
  \put(62,12){\circle*{2}}
  \put(72,12){\circle*{2}}
  \put(82,12){\circle*{2}}
  \put(52,12){\line(1,0){10}}
  \put(72,12){\line(1,0){10}}
  \put(62,12.5){\line(1,0){10}}
  \put(62,11.5){\line(1,0){10}}
  \put(50.3,7.5){$1$}
  \put(60.3,7.5){$2$}
  \put(70.3,7.5){$3$}
  \put(80.3,7.5){$4$}
  \put(65.3,10.9){\large $>$}
  
   \end{picture}
   \vspace{-15pt}
   \caption{The Dynkin diagrams for the Lie algebras $A_4$, $B_4$, $C_4$, $D_4$, $F_4$, respectively.}
  \end{figure}

        It  follows from (\ref{2.16}), (\ref{2.18})  that 
        \beq{4.3.26}
        \frac{h_s}{h_l} = \frac{K_{l}}{K_{s}} = \frac{B_{ll}}{B_{ss}} = 
           \frac{B_{ls}}{B_{ss}} \frac{B_{ll}}{B_{sl}} = \frac{A_{ls}}{A_{sl}}
         \eeq
        for any $s \neq l$ obeying $A_{sl}, A_{ls} \neq 0$. This
        implies
        \beq{4.3.26K}
         K_1 =K_2 = K_3 = K, \qquad K_4 = K, \ \frac{1}{2} K, \ 2K, \ K, 
         \eeq
         or 
         \beq{4.3.26hF}
         h_1 = h_2 = h_3 =  h, \qquad h_4 = h, \ 2h, \ \frac{1}{2} h , \ h   , 
         \eeq
         ($h = K^{-1}$) for ${\cal G} =  A_4,  B_4,  C_4, D_4$,  respectively,
         and 
         \beq{4.3.26KF}
                K_1 =K_2 =  K, \qquad K_3 = K_4 = \frac{1}{2} K,  
                 \eeq
                 or 
                 \beq{4.3.26h}
                 h_1 = h_2 =  h, \qquad h_3 = h_4 =  2h, 
                                 \eeq
         ($h = K^{-1}$) for ${\cal G} =  F_4$.

      {\bf Polynomials.}
      Due to  polynomial conjecture, the  functions  $H_1(z), ..., H_4(z)$, 
      obeying Eqs. (\ref{1.1}) and (\ref{1.2})
      with the Cartan matrix $A =  (A_{sl})$ from (\ref{4.A.5}) 
      should be polynomials with powers 
      $(n_1,n_2, n_3, n_4) = (4,6,6,4)$, $(8,14,18,10)$, $(7,12,15,16)$, $(6,10,6,6)$, $(22,42,30,16)$ 
       (see (\ref{1.4})) for
       Lie algebras $A_4$, $B_4$, $C_4$, $D_4$, $F_4$, respectively.

      One can verify this conjecture by using appropriate MATHEMATICA algorithm  
      which follows from master equations (\ref{1.1})). Below we present a list of the  obtained polynomials \cite{BolIvas_R4a-18a,BolIvas_R4-21}.

           {\bf $A_4$-case.} For the Lie algebra $A_4 \cong sl(5)$  we find
 \bearr{4.A.4}
 H_1=1+4 p_1 \textcolor{red}{z}+6 p_1 p_2 \textcolor{red}{z^2}+4 p_1 p_2 p_3 \textcolor{red}{z^3}+p_1 p_2 p_3 p_4 \textcolor{red}{z^4},
 \nl
 H_2=1+6 p_2 \textcolor{red}{z}+\left(6 p_1 p_2+9 p_2 p_3\right) \textcolor{red}{z^2}+\left(16 p_1 p_2 p_3+4 p_2 p_3 p_4\right) \textcolor{red}{z^3}\nll
 +\left(6 p_1 p_2^2 p_3+9 p_1 p_2 p_3 p_4\right) \textcolor{red}{z^4}+6 p_1 p_2^2 p_3 p_4 \textcolor{red}{z^5}+p_1 p_2^2 p_3^2 p_4 \textcolor{red}{z^6},
 \nl
 H_3=1+6 p_3 \textcolor{red}{z}+\left(9 p_2 p_3+6 p_3 p_4\right) \textcolor{red}{z^2}+\left(4 p_1 p_2 p_3+16 p_2 p_3 p_4\right) \textcolor{red}{z^3}\nll
 +\left(9 p_1 p_2 p_3 p_4+6 p_2 p_3^2 p_4\right) \textcolor{red}{z^4}+6 p_1 p_2 p_3^2 p_4 \textcolor{red}{z^5}+p_1 p_2^2 p_3^2 p_4 \textcolor{red}{z^6},
 \nl
 H_4=1+4 p_4 \textcolor{red}{z}+6 p_3 p_4 \textcolor{red}{z^2}+4 p_2 p_3 p_4 \textcolor{red}{z^3}+p_1 p_2 p_3 p_4 \textcolor{red}{z^4}.
 \ear
                      {\bf $B_4$-case.}
 In case of Lie algebra $B_4 \cong so(9)$  the fluxbrane polynomials read:
 \bearr{4.B.4}
 H_1=1+8 p_1 \textcolor{red}{z}+28 p_1 p_2 \textcolor{red}{z^2}+56 p_1 p_2 p_3 \textcolor{red}{z^3}+70 p_1 p_2 p_3 p_4 \textcolor{red}{z^4}+56 p_1 p_2 p_3 p_4^2 \textcolor{red}{z^5}\nll
 +28 p_1 p_2 p_3^2 p_4^2 \textcolor{red}{z^6}+8 p_1 p_2^2 p_3^2 p_4^2 \textcolor{red}{z^7}+p_1^2 p_2^2 p_3^2 p_4^2 \textcolor{red}{z^8},
 \nl
 H_2=1+14 p_2 \textcolor{red}{z}+\left(28 p_1 p_2+63 p_2 p_3\right) \textcolor{red}{z^2}+\left(224 p_1 p_2 p_3+140 p_2 p_3 p_4\right) \textcolor{red}{z^3}\nll
 +(196 p_1 p_2^2 p_3+630 p_1 p_2 p_3 p_4+175 p_2 p_3 p_4^2) \textcolor{red}{z^4}\nll
 +\left(980 p_1 p_2^2 p_3 p_4+896 p_1 p_2 p_3 p_4^2+126 p_2 p_3^2 p_4^2\right) \textcolor{red}{z^5}\nll
 +\left(490 p_1 p_2^2 p_3^2 p_4+1764 p_1 p_2^2 p_3 p_4^2+700 p_1 p_2 p_3^2 p_4^2+49 p_2^2 p_3^2 p_4^2\right) \textcolor{red}{z^6}+3432 p_1 p_2^2 p_3^2 p_4^2 \textcolor{red}{z^7}\nll
 +\left(49 p_1^2 p_2^2 p_3^2 p_4^2+700 p_1 p_2^3 p_3^2 p_4^2+1764 p_1 p_2^2 p_3^3 p_4^2+490 p_1 p_2^2 p_3^2 p_4^3\right) \textcolor{red}{z^8}\nll
 +\left(126 p_1^2 p_2^3 p_3^2 p_4^2+896 p_1 p_2^3 p_3^3 p_4^2+980 p_1 p_2^2 p_3^3 p_4^3\right) \textcolor{red}{z^9}\nll
 +\left(175 p_1^2 p_2^3 p_3^3 p_4^2+630 p_1 p_2^3 p_3^3 p_4^3+196 p_1 p_2^2 p_3^3 p_4^4\right) \textcolor{red}{z^{10}}+\left(140 p_1^2 p_2^3 p_3^3 p_4^3+224 p_1 p_2^3 p_3^3 p_4^4\right) \textcolor{red}{z^{11}}\nll
 +\left(63 p_1^2 p_2^3 p_3^3 p_4^4+28 p_1 p_2^3 p_3^4 p_4^4\right) \textcolor{red}{z^{12}}+14 p_1^2 p_2^3 p_3^4 p_4^4 \textcolor{red}{z^{13}}+p_1^2 p_2^4 p_3^4 p_4^4 \textcolor{red}{z^{14}},
 \nl
 H_3=1+18 p_3 \textcolor{red}{z}+\left(63 p_2 p_3+90 p_3 p_4\right) \textcolor{red}{z^2}+\left(56 p_1 p_2 p_3+560 p_2 p_3 p_4+200 p_3 p_4^2\right) \textcolor{red}{z^3}\nll
 +\left(630 p_1 p_2 p_3 p_4+630 p_2 p_3^2 p_4+1575 p_2 p_3 p_4^2+225 p_3^2 p_4^2\right) \textcolor{red}{z^4}\nll
 +\left(1260 p_1 p_2 p_3^2 p_4+2016 p_1 p_2 p_3 p_4^2+5292 p_2 p_3^2 p_4^2\right) \textcolor{red}{z^5}\nll
 +\left(490 p_1 p_2^2 p_3^2 p_4+9996 p_1 p_2 p_3^2 p_4^2+1225 p_2^2 p_3^2 p_4^2+5103 p_2 p_3^3 p_4^2+1750 p_2 p_3^2 p_4^3\right) \textcolor{red}{z^6}\nll
 +\left(5616 p_1 p_2^2 p_3^2 p_4^2+12600 p_1 p_2 p_3^3 p_4^2+3528 p_2^2 p_3^3 p_4^2+5040 p_1 p_2 p_3^2 p_4^3+5040 p_2 p_3^3 p_4^3\right) \textcolor{red}{z^7}\nll
 +\left(441 p_1^2 p_2^2 p_3^2 p_4^2+17172 p_1 p_2^2 p_3^3 p_4^2+4410 p_1 p_2^2 p_3^2 p_4^3+15750 p_1 p_2 p_3^3 p_4^3+4410 p_2^2 p_3^3 p_4^3+1575 p_2 p_3^3 p_4^4\right) \textcolor{red}{z^8}\nll
 +\left(2450 p_1^2 p_2^2 p_3^3 p_4^2+5600 p_1 p_2^3 p_3^3 p_4^2+32520 p_1 p_2^2 p_3^3 p_4^3+5600 p_1 p_2 p_3^3 p_4^4+2450 p_2^2 p_3^3 p_4^4\right) \textcolor{red}{z^9}\nll
 +\left(1575 p_1^2 p_2^3 p_3^3 p_4^2+4410 p_1^2 p_2^2 p_3^3 p_4^3+15750 p_1 p_2^3 p_3^3 p_4^3+4410 p_1 p_2^2 p_3^4 p_4^3+17172 p_1 p_2^2 p_3^3 p_4^4+441 p_2^2 p_3^4 p_4^4\right) \textcolor{red}{z^{10}}\nll
 +\left(5040 p_1^2 p_2^3 p_3^3 p_4^3+5040 p_1 p_2^3 p_3^4 p_4^3+3528 p_1^2 p_2^2 p_3^3 p_4^4+12600 p_1 p_2^3 p_3^3 p_4^4+5616 p_1 p_2^2 p_3^4 p_4^4\right) \textcolor{red}{z^{11}}\nll
 +\left(1750 p_1^2 p_2^3 p_3^4 p_4^3+5103 p_1^2 p_2^3 p_3^3 p_4^4+1225 p_1^2 p_2^2 p_3^4 p_4^4+9996 p_1 p_2^3 p_3^4 p_4^4+490 p_1 p_2^2 p_3^4 p_4^5\right) \textcolor{red}{z^{12}}\nll
 +\left(5292 p_1^2 p_2^3 p_3^4 p_4^4+2016 p_1 p_2^3 p_3^5 p_4^4+1260 p_1 p_2^3 p_3^4 p_4^5\right) \textcolor{red}{z^{13}}\nll
 +\left(225 p_1^2 p_2^4 p_3^4 p_4^4+1575 p_1^2 p_2^3 p_3^5 p_4^4+630 p_1^2 p_2^3 p_3^4 p_4^5+630 p_1 p_2^3 p_3^5 p_4^5\right) \textcolor{red}{z^{14}}\nll
 +\left(200 p_1^2 p_2^4 p_3^5 p_4^4+560 p_1^2 p_2^3 p_3^5 p_4^5+56 p_1 p_2^3 p_3^5 p_4^6\right) \textcolor{red}{z^{15}}\nll
 +\left(90 p_1^2 p_2^4 p_3^5 p_4^5+63 p_1^2 p_2^3 p_3^5 p_4^6\right) \textcolor{red}{z^{16}}+18 p_1^2 p_2^4 p_3^5 p_4^6 \textcolor{red}{z^{17}}+p_1^2 p_2^4 p_3^6 p_4^6 \textcolor{red}{z^{18}},
 \nl
 H_4=1+10 p_4 \textcolor{red}{z}+45 p_3 p_4 \textcolor{red}{z^2}+\left(70 p_2 p_3 p_4+50 p_3 p_4^2\right) \textcolor{red}{z^3}+\left(35 p_1 p_2 p_3 p_4+175 p_2 p_3 p_4^2\right) \textcolor{red}{z^4}\nll
 +\left(126 p_1 p_2 p_3 p_4^2+126 p_2 p_3^2 p_4^2\right) \textcolor{red}{z^5}+\left(175 p_1 p_2 p_3^2 p_4^2+35 p_2 p_3^2 p_4^3\right) \textcolor{red}{z^6}\nll
 +\left(50 p_1 p_2^2 p_3^2 p_4^2+70 p_1 p_2 p_3^2 p_4^3\right) \textcolor{red}{z^7}+45 p_1 p_2^2 p_3^2 p_4^3 \textcolor{red}{z^8}+10 p_1 p_2^2 p_3^3 p_4^3 \textcolor{red}{z^9}+p_1 p_2^2 p_3^3 p_4^4 \textcolor{red}{z^{10}}.
 \ear
    
  \newpage                         
 { \bf $C_4$-case. }
            For the Lie algebra $C_4 \cong sp(6)$ we have the following polynomials
            \vspace{10pt}
 \bearr{4.C.4}
 H_1=1+7 p_1 \textcolor{red}{z}+21 p_1 p_2 \textcolor{red}{z^2}+35 p_1 p_2 p_3 \textcolor{red}{z^3}+35 p_1 p_2 p_3 p_4 \textcolor{red}{z^4}\nll
 +21 p_1 p_2 p_3^2 p_4 \textcolor{red}{z^5}+7 p_1 p_2^2 p_3^2 p_4 \textcolor{red}{z^6}+p_1^2 p_2^2 p_3^2 p_4 \textcolor{red}{z^7},
 \nl
 H_2=1+12 p_2 \textcolor{red}{z}+\left(21 p_1 p_2+45 p_2 p_3\right) \textcolor{red}{z^2}+\left(140 p_1 p_2 p_3+80 p_2 p_3 p_4\right) \textcolor{red}{z^3}\nll
 +\left(105 p_1 p_2^2 p_3+315 p_1 p_2 p_3 p_4+75 p_2 p_3^2 p_4\right) \textcolor{red}{z^4}+\left(420 p_1 p_2^2 p_3 p_4+336 p_1 p_2 p_3^2 p_4+36 p_2^2 p_3^2 p_4\right) \textcolor{red}{z^5}\nll
 +924 p_1 p_2^2 p_3^2 p_4 \textcolor{red}{z^6}+\left(36 p_1^2 p_2^2 p_3^2 p_4+336 p_1 p_2^3 p_3^2 p_4+420 p_1 p_2^2 p_3^3 p_4\right) \textcolor{red}{z^7}\nll
 +\left(75 p_1^2 p_2^3 p_3^2 p_4+315 p_1 p_2^3 p_3^3 p_4+105 p_1 p_2^2 p_3^3 p_4^2\right) \textcolor{red}{z^8}+\left(80 p_1^2 p_2^3 p_3^3 p_4+140 p_1 p_2^3 p_3^3 p_4^2\right) \textcolor{red}{z^9}\nll
 +\left(45 p_1^2 p_2^3 p_3^3 p_4^2+21 p_1 p_2^3 p_3^4 p_4^2\right) \textcolor{red}{z^{10}}+12 p_1^2 p_2^3 p_3^4 p_4^2 \textcolor{red}{z^{11}}+p_1^2 p_2^4 p_3^4 p_4^2 \textcolor{red}{z^{12}},
 \nl
 H_3=1+15 p_3 \textcolor{red}{z}+\left(45 p_2 p_3+60 p_3 p_4\right) \textcolor{red}{z^2}+\left(35 p_1 p_2 p_3+320 p_2 p_3 p_4+100 p_3^2 p_4\right) \textcolor{red}{z^3}\nll
 +\left(315 p_1 p_2 p_3 p_4+1050 p_2 p_3^2 p_4\right) \textcolor{red}{z^4}+\left(1302 p_1 p_2 p_3^2 p_4+576 p_2^2 p_3^2 p_4+1125 p_2 p_3^3 p_4\right) \textcolor{red}{z^5}\nll
 +\left(1050 p_1 p_2^2 p_3^2 p_4+2240 p_1 p_2 p_3^3 p_4+1215 p_2^2 p_3^3 p_4+500 p_2 p_3^3 p_4^2\right)\textcolor{red}{z^6}
 \nll
 +\left(225 p_1^2 p_2^2 p_3^2 p_4+3990 p_1 p_2^2 p_3^3 p_4+1260 p_1 p_2 p_3^3 p_4^2+960 p_2^2 p_3^3 p_4^2\right) \textcolor{red}{z^7}\nll
 +\left(960 p_1^2 p_2^2 p_3^3 p_4+1260 p_1 p_2^3 p_3^3 p_4+3990 p_1 p_2^2 p_3^3 p_4^2+225 p_2^2 p_3^4 p_4^2\right) \textcolor{red}{z^8}\nll
 +\left(500 p_1^2 p_2^3 p_3^3 p_4+1215 p_1^2 p_2^2 p_3^3 p_4^2+2240 p_1 p_2^3 p_3^3 p_4^2+1050 p_1 p_2^2 p_3^4 p_4^2\right) \textcolor{red}{z^9}\nll
 +\left(1125 p_1^2 p_2^3 p_3^3 p_4^2+576 p_1^2 p_2^2 p_3^4 p_4^2+1302 p_1 p_2^3 p_3^4 p_4^2\right) \textcolor{red}{z^{10}}+\left(1050 p_1^2 p_2^3 p_3^4 p_4^2+315 p_1 p_2^3 p_3^5 p_4^2\right) \textcolor{red}{z^{11}}\nll
 +\left(100 p_1^2 p_2^4 p_3^4 p_4^2+320 p_1^2 p_2^3 p_3^5 p_4^2+35 p_1 p_2^3 p_3^5 p_4^3\right) \textcolor{red}{z^{12}}+\left(60 p_1^2 p_2^4 p_3^5 p_4^2+45 p_1^2 p_2^3 p_3^5 p_4^3\right) \textcolor{red}{z^{13}}\nll
 +15 p_1^2 p_2^4 p_3^5 p_4^3 \textcolor{red}{z^{14}}+p_1^2 p_2^4 p_3^6 p_4^3 \textcolor{red}{z^{15}},
 \nl
 H_4=1+16 p_4 \textcolor{red}{z}+120 p_3 p_4 \textcolor{red}{z^2}+\left(160 p_2 p_3 p_4+400 p_3^2 p_4\right) \textcolor{red}{z^3}\nll
 +\left(70 p_1 p_2 p_3 p_4+1350 p_2 p_3^2 p_4+400 p_3^2 p_4^2\right) \textcolor{red}{z^4}+\left(672 p_1 p_2 p_3^2 p_4+1296 p_2^2 p_3^2 p_4+2400 p_2 p_3^2 p_4^2\right) \textcolor{red}{z^5}\nll
 +\left(1400 p_1 p_2^2 p_3^2 p_4+1512 p_1 p_2 p_3^2 p_4^2+4096 p_2^2 p_3^2 p_4^2+1000 p_2 p_3^3 p_4^2\right) \textcolor{red}{z^6}\nll
 +\left(400 p_1^2 p_2^2 p_3^2 p_4+5600 p_1 p_2^2 p_3^2 p_4^2+1120 p_1 p_2 p_3^3 p_4^2+4320 p_2^2 p_3^3 p_4^2\right) \textcolor{red}{z^7}\nll
 +\left(2025 p_1^2 p_2^2 p_3^2 p_4^2+8820 p_1 p_2^2 p_3^3 p_4^2+2025 p_2^2 p_3^4 p_4^2\right) \textcolor{red}{z^8}\nll
 +\left(4320 p_1^2 p_2^2 p_3^3 p_4^2+1120 p_1 p_2^3 p_3^3 p_4^2+5600 p_1 p_2^2 p_3^4 p_4^2+400 p_2^2 p_3^4 p_4^3\right) \textcolor{red}{z^9}\nll
 +\left(1000 p_1^2 p_2^3 p_3^3 p_4^2+4096 p_1^2 p_2^2 p_3^4 p_4^2+1512 p_1 p_2^3 p_3^4 p_4^2+1400 p_1 p_2^2 p_3^4 p_4^3\right) \textcolor{red}{z^{10}}\nll
 +\left(2400 p_1^2 p_2^3 p_3^4 p_4^2+1296 p_1^2 p_2^2 p_3^4 p_4^3+672 p_1 p_2^3 p_3^4 p_4^3\right) \textcolor{red}{z^{11}}\nll
 +\left(400 p_1^2 p_2^4 p_3^4 p_4^2+1350 p_1^2 p_2^3 p_3^4 p_4^3+70 p_1 p_2^3 p_3^5 p_4^3\right) \textcolor{red}{z^{12}}+\left(400 p_1^2 p_2^4 p_3^4 p_4^3+160 p_1^2 p_2^3 p_3^5 p_4^3\right) \textcolor{red}{z^{13}}\nll
 +120 p_1^2 p_2^4 p_3^5 p_4^3 \textcolor{red}{z^{14}}+16 p_1^2 p_2^4 p_3^6 p_4^3 \textcolor{red}{z^{15}}+p_1^2 p_2^4 p_3^6 p_4^4 \textcolor{red}{z^{16}}.
 \ear
   
    { \bf $D_4$-case. }
             In case of  Lie algebra $D_4 \cong so(8)$ we obtain the following polynomials
             \vspace{10pt}
  \bearr{4.D.4}
 H_1=1+6 p_1 \textcolor{red}{z}+15 p_1 p_2 \textcolor{red}{z^2}+\left(10 p_1 p_2 p_3+10 p_1 p_2 p_4\right) \textcolor{red}{z^3}+15 p_1 p_2 p_3 p_4 \textcolor{red}{z^4}\nll
 +6 p_1 p_2^2 p_3 p_4 \textcolor{red}{z^5}+p_1^2 p_2^2 p_3 p_4 \textcolor{red}{z^6},
 \nl
 H_2=1+10 p_2 \textcolor{red}{z}+\left(15 p_1 p_2+15 p_2 p_3+15 p_2 p_4\right) \textcolor{red}{z^2}+\left(40 p_1 p_2 p_3+40 p_1 p_2 p_4+40 p_2 p_3 p_4\right) \textcolor{red}{z^3}\nll
 +\left(25 p_1 p_2^2 p_3+25 p_1 p_2^2 p_4+135 p_1 p_2 p_3 p_4+25 p_2^2 p_3 p_4\right) \textcolor{red}{z^4}+252 p_1 p_2^2 p_3 p_4 \textcolor{red}{z^5}\nll
 +\left(25 p_1^2 p_2^2 p_3 p_4+135 p_1 p_2^3 p_3 p_4+25 p_1 p_2^2 p_3^2 p_4+25 p_1 p_2^2 p_3 p_4^2\right) \textcolor{red}{z^6}\nll
 +\left(40 p_1^2 p_2^3 p_3 p_4+40 p_1 p_2^3 p_3^2 p_4+40 p_1 p_2^3 p_3 p_4^2\right) \textcolor{red}{z^7}+\left(15 p_1^2 p_2^3 p_3^2 p_4+15 p_1^2 p_2^3 p_3 p_4^2+15 p_1 p_2^3 p_3^2 p_4^2\right) \textcolor{red}{z^8}\nll+10 p_1^2 p_2^3 p_3^2 p_4^2 \textcolor{red}{z^9}+p_1^2 p_2^4 p_3^2 p_4^2 \textcolor{red}{z^{10}},
 \nl
 H_3=1+6 p_3 \textcolor{red}{z}+15 p_2 p_3 \textcolor{red}{z^2}+\left(10 p_1 p_2 p_3+10 p_2 p_3 p_4\right) \textcolor{red}{z^3}+15 p_1 p_2 p_3 p_4 \textcolor{red}{z^4}\nll
 +6 p_1 p_2^2 p_3 p_4 \textcolor{red}{z^5}+p_1 p_2^2 p_3^2 p_4 \textcolor{red}{z^6},
 \nl
 H_4=1+6 p_4 \textcolor{red}{z}+15 p_2 p_4 \textcolor{red}{z^2}+\left(10 p_1 p_2 p_4+10 p_2 p_3 p_4\right) \textcolor{red}{z^3}+15 p_1 p_2 p_3 p_4 \textcolor{red}{z^4}\nll
 +6 p_1 p_2^2 p_3 p_4 \textcolor{red}{z^5}+p_1 p_2^2 p_3 p_4^2 z^6.
 \ear 
  
  \newpage
  
  { \bf $F_4$-case.}      
      Now we consider the exceptional Lie algebra $F_4$. We obtain the following polynomials 
               \vspace{15pt}         
 \begin{itemize}
 \item[$H_1=$]\(
 1+22 p_1 \textcolor{red}{z}+231 p_1 p_2 \textcolor{red}{z^2}+1540 p_1 p_2 p_3 \textcolor{red}{z^3}+(5775 p_1 p_2 p_3^2+1540 p_1 p_2 p_3 p_4) \textcolor{red}{z^4}+(9702 p_1 p_2^2 p_3^2+16632 p_1 p_2 p_3^2 p_4) \textcolor{red}{z^5} \\
 +(5929 p_1^2 p_2^2 p_3^2+53900 p_1 p_2^2 p_3^2 p_4+14784 p_1 p_2 p_3^2 p_4^2) \textcolor{red}{z^6}+(47432 p_1^2 p_2^2 p_3^2 p_4+33000 p_1 p_2^2 p_3^3 p_4+90112 p_1 p_2^2 p_3^2 p_4^2) \textcolor{red}{z^7} \\
 +(65340 p_1^2 p_2^2 p_3^3 p_4+108900 p_1^2 p_2^2 p_3^2 p_4^2+145530 p_1 p_2^2 p_3^3 p_4^2) \textcolor{red}{z^8}+(33880 p_1^2 p_2^3 p_3^3 p_4+355740 p_1^2 p_2^2 p_3^3 p_4^2+107800 p_1 p_2^2 p_3^4 p_4^2) \textcolor{red}{z^9} \\
 +(10164 p_1^2 p_2^3 p_3^4 p_4+211750 p_1^2 p_2^3 p_3^3 p_4^2+379456 p_1^2 p_2^2 p_3^4 p_4^2+45276 p_1 p_2^3 p_3^4 p_4^2) \textcolor{red}{z^{10}}+705432 p_1^2 p_2^3 p_3^4 p_4^2 \textcolor{red}{z^{11}} \\
 +(45276 p_1^3 p_2^3 p_3^4 p_4^2+379456 p_1^2 p_2^4 p_3^4 p_4^2+211750 p_1^2 p_2^3 p_3^5 p_4^2+10164 p_1^2 p_2^3 p_3^4 p_4^3) \textcolor{red}{z^{12}} \\
 +(107800 p_1^3 p_2^4 p_3^4 p_4^2+355740 p_1^2 p_2^4 p_3^5 p_4^2+33880 p_1^2 p_2^3 p_3^5 p_4^3) \textcolor{red}{z^{13}} \\
 +(145530 p_1^3 p_2^4 p_3^5 p_4^2+108900 p_1^2 p_2^4 p_3^6 p_4^2+65340 p_1^2 p_2^4 p_3^5 p_4^3) \textcolor{red}{z^{14}} \\
 +(90112 p_1^3 p_2^4 p_3^6 p_4^2+33000 p_1^3 p_2^4 p_3^5 p_4^3+47432 p_1^2 p_2^4 p_3^6 p_4^3) \textcolor{red}{z^{15}}+(14784 p_1^3 p_2^5 p_3^6 p_4^2+53900 p_1^3 p_2^4 p_3^6 p_4^3+5929 p_1^2 p_2^4 p_3^6 p_4^4) \textcolor{red}{z^{16}} \\
 +(16632 p_1^3 p_2^5 p_3^6 p_4^3+9702 p_1^3 p_2^4 p_3^6 p_4^4) \textcolor{red}{z^{17}}+(1540 p_1^3 p_2^5 p_3^7 p_4^3+5775 p_1^3 p_2^5 p_3^6 p_4^4) \textcolor{red}{z^{18}}+1540 p_1^3 p_2^5 p_3^7 p_4^4 \textcolor{red}{z^{19}}+231 p_1^3 p_2^5 p_3^8 p_4^4 \textcolor{red}{z^{20}} \\
 +22 p_1^3 p_2^6 p_3^8 p_4^4 \textcolor{red}{z^{21}}+p_1^4 p_2^6 p_3^8 p_4^4 \textcolor{red}{z^{22}},
 \)
 \begin{equation}\label{4.H1}\end{equation}
 
 \item[$H_2=$]\(
 1+42 p_2 \textcolor{red}{z}+(231 p_1 p_2+630 p_2 p_3) \textcolor{red}{z^2}+(6160 p_1 p_2 p_3+4200 p_2 p_3^2+1120 p_2 p_3 p_4) \textcolor{red}{z^3} \\  +(16170 p_1 p_2^2 p_3+51975 p_1 p_2 p_3^2+11025 p_2^2 p_3^2+13860 p_1 p_2 p_3 p_4+18900 p_2 p_3^2 p_4) \textcolor{red}{z^4}\\
 +(407484 p_1 p_2^2 p_3^2+64680 p_1 p_2^2 p_3 p_4+266112 p_1 p_2 p_3^2 p_4+88200 p_2^2 p_3^2 p_4+24192 p_2 p_3^2 p_4^2) \textcolor{red}{z^5}\\
 +(148225 p_1^2 p_2^2 p_3^2+916839 p_1 p_2^3 p_3^2+404250 p_1 p_2^2 p_3^3+3132668 p_1 p_2^2 p_3^2 p_4 +73500 p_2^2 p_3^3 p_4+369600 p_1 p_2 p_3^2 p_4^2+200704 p_2^2 p_3^2 p_4^2) \textcolor{red}{z^6} \\
 +(996072 p_1^2 p_2^3 p_3^2+2716560 p_1 p_2^3 p_3^3+1707552 p_1^2 p_2^2 p_3^2 p_4+9055200 p_1 p_2^3 p_3^2 p_4 +6035040 p_1 p_2^2 p_3^3 p_4+6044544 p_1 p_2^2 p_3^2 p_4^2+423360 p_2^2 p_3^3 p_4^2) \textcolor{red}{z^7}\\
 +(3735270 p_1^2 p_2^3 p_3^3+2546775 p_1 p_2^3 p_3^4+12450900 p_1^2 p_2^3 p_3^2 p_4+3201660 p_1^2 p_2^2 p_3^3 p_4+43423380 p_1 p_2^3 p_3^3 p_4+4365900 p_1 p_2^2 p_3^4 p_4 +5336100 p_1^2 p_2^2 p_3^2 p_4^2+23654400 p_1 p_2^3 p_3^2 p_4^2+18918900 p_1 p_2^2 p_3^3 p_4^2+396900 p_2^2 p_3^4 p_4^2) \textcolor{red}{z^8} \\
 +(6225450 p_1^2 p_2^3 p_3^4+81650800 p_1^2 p_2^3 p_3^3 p_4+93601200 p_1 p_2^3 p_3^4 p_4+41164200 p_1^2 p_2^3 p_3^2 p_4^2+22767360 p_1^2 p_2^2 p_3^3 p_4^2+171990280 p_1 p_2^3 p_3^3 p_4^2+24147200 p_1 p_2^2 p_3^4 p_4^2+205800 p_2^3 p_3^4 p_4^2+4139520 p_1 p_2^2 p_3^3 p_4^3) \textcolor{red}{z^9}\\
 +(2614689 p_1^2 p_2^4 p_3^4+17431260 p_1^2 p_2^4 p_3^3 p_4+231708708 p_1^2 p_2^3 p_3^4 p_4+23769900 p_1 p_2^4 p_3^4 p_4+77962500 p_1 p_2^3 p_3^5 p_4 +420637140 p_1^2 p_2^3 p_3^3 p_4^2+30735936 p_1^2 p_2^2 p_3^4 p_4^2+598635576 p_1 p_2^3 p_3^4 p_4^2+56770560 p_1 p_2^3 p_3^3 p_4^3+11176704 p_1 p_2^2 p_3^4 p_4^3) \textcolor{red}{z^{10}}\\
 +(175877856 p_1^2 p_2^4 p_3^4 p_4+274428000 p_1^2 p_2^3 p_3^5 p_4+58212000 p_1 p_2^4 p_3^5 p_4+142296000 p_1^2 p_2^4 p_3^3 p_4^2+1896293952 p_1^2 p_2^3 p_3^4 p_4^2 +191866752 p_1 p_2^4 p_3^4 p_4^2+984060000 p_1 p_2^3 p_3^5 p_4^2+121968000 p_1^2 p_2^3 p_3^3 p_4^3+435558816 p_1 p_2^3 p_3^4 p_4^3) \textcolor{red}{z^{11}}\\
 +(12782924 p_1^3 p_2^4 p_3^4 p_4+525427980 p_1^2 p_2^4 p_3^5 p_4+5478396 p_1^3 p_2^3 p_3^4 p_4^2+2005022376 p_1^2 p_2^4 p_3^4 p_4^2 +4106272940 p_1^2 p_2^3 p_3^5 p_4^2+816487980 p_1 p_2^4 p_3^5 p_4^2+707437500 p_1 p_2^3 p_3^6 p_4^2+1396604748 p_1^2 p_2^3 p_3^4 p_4^3 +220774400 p_1 p_2^4 p_3^4 p_4^3+1201272380 p_1 p_2^3 p_3^5 p_4^3+60555264 p_1 p_2^3 p_3^4 p_4^4) \textcolor{red}{z^{12}}\\
 +(70436520 p_1^3 p_2^4 p_3^5 p_4+239057280 p_1^2 p_2^5 p_3^5 p_4+96049800 p_1^2 p_2^4 p_3^6 p_4+180457200 p_1^3 p_2^4 p_3^4 p_4^2 \\+398428800 p_1^2 p_2^5 p_3^4 p_4^2+9178974000 p_1^2 p_2^4 p_3^5 p_4^2+3585859200 p_1^2 p_2^3 p_3^6 p_4^2+1189465200 p_1 p_2^4 p_3^6 p_4^2 +1611502200 p_1^2 p_2^4 p_3^4 p_4^3+5439772800 p_1^2 p_2^3 p_3^5 p_4^3+1540871640 p_1 p_2^4 p_3^5 p_4^3+1303948800 p_1 p_2^3 p_3^6 p_4^3 +292723200 p_1^2 p_2^3 p_3^4 p_4^4+391184640 p_1 p_2^3 p_3^5 p_4^4) \textcolor{red}{z^{13}}\\
 +(82175940 p_1^3 p_2^5 p_3^5 p_4+112058100 p_1^2 p_2^5 p_3^6 p_4+136959900 p_1^3 p_2^5 p_3^4 p_4^2+1285029900 p_1^3 p_2^4 p_3^5 p_4^2 +5685080940 p_1^2 p_2^5 p_3^5 p_4^2+15028648200 p_1^2 p_2^4 p_3^6 p_4^2+499167900 p_1 p_2^5 p_3^6 p_4^2+234788400 p_1^3 p_2^4 p_3^4 p_4^3 +15327479700 p_1^2 p_2^4 p_3^5 p_4^3+7171718400 p_1^2 p_2^3 p_3^6 p_4^3+3451486500 p_1 p_2^4 p_3^6 p_4^3+446054400 p_1^2 p_2^4 p_3^4 p_4^4+2151515520 p_1^2 p_2^3 p_3^5 p_4^4+596090880 p_1 p_2^4 p_3^5 p_4^4+651974400 p_1 p_2^3 p_3^6 p_4^4) \textcolor{red}{z^{14}} \\
 +(43827168 p_1^3 p_2^5 p_3^6 p_4+2179888480 p_1^3 p_2^5 p_3^5 p_4^2+2414513024 p_1^3 p_2^4 p_3^6 p_4^2+21026246976 p_1^2 p_2^5 p_3^6 p_4^2 +3557400000 p_1^2 p_2^4 p_3^7 p_4^2+3277206240 p_1^3 p_2^4 p_3^5 p_4^3+10654446880 p_1^2 p_2^5 p_3^5 p_4^3 +38613582112 p_1^2 p_2^4 p_3^6 p_4^3+1774819200 p_1 p_2^5 p_3^6 p_4^3+646800000 p_1 p_2^4 p_3^7 p_4^3+8150714880 p_1^2 p_2^4 p_3^5 p_4^4 +4079910912 p_1^2 p_2^3 p_3^6 p_4^4+2253071744 p_1 p_2^4 p_3^6 p_4^4) \textcolor{red}{z^{15}}\\
 +(9717029784 p_1^3 p_2^5 p_3^6 p_4^2+8199664704 p_1^2 p_2^6 p_3^6 p_4^2+13199224500 p_1^2 p_2^5 p_3^7 p_4^2+4946287500 p_1^3 p_2^5 p_3^5 p_4^3+10108843668 p_1^3 p_2^4 p_3^6 p_4^3+64474736508 p_1^2 p_2^5 p_3^6 p_4^3+14007262500 p_1^2 p_2^4 p_3^7 p_4^3+611226000 p_1 p_2^5 p_3^7 p_4^3+1760913000 p_1^3 p_2^4 p_3^5 p_4^4+7805952000 p_1^2 p_2^5 p_3^5 p_4^4+29296429974 p_1^2 p_2^4 p_3^6 p_4^4+1669054464 p_1 p_2^5 p_3^6 p_4^4+713097000 p_1 p_2^4 p_3^7 p_4^4) \textcolor{red}{z^{16}}\\
 +(439267752 p_1^4 p_2^5 p_3^6 p_4^2+6754454784 p_1^3 p_2^6 p_3^6 p_4^2+6903638280 p_1^3 p_2^5 p_3^7 p_4^2+10040405760 p_1^2 p_2^6 p_3^7 p_4^2+2858625000 p_1^2 p_2^5 p_3^8 p_4^2+37825702992 p_1^3 p_2^5 p_3^6 p_4^3+33468019200 p_1^2 p_2^6 p_3^6 p_4^3+4507937280 p_1^3 p_2^4 p_3^7 p_4^3+57537501840 p_1^2 p_2^5 p_3^7 p_4^3+4192650000 p_1^3 p_2^5 p_3^5 p_4^4+8611029504 p_1^3 p_2^4 p_3^6 p_4^4+63276492636 p_1^2 p_2^5 p_3^6 p_4^4+16802311680 p_1^2 p_2^4 p_3^7 p_4^4+1198002960 p_1 p_2^5 p_3^7 p_4^4+245887488 p_1^2 p_2^4 p_3^6 p_4^5) \textcolor{red}{z^{17}} \\
 +(1423552900 p_1^4 p_2^6 p_3^6 p_4^2+10086748980 p_1^3 p_2^6 p_3^7 p_4^2+2862182400 p_1^3 p_2^5 p_3^8 p_4^2+3890016900 p_1^2 p_2^6 p_3^8 p_4^2+2440376400 p_1^4 p_2^5 p_3^6 p_4^3+33759456500 p_1^3 p_2^6 p_3^6 p_4^3+44524657100 p_1^3 p_2^5 p_3^7 p_4^3+59339922180 p_1^2 p_2^6 p_3^7 p_4^3+16165587900 p_1^2 p_2^5 p_3^8 p_4^3+43888833450 p_1^3 p_2^5 p_3^6 p_4^4+38856294400 p_1^2 p_2^6 p_3^6 p_4^4+6135803520 p_1^3 p_2^4 p_3^7 p_4^4+86086107380 p_1^2 p_2^5 p_3^7 p_4^4+1859334400 p_1^2 p_2^4 p_3^8 p_4^4+221852400 p_1 p_2^5 p_3^8 p_4^4+1040793600 p_1^2 p_2^5 p_3^6 p_4^5+1115600640 p_1^2 p_2^4 p_3^7 p_4^5) \textcolor{red}{z^{18}} \\
 +(2510101440 p_1^4 p_2^6 p_3^7 p_4^2+6411081600 p_1^3 p_2^6 p_3^8 p_4^2+8367004800 p_1^4 p_2^6 p_3^6 p_4^3+2151515520 p_1^4 p_2^5 p_3^7 p_4^3+81592267680 p_1^3 p_2^6 p_3^7 p_4^3+18912247200 p_1^3 p_2^5 p_3^8 p_4^3+38377231200 p_1^2 p_2^6 p_3^8 p_4^3+3585859200 p_1^4 p_2^5 p_3^6 p_4^4+45964195200 p_1^3 p_2^6 p_3^6 p_4^4+79733253600 p_1^3 p_2^5 p_3^7 p_4^4+102862932480 p_1^2 p_2^6 p_3^7 p_4^4+46561158000 p_1^2 p_2^5 p_3^8 p_4^4+804988800 p_1^3 p_2^5 p_3^6 p_4^5+8941474080 p_1^2 p_2^5 p_3^7 p_4^5) \textcolor{red}{z^{19}}\\
 +(1967099904 p_1^4 p_2^6 p_3^8 p_4^2+788889024 p_1^3 p_2^7 p_3^8 p_4^2+24726420180 p_1^4 p_2^6 p_3^7 p_4^3+5259260160 p_1^3 p_2^7 p_3^7 p_4^3+75784320612 p_1^3 p_2^6 p_3^8 p_4^3+8004150000 p_1^2 p_2^6 p_3^9 p_4^3+13340250000 p_1^4 p_2^6 p_3^6 p_4^4+6589016280 p_1^4 p_2^5 p_3^7 p_4^4+166955605740 p_1^3 p_2^6 p_3^7 p_4^4+57761551386 p_1^3 p_2^5 p_3^8 p_4^4+113404704966 p_1^2 p_2^6 p_3^8 p_4^4+9338175000 p_1^2 p_2^5 p_3^9 p_4^4+7582847580 p_1^3 p_2^5 p_3^7 p_4^5+13113999360 p_1^2 p_2^6 p_3^7 p_4^5+9175317228 p_1^2 p_2^5 p_3^8 p_4^5) \textcolor{red}{z^{20}}\\
 +(398428800 p_1^4 p_2^7 p_3^8 p_4^2+2656192000 p_1^4 p_2^7 p_3^7 p_4^3+29530356856 p_1^4 p_2^6 p_3^8 p_4^3+14144946816 p_1^3 p_2^7 p_3^8 p_4^3+20764887000 p_1^3 p_2^6 p_3^9 p_4^3+60120060000 p_1^4 p_2^6 p_3^7 p_4^4+14609056000 p_1^3 p_2^7 p_3^7 p_4^4+3123681792 p_1^4 p_2^5 p_3^8 p_4^4+247562655912 p_1^3 p_2^6 p_3^8 p_4^4+3123681792 p_1^2 p_2^7 p_3^8 p_4^4+14609056000 p_1^3 p_2^5 p_3^9 p_4^4+60120060000 p_1^2 p_2^6 p_3^9 p_4^4+20764887000 p_1^3 p_2^6 p_3^7 p_4^5+14144946816 p_1^3 p_2^5 p_3^8 p_4^5+29530356856 p_1^2 p_2^6 p_3^8 p_4^5+2656192000 p_1^2 p_2^5 p_3^9 p_4^5+398428800 p_1^2 p_2^5 p_3^8 p_4^6) \textcolor{red}{z^{21}}\\
 +(9175317228 p_1^4 p_2^7 p_3^8 p_4^3+13113999360 p_1^4 p_2^6 p_3^9 p_4^3+7582847580 p_1^3 p_2^7 p_3^9 p_4^3+9338175000 p_1^4 p_2^7 p_3^7 p_4^4+113404704966 p_1^4 p_2^6 p_3^8 p_4^4+57761551386 p_1^3 p_2^7 p_3^8 p_4^4+166955605740 p_1^3 p_2^6 p_3^9 p_4^4+6589016280 p_1^2 p_2^7 p_3^9 p_4^4+13340250000 p_1^2 p_2^6 p_3^{10} p_4^4+8004150000 p_1^4 p_2^6 p_3^7 p_4^5+75784320612 p_1^3 p_2^6 p_3^8 p_4^5+5259260160 p_1^3 p_2^5 p_3^9 p_4^5+24726420180 p_1^2 p_2^6 p_3^9 p_4^5+788889024 p_1^3 p_2^5 p_3^8 p_4^6+1967099904 p_1^2 p_2^6 p_3^8 p_4^6) \textcolor{red}{z^{22}}\\
 +(8941474080 p_1^4 p_2^7 p_3^9 p_4^3+804988800 p_1^3 p_2^7 p_3^{10} p_4^3+46561158000 p_1^4 p_2^7 p_3^8 p_4^4+102862932480 p_1^4 p_2^6 p_3^9 p_4^4+79733253600 p_1^3 p_2^7 p_3^9 p_4^4+45964195200 p_1^3 p_2^6 p_3^{10} p_4^4+3585859200 p_1^2 p_2^7 p_3^{10} p_4^4+38377231200 p_1^4 p_2^6 p_3^8 p_4^5+18912247200 p_1^3 p_2^7 p_3^8 p_4^5+81592267680 p_1^3 p_2^6 p_3^9 p_4^5+2151515520 p_1^2 p_2^7 p_3^9 p_4^5+8367004800 p_1^2 p_2^6 p_3^{10} p_4^5+6411081600 p_1^3 p_2^6 p_3^8 p_4^6+2510101440 p_1^2 p_2^6 p_3^9 p_4^6) \textcolor{red}{z^{23}}\\
 +(1115600640 p_1^4 p_2^8 p_3^9 p_4^3+1040793600 p_1^4 p_2^7 p_3^{10} p_4^3+221852400 p_1^5 p_2^7 p_3^8 p_4^4+1859334400 p_1^4 p_2^8 p_3^8 p_4^4+86086107380 p_1^4 p_2^7 p_3^9 p_4^4+6135803520 p_1^3 p_2^8 p_3^9 p_4^4+38856294400 p_1^4 p_2^6 p_3^{10} p_4^4+43888833450 p_1^3 p_2^7 p_3^{10} p_4^4+16165587900 p_1^4 p_2^7 p_3^8 p_4^5+59339922180 p_1^4 p_2^6 p_3^9 p_4^5+44524657100 p_1^3 p_2^7 p_3^9 p_4^5+33759456500 p_1^3 p_2^6 p_3^{10} p_4^5+2440376400 p_1^2 p_2^7 p_3^{10} p_4^5+3890016900 p_1^4 p_2^6 p_3^8 p_4^6+2862182400 p_1^3 p_2^7 p_3^8 p_4^6+10086748980 p_1^3 p_2^6 p_3^9 p_4^6+1423552900 p_1^2 p_2^6 p_3^{10} p_4^6) \textcolor{red}{z^{24}}\\
 +(245887488 p_1^4 p_2^8 p_3^{10} p_4^3+1198002960 p_1^5 p_2^7 p_3^9 p_4^4+16802311680 p_1^4 p_2^8 p_3^9 p_4^4+63276492636 p_1^4 p_2^7 p_3^{10} p_4^4+8611029504 p_1^3 p_2^8 p_3^{10} p_4^4+4192650000 p_1^3 p_2^7 p_3^{11} p_4^4+57537501840 p_1^4 p_2^7 p_3^9 p_4^5+4507937280 p_1^3 p_2^8 p_3^9 p_4^5+33468019200 p_1^4 p_2^6 p_3^{10} p_4^5+37825702992 p_1^3 p_2^7 p_3^{10} p_4^5+2858625000 p_1^4 p_2^7 p_3^8 p_4^6+10040405760 p_1^4 p_2^6 p_3^9 p_4^6+6903638280 p_1^3 p_2^7 p_3^9 p_4^6+6754454784 p_1^3 p_2^6 p_3^{10} p_4^6+439267752 p_1^2 p_2^7 p_3^{10} p_4^6) \textcolor{red}{z^{25}}\\
 +(713097000 p_1^5 p_2^8 p_3^9 p_4^4+1669054464 p_1^5 p_2^7 p_3^{10} p_4^4+29296429974 p_1^4 p_2^8 p_3^{10} p_4^4+7805952000 p_1^4 p_2^7 p_3^{11} p_4^4+1760913000 p_1^3 p_2^8 p_3^{11} p_4^4+611226000 p_1^5 p_2^7 p_3^9 p_4^5+14007262500 p_1^4 p_2^8 p_3^9 p_4^5+64474736508 p_1^4 p_2^7 p_3^{10} p_4^5+10108843668 p_1^3 p_2^8 p_3^{10} p_4^5+4946287500 p_1^3 p_2^7 p_3^{11} p_4^5+13199224500 p_1^4 p_2^7 p_3^9 p_4^6+8199664704 p_1^4 p_2^6 p_3^{10} p_4^6+9717029784 p_1^3 p_2^7 p_3^{10} p_4^6) \textcolor{red}{z^{26}} \\
 +(2253071744 p_1^5 p_2^8 p_3^{10} p_4^4+4079910912 p_1^4 p_2^9 p_3^{10} p_4^4+8150714880 p_1^4 p_2^8 p_3^{11} p_4^4+646800000 p_1^5 p_2^8 p_3^9 p_4^5+1774819200 p_1^5 p_2^7 p_3^{10} p_4^5+38613582112 p_1^4 p_2^8 p_3^{10} p_4^5+10654446880 p_1^4 p_2^7 p_3^{11} p_4^5+3277206240 p_1^3 p_2^8 p_3^{11} p_4^5+3557400000 p_1^4 p_2^8 p_3^9 p_4^6+21026246976 p_1^4 p_2^7 p_3^{10} p_4^6+2414513024 p_1^3 p_2^8 p_3^{10} p_4^6+2179888480 p_1^3 p_2^7 p_3^{11} p_4^6+43827168 p_1^3 p_2^7 p_3^{10} p_4^7) \textcolor{red}{z^{27}} \\
 +(651974400 p_1^5 p_2^9 p_3^{10} p_4^4+596090880 p_1^5 p_2^8 p_3^{11} p_4^4+2151515520 p_1^4 p_2^9 p_3^{11} p_4^4+446054400 p_1^4 p_2^8 p_3^{12} p_4^4+3451486500 p_1^5 p_2^8 p_3^{10} p_4^5+7171718400 p_1^4 p_2^9 p_3^{10} p_4^5+15327479700 p_1^4 p_2^8 p_3^{11} p_4^5+234788400 p_1^3 p_2^8 p_3^{12} p_4^5+499167900 p_1^5 p_2^7 p_3^{10} p_4^6+15028648200 p_1^4 p_2^8 p_3^{10} p_4^6+5685080940 p_1^4 p_2^7 p_3^{11} p_4^6+1285029900 p_1^3 p_2^8 p_3^{11} p_4^6+136959900 p_1^3 p_2^7 p_3^{12} p_4^6+112058100 p_1^4 p_2^7 p_3^{10} p_4^7+82175940 p_1^3 p_2^7 p_3^{11} p_4^7) \textcolor{red}{z^{28}} \\
 +(391184640 p_1^5 p_2^9 p_3^{11} p_4^4+292723200 p_1^4 p_2^9 p_3^{12} p_4^4+1303948800 p_1^5 p_2^9 p_3^{10} p_4^5+1540871640 p_1^5 p_2^8 p_3^{11} p_4^5+5439772800 p_1^4 p_2^9 p_3^{11} p_4^5+1611502200 p_1^4 p_2^8 p_3^{12} p_4^5+1189465200 p_1^5 p_2^8 p_3^{10} p_4^6+3585859200 p_1^4 p_2^9 p_3^{10} p_4^6+9178974000 p_1^4 p_2^8 p_3^{11} p_4^6+398428800 p_1^4 p_2^7 p_3^{12} p_4^6+180457200 p_1^3 p_2^8 p_3^{12} p_4^6+96049800 p_1^4 p_2^8 p_3^{10} p_4^7+239057280 p_1^4 p_2^7 p_3^{11} p_4^7+70436520 p_1^3 p_2^8 p_3^{11} p_4^7) \textcolor{red}{z^{29}} \\
 +(60555264 p_1^5 p_2^9 p_3^{12} p_4^4+1201272380 p_1^5 p_2^9 p_3^{11} p_4^5+220774400 p_1^5 p_2^8 p_3^{12} p_4^5+1396604748 p_1^4 p_2^9 p_3^{12} p_4^5+707437500 p_1^5 p_2^9 p_3^{10} p_4^6+816487980 p_1^5 p_2^8 p_3^{11} p_4^6+4106272940 p_1^4 p_2^9 p_3^{11} p_4^6+2005022376 p_1^4 p_2^8 p_3^{12} p_4^6+5478396 p_1^3 p_2^9 p_3^{12} p_4^6+525427980 p_1^4 p_2^8 p_3^{11} p_4^7+12782924 p_1^3 p_2^8 p_3^{12} p_4^7) \textcolor{red}{z^{30}} \\
 +(435558816 p_1^5 p_2^9 p_3^{12} p_4^5+121968000 p_1^4 p_2^9 p_3^{13} p_4^5+984060000 p_1^5 p_2^9 p_3^{11} p_4^6+191866752 p_1^5 p_2^8 p_3^{12} p_4^6+1896293952 p_1^4 p_2^9 p_3^{12} p_4^6+142296000 p_1^4 p_2^8 p_3^{13} p_4^6+58212000 p_1^5 p_2^8 p_3^{11} p_4^7+274428000 p_1^4 p_2^9 p_3^{11} p_4^7+175877856 p_1^4 p_2^8 p_3^{12} p_4^7) \textcolor{red}{z^{31}} \\
 +(11176704 p_1^5 p_2^{10} p_3^{12} p_4^5+56770560 p_1^5 p_2^9 p_3^{13} p_4^5+598635576 p_1^5 p_2^9 p_3^{12} p_4^6+30735936 p_1^4 p_2^{10} p_3^{12} p_4^6+420637140 p_1^4 p_2^9 p_3^{13} p_4^6+77962500 p_1^5 p_2^9 p_3^{11} p_4^7+23769900 p_1^5 p_2^8 p_3^{12} p_4^7+231708708 p_1^4 p_2^9 p_3^{12} p_4^7+17431260 p_1^4 p_2^8 p_3^{13} p_4^7+2614689 p_1^4 p_2^8 p_3^{12} p_4^8) \textcolor{red}{z^{32}} \\
 +(4139520 p_1^5 p_2^{10} p_3^{13} p_4^5+205800 p_1^6 p_2^9 p_3^{12} p_4^6+24147200 p_1^5 p_2^{10} p_3^{12} p_4^6+171990280 p_1^5 p_2^9 p_3^{13} p_4^6+22767360 p_1^4 p_2^{10} p_3^{13} p_4^6+41164200 p_1^4 p_2^9 p_3^{14} p_4^6+93601200 p_1^5 p_2^9 p_3^{12} p_4^7+81650800 p_1^4 p_2^9 p_3^{13} p_4^7+6225450 p_1^4 p_2^9 p_3^{12} p_4^8) \textcolor{red}{z^{33}} \\
 +(396900 p_1^6 p_2^{10} p_3^{12} p_4^6+18918900 p_1^5 p_2^{10} p_3^{13} p_4^6+23654400 p_1^5 p_2^9 p_3^{14} p_4^6+5336100 p_1^4 p_2^{10} p_3^{14} p_4^6+4365900 p_1^5 p_2^{10} p_3^{12} p_4^7+43423380 p_1^5 p_2^9 p_3^{13} p_4^7+3201660 p_1^4 p_2^{10} p_3^{13} p_4^7+12450900 p_1^4 p_2^9 p_3^{14} p_4^7+2546775 p_1^5 p_2^9 p_3^{12} p_4^8+3735270 p_1^4 p_2^9 p_3^{13} p_4^8) \textcolor{red}{z^{34}} \\
 +(423360 p_1^6 p_2^{10} p_3^{13} p_4^6+6044544 p_1^5 p_2^{10} p_3^{14} p_4^6+6035040 p_1^5 p_2^{10} p_3^{13} p_4^7+9055200 p_1^5 p_2^9 p_3^{14} p_4^7+1707552 p_1^4 p_2^{10} p_3^{14} p_4^7+2716560 p_1^5 p_2^9 p_3^{13} p_4^8+996072 p_1^4 p_2^9 p_3^{14} p_4^8) \textcolor{red}{z^{35}} \\
 +(200704 p_1^6 p_2^{10} p_3^{14} p_4^6+369600 p_1^5 p_2^{11} p_3^{14} p_4^6+73500 p_1^6 p_2^{10} p_3^{13} p_4^7+3132668 p_1^5 p_2^{10} p_3^{14} p_4^7+404250 p_1^5 p_2^{10} p_3^{13} p_4^8+916839 p_1^5 p_2^9 p_3^{14} p_4^8+148225 p_1^4 p_2^{10} p_3^{14} p_4^8) \textcolor{red}{z^{36}} \\
 +(24192 p_1^6 p_2^{11} p_3^{14} p_4^6+88200 p_1^6 p_2^{10} p_3^{14} p_4^7+266112 p_1^5 p_2^{11} p_3^{14} p_4^7+64680 p_1^5 p_2^{10} p_3^{15} p_4^7+407484 p_1^5 p_2^{10} p_3^{14} p_4^8) \textcolor{red}{z^{37}} \\
 +(18900 p_1^6 p_2^{11} p_3^{14} p_4^7+13860 p_1^5 p_2^{11} p_3^{15} p_4^7+11025 p_1^6 p_2^{10} p_3^{14} p_4^8+51975 p_1^5 p_2^{11} p_3^{14} p_4^8+16170 p_1^5 p_2^{10} p_3^{15} p_4^8) \textcolor{red}{z^{38}} \\
 +(1120 p_1^6 p_2^{11} p_3^{15} p_4^7+4200 p_1^6 p_2^{11} p_3^{14} p_4^8+6160 p_1^5 p_2^{11} p_3^{15} p_4^8) \textcolor{red}{z^{39}} \\
 +(630 p_1^6 p_2^{11} p_3^{15} p_4^8+231 p_1^5 p_2^{11} p_3^{16} p_4^8) \textcolor{red}{z^{40}} +42 p_1^6 p_2^{11} p_3^{16} p_4^8 \textcolor{red}{z^{41}}+p_1^6 p_2^{12} p_3^{16} p_4^8 \textcolor{red}{z^{42}},
 \)
 \begin{equation}\label{4.H2}\end{equation}

 \item[$H_3=$]\(
 1+30 p_3 \textcolor{red}{z}+(315 p_2 p_3+120 p_3 p_4) \textcolor{red}{z^2}+(770 p_1 p_2 p_3+1050 p_2 p_3^2+2240 p_2 p_3 p_4) \textcolor{red}{z^3} \\
 +(5775 p_1 p_2 p_3^2+6930 p_1 p_2 p_3 p_4+14700 p_2 p_3^2 p_4) \textcolor{red}{z^4}+(9702 p_1 p_2^2 p_3^2+90552 p_1 p_2 p_3^2 p_4+31500 p_2 p_3^3 p_4+10752 p_2 p_3^2 p_4^2) \textcolor{red}{z^5} \\
 +(8085 p_1 p_2^2 p_3^3+161700 p_1 p_2^2 p_3^2 p_4+249480 p_1 p_2 p_3^3 p_4+36750 p_2^2 p_3^3 p_4+92400 p_1 p_2 p_3^2 p_4^2+45360 p_2 p_3^3 p_4^2) \textcolor{red}{z^6} \\
 +(1181400 p_1 p_2^2 p_3^3 p_4+316800 p_1 p_2^2 p_3^2 p_4^2+443520 p_1 p_2 p_3^3 p_4^2+94080 p_2^2 p_3^3 p_4^2) \textcolor{red}{z^7} \\
 +(177870 p_1^2 p_2^2 p_3^3 p_4+1358280 p_1 p_2^3 p_3^3 p_4+782100 p_1 p_2^2 p_3^4 p_4+3490575 p_1 p_2^2 p_3^3 p_4^2+44100 p_2^2 p_3^4 p_4^2) \textcolor{red}{z^8} \\
 +(830060 p_1^2 p_2^3 p_3^3 p_4+2633400 p_1 p_2^3 p_3^4 p_4+711480 p_1^2 p_2^2 p_3^3 p_4^2+4928000 p_1 p_2^3 p_3^3 p_4^2+5035250 p_1 p_2^2 p_3^4 p_4^2+168960 p_1 p_2^2 p_3^3 p_4^3) \textcolor{red}{z^9} \\
 +(2144604 p_1^2 p_2^3 p_3^4 p_4+1559250 p_1 p_2^3 p_3^5 p_4+3811500 p_1^2 p_2^3 p_3^3 p_4^2+853776 p_1^2 p_2^2 p_3^4 p_4^2+16967181 p_1 p_2^3 p_3^4 p_4^2 \\
 +3234000 p_1 p_2^2 p_3^5 p_4^2+1474704 p_1 p_2^2 p_3^4 p_4^3) \textcolor{red}{z^{10}} \\
 +(2439360 p_1^2 p_2^3 p_3^5 p_4+18117750 p_1^2 p_2^3 p_3^4 p_4^2+26826030 p_1 p_2^3 p_3^5 p_4^2 +5174400 p_1 p_2^3 p_3^4 p_4^3+2069760 p_1 p_2^2 p_3^5 p_4^3) \textcolor{red}{z^{11}} \\
 +(711480 p_1^2 p_2^4 p_3^5 p_4+2371600 p_1^2 p_2^4 p_3^4 p_4^2+38368225 p_1^2 p_2^3 p_3^5 p_4^2+6338640 p_1 p_2^4 p_3^5 p_4^2 \\
 +14437500 p_1 p_2^3 p_3^6 p_4^2+5336100 p_1^2 p_2^3 p_3^4 p_4^3+18929680 p_1 p_2^3 p_3^5 p_4^3) \textcolor{red}{z^{12}} \\
 +(21783930 p_1^2 p_2^4 p_3^5 p_4^2+32524800 p_1^2 p_2^3 p_3^6 p_4^2+8731800 p_1 p_2^4 p_3^6 p_4^2+29988000 p_1^2 p_2^3 p_3^5 p_4^3+8279040 p_1 p_2^4 p_3^5 p_4^3 \\
 +16678200 p_1 p_2^3 p_3^6 p_4^3+1774080 p_1 p_2^3 p_3^5 p_4^4) \textcolor{red}{z^{13}} \\
 +(1584660 p_1^3 p_2^4 p_3^5 p_4^2+46973475 p_1^2 p_2^4 p_3^6 p_4^2+25194480 p_1^2 p_2^4 p_3^5 p_4^3+43705200 p_1^2 p_2^3 p_3^6 p_4^3 \\
 +17948700 p_1 p_2^4 p_3^6 p_4^3+5488560 p_1^2 p_2^3 p_3^5 p_4^4+4527600 p_1 p_2^3 p_3^6 p_4^4) \textcolor{red}{z^{14}} \\
 +(5588352 p_1^3 p_2^4 p_3^6 p_4^2+15937152 p_1^2 p_2^5 p_3^6 p_4^2+5808000 p_1^2 p_2^4 p_3^7 p_4^2+3234000 p_1^3 p_2^4 p_3^5 p_4^3+93982512 p_1^2 p_2^4 p_3^6 p_4^3 \\+3234000 p_1 p_2^4 p_3^7 p_4^3+5808000 p_1^2 p_2^4 p_3^5 p_4^4+15937152 p_1^2 p_2^3 p_3^6 p_4^4+5588352 p_1 p_2^4 p_3^6 p_4^4) \textcolor{red}{z^{15}} \\
 +(4527600 p_1^3 p_2^5 p_3^6 p_4^2+5488560 p_1^2 p_2^5 p_3^7 p_4^2+17948700 p_1^3 p_2^4 p_3^6 p_4^3+43705200 p_1^2 p_2^5 p_3^6 p_4^3 \\+25194480 p_1^2 p_2^4 p_3^7 p_4^3+46973475 p_1^2 p_2^4 p_3^6 p_4^4+1584660 p_1 p_2^4 p_3^7 p_4^4) \textcolor{red}{z^{16}} \\
 +(1774080 p_1^3 p_2^5 p_3^7 p_4^2+16678200 p_1^3 p_2^5 p_3^6 p_4^3+8279040 p_1^3 p_2^4 p_3^7 p_4^3+29988000 p_1^2 p_2^5 p_3^7 p_4^3 \\+8731800 p_1^3 p_2^4 p_3^6 p_4^4+32524800 p_1^2 p_2^5 p_3^6 p_4^4+21783930 p_1^2 p_2^4 p_3^7 p_4^4) \textcolor{red}{z^{17}} \\
 +(18929680 p_1^3 p_2^5 p_3^7 p_4^3+5336100 p_1^2 p_2^5 p_3^8 p_4^3+14437500 p_1^3 p_2^5 p_3^6 p_4^4+6338640 p_1^3 p_2^4 p_3^7 p_4^4 \\+38368225 p_1^2 p_2^5 p_3^7 p_4^4+2371600 p_1^2 p_2^4 p_3^8 p_4^4+711480 p_1^2 p_2^4 p_3^7 p_4^5) \textcolor{red}{z^{18}} \\
 +(2069760 p_1^3 p_2^6 p_3^7 p_4^3+5174400 p_1^3 p_2^5 p_3^8 p_4^3+26826030 p_1^3 p_2^5 p_3^7 p_4^4+18117750 p_1^2 p_2^5 p_3^8 p_4^4+2439360 p_1^2 p_2^5 p_3^7 p_4^5) \textcolor{red}{z^{19}} \\
 +(1474704 p_1^3 p_2^6 p_3^8 p_4^3+3234000 p_1^3 p_2^6 p_3^7 p_4^4+16967181 p_1^3 p_2^5 p_3^8 p_4^4+853776 p_1^2 p_2^6 p_3^8 p_4^4 \\+3811500 p_1^2 p_2^5 p_3^9 p_4^4+1559250 p_1^3 p_2^5 p_3^7 p_4^5+2144604 p_1^2 p_2^5 p_3^8 p_4^5) \textcolor{red}{z^{20}} \\
 +(168960 p_1^3 p_2^6 p_3^9 p_4^3+5035250 p_1^3 p_2^6 p_3^8 p_4^4+4928000 p_1^3 p_2^5 p_3^9 p_4^4+711480 p_1^2 p_2^6 p_3^9 p_4^4+2633400 p_1^3 p_2^5 p_3^8 p_4^5+830060 p_1^2 p_2^5 p_3^9 p_4^5) \textcolor{red}{z^{21}} \\
 +(44100 p_1^4 p_2^6 p_3^8 p_4^4+3490575 p_1^3 p_2^6 p_3^9 p_4^4+782100 p_1^3 p_2^6 p_3^8 p_4^5+1358280 p_1^3 p_2^5 p_3^9 p_4^5+177870 p_1^2 p_2^6 p_3^9 p_4^5) \textcolor{red}{z^{22}} \\
 +(94080 p_1^4 p_2^6 p_3^9 p_4^4+443520 p_1^3 p_2^7 p_3^9 p_4^4+316800 p_1^3 p_2^6 p_3^{10} p_4^4+1181400 p_1^3 p_2^6 p_3^9 p_4^5) \textcolor{red}{z^{23}} \\
 +(45360 p_1^4 p_2^7 p_3^9 p_4^4+92400 p_1^3 p_2^7 p_3^{10} p_4^4+36750 p_1^4 p_2^6 p_3^9 p_4^5+249480 p_1^3 p_2^7 p_3^9 p_4^5+161700 p_1^3 p_2^6 p_3^{10} p_4^5+8085 p_1^3 p_2^6 p_3^9 p_4^6) \textcolor{red}{z^{24}} \\
 +(10752 p_1^4 p_2^7 p_3^{10} p_4^4+31500 p_1^4 p_2^7 p_3^9 p_4^5+90552 p_1^3 p_2^7 p_3^{10} p_4^5+9702 p_1^3 p_2^6 p_3^{10} p_4^6) \textcolor{red}{z^{25}} \\
 +(14700 p_1^4 p_2^7 p_3^{10} p_4^5+6930 p_1^3 p_2^7 p_3^{11} p_4^5+5775 p_1^3 p_2^7 p_3^{10} p_4^6) \textcolor{red}{z^{26}}
 +(2240 p_1^4 p_2^7 p_3^{11} p_4^5+1050 p_1^4 p_2^7 p_3^{10} p_4^6+770 p_1^3 p_2^7 p_3^{11} p_4^6) \textcolor{red}{z^{27}} \\ 
 +(120 p_1^4 p_2^8 p_3^{11} p_4^5+315 p_1^4 p_2^7 p_3^{11} p_4^6) \textcolor{red}{z^{28}} 
 +30 p_1^4 p_2^8 p_3^{11} p_4^6 \textcolor{red}{z^{29}} +p_1^4 p_2^8 p_3^{12} p_4^6 \textcolor{red}{z^{30}},
 \)
 \begin{equation}\label{H3}\end{equation}

 \item[$H_4=$]\(
 1+16 p_4 \textcolor{red}{z}+120 p_3 p_4 \textcolor{red}{z^2}+560 p_2 p_3 p_4 \textcolor{red}{z^3}+(770 p_1 p_2 p_3 p_4+1050 p_2 p_3^2 p_4) \textcolor{red}{z^4}+(3696 p_1 p_2 p_3^2 p_4+672 p_2 p_3^2 p_4^2) \textcolor{red}{z^5} \\
 +(4312 p_1 p_2^2 p_3^2 p_4+3696 p_1 p_2 p_3^2 p_4^2) \textcolor{red}{z^6}+(2640 p_1 p_2^2 p_3^3 p_4+8800 p_1 p_2^2 p_3^2 p_4^2) \textcolor{red}{z^7}+12870 p_1 p_2^2 p_3^3 p_4^2 \textcolor{red}{z^8} \\
 +(8800 p_1 p_2^2 p_3^4 p_4^2+2640 p_1 p_2^2 p_3^3 p_4^3) \textcolor{red}{z^9}+(3696 p_1 p_2^3 p_3^4 p_4^2+4312 p_1 p_2^2 p_3^4 p_4^3) \textcolor{red}{z^{10}}+(672 p_1^2 p_2^3 p_3^4 p_4^2+3696 p_1 p_2^3 p_3^4 p_4^3) \textcolor{red}{z^{11}} \\
 +(1050 p_1^2 p_2^3 p_3^4 p_4^3+770 p_1 p_2^3 p_3^5 p_4^3) \textcolor{red}{z^{12}}+560 p_1^2 p_2^3 p_3^5 p_4^3 \textcolor{red}{z^{13}}+120 p_1^2 p_2^4 p_3^5 p_4^3 \textcolor{red}{z^{14}}+16 p_1^2 p_2^4 p_3^6 p_4^3 \textcolor{red}{z^{15}}+p_1^2 p_2^4 p_3^6 p_4^4 \textcolor{red}{z^{16}}.
 \)
 \begin{equation}\label{4.H4}\end{equation}              
 \end{itemize}              
      Now we denote
       \begin{equation}
         \label{4.3e.5}
          H_s \equiv H_s(z) = H_s(z, (p_i) ), \quad (p_i) \equiv (p_1,p_2,p_3,p_4).
        \end{equation}
              
       The asymptotic formulae for the polynomials read:
       \begin{equation}
         \label{4.3e.6}
          H_s = H_s(z, (p_i) )  \sim \left( \prod_{l=1}^{4} (p_l)^{\nu^{sl}} \right) z^{n_s} \equiv 
          H_s^{as}(z, (p_i)), \quad \text{ as } z \to \infty,
        \end{equation}
      where the  matricies $\nu = (\nu^{sl})$ have the following form 
 \begin{align}
          \label{4.3e.7}
         \nu &=  
          \begin{pmatrix}
           1 & 1 & 1 & 1 \\
           1 & 2 & 2 & 1 \\
           1 & 2 & 2 & 1 \\
           1 & 1 & 1 & 1 \\
          \end{pmatrix},
          \quad  
          \begin{pmatrix}
           2 & 2 & 2 & 2 \\
           2 & 4 & 4 & 4 \\
           2 & 4 & 6 & 6 \\
           1 & 2 & 3 & 4 \\
          \end{pmatrix},
          \quad 
          \begin{pmatrix}
           2 & 2 & 2 & 1 \\
           2 & 4 & 4 & 2 \\
           2 & 4 & 6 & 3 \\
           2 & 4 & 6 & 4 \\ 
          \end{pmatrix},
          \quad \nonumber \\
         &\hspace{5em} \begin{pmatrix}
           2 & 2 & 1 & 1 \\
           2 & 4 & 2 & 2 \\
           1 & 2 & 2 & 1 \\
           1 & 2 & 1 & 2 \\ 
          \end{pmatrix},
          \quad 
          \begin{pmatrix}
                    4 & 6 & 8 & 4 \\
                    6 & 12 & 16 & 8 \\
                    4 & 8 & 12 & 6 \\
                    2 & 4 & 6 & 4 \\ 
                   \end{pmatrix}
 \end{align}
       for Lie algebras $A_4,  B_4,  C_4, D_4, F_4$, respectively.
       In all these cases we are led to the relations 
                    \begin{equation}
                      \label{4e.10a}
                       \sum_{l= 1}^4 \nu^{sl} = n_s, \quad s = 1, 2, 3, 4.
                     \end{equation}
                                        
      In the case of Lie algebras  $B_4$, $C_4$, $D_4$ and $F_4$ we have 
       \begin{equation}
         \label{4.3e.8a}
       \nu({\cal G}) = 2 A^{-1}, \quad {\cal G}=B_4, C_4, D_4, F_4,
       \end{equation}
       where $A^{-1}$ is inverse Cartan matrix, whereas in the $A_4$-case   
       the matrix $\nu$ reads as follows
       \begin{equation}
         \label{4.3e.8b}
          \nu({\cal G})  = A^{-1} (I + P), \quad {\cal G}=A_4.
        \end{equation}
        Here $I$ is $4 \times 4$ identity matrix and
       \begin{equation}
          \label{4.3e.9}
          P  = 
          \begin{pmatrix}
           0 & 0 & 0 & 1 \\
           0 & 0 & 1 & 0 \\
           0 & 1 & 0 & 0 \\
           1 & 0 & 0 & 0 \\
           \end{pmatrix}
          \end{equation}
       is a  matrix which corresponds to the permutation $\sigma \in S_4$ ($S_4$ is symmetric group)
       \begin{equation}
         \label{4.3.10}
          \sigma: (1,2,3,4) \mapsto (4,3,2,1),
        \end{equation}
         due to  relation  $P = (P^i_j) = (\delta^i_{\sigma(j)})$. 
        Here $\sigma$   is the generator of the group of symmetry of the Dynkin diagram for $A_4$:
        $G = \{ \sigma, {\rm id} \}$. This group is  isomorphic to the group $\mathbb{Z}_2$.      
         
 
  For the Lie algebra $D_4$ we are led to the  group of symmetry of the Dynkin diagram 
  $G'$ which is isomorphic to the symmetric group $S_3$. This group is acting on the 
  set of three vertices of the diagram $\{1, 3, 4 \}$  by their  permutations. 
  The  groups  symmetry $G \cong \mathbb{Z}_2$ 
  and $G' \cong S_3$  imply certain identity properties for the polynomials $H_s(z)$.
   
  Now we introduce the dual (ordered) set: $(\hat{p}_i) = (p_{\sigma(i)})$ for the algebra $A_4$ , and
  $(\hat{p}_i) = (p_{i})$ for algebras $B_4$, $C_4$, $D_4$, $F_4$  ($i= 1,2,3,4$).         
  The dual set for $A_4$ case is a result of action  of the generator $\sigma$ 
   of the group $G \cong \mathbb{Z}_2$   on vertices of the Dynkin diagram. 
       
    Afterwards we obtain symmetry and duality  identities. 
    They were verified by using certain MATHEMATICA algorithm. 
    \\[1em] 
        {\bf Symmetry relations.}
    \\
        {\bf Proposition 4.1.} {\em  
          The fluxbrane polynomials satisfy for all $p_i$ and $z$
                  the following identities:
         \bearr{4.3.11}
         H_{\sigma(s)}(z, (p_i) )\, = H_s(z, (\hat{p}_i)) \qquad \text{ for $A_4$ case},\nnn
         H_{\sigma'(s)}(z, (p_i) ) = H_s(z, (p_{\sigma'(i)})\qquad\! \text{ for $D_4$ case},
          \ear
          for any $\sigma' \in S_3$, $s= 1, \dots, 4$.  }
    \\[1em]
        {\bf Duality relations.}
    \\
        {\bf Proposition 4.2.} {\em  The fluxbrane polynomials
         corresponding to Lie algebras $A_4$,
         $B_4$, $C_4$, $D_4$ and $F_4$ obey for all $p_i > 0$ and $z > 0$
         the identities:
         \begin{equation}
           \label{4.3.12}
            H_{s}(z, (p_i) ) = H_s^{as}(z, (p_i)) H_s(z^{-1}, (\hat{p}_i^{-1})),
          \end{equation}
          $s = 1, 2, 3, 4$. }

  \subsection{Rank-$5$ algebras  }

  Now turn our attention to the solutions corresponding to  Lie algebras  of rank $5$ 
  when the matrix  $A = (A_{sl})$ is coinciding with one of the Cartan matrices 
  \begin{gather} 
       \left(A_{ss'}\right) =
       \begin{pmatrix}
       2 & -1 & 0 & 0 & 0 \\
        -1 & 2 & -1 & 0 & 0 \\
        0 & -1 & 2 & -1 & 0 \\
        0 & 0 & -1 & 2 & -1 \\
        0 & 0 & 0 & -1 & 2 \\
       \end{pmatrix},\!
    \quad
     \begin{pmatrix}
      2 & -1 & 0 & 0 & 0 \\
       -1 & 2 & -1 & 0 & 0 \\
       0 & -1 & 2 & -1 & 0 \\
       0 & 0 & -1 & 2 & -2 \\
       0 & 0 & 0 & -1 & 2 \\
    \end{pmatrix}, 
    \quad \nonumber \\
    \begin{pmatrix}
        2 & -1 & 0 & 0 & 0 \\
        -1 & 2 & -1 & 0 & 0 \\
        0 & -1 & 2 & -1 & 0 \\
        0 & 0 & -1 & 2 & -1 \\
        0 & 0 & 0 & -2 & 2 \\
      \end{pmatrix}, 
     \quad 
    \begin{pmatrix}
          2 & -1 & 0 & 0 & 0 \\
           -1 & 2 & -1 & 0 & 0 \\
           0 & -1 & 2 & -1 & -1 \\
           0 & 0 & -1 & 2 & 0 \\
           0 & 0 & -1 & 0 & 2 \\
      \end{pmatrix}.
    \label{5.A.5}
  \end{gather}
  for ${\cal G} = A_5, \, B_5, \, C_5, \, D_5$, respectively.
  
  Figure 4  gives us graphical presentations of these matrices by Dynkin diagrams.
   \vspace{25pt}
    
    \setlength{\unitlength}{1mm}
   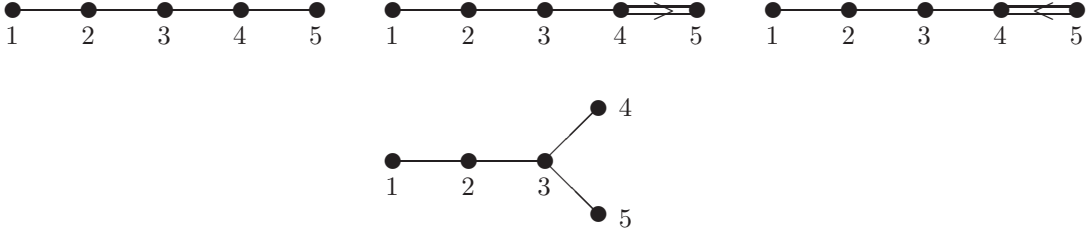
\begin{figure}[h]
   \centering
   \begin{picture}(140, 25)
   \put(2,25){\circle*{2}}
   \put(12,25){\circle*{2}}
   \put(22,25){\circle*{2}}
   \put(32,25){\circle*{2}}
   \put(42,25){\circle*{2}}
   \put(1,25){\line(1,0){40}}
    \put(0.3,20.5){$1$}
   \put(10.3,20.5){$2$}
   \put(20.3,20.5){$3$}
   \put(30.3,20.5){$4$}
   \put(40.3,20.5){$5$}
   \put(52,25){\circle*{2}}
   \put(62,25){\circle*{2}}
   \put(72,25){\circle*{2}}
   \put(82,25){\circle*{2}}
   \put(92,25){\circle*{2}}
   \put(52,25){\line(1,0){30}}
   \put(82,25.5){\line(1,0){10}}
   \put(82,24.5){\line(1,0){10}}
   \put(50.3,20.5){$1$}
   \put(60.3,20.5){$2$}
   \put(70.3,20.5){$3$}
   \put(80.3,20.5){$4$}
   \put(90.3,20.5){$5$}
   \put(85.3,23.9){\large $>$}
   \put(102,25){\circle*{2}}
   \put(112,25){\circle*{2}}
   \put(122,25){\circle*{2}}
   \put(132,25){\circle*{2}}
   \put(142,25){\circle*{2}}
   \put(102,25){\line(1,0){30}}
   \put(132,25.5){\line(1,0){10}}
   \put(132,24.5){\line(1,0){10}}
   \put(100.3,20.5){$1$}
   \put(110.3,20.5){$2$}
   \put(120.3,20.5){$3$}
   \put(130.3,20.5){$4$}
   \put(140.3,20.5){$5$}
   \put(135.3,23.9){\large $<$}
   \put(52,05){\circle*{2}}
   \put(62,05){\circle*{2}}
   \put(72,05){\circle*{2}}
   \put(79,12){\circle*{2}}
   \put(79,-2){\circle*{2}}
   \put(51,05){\line(1,0){20}}
   \put(71.3,04.5){\line(1,1){7.5}}
   \put(72,05){\line(1,-1){7.5}}
   \put(50.3,0.5){$1$}
   \put(60.3,0.5){$2$}
   \put(70.3,0.5){$3$}
   \put(81,11){$4$}
   \put(81,-3.7){$5$}
  \end{picture}
    \vspace{10pt}
    \caption{Dynkin diagrams for the Lie algebras $A_5, \, B_5, \, C_5, \, D_5$, respectively.}
   \end{figure}

       {\bf Polynomials.}
      According  to the conjecture from Ref. \cite{Iflux}, 
      the  functions  $H_1(z), ..., H_5(z)$
       obeying Eqs. (\ref{1.1}) and (\ref{1.2})
           with any matrix $A =  (A_{sl})$ from (\ref{5.A.5}), 
           should be polynomials. Relation (\ref{1.4}) gives us the powers of these polynomials: 
           $(n_1, n_2, n_3, n_4, n_5) = (5,8,9,8,5), (10,18,24,28,15), 
           (9,16,21,24,25), (8,14,18,10,10)$  
           for Lie algebras $A_5$, $B_5$, $C_5$, $D_5$, respectively.
           
           In this case the verification (or proof) of the polynomial conjecture \cite{Iflux} is following: 
           we  solve the set of algebraic equations for the coefficients of the polynomials (\ref{1.3}) 
           which follow from the master equations (\ref{1.1}). 
           
          In this subsection we present the structures (or ``truncated versions'')  of these polynomials. 
          The total list of the polynomials   is presented in Appendix A.
          The polynomials were obtained by using a certain MATHEMATICA algorithm in Ref. \cite{BolIvas_R5-22}.

  \noindent{\bf $A_5$-case.} 
  For the Lie algebra $A_5 $ the polynomials have the following structure
    \begin{itemize} 
  \item[$H_1=$]\(
  1+5 p_1 \textcolor{red}{z}+10 p_1 p_2 \textcolor{red}{{z}^2}+10 p_1 p_2 p_3 \textcolor{red}{{z}^3}+5 p_1 p_2 p_3 p_4 \textcolor{red}{{z}^4}+p_1 p_2 p_3 p_4 p_5 \textcolor{red}{{z}^5},
  \)
  \item[$H_2=$]\(
  1+8 p_2 \textcolor{red}{z}+(10 p_1 p_2 + 18 p_2 p_3) \textcolor{red}{{z}^2}+
  \dots
  +(10 p_1 p_2^2 p_3^2 p_4+18 p_1 p_2^2 p_3 p_4 p_5) \textcolor{red}{{z}^6}+8 p_1 p_2^2 p_3^2 p_4 p_5 \textcolor{red}{{z}^7}+p_1 p_2^2 p_3^2 p_4^2 p_5 \textcolor{red}{{z}^8}\),
  \item[$H_3=$]\(
  1+9 p_3 \textcolor{red}{z}+(18 p_2 p_3+18 p_3 p_4) \textcolor{red}{{z}^2}+
  \dots
  +(18 p_1 p_2^2 p_3^2 p_4 p_5+18 p_1 p_2 p_3^2 p_4^2 p_5) \textcolor{red}{{z}^7}
  +9 p_1 p_2^2 p_3^2 p_4^2 p_5 \textcolor{red}{{z}^8}+p_1 p_2^2 p_3^3 p_4^2 p_5 \textcolor{red}{{z}^9}\),
  \item[$H_4=$]\(
  1+8 p_4 \textcolor{red}{z}+(18 p_3 p_4+10 p_4 p_5) \textcolor{red}{{z}^2}+ \dots 
  +(18 p_1 p_2 p_3 p_4^2 p_5+10 p_2 p_3^2 p_4^2 p_5) \textcolor{red}{{z}^6}+8 p_1 p_2 p_3^2 p_4^2 p_5 \textcolor{red}{{z}^7}+p_1 p_2^2 p_3^2 p_4^2 p_5 \textcolor{red}{{z}^8}\),
  \item[$H_5=$]\(
  1+5 p_5 \textcolor{red}{z}+10 p_4 p_5 \textcolor{red}{{z}^2}+10 p_3 p_4 p_5 \textcolor{red}{{z}^3}+5 p_2 p_3 p_4 p_5 \textcolor{red}{{z}^4}+p_1 p_2 p_3 p_4 p_5 \textcolor{red}{{z}^5}.
  \)
  \end{itemize}

  \bigskip
  \noindent {\bf $B_5$-case.}
  In the case of Lie algebra $B_5 $ the polynomial structure is following
  \begin{itemize}
  \item[$H_1=$]\(
  1+10 p_1 \textcolor{red}{z}+45 p_1 p_2 \textcolor{red}{{z}^2}+ \dots
  +45 p_1 p_2 p_3^2 p_4^2 p_5^2 \textcolor{red}{{z}^8}+10 p_1 p_2^2 p_3^2 p_4^2 p_5^2 \textcolor{red}{{z}^9}+p_1^2 p_2^2 p_3^2 p_4^2 p_5^2 \textcolor{red}{{z}^{10}},
  \)
  \item[$H_2=$]\(
  1+18 p_2 \textcolor{red}{z}+(45 p_1 p_2+108 p_2 p_3) \textcolor{red}{{z}^2}+ \dots
  +(108 p_1^2 p_2^3 p_3^3 p_4^4 p_5^4+45 p_1 p_2^3 p_3^4 p_4^4 p_5^4) \textcolor{red}{{z}^{16}}
  +18 p_1^2 p_2^3 p_3^4 p_4^4 p_5^4 \textcolor{red}{{z}^{17}}+p_1^2 p_2^4 p_3^4 p_4^4 p_5^4 \textcolor{red}{{z}^{18}},
  \)
  \item[$H_3=$]\(
  1+24 p_3 \textcolor{red}{z}+(108 p_2 p_3+168 p_3 p_4) \textcolor{red}{{z}^2}+
  \dots
  +(168 p_1^2 p_2^4 p_3^5 p_4^5 p_5^6+108 p_1^2 p_2^3 p_3^5 p_4^6 p_5^6) \textcolor{red}{{z}^{22}}+24 p_1^2 p_2^4 p_3^5 p_4^6 p_5^6 \textcolor{red}{{z}^{23}}+p_1^2 p_2^4 p_3^6 p_4^6 p_5^6 \textcolor{red}{{z}^{24}},
  \)
  \item[$H_4=$]\(
  1+28 p_4 \textcolor{red}{z}+(168 p_3 p_4+210 p_4 p_5) \textcolor{red}{{z}^2}+ \dots
  +(210 p_1^2 p_2^4 p_3^6 p_4^7 p_5^7+168 p_1^2 p_2^4 p_3^5 p_4^7 p_5^8) \textcolor{red}{{z}^{26}}+28 p_1^2 p_2^4 p_3^6 p_4^7 p_5^8 \textcolor{red}{{z}^{27}}+p_1^2 p_2^4 p_3^6 p_4^8 p_5^8 \textcolor{red}{{z}^{28}},
  \)
  \item[$H_5=$]\(
  1+15 p_5 \textcolor{red}{z}+105 p_4 p_5 \textcolor{red}{{z}^2}+ \dots
  +105 p_1 p_2^2 p_3^3 p_4^3 p_5^4 \textcolor{red}{{z}^{13}}+15 p_1 p_2^2 p_3^3 p_4^4 p_5^4 \textcolor{red}{{z}^{14}}+p_1 p_2^2 p_3^3 p_4^4 p_5^5 \textcolor{red}{{z}^{15}}.
  \)
  \end{itemize}

  \bigskip 
  \noindent { \bf $C_5$-case. }
   The polynomials corresponding to Lie algebra $C_5 $ have the following structure 
  \begin{itemize}
  \item[$H_1=$]\(
  1+9 p_1 \textcolor{red}{z}+36 p_1 p_2 \textcolor{red}{{z}^2}+ 
  \dots +36 p_1 p_2 p_3^2 p_4^2 p_5 \textcolor{red}{{z}^7}+9 p_1 p_2^2 p_3^2 p_4^2 p_5 \textcolor{red}{{z}^8}+p_1^2 p_2^2 p_3^2 p_4^2 p_5 \textcolor{red}{{z}^9},
  \)
  \item[$H_2=$]\(
  1+16 p_2 \textcolor{red}{z}+(36 p_1 p_2+84 p_2 p_3) \textcolor{red}{{z}^2}+ \dots
  +(84 p_1^2 p_2^3 p_3^3 p_4^4 p_5^2+36 p_1 p_2^3 p_3^4 p_4^4 p_5^2) \textcolor{red}{{z}^{14}}+16 p_1^2 p_2^3 p_3^4 p_4^4 p_5^2 \textcolor{red}{{z}^{15}}+p_1^2 p_2^4 p_3^4 p_4^4 p_5^2 \textcolor{red}{{z}^{16}},
  \)
  \item[$H_3=$]\(
  1+21 p_3 \textcolor{red}{z}+(84 p_2 p_3+126 p_3 p_4) \textcolor{red}{{z}^2}+ \dots +(126 p_1^2 p_2^4 p_3^5 p_4^5 p_5^3+84 p_1^2 p_2^3 p_3^5 p_4^6 p_5^3) \textcolor{red}{{z}^{19}}+21 p_1^2 p_2^4 p_3^5 p_4^6 p_5^3 \textcolor{red}{{z}^{20}}+p_1^2 p_2^4 p_3^6 p_4^6 p_5^3 \textcolor{red}{{z}^{21}},
  \)
  \item[$H_4=$]\(
  1+24 p_4 \textcolor{red}{z}+(126 p_3 p_4+150 p_4 p_5) \textcolor{red}{{z}^2}+ \dots +(150 p_1^2 p_2^4 p_3^6 p_4^7 p_5^3+126 p_1^2 p_2^4 p_3^5 p_4^7 p_5^4) \textcolor{red}{{z}^{22}}+24 p_1^2 p_2^4 p_3^6 p_4^7 p_5^4 \textcolor{red}{{z}^{23}}+p_1^2 p_2^4 p_3^6 p_4^8 p_5^4 \textcolor{red}{{z}^{24}},
  \)
  \item[$H_5=$]\(
  1+25 p_5 \textcolor{red}{z}+300 p_4 p_5 \textcolor{red}{{z}^2}+ \dots +300 p_1^2 p_2^4 p_3^6 p_4^7 p_5^4 \textcolor{red}{{z}^{23}}+25 p_1^2 p_2^4 p_3^6 p_4^8 p_5^4 \textcolor{red}{{z}^{24}}+p_1^2 p_2^4 p_3^6 p_4^8 p_5^5 \textcolor{red}{{z}^{25}}.
  \)
  \end{itemize}

  \bigskip
  \noindent { \bf $D_5$-case. }
   In the case of  Lie algebra $D_5 $ we are led to the following structure of polynomials  
  \begin{itemize}
  \item[$H_1=$]\(
  1+8 p_1 \textcolor{red}{z}+28 p_1 p_2 \textcolor{red}{{z}^2}+ \dots +28 p_1 p_2 p_3^2 p_4 p_5 \textcolor{red}{{z}^6}+8 p_1 p_2^2 p_3^2 p_4 p_5 \textcolor{red}{{z}^7}+p_1^2 p_2^2 p_3^2 p_4 p_5 \textcolor{red}{{z}^8},
  \)
  \item[$H_2=$]\(
  1+14 p_2 \textcolor{red}{z}+(28 p_1 p_2+63 p_2 p_3) \textcolor{red}{{z}^2}+ \dots +(63 p_1^2 p_2^3 p_3^3 p_4^2 p_5^2+28 p_1 p_2^3 p_3^4 p_4^2 p_5^2) \textcolor{red}{{z}^{12}}+14 p_1^2 p_2^3 p_3^4 p_4^2 p_5^2 \textcolor{red}{{z}^{13}}+p_1^2 p_2^4 p_3^4 p_4^2 p_5^2 \textcolor{red}{{z}^{14}},
  \) %
  \item[$H_3=$]\(
  1+18 p_3 \textcolor{red}{z}+(63 p_2 p_3+45 p_3 p_4+45 p_3 p_5) \textcolor{red}{{z}^2}+ \dots +(45 p_1^2 p_2^4 p_3^5 p_4^3 p_5^2+45 p_1^2 p_2^4 p_3^5 p_4^2 p_5^3+63 p_1^2 p_2^3 p_3^5 p_4^3 p_5^3) \textcolor{red}{{z}^{16}}+18 p_1^2 p_2^4 p_3^5 p_4^3 p_5^3 \textcolor{red}{{z}^{17}}+p_1^2 p_2^4 p_3^6 p_4^3 p_5^3 \textcolor{red}{{z}^{18}},
  \)
  \item[$H_4=$]\(
  1+10 p_4 \textcolor{red}{z}+45 p_3 p_4 \textcolor{red}{{z}^2}+ \dots +45 p_1 p_2^2 p_3^2 p_4^2 p_5 \textcolor{red}{{z}^8}+10 p_1 p_2^2 p_3^3 p_4^2 p_5 \textcolor{red}{{z}^9}+p_1 p_2^2 p_3^3 p_4^2 p_5^2 \textcolor{red}{{z}^{10}},
  \)
  \item[$H_5=$]\(
  1+10 p_5 \textcolor{red}{z}+45 p_3 p_5 \textcolor{red}{{z}^2}+ \dots +45 p_1 p_2^2 p_3^2 p_4 p_5^2 \textcolor{red}{{z}^8}+10 p_1 p_2^2 p_3^3 p_4 p_5^2 \textcolor{red}{{z}^9}+p_1 p_2^2 p_3^3 p_4^2 p_5^2 \textcolor{red}{{z}^{10}}.
  \)
  \end{itemize}
       
  By using  notations
        \begin{equation}
          \label{5.3e.5}
           H_s \equiv H_s(z) = H_s(z, (p_i) ), \quad (p_i) \equiv (p_1,p_2,p_3,p_4,p_5)
         \end{equation}
  we write  asymptotic relations for polynomials
  
  \begin{equation}
          \label{5.3e.6}
           H_s = H_s(z, (p_i) )  \sim \left( \prod_{l=1}^{5} (p_l)^{\nu^{sl}} \right) z^{n_s} \equiv 
           H_s^{as}(z, (p_i)), \quad \text{ as } z \to \infty.
         \end{equation}
       Here  we denote by $\nu = (\nu^{sl})$ the integer valued matrix. It has the form 
  \begin{align}
           \label{5.3e.7}
          \nu &=  
           \begin{pmatrix}
            1 & 1 & 1 & 1 & 1 \\
            1 & 2 & 2 & 2 & 1 \\
            1 & 2 & 3 & 2 & 1 \\
            1 & 2 & 2 & 2 & 1 \\
            1 & 1 & 1 & 1 & 1 \\
           \end{pmatrix},
           \quad  
           \begin{pmatrix}
             2 & 2 & 2 & 2 & 2 \\
             2 & 4 & 4 & 4 & 4 \\
             2 & 4 & 6 & 6 & 6 \\
             2 & 4 & 6 & 8 & 8 \\
             1 & 2 & 3 & 4 & 5 \\
           \end{pmatrix},
           \quad 
           \begin{pmatrix}
            2 & 2 & 2 & 2 & 1 \\
            2 & 4 & 4 & 4 & 2 \\
            2 & 4 & 6 & 6 & 3 \\
            2 & 4 & 6 & 8 & 4 \\
            2 & 4 & 6 & 8 & 5 \\
           \end{pmatrix},
           \quad \nonumber \\
          &\hspace{5em}
           \begin{pmatrix}
             2 & 2 & 2 & 1 & 1 \\
             2 & 4 & 4 & 2 & 2 \\
             2 & 4 & 6 & 3 & 3 \\
             1 & 2 & 3 & 2 & 2 \\
             1 & 2 & 3 & 2 & 2 \\
           \end{pmatrix}
  \end{align}
        for Lie algebras $A_5,  B_5,  C_5, D_5$, respectively.
        
      One can readily verify that any matrix $\nu = (\nu^{sl})$ obeys 
       the following identity 
                     \begin{equation}
                       \label{3e.10a}
                        \sum_{l= 1}^5 \nu^{sl} = n_s, \quad s = 1, 2, 3, 4, 5.
                      \end{equation}
  \\
  For Lie algebras  $B_5$, $C_5$ 
   the  matrix $\nu$ is  twice inverse Cartan matrix $A^{-1}$:
           \begin{equation}
          \label{5.3e.8a}
        \nu({\cal G}) = 2 A^{-1}, \quad {\cal G}=B_5, C_5,
        \end{equation}
       and in the $A_5$ and $D_5$ cases   
        we are led to another relation:
         \begin{equation}
          \label{5.3e.8b}
           \nu({\cal G})  = A^{-1} (I + P({\cal G})), \quad {\cal G}=A_5, D_5.
         \end{equation}
         Here we use the  notations: $I$ - for $5 \times 5$ identity matrix and   $P({\cal G})$
          - for a matrix corresponding to a certain permutation $\sigma \in S_5$ ($S_5$ is symmetric group), 
           $P = (P^i_j) = (\delta^i_{\sigma(j)})$,  where $\sigma$ is the generator of the group 
         $G = \{ \sigma, {\rm id} \}$. The group $G$  is isomorphic to the group $\mathbb{Z}_2$. 
        For $A_5$ and $D_5$  the group of symmetry of the Dynkin diagram  acts on the set 
         of corresponding five vertices via their permutations. 
         
  The explicit forms for the permutation matrix $P$ and the generator $\sigma$ for both Lie algebras 
  $A_5, D_5$ read as follows:
        \begin{equation}
           \label{5.3e.9a}
           P({A_5})  = 
           \begin{pmatrix}
             0 & 0 & 0 & 0 & 1 \\
             0 & 0 & 0 & 1 & 0 \\
             0 & 0 & 1 & 0 & 0 \\
             0 & 1 & 0 & 0 & 0 \\
             1 & 0 & 0 & 0 & 0 \\
           \end{pmatrix},\quad \sigma: (1,2,3,4,5) \mapsto (5,4,3,2,1);
           \end{equation}
        \begin{equation}
           \label{5.3e.9b}
           P({D_5})  = 
           \begin{pmatrix}
              1 & 0 & 0 & 0 & 0 \\
              0 & 1 & 0 & 0 & 0 \\
              0 & 0 & 1 & 0 & 0 \\
              0 & 0 & 0 & 0 & 1 \\
              0 & 0 & 0 & 1 & 0 \\
           \end{pmatrix},\quad \sigma: (1,2,3,4,5) \mapsto (1,2,3,5,4).
           \end{equation} \\ \indent       
      
      The above symmetry groups control certain identity properties for polynomials $H_s(z)$.
    \\ \indent
           We also denote $\hat{p}_i = p_{\sigma(i)} $ for the $A_5$ and $D_5$ cases, and
         $\hat{p}_i = p_{i}$ for $B_5$ and $C_5$  cases ($i= 1,2,3,4,5$).

        By using MATHEMATICA algorithms we were able to verify the validity of the following identities.
     \\[1em] 
         {\bf Symmetry relations.}
     \\
         {\bf Proposition 5.1} {\em  
           The fluxbrane polynomials, corresponding to 
           Lie algebras  $A_5$ and $D_5$, obey for all $p_i$ and $z$
           the following identities:
          \bearr{5.3.11}
          H_{\sigma(s)}(z, (p_i) )\, = H_s(z, (\hat{p}_i)),
           \ear
           where $\sigma\in S_5$, $s= 1, \dots, 5$, is defined for each algebra by Eqs.
            \eqref{5.3e.9a}, \eqref{5.3e.9b}}.
     \\[1em]
         {\bf Duality relations.}
     \\
         {\bf Proposition 5.2} {\em  The fluxbrane polynomials which
          correspond to Lie algebras $A_5$,
          $B_5$, $C_5$, $D_5$, satisfy for all $p_i > 0$ and $z > 0$
          the following identities
          \begin{equation}
            \label{5.3.12}
             H_{s}(z, (p_i) ) = H_s^{as}(z, (p_i)) H_s(z^{-1}, (\hat{p}_i^{-1})),
           \end{equation}
           $s = 1, 2, 3, 4,5$. }

      \subsection{$E_6$ algebra  }      
  
    Now we deal with the exceptional Lie algebra $E_6$. 
    The matrix  $A $ is coinciding with the Cartan matrix 
    (for $E_6$) 	
    
     \begin{equation}
   \label{3.1}
   A=  (A_{ss'}) = \left(
   \begin{array}{cccccc}
    2 & -1 & 0 & 0 & 0 & 0 \\
    -1 & 2 & -1 & 0 & 0 & 0 \\
    0 & -1 & 2 & -1 & 0 & -1 \\
    0 & 0 & -1 & 2 & -1 & 0 \\
    0 & 0 & 0 & -1 & 2 & 0 \\
    0 & 0 & -1 & 0 & 0 & 2
   \end{array}
   \right).
   \end{equation}
   
   \newpage
   
   This matrix  is graphically depicted   by the Dynkin diagram in   Figure  5.
    
    \vspace{65pt}
    
    \setlength{\unitlength}{1mm}
   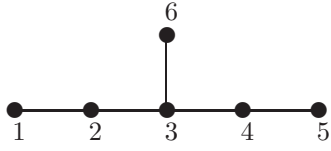
\begin{figure}[h]
   \centering
   \begin{picture}(40, 0)
   \put(1,5){\circle*{2}}
   \put(11,5){\circle*{2}}
   \put(21,5){\circle*{2}}
   \put(31,5){\circle*{2}}
   \put(41,5){\circle*{2}}
   \put(21,15){\circle*{2}}
   \put(0,1){$1$}
   \put(10,1){$2$}
   \put(20,1){$3$}
   \put(30,1){$4$}
   \put(40,1){$5$}
   \put(20,17){$6$}
   \put(1,5){\line(10,00){40}}
   \put(20.9,5){\line(0,10){10}}
   \end{picture}
   \caption{The Dynkin diagram for the Lie algebra $E_6$.}
   \end{figure}

    The inverse Cartan matrix for $E_6$
   \begin{equation}
   \label{3.2}
    A^{-1}= (A^{ss'}) = 
    \begin{pmatrix} \vspace{0.3em}
    \frac{4}{3} & \frac{5}{3} & 2 & \frac{4}{3} & \frac{2}{3} & 1 \\ \vspace{0.3em}
    \frac{5}{3} & \frac{10}{3} & 4 & \frac{8}{3} & \frac{4}{3} & 2 \\ \vspace{0.3em}
    2 & 4 & 6 & 4 & 2 & 3 \\ \vspace{0.3em}
    \frac{4}{3} & \frac{8}{3} & 4 & \frac{10}{3} & \frac{5}{3} & 2 \\ \vspace{0.3em}
    \frac{2}{3} & \frac{4}{3} & 2 & \frac{5}{3} & \frac{4}{3} & 1 \\ \vspace{0.3em}
    1 & 2 & 3 & 2 & 1 & 2
    \end{pmatrix}
   \end{equation}
    
   implies due to (\ref{1.4}) 
   \begin{equation}
    \label{3.3}
   (n_1,n_2,n_3,n_4,n_5,n_6) =  (16,30,42,30,16,22).
   \end{equation}
    
   For the Lie algebra $E_6$ we find the  set of six fluxbrane polynomials \cite{BolIvas-17}, which are
   listed in Appendix B. 
   
    The polynomials have the following structure:
  \begin{multline*}
  H_1 = 1 +
  16 p_1 z+ 120 p_1 p_2 z^2+ \dots +120 p_1^2 p_2^3 p_3^4 p_4^2 p_5 p_6^2 z^{14}\\
  +16 p_1^2 p_2^3 p_3^4 p_4^3 p_5 p_6^2 z^{15}+p_1^2 p_2^3 p_3^4 p_4^3 p_5^2 p_6^2 z^{16},
  \end{multline*}\vspace{-2.5em}
  \begin{multline*}
  H_2 = 1 +
  30 p_2 z+(120 p_1 p_2+315 p_3 p_2) z^2 \\ \dots +(120 p_1^3 p_2^6 p_4^5 p_5^2
  p_6^4 p_3^8 +315 p_1^3 p_2^6 p_4^5 p_5^3 p_6^4 p_3^7) z^{28} \\
  + 30 p_1^3 p_2^6 p_3^8 p_4^5 p_5^3 p_6^4 z^{29}+p_1^3 p_2^6 p_3^8 p_4^6 p_5^3 p_6^4 z^{30},
  \end{multline*}\vspace{-2.5em}
  \begin{multline*}
  H_3 = 1 + 42 p_3 z+(315 p_2 p_3+315 p_4 p_3+231 p_6 p_3) z^2 \\ \dots +
  (315 p_1^4 p_2^7 p_4^8 p_5^4 p_6^6 p_3^{11}
  +315 p_1^4 p_2^8 p_4^7 p_5^4 p_6^6 p_3^{11}+231 p_1^4 p_2^8 p_4^8 p_5^4 p_6^5 p_3^{11}) z^{40} \\
  +42 p_1^4 p_2^8 p_3^{11} p_4^8 p_5^4 p_6^6 z^{41}
  +p_1^4 p_2^8 p_3^{12} p_4^8 p_5^4 p_6^6 z^{42},
  \end{multline*}\vspace{-2.5em}
  \begin{multline*}
  H_4 = 1+
  30 p_4 z+(315 p_3 p_4+120 p_5 p_4) z^2 \\ \dots +(120 p_1^2 p_2^5 p_4^6 p_5^3
  p_6^4 p_3^8+315 p_1^3 p_2^5 p_4^6 p_5^3 p_6^4 p_3^7) z^{28} \\
  +30 p_1^3 p_2^5 p_3^8 p_4^6 p_5^3 p_6^4 z^{29}+p_1^3 p_2^6 p_3^8 p_4^6 p_5^3 p_6^4 z^{30},
  \end{multline*}\vspace{-2.5em}
  \begin{multline*}
  H_5 = 1 +
  16 p_5 z+120 p_4 p_5 z^2+ \dots +120 p_1 p_2^2 p_3^4 p_4^3 p_5^2 p_6^2z^{14}\\
  +16 p_1 p_2^3 p_3^4 p_4^3 p_5^2 p_6^2 z^{15}+p_1^2 p_2^3 p_3^4 p_4^3 p_5^2 p_6^2 z^{16},
  \end{multline*}\vspace{-2.5em}
  \begin{multline}
  H_6 = 1 +
   22 p_6 z+231 p_3 p_6 z^2+ \dots + 231 p_1^2 p_2^4 p_3^5 p_4^4 p_5^2
   p_6^3 z^{20}\\
   +22 p_1^2 p_2^4 p_3^6 p_4^4 p_5^2 p_6^3 z^{21}+p_1^2 p_2^4 p_3^6 p_4^4 p_5^2 p_6^4 z^{22}.
  \label{3.4}
  \end{multline}
  
  The powers of polynomials are in agreement with the relation (\ref{3.3}).
  In what follows we denote
  \begin{equation}
    \label{3.5}
     H_s = H_s(z) = H_s(z, (p_i) ),
   \end{equation}
  $s = 1, \dots, 6$; where $(p_i) = (p_1,p_2,p_3,p_4,p_5,p_6)$.
  
  Due to  (\ref{3.4})  the  asymptotical relations for the polynomials read as follows
  \begin{equation}
    \label{3.6}
     H_s = H_s(z, (p_i) )  \sim \left( \prod_{l=1}^{6} (p_l)^{\nu^{sl}} \right) z^{n_s} \equiv 
     H_s^{as}(z, (p_i)),
   \end{equation}
  $s = 1, \dots, 6$, as $z \to \infty$. Here
  \begin{equation}
     \label{3.7}
     \nu = (\nu^{sl})  = \left(
     \begin{array}{cccccc}
      2 & 3 & 4 & 3 & 2 & 2 \\
      3 & 6 & 8 & 6 & 3 & 4 \\
      4 & 8 & 12& 8 & 4 & 6 \\
      3 & 6 & 8 & 6 & 3 & 4 \\
      2 & 3 & 4 & 3 & 2 & 2 \\
      2 & 4 & 6 & 4 & 2 & 4
     \end{array}
     \right).
     \end{equation}
  
  This matrix reads 
  \begin{equation}
    \label{3.8}
    \nu = A^{-1} (I + P),
   \end{equation}
   where $A^{-1}$ is inverse Cartan matrix, $I$ is $6 \times 6$ identity matrix and
  \begin{equation}
     \label{3.9}
     P  = \left(
     \begin{array}{cccccc}
      0 & 0 & 0 & 0 & 1 & 0 \\
      0 & 0 & 0 & 1 & 0 & 0 \\
      0 & 0 & 1 & 0 & 0 & 0 \\
      0 & 1 & 0 & 0 & 0 & 0 \\
      1 & 0 & 0 & 0 & 0 & 0 \\
      0 & 0 & 0 & 0 & 0 & 1
     \end{array}
     \right).
     \end{equation}
  is permutation matrix. This matrix is related to the permutation $\sigma \in S_6$ ($S_6$ is symmetric group)
  \begin{equation}
    \label{3.10}
     \sigma: (1,2,3,4,5,6) \mapsto (5,4,3,2,1,6),
   \end{equation}
   by the following formula  $P = (P^i_j) = (\delta^i_{\sigma(j)})$.
   Here $\sigma$   is the generator of the group 
   of symmetry of the Dynkin diagram $G = \{ \sigma, id \}$. 
   ($G$ is isomorphic to the group $Z_2$.)
   $\sigma$ is a composition of two transpositions: $(1 \ 5)$ and $(2 \ 4)$.
   
   We note that the matrix $\nu$ is symmetric one and 
   \begin{equation}
     \label{3.10a}
      \sum_{s= 1}^6 \nu^{sl} = n_l,
    \end{equation}
    $l = 1, \dots, 6$.
    
   Now we introduce the dual (ordered) set $(\hat{p}_i) = (p_{\sigma(i)}) $, $i= 1, \dots, 6$.
    By using the relations for  polynomials from Appendix   we are 
   led  to the following two  identities which are verified with the aid of MATHEMATICA.
   
   {\bf Symmetry relations.}
   
   {\bf Proposition 6.1} {\em  
    For all $p_i$ and $z$
   \begin{equation}
     \label{3.11}
      H_{\sigma(s)}(z, (p_i) ) = H_s(z, (\hat{p}_i)),
    \end{equation}
    $s= 1, \dots, 6$.  }
    
     {\bf Duality relations.}
    
   {\bf Proposition 6.2} {\em  
     For all $p_i > 0$ and $z > 0$
    \begin{equation}
      \label{3.12}
       H_{s}(z, (p_i) ) = H_s^{as}(z, (p_i)) H_s(z^{-1}, (\hat{p}_i^{-1})),
     \end{equation}
     $s = 1, \dots, 6$. }
    
          
    The solution  (\ref{2.30})- (\ref{2.32a})  reads in our case
    
       \bear{6.3.13}
        g= \Bigl(\prod_{s = 1}^{6} H_s^{2 h /(D-2)} \Bigr)
        \biggl\{ d\rho \otimes d \rho  +
        \Bigl(\prod_{s = 1}^{6} H_s^{-2 h} \Bigr) \rho^2 d\phi \otimes d\phi +
          g^2  \biggr\},
       \\  \label{6.3.14}
        \exp(\varphi^a)=
        \prod_{s = 1}^{6} H_s^{h \lambda_{s}^a},
       \\  \label{6.3.15}
        F^s= {\cal B}^s \rho d\rho \wedge d \phi,
        \ear
       $a, s = 1,\dots, 6$, where  $g^1 = d \phi \otimes d \phi$ is a
        metric on $M_1 = S^1$ ($0 < \phi < 2 \pi$), 
         $g^2$ is a  Ricci-flat metric on $M_{2}$ of signature $(-,+,\dots,+)$.
     Here 
     \begin{equation}
         \label{6.3.16}
     {\cal B}^s =  - Q_s 
     \left( \prod_{l = 1}^{6}  H_{l}^{- A_{s l}} \right)
       \end{equation} 
    and due to (\ref{2.16})-(\ref{2.18})
      \begin{equation}
             \label{6.3.17}
      K = K_s =   \frac{D - 3}{D -2} +  \vec{\lambda}_{s}^2,
        \end{equation} 
      $h_s = h = K^{-1}$,    
        \beq{6.3.18}
          \vec{\lambda}_{s} \vec{\lambda}_{s'} = 
              \frac{1}{2} K A_{s s'}  - \frac{D - 3}{D -2} \equiv G_{s s'},
        \eeq    
       $s,s' = 1, \dots, 6$.

   \subsection{Some relations between polynomials}
   
    Here we  denote the set of polynomials corresponding
    to a set of parameters $p_1 > 0$, ..., $p_n > 0$ as
    following
    \beq{6.1}
      H_s = H_s(z,p_1, ...,p_n; A),
     \eeq
      $s = 1, \dots, n$,
      where $A = A[{\cal G}]$ is the Cartan matrix corresponding to
      a (semi)simple Lie algebra $\cal G$ of rank $n$.
   
      \subsubsection{$C_{n+1}$-polynomials from $A_{2n+1}$-ones}
   
      It was conjected in Ref. \cite{GI} that 
      the set of polynomials corresponding to the Lie algebra
      $C_{n+1}$ may be obtained from the set of polynomials
      corresponding to the Lie algebra $A_{2n+1}$
      according to the following relations
       \beq{6.AC}
      H_s(z,p_1, ...,p_{n+1}; A[C_{n+1}]) =
      H_s(z,p_1, ...,p_{n+1}, p_{n+2}= p_{n},..., p_{2n+1} = p_{1}; A[A_{2n+1}]),
     \eeq
      $s = 1, \dots, n+1$, i.e. the parameters $p_1, ...,p_{n+1}, p_{n+2},...,
      p_{2n+1}$ are identified symmetrically with respect to $p_{n+1}$.
      
      Relation (\ref{6.AC}) may be readily verified at least for $n= 1,2$
      by using explicit relations for corresponding polynomials presented above.

      \subsubsection{$B_{n}$-polynomials from $D_{n+1}$-ones}
   
      Due to conjecter from Ref. \cite{GI} 
      the set polynomials corresponding to the Lie algebra
      $B_{n}$ can be obtained from the set of polynomials
      corresponding to the Lie algebra $D_{n+1}$
      according to the following relation
       \beq{6.BD}
      H_s(z,p_1, ...,p_{n}; A[B_{n}]) =
      H_s(z,p_1, ...,p_{n}, p_{n+1}= p_{n}; A[D_{n+1}]),
     \eeq
      $s = 1, \dots, n$, i.e. the parameters $p_{n}$ and $p_{n+1}$
      are identified.  
      
      
      Relation (\ref{6.BD}) can be readily verified at least for $n= 3,4$
      by using explicit relations for corresponding
      polynomials presented above.

      \subsubsection{$G_{2}$-polynomials from $D_{4}$-ones}
      
      It can be readily checked  that $G_{2}$-polynomials may be obtained just by
           imposing the following relations on parameters of $D_{4}$-polynomials:
           $p_1 = p_3 = p_4$. We obtain
      \begin{eqnarray}
            \label{6.GD1}
             H_1(z,p_1, p_2, p_1, p_{1}; A[D_{4}]) =
             H_1(z, p_1, p_2; A[G_{2}]), \\
             \label{6.GD2}
              H_2(z,p_1, p_2, p_1, p_{1}; A[D_{4}]) =
                          H_2(z, p_1, p_2; A[G_{2}]).             
              \end{eqnarray} 
      
       It looks like  we glue the symmetric points ($1$, $3$  and $4$ )
          at the Dynkin graph for $D_4$ algebra (see Figure 3) 
          in order to obtain the Dynkin graph for $G_2$ algebra (see Fifure 1).
      
      \subsubsection{$F_{4}$-polynomials from  $E_{6}$-ones}
   
     It can be verified that $F_{4}$-polynomials may be obtained by
     imposing the following relations on parameters of $E_{6}$-polynomials:
     $p_1 = p_5 = \bar{p}_4$,  $p_2 = p_4 = \bar{p}_3$, $p_3 = \bar{p}_2$, $p_6 = \bar{p}_1$. We obtain 
     
     \begin{eqnarray}
      \label{6.FE1}
       H_1(z,p_1, p_2, p_3, p_2,  p_1, p_{6}; A[E_{6}]) =
       H_4(z, \bar{p}_1, \bar{p}_2, \bar{p}_3, \bar{p}_4; A[F_{4}]), \\
       \label{6.FE2}
       H_2(z,p_1, p_2, p_3,  p_2,  p_1, p_{6}; A[E_{6}]) =
       H_3(z, \bar{p}_1, \bar{p}_2, \bar{p}_3, \bar{p}_4; A[F_{4}]), \\              
        \label{6.FE3}
       H_3(z,p_1, p_2, p_3,  p_2,  p_1, p_{6}; A[E_{6}]) =
       H_2(z, \bar{p}_1, \bar{p}_2, \bar{p}_3, \bar{p}_4; A[F_{4}]), \\
       \label{6.F6}
       H_6(z,p_1, p_2, p_3,  p_2,  p_1, p_{6}; A[E_{6}]) =
       H_1(z, \bar{p}_1, \bar{p}_2, \bar{p}_3, \bar{p}_4; A[F_{4}]). 
       \end{eqnarray}  
  
   It looks like  we glue the symmetric points ($1$ and $5$, $2$ and $4$ )
    at the Dynkin graph for $E_6$ algebra (see Figure 5) 
    in order to obtain the Dynkin graph for $F_4$ algebra (see Fifure 3).

     \subsubsection{Reduction formulas}
        Here, we denote the Cartan matrix in the following way: $A = A_{\Gamma}$,
        where $\Gamma$ is the related Dynkin graph. Let $i$ be a node of
        $\Gamma$. We denote by $\Gamma_i$ a Dynkin graph (which corresponds to a
        certain semi-simple   Lie algebra) that is obtained
        from $\Gamma$ by erasing all lines that have endpoints at $i$.
        It can be verified (e.g., by using MATHEMATICA) that 
        for the polynomials presented above the
        following reduction formulae hold
       
     \beq{6.R1}
      H_s(z,p_1, ...,p_i=0 ,...,p_{n}; A_{\Gamma}) =
      H_s(z,p_1, ...,p_{i-1},p_{i+1},...,p_{n}; A_{\Gamma_i}),
     \eeq
      $s = 1, \dots, i-1,i+1, \dots, n$, 
      and
            \beq{6.R2}
      H_i(z,p_1, ...,p_i=0 ,...,p_{n}; A_{\Gamma}) = 1.
       \eeq
     This means that by setting $p_i = 0$ we
     reduce the set of polynomials by replacing 
     the Cartan matrix $A_{\Gamma}$ with the
     Cartan matrix $A_{\Gamma_i}$. In this case the polynomial
      $H_i = 1$ corresponds to
      $A_1$-subalgebra (represented by  the node $i$) and the parameter $p_i = 0$.
      
      As an example of reduction formulas we
      present the following relations
      \beq{6.R3}
      H_s(z,p_1, ...,p_n, p_{n+1} = 0; A[{\cal G}]) =
      H_s(z,p_1, ...,p_{n}; A[A_n]),
     \eeq
      $s = 1, \dots, n$, for ${\cal G} = A_{n+1}, B_{n+1}, C_{n+1}, D_{n+1}$
      with appropriate restrictions on $n$. These relations are valid at least for
      all $ABCD$-polynomials presented above.  In
      writing relation (\ref{6.R1}) we use the numbering of nodes in
      agreement with the Dynkin diagrams depicted on figures presented above.

      The reduction formulas  (\ref{6.R2}) for $A_5$-polynomials
      with $p_3 = 0$ are shown in Figure 6. The reduced polynomials
      coincide with those corresponding to semisimple
      Lie algebra $A_2 \bigoplus A_1 \bigoplus A_2$.  
  
     \begin{center}
  \bigskip
  \begin{picture}(66,11)
  \put(5,10){\circle*{2}} \put(15,10){\circle*{2}}
  \put(25,10){\circle*{2}} \put(35,10){\circle*{2}}
  \put(45,10){\circle*{2}} \put(5,10){\line(1,0){10}}
  \put(5,5){\makebox(0,0)[cc]{1}}
  \put(15,5){\makebox(0,0)[cc]{2}}
  \put(25,5){\makebox(0,0)[cc]{3}}
  \put(35,5){\makebox(0,0)[cc]{$4$}}
  \put(45,5){\makebox(0,0)[cc]{$5$}} \put(35,10){\line(1,0){10}}
  \end{picture} \\[5pt]
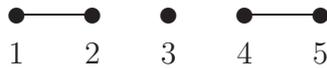
   Figure 6.   Dynkin diagram for semisimple
   Lie algebra $A_2 \bigoplus A_1 \bigoplus A_2$
    describing the set of $A_5$-polynomials with $p_3 =0$ \cite{GI}.
  \end{center}
  
\section{Flux integrals  }

As in a previous section  we deal here with the solution (\ref{2.30})-(\ref{2.32a})
with  $n = l$, $w = + 1$ and $M_1 = S^1$ and  $h_{\alpha\beta} = \delta_{\alpha \beta}$. 
 
 We denote  $(\lambda_{s a }) =  (\lambda_{s}^{a}) = \vec{\lambda}_{s}$, $s = 1, \dots, n$. 
 The solution corresponds to a simple finite-dimensional Lie algebra ${\cal G}$, i.e. 
the matrix   $A =  (A_{ss'})$ is coinciding with the Cartan matrix of this Lie algebra.

 In this case we get relations \ref{2.3.17}), (\ref{2.3.18}) 
 for  $h_s =  K_s^{-1} > 0$ and scalar products 
 $\vec{\lambda}_{s} \vec{\lambda}_{l}$ which follow from 
  (\ref{2.16})-(\ref{2.18}). 
         
          Due to  (\ref{2.16})-(\ref{2.18}) that 
         \beq{3.26}
          \frac{h_i}{h_j} = \frac{K_{j}}{K_{i}} = \frac{B_{jj}}{B_{ii}} = 
          \frac{B_{ji}}{B_{ii}} \frac{B_{jj}}{B_{ij}} = \frac{A_{ji}}{A_{ij}}
           \eeq
         for any $i \neq j$ obeying $A_{ij} = A_{ji} \neq 0$; $i,j = 1, \dots,n$.

      It follows from (\ref{3.26}) and the  connectedness of the Dynkin diagram for a simple Lie algebra 
      that
       \beq{3.27}
        \frac{h_i}{h_j} = \frac{K_{j}}{K_{i}} = \frac{r_j }{r_i} 
         \eeq
        $i \neq j$, where $r_i = (\alpha_{i}, \alpha_{i})$  
        is length squared  of a simple root $\alpha_{i}$ of the Lie algebra ${\cal G}$. 
        Here  ($(.,.)$ is dual Killing-Cartan form and
        \beq{3.27C}
          A_{ij} = 2 (\alpha_{i}, \alpha_{j})/(\alpha_{j}, \alpha_{j}),
         \eeq
        $i,j = 1, \dots,n$. Due to (\ref{3.27}) we obtain 
      \beq{3.27a}
       K_{i} = \frac12 K r_i,  
      \eeq
     $i = 1, \dots,n$, where $K > 0$.
                      
      Now we consider the oriented $2$-dimensional manifold 
      $M_{*} =(0, + \infty)  \times S^1$. The flux integrals read
       \beq{3.19}
       \Phi^s = \int_{M_{*}} F^s =  
        \int_{0}^{+ \infty} d \rho \int_{0}^{2 \pi} d \phi \ \rho {\cal B}^s(\rho^2)
        = 2 \pi \int_{0}^{+ \infty} d \rho \ \rho {\cal B}^s(\rho^2) ,
       \eeq    
       where 
        \begin{equation}
           \label{3.16}
           {\cal B}^s(\rho^2) =   q_s  \prod_{l = 1}^{n}  (H_{l}(\rho^2))^{- A_{s l}}.
                   \end{equation} 
       Due to (\ref{2.21}) and (\ref{4.g}) we have
        \begin{equation}
                   \label{3.16a}
                    p_s = \frac{P_s}{n_s} = \frac{K_s}{4n_s}  q_s^2.
          \end{equation} 
       
       The integrals (\ref{3.19})
       are convergent for all $s = 1, \dots, n$, if the conjecture 
       on polynomial structure  of moduli functions $H_s$  is satisfied 
       for the  Lie algebra ${\cal G}$ under consideration.
       
        Indeed,  polynomial assumption  (\ref{1.3}) implies 
      \beq{3.20}
       H_s(\rho^2) \sim C_s \rho^{2n_s},   \qquad C_s = P_s^{(n_s)},
            \eeq 
      as $\rho \to + \infty$; $s =1, \dots, n$. From (\ref{3.16}), (\ref{3.20}) and 
      the identities $\sum_{1}^{n} A_{s l} n_l = 2$,     following from (\ref{1.4}), we obtain
      \beq{3.21}
           {\cal B}^s(\rho^2) \sim  q_s C^s \rho^{-4},  \quad  C^s =  \prod_{l = 1}^{n} C_l^{-A_{sl}}.              
       \eeq 
       Hence the integral (\ref{3.19}) is convergent for any $s =1, \dots, n$.

        From  master equations (\ref{1.1})    we obtain
      \bear{3.24}
      \int_{0}^{+ \infty} d \rho \rho {\cal B}^s(\rho^2) =  q_s P_s^{-1}
      \frac12 \int_{0}^{+ \infty} d z  
      \frac{d}{dz} \left( \frac{ z}{H_s} \frac{d}{dz} H_s \right)
      \\ \nonumber
      =     \frac12 q_s P_s^{-1} \lim_{z \to + \infty}   
              \left( \frac{ z}{H_s} \frac{d}{dz} H_s \right) =
               \frac12 n_s q_s P_s^{-1},          
      \ear
   which implies (see (\ref{2.21})) \cite{Ivas-17-flux}
            \beq{3.25}
         \Phi^s =   4 \pi n_s q_s^{-1} h_s,  
         \eeq
   $s =1, \dots, n $.

  Thus, we see that any flux $\Phi^s$ depends only upon one integration constant  $q_s \neq 0$. 
  This  is a nontrivial fact since the integrand form $F^s$ depends upon all constants: $q_1, \dots, q_n$.
   
  It should be noted that for $D =4$ and 
   $g^2 = -  dt \otimes dt + d x \otimes d x$, $q_s$ is coinciding  with 
  the value of the $x$-component of the magnetic field on the axis of symmetry. 
  
  For the case of Gibbons-Maeda dilatonic generalization of the Melvin solution,   corresponding to  
  $D = 4$, $n = l= 1$   and ${\cal G} = A_1$ \cite{GM}, the flux from (\ref{3.25}) ($s=1$)  is in agreement  
  with that obtained in Ref. \cite{KKR}. For the Melvin's solution and some higher dimensional 
  extensions (with ${\cal G} = A_1$) see also Ref. \cite{GutSt}.
   
 Owing to (\ref{3.27})  the ratios 
    \beq{3.28}
    \frac{q_i \Phi^i}{q_j \Phi^j} =  \frac{n_i h_i}{n_j h_j} =  \frac{n_i r_{j} }{n_j r_{i}}  
    \eeq
  are just fixed numbers which depend upon the  Cartan  matrix $({A_{ij}})$ 
  of a simple finite-dimensional Lie algebra ${\cal G}$.

   Since the manifold $M_{*} =(0, + \infty)  \times S^1$ is isomorphic to
   the manifold $\R^2_{*} = \R^2 \setminus \{ 0 \}$,  
   the solution (\ref{2.30})-(\ref{2.32a}) may be rewritten 
   (by pull-backs)  onto  manifold $\R^2_{*} \times M_2$.  
   In this new presentation of the solution  coordinates $\rho$, $\phi$ are 
   understood as polar coordinates on  $\R^2_{*}$. Since they  are not globally defined 
   one may consider two charts   with  coordinates $\rho$, 
   $\phi= \phi_1$ and $\rho$, $\phi= \phi_2$, where $\rho > 0$, $0 < \phi_1 < 2 \pi$
    and  $- \pi < \phi_2 <  \pi$. Here $\exp(i \phi_1 ) = \exp(i \phi_2)$. 
    For both charts we have  $x = \rho \cos \phi$ and
    $y = \rho \sin \phi $, where $x, y$ are standard (Euclidean) coordinates of $\R^2$.
    By using the identity $ \rho  d\rho \wedge  d\phi = d x \wedge  dy$ we obtain
    \beq{3.30}
     F^s=  q_s  \prod_{l = 1}^{n}  
      (H_{l}(x^2 + y^2))^{- A_{s l}}   dx \wedge d y,
    \eeq
    $s =1, \dots, n $.   
    
    Now, let us show that 2-forms (\ref{3.30}) are well-defined on $\R^2$. Indeed, 
    due to conjecture from Ref. \cite{Iflux} any  polynomial $H_{s}(z)$ is a smooth function
    on $\R = (- \infty, + \infty)$  obeying $H_{s}(z) > 0$ for $z \in (- \varepsilon_s, + \infty)$,
     where     $\varepsilon_s > 0$. This is so since due to polynomial conjecture 
     from Ref. \cite{Iflux} we have $H_{s}(z) > 0$ for $z > 0$    and $H_{s}(+0) = 1$. 
    Hence, $\left( \prod_{l = 1}^{n} (H_{l}(x^2 + y^2))^{- A_{s l}} \right)$
    is a smooth function since it is just a composition of two well-defined smooth functions:  
    $\left( \prod_{l = 1}^{n} (H_{l}(z))^{- A_{s l}} \right)$ and $z = x^2 + y^2$. 
           
    Now we show that there exist 1-forms $A^s$ obeying $F^s = dA^s$ which are globally defined on $\R^2$.
    Let us consider the open submanifold  $\R^2_{*}$. The  1-forms 
    \beq{3.31}
     A^s = \left( \int_{0}^{\rho}  d \bar{\rho} \bar{\rho} {\cal B}^s(\bar{\rho}^2)  \right) d \phi
         = \frac{1}{2}  \left( \int_{0}^{\rho^2} d \bar{z}  {\cal B}^s(\bar{z})  \right) d \phi
    \eeq
    are well defined on $\R^2_{*}$  and obey $F^s = dA^s$,
    $s = 1, \dots, n $.  It is obvious that 
        here 
         \beq{3.31p}
        d\phi = (x^2 + y^2)^{-1} ( - y dx + x dy).
         \eeq           
    
    It follows from the master equation (\ref{1.1}) that
     \bear{3.32}
         A^s = \frac{q_s}{2 P_s}  \left( \int_{0}^{\rho^2} d \bar{z} 
    \frac{d}{d \bar{z}}  \left( \frac{ \bar{z}}{H_s(\bar{z})} \frac{d}{d\bar{z}} H_s ( \bar{z}) 
    \right) \right) d \phi
         \\ \nonumber
     =  \frac{2 h_s}{q_s} \frac{H^{'}_{s} (\rho^2)}{H_s(\rho^2)} \rho^2 d \phi,  
        \ear
    $s = 1, \dots, n $. Here $H{'}_s = \frac{d}{dz} H_s$. Due to relation $\rho^2 d \phi = - y dx + x dy $,
    we obtain 
     \beq{3.33}
             A^s   =  \frac{2 h_s}{q_s} \frac{H^{'}_{s} (x^2 + y^2)}{H_s(x^2 + y^2)} (- y dx + x dy),  
     \eeq
      $s = 1, \dots, n $. Thus we are led to 1-forms (\ref{3.33}) 
      which are well-defined (smooth)  1-forms on $\R^2$.
      
      It should be noted that in  case of Gibbons-Maeda solution \cite{GM} (with $D = 4$, $n = l= 1$ 
      and ${\cal G} = A_1$) the gauge potential from (\ref{3.32}) coincides (up to notations) 
      with that considered in Ref. \cite{DGGidH}.
                         
     Now we verify the formula (\ref{3.25}) for flux integrals
     by using the relation (\ref{3.33}) for  1-forms $A^s$. 
     Let us consider a $2d$ oriented manifold (disk) 
     \beq{3.33D}
     D_R = \{ (x,y): x^2 + y^2 \leq R^2 \} 
     \eeq
     with the boundary
     \beq{3.33C}
     \partial D_R = C_R = \{ (x,y): x^2 + y^2 = R^2 \}.
      \eeq
     Here $C_R$ is a circle of radius $R$. It is just
      an $1d$ oriented manifold  with the orientation (inherited from that of $D_R$)
      obeying the relation $\int_{C_R} d\phi = 2 \pi$. 
     Using the Stokes-Cartan theorem we obtain
     \beq{3.34}
       \Phi^s(R) =  \int_{D_{R}} F^s=    \int_{D_{R}} d A^s=   \int_{C_{R}}  A^s
          =  \frac{4 \pi h_s}{q_s} \frac{H^{'}_{s} (R^2)}{H_s(R^2)} R^2,  
     \eeq
     $s = 1, \dots, n $. 
     By using the asymptotic relation (\ref{3.20}) we get
     \beq{3.35}
         \lim_{R \to + \infty}   \Phi^s(R) = \lim_{R \to + \infty}  \int_{D_{R}} F^s
      =  \frac{4 \pi h_s n_s}{q_s} = \Phi^s,  
          \eeq
          $s = 1, \dots, n $, in agreeement with (\ref{3.25}).

      We note that according to the definition of Abelian Wilson loop (factor), we have 
            \beq{3.W}
                   W^s(C_R)  =  \exp (i \int_{C_{R}}  A^s ) = \exp (i  \Phi^s(R) ),          
            \eeq
       $s = 1, \dots, n $.         
       As a consequence (see (\ref{3.35})), we obtain finite limits
       \beq{3.WL}
         \lim_{R \to + \infty} W^s(C_R)  =  \exp (i \Phi^s),          
        \eeq
       $s = 1, \dots, n $.
     
         Finally, we note  that the metric and scalar fields
     for our solution with $w = +1$ and $l = n$ 
     can be extended  to the manifold $\R^2 \times M_2$. 
     
     Indeed, in standard   $x, y$ coordinates on $\R^2$ 
     the metric (\ref{2.30}) and scalar fields
     (\ref{2.31}) read as follows \cite{Ivas-17-flux}         
      \bear{3.36}
        g= \Bigl(\prod_{s = 1}^{n} H_s^{2 h_s /(D-2)} \Bigr)
        \biggl\{  d x \otimes d x  + d y \otimes d y + 
         F  (- y dx + x dy)^2_{\otimes} +   g^2  \biggr\},
       \\  \label{3.37}
        \varphi^{a}=
        \sum_{s = 1}^{n} h_s  \lambda_{s}^{a} \ln H_s,
          \ear
     $a =1, \dots, n $. Here 
     $H_s = H_s(x^2 + y^2)$, $s = 1, \dots, n $,  
      $ \omega^2_{\otimes} = \omega \otimes \omega $ and
     $F = F(x^2 + y^2)$,
     where
     \beq{3.38}
     F(z) = \left(\Bigl(\prod_{s = 1}^{n} (H_s(z))^{-2 h_s} \Bigr) - 1 \right) z^{-1},
     \eeq
     for $z \neq 0$ and 
     \beq{3.38a}
          F(0) =  \lim_{z \to 0} F(z) = \sum_{s = 1}^{n} (-2 h_s p_s).
          \eeq
      
      The  metric and scalar fields are  smooth on the manifold $\R^2 \times M_2$ \cite{Ivas-17-flux}.

      \section{Dilatonic black holes}
          
          Relations  on dilatonic coupling vectors  
          (\ref{2.16}),  (\ref{2.17}), (\ref{2.18}) appear also 
          for dilatonic black hole (DBH) solutions  defined on the manifold
         \beq{5.1}
           M = (R_0, + \infty) \times (M_0 = S^2) \times (M_1 = \R) \times M_2,
          \eeq
          where $R_0 = 2 \mu > 0$ and $M_2$ is a  Ricci-flat manifold. These DBH solutions on $M$ from 
          (\ref{5.1})
          for the model under consideration may be extracted just from more general black brane solutions, see  
          Refs. \cite{I-14,Ivas-Symmetry-17,IMp3}. They read: 
          \bear{5.1g}
           g= \Bigl(\prod_{s = 1}^{n} {\bf H}_s^{2 h_s /(D-2)} \Bigr)
           \biggl\{  f^{-1} dR \otimes d R  + R^2 g^0  -
           \Bigl(\prod_{s = 1}^{n} {\bf H}^{-2 h_s} \Bigr) f dt \otimes dt +
             g^2  \biggr\},
          \\  \label{5.2p}
           \exp(\varphi^a)=
           \prod_{s = 1}^{n} {\bf H}^{h_s  \lambda_{s}^a},
          \\  \label{5.2f}
           F^s = - Q_s R^{-2} \left( \prod_{l = 1}^{n}  {\bf H}_{l}^{- A_{s
           l}} \right)  dR \wedge d t,
           \ear
          $s,a = 1, \dots ,n$, where $f = 1 - 2\mu R^{-1}$, $g^0$ is the standard 
           metric on $M_0 = S^2$ and $g^2$ is a  Ricci-flat metric of 
           signatute $(+, \dots, +)$ on $M_{2}$.  Here $Q_s \neq 0$ are integration constants  
          (charges).
         
          The functions ${\bf H}_s = {\bf H}_s (R) > 0$ obey the master equations
         \beq{5.3}
           R^2 \frac{d}{dR} \left( f \frac{ R^2}{{\bf H}_s} \frac{d}{dR} {\bf H}_s \right) =
            B_s \prod_{l = 1}^{n}  {\bf H}_{l}^{- A_{s l}},
           \eeq
          with  the following boundary conditions (on the horizon and at infinity) imposed:
          \beq{5.4}
            {\bf H}_{s}(R_0 + 0) =  {\bf H}_{s0} > 0, \qquad {\bf H}_{s}(+ \infty) = 1,
          \eeq
          where
          \beq{5.5}
           B_s =  - K_s Q_s^2,
          \eeq
          $s = 1, \dots ,n$. 
          
          It was shown in Ref. \cite{I-14} that these polynomials may be 
          obtained (at least for small enough $Q_s$) from fluxbrane polynomials $H_s(z)$ 
          presented in this paper.  
          
           Indeed, let us denote $f = f(z)= 1 - 2\mu z$, $z = 1/R$.
          Then the relations (\ref{5.3}) may be rewritten as follows
          \beq{5.6}
           \frac{d}{df} \left( \frac{f}{{\bf H}_s}
           \frac{d}{df} {\bf H}_s \right) =  B_s (2 \mu )^{-2}
           \prod_{l =1}^{n}  {\bf H}_{l}^{- A_{s l}}, 
           \eeq
            $s = 1,\dots,n$.
           These relations could be solved (at least for small enough $Q_s$) by using fluxbrane
          polynomials $H_{s}(f) = H_{s}(f; p)$, given by
           $n \times n$ Cartan matrix $(A_{s l})$, where $p =
          (p_1,\dots ,p_n)$ is the set of  parameters \cite{I-14}. Here
          we impose the restrictions $p_s \neq 0$ for all $s$.
          
          Due to approach of Ref. \cite{I-14} we put 
          \beq{5.7}
            {\bf H}_s(z) =  H_{s}(f(z);p)/H_{s}(1;p)
          \eeq 
          for  $s = 1, \dots ,n$. Then the relations (\ref{5.6}), 
          are satisfied identically if \cite{I-14}
          
          \beq{5.7a}
           n_s p_s  \prod_{l =1}^{n}  (H_{l}(1;p))^{- A_{s l}} = B_s /(2 \mu)^2,
           \eeq 
          $s = 1, ...,n$.
          
          We call the set of parameters $p = (p_1,\dots,p_n)$ ($p_i \neq 0$)  as proper one if \cite{I-14}
            \beq{5.8}
            H_{s}(f;p) > 0
           \eeq
           for all $f \in [0,1]$ and $s = 1,\dots,n$.  Here we consider only proper $p$.
           Due to  relations (\ref{5.7a}) and $B_s < 0$  we have  $p_s < 0$ for all $s = 1,\dots ,n$, 
           i.e. one should use fluxbrane polynomials with negative parameters $p_s$ 
           for a description of black hole solutions under consideration. 
          
          The  boundary conditions  (\ref{5.4}) are valid since
            \beq{5.9}
            {\bf H}_{s}((2\mu)^{-1} -0) = 1/ H_{s}(1;p) > 0,
            \eeq
           $s = 1,\dots,n$  (see  definition (\ref{5.7})).
          
           For small enough
           $p_i$  the set $(p_1, \dots,p_n)$ is proper 
           and relation  (\ref{5.7a}) defines one-to-one correspondence between
           the sets of parameters $(p_1, \dots,p_n)$ and $(Q_1^2, \dots, Q_n^2 )$. 
          
           Relations (\ref{5.9}) imply the following formula for
            the Hawking temperature  \cite{I-14}
           \beq{5.7T}
           T_H =   \frac{1}{8 \pi \mu}
           \prod_{s = 1}^{n} (H_{s}(1;p))^{h_s}.  \eeq
          
        It should be noted that fluxbrane polynomials were used  in Refs.
        \cite{Dav,ZGCh,AbIMT} in a context of dyon-like black hole solutions.
        (To a certain extent  these papers were inspired by Ref. \cite{ABI}.)

        \section{\bf Conclusions} 
          Here, we have explored a multidimensional generalization of the Melvin's
         solution corresponding to a simple finite-dimensional Lie algebra ${\cal G}$.
         It takes place in a $D$-dimensional model which contains metric $n$ Abelian 2-forms $ F^s = dA^s$ 
         and  $l \geq n$ scalar fields  $\varphi^{\alpha}$. (Here, we have put $l = n$ 
         for simplicity.)  
          
         The solution is governed by a set of $n$  moduli functions $H_s(z)$, $s =1,\dots,n$,
         which were  assumed earlier to be polynomials---so-called fluxbrane polynomials.
         These polynomials  define special solutions to open Toda chain equations corresponding 
         to the Lie algebra  ${\cal G}$.     
         The polynomials $H_s(z)$ also depend upon parameters $p_s \sim q_s^2$, where   
         $ q_s$ are coinciding for $D =4$  (up to a sign) with the values  of  colored 
         magnetic fields on the axis of symmetry. 
           
         Here, we have presented  examples of polynomials corresponding to  Lie algebras of 
          rank $r = 1,2,3,4,5$ and  exceptional Lie algebra $E_6$
         and have outlined the symmetry and duality relations for these polynomials. 
         The so-called duality relations for fluxbrane polynomials  describe
          a behavior of the solutions under the inversion: $\rho \to 1/\rho$, 
          which makes the model in tune with so-called $T$-duality in 
           string models. These relations  can be also mathematically understood in terms
          of the discrete groups of symmetry  of Dynkin diagrams for the corresponding Lie algebras.  
                 
         We have presented  calculations of $2d$ flux integrals
         $\Phi^s = \int F^s$,  where  $F^s$ are 2-forms, 
         $s =1, \dots, n$. It is remarkable that 
         any flux  $\Phi^s$ depends only upon one parameter $q_s$, 
         while the integrand  $F^s$ depends upon all parameters $q_1, \dots, q_n$.
            
         Here, we have also outlined  possible applications of fluxbrane polynomials for seeking 
              dilatonic black hole solutions (e.g., $4d$ ones) in the model under consideration.
          
          It should be noted that fluxbrane polynomials may also describe certain 
          subclass of  self-dual solutions to Yang-Mills equations in a flat space of signature $(+,+,-,-)$ 
         ~\cite{IvPop}. This subclass belong to a more general  Toda chain class of self-dual solutions.
          
          Here, an interesting question occurs:  how can the nice properties of polynomials (symmetries, duality relations, reduction formulas)  be used for a description of geometries of Melvin-like solutions (and other fluxbrane ones)
          and possible generating of new solutions? This and other related questions may be a subject of  future  publications. 
              

  \section{Appendix}
  
   \renewcommand{\theequation}{\Alph{section}.\arabic{equation}}
       \renewcommand{\thesection}{\Alph{section}}

  \renewcommand{\thesubsection}{\Alph{subsection}}
  
  \subsection{Polynomials for $A_5,B_5,C_5,D_5$}
  
  Here we present the polynomials corresponding to Lie algebras of rank $5$
   \cite{BolIvas_R5-22}.

  \noindent{\bf $A_5$-case.} For the Lie algebra $A_5 \cong sl(6)$ the polynomials read
   \begin{itemize} 
  \item[$H_1=$]\(
  1+5 p_1 \textcolor{red}{z}+10 p_1 p_2 \textcolor{red}{{z}^2}+10 p_1 p_2 p_3 \textcolor{red}{{z}^3}+5 p_1 p_2 p_3 p_4 \textcolor{red}{{z}^4}+p_1 p_2 p_3 p_4 p_5 \textcolor{red}{{z}^5}
  \)
  \item[$H_2=$]\(
  1+8 p_2 \textcolor{red}{z}+(10 p_1 p_2+18 p_2 p_3) \textcolor{red}{{z}^2}+(40 p_1 p_2 p_3+16 p_2 p_3 p_4) \textcolor{red}{{z}^3}
  +(20 p_1 p_2^2 p_3+45 p_1 p_2 p_3 p_4+5 p_2 p_3 p_4 p_5) \textcolor{red}{{z}^4}+(40 p_1 p_2^2 p_3 p_4+16 p_1 p_2 p_3 p_4 p_5) \textcolor{red}{{z}^5}
  +(10 p_1 p_2^2 p_3^2 p_4+18 p_1 p_2^2 p_3 p_4 p_5) \textcolor{red}{{z}^6}+8 p_1 p_2^2 p_3^2 p_4 p_5 \textcolor{red}{{z}^7}+p_1 p_2^2 p_3^2 p_4^2 p_5 \textcolor{red}{{z}^8}\)
  \item[$H_3=$]\(
  1+9 p_3 \textcolor{red}{z}+(18 p_2 p_3+18 p_3 p_4) \textcolor{red}{{z}^2}+(10 p_1 p_2 p_3+64 p_2 p_3 p_4+10 p_3 p_4 p_5) \textcolor{red}{{z}^3}
  +(45 p_1 p_2 p_3 p_4+36 p_2 p_3^2 p_4+45 p_2 p_3 p_4 p_5) \textcolor{red}{{z}^4}+(45 p_1 p_2 p_3^2 p_4+36 p_1 p_2 p_3 p_4 p_5+45 p_2 p_3^2 p_4 p_5) \textcolor{red}{{z}^5}
  +(10 p_1 p_2^2 p_3^2 p_4+64 p_1 p_2 p_3^2 p_4 p_5+10 p_2 p_3^2 p_4^2 p_5) \textcolor{red}{{z}^6}+(18 p_1 p_2^2 p_3^2 p_4 p_5+18 p_1 p_2 p_3^2 p_4^2 p_5) \textcolor{red}{{z}^7}
  +9 p_1 p_2^2 p_3^2 p_4^2 p_5 \textcolor{red}{{z}^8}+p_1 p_2^2 p_3^3 p_4^2 p_5 \textcolor{red}{{z}^9}\)
  \item[$H_4=$]\(
  1+8 p_4 \textcolor{red}{z}+(18 p_3 p_4+10 p_4 p_5) \textcolor{red}{{z}^2}+(16 p_2 p_3 p_4+40 p_3 p_4 p_5) \textcolor{red}{{z}^3}
  +(5 p_1 p_2 p_3 p_4+45 p_2 p_3 p_4 p_5+20 p_3 p_4^2 p_5) \textcolor{red}{{z}^4}+(16 p_1 p_2 p_3 p_4 p_5+40 p_2 p_3 p_4^2 p_5) \textcolor{red}{{z}^5}
  +(18 p_1 p_2 p_3 p_4^2 p_5+10 p_2 p_3^2 p_4^2 p_5) \textcolor{red}{{z}^6}+8 p_1 p_2 p_3^2 p_4^2 p_5 \textcolor{red}{{z}^7}+p_1 p_2^2 p_3^2 p_4^2 p_5 \textcolor{red}{{z}^8}\)
  \item[$H_5=$]\(
  1+5 p_5 \textcolor{red}{z}+10 p_4 p_5 \textcolor{red}{{z}^2}+10 p_3 p_4 p_5 \textcolor{red}{{z}^3}+5 p_2 p_3 p_4 p_5 \textcolor{red}{{z}^4}+p_1 p_2 p_3 p_4 p_5 \textcolor{red}{{z}^5}
  \)
  \end{itemize}

  \bigskip
  \noindent {\bf $B_5$-case.}
  For the Lie algebra $B_5 \cong so(11)$ we find
  \begin{itemize}
  \item[$H_1=$]\(
  1+10 p_1 \textcolor{red}{z}+45 p_1 p_2 \textcolor{red}{{z}^2}+120 p_1 p_2 p_3 \textcolor{red}{{z}^3}+210 p_1 p_2 p_3 p_4 \textcolor{red}{{z}^4}+252 p_1 p_2 p_3 p_4 p_5 \textcolor{red}{{z}^5}+210 p_1 p_2 p_3 p_4 p_5^2 \textcolor{red}{{z}^6}
  +120 p_1 p_2 p_3 p_4^2 p_5^2 \textcolor{red}{{z}^7}+45 p_1 p_2 p_3^2 p_4^2 p_5^2 \textcolor{red}{{z}^8}+10 p_1 p_2^2 p_3^2 p_4^2 p_5^2 \textcolor{red}{{z}^9}+p_1^2 p_2^2 p_3^2 p_4^2 p_5^2 \textcolor{red}{{z}^{10}}
  \)
  \item[$H_2=$]\(
  1+18 p_2 \textcolor{red}{z}+(45 p_1 p_2+108 p_2 p_3) \textcolor{red}{{z}^2}+(480 p_1 p_2 p_3+336 p_2 p_3 p_4) \textcolor{red}{{z}^3}
  +(540 p_1 p_2^2 p_3+1890 p_1 p_2 p_3 p_4+630 p_2 p_3 p_4 p_5) \textcolor{red}{{z}^4}+(3780 p_1 p_2^2 p_3 p_4+4032 p_1 p_2 p_3 p_4 p_5 
  +756 p_2 p_3 p_4 p_5^2) \textcolor{red}{{z}^5}+(2520 p_1 p_2^2 p_3^2 p_4+10206 p_1 p_2^2 p_3 p_4 p_5+5250 p_1 p_2 p_3 p_4 p_5^2+588 p_2 p_3 p_4^2 p_5^2) \textcolor{red}{{z}^6}
  +(12096 p_1 p_2^2 p_3^2 p_4 p_5+15120 p_1 p_2^2 p_3 p_4 p_5^2+4320 p_1 p_2 p_3 p_4^2 p_5^2+288 p_2 p_3^2 p_4^2 p_5^2) \textcolor{red}{{z}^7}
  +(5292 p_1 p_2^2 p_3^2 p_4^2 p_5+22680 p_1 p_2^2 p_3^2 p_4 p_5^2+13500 p_1 p_2^2 p_3 p_4^2 p_5^2+2205 p_1 p_2 p_3^2 p_4^2 p_5^2+81 p_2^2 p_3^2 p_4^2 p_5^2) \textcolor{red}{{z}^8}
  +48620 p_1 p_2^2 p_3^2 p_4^2 p_5^2 \textcolor{red}{{z}^9}+(81 p_1^2 p_2^2 p_3^2 p_4^2 p_5^2+2205 p_1 p_2^3 p_3^2 p_4^2 p_5^2+13500 p_1 p_2^2 p_3^3 p_4^2 p_5^2
  +22680 p_1 p_2^2 p_3^2 p_4^3 p_5^2+5292 p_1 p_2^2 p_3^2 p_4^2 p_5^3) \textcolor{red}{{z}^{10}}+(288 p_1^2 p_2^3 p_3^2 p_4^2 p_5^2+4320 p_1 p_2^3 p_3^3 p_4^2 p_5^2
  +15120 p_1 p_2^2 p_3^3 p_4^3 p_5^2+12096 p_1 p_2^2 p_3^2 p_4^3 p_5^3) \textcolor{red}{{z}^{11}}+(588 p_1^2 p_2^3 p_3^3 p_4^2 p_5^2+5250 p_1 p_2^3 p_3^3 p_4^3 p_5^2
  +10206 p_1 p_2^2 p_3^3 p_4^3 p_5^3+2520 p_1 p_2^2 p_3^2 p_4^3 p_5^4) \textcolor{red}{{z}^{12}}+(756 p_1^2 p_2^3 p_3^3 p_4^3 p_5^2+4032 p_1 p_2^3 p_3^3 p_4^3 p_5^3
  +3780 p_1 p_2^2 p_3^3 p_4^3 p_5^4) \textcolor{red}{{z}^{13}}+(630 p_1^2 p_2^3 p_3^3 p_4^3 p_5^3+1890 p_1 p_2^3 p_3^3 p_4^3 p_5^4+540 p_1 p_2^2 p_3^3 p_4^4 p_5^4) \textcolor{red}{{z}^{14}}
  +(336 p_1^2 p_2^3 p_3^3 p_4^3 p_5^4+480 p_1 p_2^3 p_3^3 p_4^4 p_5^4) \textcolor{red}{{z}^{15}}+(108 p_1^2 p_2^3 p_3^3 p_4^4 p_5^4+45 p_1 p_2^3 p_3^4 p_4^4 p_5^4) \textcolor{red}{{z}^{16}}
  +18 p_1^2 p_2^3 p_3^4 p_4^4 p_5^4 \textcolor{red}{{z}^{17}}+p_1^2 p_2^4 p_3^4 p_4^4 p_5^4 \textcolor{red}{{z}^{18}}
  \)
  \item[$H_3=$]\(
  1+24 p_3 \textcolor{red}{z}+(108 p_2 p_3+168 p_3 p_4) \textcolor{red}{{z}^2}+(120 p_1 p_2 p_3+1344 p_2 p_3 p_4+560 p_3 p_4 p_5) \textcolor{red}{{z}^3}
  +(1890 p_1 p_2 p_3 p_4+2016 p_2 p_3^2 p_4+5670 p_2 p_3 p_4 p_5+1050 p_3 p_4 p_5^2) \textcolor{red}{{z}^4}
  +(5040 p_1 p_2 p_3^2 p_4+9072 p_1 p_2 p_3 p_4 p_5+15120 p_2 p_3^2 p_4 p_5+12096 p_2 p_3 p_4 p_5^2+1176 p_3 p_4^2 p_5^2) \textcolor{red}{{z}^5}
  +(2520 p_1 p_2^2 p_3^2 p_4+43008 p_1 p_2 p_3^2 p_4 p_5+11760 p_2 p_3^2 p_4^2 p_5+21000 p_1 p_2 p_3 p_4 p_5^2
  +40824 p_2 p_3^2 p_4 p_5^2+14700 p_2 p_3 p_4^2 p_5^2+784 p_3^2 p_4^2 p_5^2) \textcolor{red}{{z}^6}+(27216 p_1 p_2^2 p_3^2 p_4 p_5
  +42336 p_1 p_2 p_3^2 p_4^2 p_5+126000 p_1 p_2 p_3^2 p_4 p_5^2+27000 p_1 p_2 p_3 p_4^2 p_5^2+123552 p_2 p_3^2 p_4^2 p_5^2) \textcolor{red}{{z}^7}
  +(47628 p_1 p_2^2 p_3^2 p_4^2 p_5+90720 p_1 p_2^2 p_3^2 p_4 p_5^2+424710 p_1 p_2 p_3^2 p_4^2 p_5^2
  +3969 p_2^2 p_3^2 p_4^2 p_5^2+43200 p_2 p_3^3 p_4^2 p_5^2+98784 p_2 p_3^2 p_4^3 p_5^2+26460 p_2 p_3^2 p_4^2 p_5^3) \textcolor{red}{{z}^8}
  +(14112 p_1 p_2^2 p_3^3 p_4^2 p_5+434720 p_1 p_2^2 p_3^2 p_4^2 p_5^2+147000 p_1 p_2 p_3^3 p_4^2 p_5^2+17496 p_2^2 p_3^3 p_4^2 p_5^2
  +408240 p_1 p_2 p_3^2 p_4^3 p_5^2+86016 p_2 p_3^3 p_4^3 p_5^2+117600 p_1 p_2 p_3^2 p_4^2 p_5^3+82320 p_2 p_3^2 p_4^3 p_5^3) \textcolor{red}{{z}^9}
  +(1296 p_1^2 p_2^2 p_3^2 p_4^2 p_5^2+291720 p_1 p_2^2 p_3^3 p_4^2 p_5^2+567000 p_1 p_2^2 p_3^2 p_4^3 p_5^2
  +370440 p_1 p_2 p_3^3 p_4^3 p_5^2+37800 p_2^2 p_3^3 p_4^3 p_5^2+190512 p_1 p_2^2 p_3^2 p_4^2 p_5^3+387072 p_1 p_2 p_3^2 p_4^3 p_5^3
  +90720 p_2 p_3^3 p_4^3 p_5^3+24696 p_2 p_3^2 p_4^3 p_5^4) \textcolor{red}{{z}^{10}}+(10584 p_1^2 p_2^2 p_3^3 p_4^2 p_5^2+52920 p_1 p_2^3 p_3^3 p_4^2 p_5^2
  +960960 p_1 p_2^2 p_3^3 p_4^3 p_5^2+127008 p_1 p_2^2 p_3^3 p_4^2 p_5^3+680400 p_1 p_2^2 p_3^2 p_4^3 p_5^3
  +444528 p_1 p_2 p_3^3 p_4^3 p_5^3+45360 p_2^2 p_3^3 p_4^3 p_5^3+126000 p_1 p_2 p_3^2 p_4^3 p_5^4+48384 p_2 p_3^3 p_4^3 p_5^4) \textcolor{red}{{z}^{11}}
  +(9408 p_1^2 p_2^3 p_3^3 p_4^2 p_5^2+30618 p_1^2 p_2^2 p_3^3 p_4^3 p_5^2+257250 p_1 p_2^3 p_3^3 p_4^3 p_5^2+252000 p_1 p_2^2 p_3^4 p_4^3 p_5^2
  +1605604 p_1 p_2^2 p_3^3 p_4^3 p_5^3+252000 p_1 p_2^2 p_3^2 p_4^3 p_5^4+257250 p_1 p_2 p_3^3 p_4^3 p_5^4
  +30618 p_2^2 p_3^3 p_4^3 p_5^4+9408 p_2 p_3^3 p_4^4 p_5^4) \textcolor{red}{{z}^{12}}+(48384 p_1^2 p_2^3 p_3^3 p_4^3 p_5^2
  +126000 p_1 p_2^3 p_3^4 p_4^3 p_5^2+45360 p_1^2 p_2^2 p_3^3 p_4^3 p_5^3+444528 p_1 p_2^3 p_3^3 p_4^3 p_5^3
  +680400 p_1 p_2^2 p_3^4 p_4^3 p_5^3+127008 p_1 p_2^2 p_3^3 p_4^4 p_5^3+960960 p_1 p_2^2 p_3^3 p_4^3 p_5^4
  +52920 p_1 p_2 p_3^3 p_4^4 p_5^4+10584 p_2^2 p_3^3 p_4^4 p_5^4) \textcolor{red}{{z}^{13}}+(24696 p_1^2 p_2^3 p_3^4 p_4^3 p_5^2+90720 p_1^2 p_2^3 p_3^3 p_4^3 p_5^3+387072 p_1 p_2^3 p_3^4 p_4^3 p_5^3+190512 p_1 p_2^2 p_3^4 p_4^4 p_5^3+37800 p_1^2 p_2^2 p_3^3 p_4^3 p_5^4+370440 p_1 p_2^3 p_3^3 p_4^3 p_5^4+567000 p_1 p_2^2 p_3^4 p_4^3 p_5^4+291720 p_1 p_2^2 p_3^3 p_4^4 p_5^4+1296 p_2^2 p_3^4 p_4^4 p_5^4) \textcolor{red}{{z}^{14}}+(82320 p_1^2 p_2^3 p_3^4 p_4^3 p_5^3+117600 p_1 p_2^3 p_3^4 p_4^4 p_5^3+86016 p_1^2 p_2^3 p_3^3 p_4^3 p_5^4+408240 p_1 p_2^3 p_3^4 p_4^3 p_5^4+17496 p_1^2 p_2^2 p_3^3 p_4^4 p_5^4+147000 p_1 p_2^3 p_3^3 p_4^4 p_5^4+434720 p_1 p_2^2 p_3^4 p_4^4 p_5^4+14112 p_1 p_2^2 p_3^3 p_4^4 p_5^5) \textcolor{red}{{z}^{15}}+(26460 p_1^2 p_2^3 p_3^4 p_4^4 p_5^3+98784 p_1^2 p_2^3 p_3^4 p_4^3 p_5^4+43200 p_1^2 p_2^3 p_3^3 p_4^4 p_5^4+3969 p_1^2 p_2^2 p_3^4 p_4^4 p_5^4+424710 p_1 p_2^3 p_3^4 p_4^4 p_5^4+90720 p_1 p_2^2 p_3^4 p_4^5 p_5^4+47628 p_1 p_2^2 p_3^4 p_4^4 p_5^5) \textcolor{red}{{z}^{16}}+(123552 p_1^2 p_2^3 p_3^4 p_4^4 p_5^4+27000 p_1 p_2^3 p_3^5 p_4^4 p_5^4+126000 p_1 p_2^3 p_3^4 p_4^5 p_5^4+42336 p_1 p_2^3 p_3^4 p_4^4 p_5^5+27216 p_1 p_2^2 p_3^4 p_4^5 p_5^5) \textcolor{red}{{z}^{17}}+(784 p_1^2 p_2^4 p_3^4 p_4^4 p_5^4+14700 p_1^2 p_2^3 p_3^5 p_4^4 p_5^4+40824 p_1^2 p_2^3 p_3^4 p_4^5 p_5^4+21000 p_1 p_2^3 p_3^5 p_4^5 p_5^4+11760 p_1^2 p_2^3 p_3^4 p_4^4 p_5^5+43008 p_1 p_2^3 p_3^4 p_4^5 p_5^5+2520 p_1 p_2^2 p_3^4 p_4^5 p_5^6) \textcolor{red}{{z}^{18}}+(1176 p_1^2 p_2^4 p_3^5 p_4^4 p_5^4+12096 p_1^2 p_2^3 p_3^5 p_4^5 p_5^4+15120 p_1^2 p_2^3 p_3^4 p_4^5 p_5^5+9072 p_1 p_2^3 p_3^5 p_4^5 p_5^5+5040 p_1 p_2^3 p_3^4 p_4^5 p_5^6) \textcolor{red}{{z}^{19}}+(1050 p_1^2 p_2^4 p_3^5 p_4^5 p_5^4+5670 p_1^2 p_2^3 p_3^5 p_4^5 p_5^5+2016 p_1^2 p_2^3 p_3^4 p_4^5 p_5^6+1890 p_1 p_2^3 p_3^5 p_4^5 p_5^6) \textcolor{red}{{z}^{20}}+(560 p_1^2 p_2^4 p_3^5 p_4^5 p_5^5+1344 p_1^2 p_2^3 p_3^5 p_4^5 p_5^6+120 p_1 p_2^3 p_3^5 p_4^6 p_5^6) \textcolor{red}{{z}^{21}}+(168 p_1^2 p_2^4 p_3^5 p_4^5 p_5^6+108 p_1^2 p_2^3 p_3^5 p_4^6 p_5^6) \textcolor{red}{{z}^{22}}+24 p_1^2 p_2^4 p_3^5 p_4^6 p_5^6 \textcolor{red}{{z}^{23}}+p_1^2 p_2^4 p_3^6 p_4^6 p_5^6 \textcolor{red}{{z}^{24}}
  \)
  \item[$H_4=$]\(
  1+28 p_4 \textcolor{red}{z}+(168 p_3 p_4+210 p_4 p_5) \textcolor{red}{{z}^2}+(336 p_2 p_3 p_4+2240 p_3 p_4 p_5+700 p_4 p_5^2) \textcolor{red}{{z}^3}
  +(210 p_1 p_2 p_3 p_4+5670 p_2 p_3 p_4 p_5+3920 p_3 p_4^2 p_5+9450 p_3 p_4 p_5^2+1225 p_4^2 p_5^2) \textcolor{red}{{z}^4}+(4032 p_1 p_2 p_3 p_4 p_5+17640 p_2 p_3 p_4^2 p_5+27216 p_2 p_3 p_4 p_5^2+49392 p_3 p_4^2 p_5^2) \textcolor{red}{{z}^5}+(15876 p_1 p_2 p_3 p_4^2 p_5+11760 p_2 p_3^2 p_4^2 p_5+21000 p_1 p_2 p_3 p_4 p_5^2+209916 p_2 p_3 p_4^2 p_5^2+19600 p_3^2 p_4^2 p_5^2+74088 p_3 p_4^3 p_5^2+24500 p_3 p_4^2 p_5^3) \textcolor{red}{{z}^6}+(18816 p_1 p_2 p_3^2 p_4^2 p_5+195120 p_1 p_2 p_3 p_4^2 p_5^2+202176 p_2 p_3^2 p_4^2 p_5^2+411600 p_2 p_3 p_4^3 p_5^2+87808 p_3^2 p_4^3 p_5^2+158760 p_2 p_3 p_4^2 p_5^3+109760 p_3 p_4^3 p_5^3) \textcolor{red}{{z}^7}+(5292 p_1 p_2^2 p_3^2 p_4^2 p_5+277830 p_1 p_2 p_3^2 p_4^2 p_5^2+35721 p_2^2 p_3^2 p_4^2 p_5^2+425250 p_1 p_2 p_3 p_4^3 p_5^2+961632 p_2 p_3^2 p_4^3 p_5^2+176400 p_1 p_2 p_3 p_4^2 p_5^3+238140 p_2 p_3^2 p_4^2 p_5^3+771750 p_2 p_3 p_4^3 p_5^3+164640 p_3^2 p_4^3 p_5^3+51450 p_3 p_4^3 p_5^4) \textcolor{red}{{z}^8}+(109760 p_1 p_2^2 p_3^2 p_4^2 p_5^2+1292760 p_1 p_2 p_3^2 p_4^3 p_5^2+308700 p_2^2 p_3^2 p_4^3 p_5^2+537600 p_2 p_3^3 p_4^3 p_5^2+470400 p_1 p_2 p_3^2 p_4^2 p_5^3+907200 p_1 p_2 p_3 p_4^3 p_5^3+2731680 p_2 p_3^2 p_4^3 p_5^3+411600 p_2 p_3 p_4^3 p_5^4+137200 p_3^2 p_4^3 p_5^4) \textcolor{red}{{z}^9}+(7056 p_1^2 p_2^2 p_3^2 p_4^2 p_5^2+666680 p_1 p_2^2 p_3^2 p_4^3 p_5^2+1029000 p_1 p_2 p_3^3 p_4^3 p_5^2+340200 p_2^2 p_3^3 p_4^3 p_5^2+190512 p_1 p_2^2 p_3^2 p_4^2 p_5^3+4484844 p_1 p_2 p_3^2 p_4^3 p_5^3+833490 p_2^2 p_3^2 p_4^3 p_5^3+2268000 p_2 p_3^3 p_4^3 p_5^3+576240 p_2 p_3^2 p_4^4 p_5^3+525000 p_1 p_2 p_3 p_4^3 p_5^4+2163672 p_2 p_3^2 p_4^3 p_5^4+38416 p_3^2 p_4^4 p_5^4) \textcolor{red}{{z}^{10}}+(81648 p_1^2 p_2^2 p_3^2 p_4^3 p_5^2+1132320 p_1 p_2^2 p_3^3 p_4^3 p_5^2+2621472 p_1 p_2^2 p_3^2 p_4^3 p_5^3+4939200 p_1 p_2 p_3^3 p_4^3 p_5^3+1632960 p_2^2 p_3^3 p_4^3 p_5^3+1524096 p_1 p_2 p_3^2 p_4^4 p_5^3+1128960 p_2 p_3^3 p_4^4 p_5^3+3591000 p_1 p_2 p_3^2 p_4^3 p_5^4+1000188 p_2^2 p_3^2 p_4^3 p_5^4+2721600 p_2 p_3^3 p_4^3 p_5^4+1100736 p_2 p_3^2 p_4^4 p_5^4) \textcolor{red}{{z}^{11}}+(166698 p_1^2 p_2^2 p_3^3 p_4^3 p_5^2+257250 p_1 p_2^3 p_3^3 p_4^3 p_5^2+272160 p_1^2 p_2^2 p_3^2 p_4^3 p_5^3+6419812 p_1 p_2^2 p_3^3 p_4^3 p_5^3+1190700 p_1 p_2^2 p_3^2 p_4^4 p_5^3+3111696 p_1 p_2 p_3^3 p_4^4 p_5^3+882000 p_2^2 p_3^3 p_4^4 p_5^3+2666720 p_1 p_2^2 p_3^2 p_4^3 p_5^4+6431250 p_1 p_2 p_3^3 p_4^3 p_5^4+2480058 p_2^2 p_3^3 p_4^3 p_5^4+2500470 p_1 p_2 p_3^2 p_4^4 p_5^4+540225 p_2^2 p_3^2 p_4^4 p_5^4+3358656 p_2 p_3^3 p_4^4 p_5^4+144060 p_2 p_3^2 p_4^4 p_5^5) \textcolor{red}{{z}^{12}}+(65856 p_1^2 p_2^3 p_3^3 p_4^3 p_5^2+987840 p_1^2 p_2^2 p_3^3 p_4^3 p_5^3+1778112 p_1 p_2^3 p_3^3 p_4^3 p_5^3+5551504 p_1 p_2^2 p_3^3 p_4^4 p_5^3+403200 p_1^2 p_2^2 p_3^2 p_4^3 p_5^4+10190880 p_1 p_2^2 p_3^3 p_4^3 p_5^4+2744000 p_1 p_2^2 p_3^2 p_4^4 p_5^4+9560880 p_1 p_2 p_3^3 p_4^4 p_5^4+4000752 p_2^2 p_3^3 p_4^4 p_5^4+1053696 p_2 p_3^3 p_4^5 p_5^4+470400 p_1 p_2 p_3^2 p_4^4 p_5^5+635040 p_2 p_3^3 p_4^4 p_5^5) \textcolor{red}{{z}^{13}}+(493920 p_1^2 p_2^3 p_3^3 p_4^3 p_5^3+714420 p_1^2 p_2^2 p_3^3 p_4^4 p_5^3+2160900 p_1 p_2^3 p_3^3 p_4^4 p_5^3+529200 p_1 p_2^2 p_3^4 p_4^4 p_5^3+1852200 p_1^2 p_2^2 p_3^3 p_4^3 p_5^4+3333960 p_1 p_2^3 p_3^3 p_4^3 p_5^4+291600 p_1^2 p_2^2 p_3^2 p_4^4 p_5^4+21364200 p_1 p_2^2 p_3^3 p_4^4 p_5^4+291600 p_2^2 p_3^4 p_4^4 p_5^4+3333960 p_1 p_2 p_3^3 p_4^5 p_5^4+1852200 p_2^2 p_3^3 p_4^5 p_5^4+529200 p_1 p_2^2 p_3^2 p_4^4 p_5^5+2160900 p_1 p_2 p_3^3 p_4^4 p_5^5+714420 p_2^2 p_3^3 p_4^4 p_5^5+493920 p_2 p_3^3 p_4^5 p_5^5) \textcolor{red}{{z}^{14}}+(635040 p_1^2 p_2^3 p_3^3 p_4^4 p_5^3+470400 p_1 p_2^3 p_3^4 p_4^4 p_5^3+1053696 p_1^2 p_2^3 p_3^3 p_4^3 p_5^4+4000752 p_1^2 p_2^2 p_3^3 p_4^4 p_5^4+9560880 p_1 p_2^3 p_3^3 p_4^4 p_5^4+2744000 p_1 p_2^2 p_3^4 p_4^4 p_5^4+10190880 p_1 p_2^2 p_3^3 p_4^5 p_5^4+403200 p_2^2 p_3^4 p_4^5 p_5^4+5551504 p_1 p_2^2 p_3^3 p_4^4 p_5^5+1778112 p_1 p_2 p_3^3 p_4^5 p_5^5+987840 p_2^2 p_3^3 p_4^5 p_5^5+65856 p_2 p_3^3 p_4^5 p_5^6) \textcolor{red}{{z}^{15}}+(144060 p_1^2 p_2^3 p_3^4 p_4^4 p_5^3+3358656 p_1^2 p_2^3 p_3^3 p_4^4 p_5^4+540225 p_1^2 p_2^2 p_3^4 p_4^4 p_5^4+2500470 p_1 p_2^3 p_3^4 p_4^4 p_5^4+2480058 p_1^2 p_2^2 p_3^3 p_4^5 p_5^4+6431250 p_1 p_2^3 p_3^3 p_4^5 p_5^4+2666720 p_1 p_2^2 p_3^4 p_4^5 p_5^4+882000 p_1^2 p_2^2 p_3^3 p_4^4 p_5^5+3111696 p_1 p_2^3 p_3^3 p_4^4 p_5^5+1190700 p_1 p_2^2 p_3^4 p_4^4 p_5^5+6419812 p_1 p_2^2 p_3^3 p_4^5 p_5^5+272160 p_2^2 p_3^4 p_4^5 p_5^5+257250 p_1 p_2 p_3^3 p_4^5 p_5^6+166698 p_2^2 p_3^3 p_4^5 p_5^6) \textcolor{red}{{z}^{16}}+(1100736 p_1^2 p_2^3 p_3^4 p_4^4 p_5^4+2721600 p_1^2 p_2^3 p_3^3 p_4^5 p_5^4+1000188 p_1^2 p_2^2 p_3^4 p_4^5 p_5^4+3591000 p_1 p_2^3 p_3^4 p_4^5 p_5^4+1128960 p_1^2 p_2^3 p_3^3 p_4^4 p_5^5+1524096 p_1 p_2^3 p_3^4 p_4^4 p_5^5+1632960 p_1^2 p_2^2 p_3^3 p_4^5 p_5^5+4939200 p_1 p_2^3 p_3^3 p_4^5 p_5^5+2621472 p_1 p_2^2 p_3^4 p_4^5 p_5^5+1132320 p_1 p_2^2 p_3^3 p_4^5 p_5^6+81648 p_2^2 p_3^4 p_4^5 p_5^6) \textcolor{red}{{z}^{17}}+(38416 p_1^2 p_2^4 p_3^4 p_4^4 p_5^4+2163672 p_1^2 p_2^3 p_3^4 p_4^5 p_5^4+525000 p_1 p_2^3 p_3^5 p_4^5 p_5^4+576240 p_1^2 p_2^3 p_3^4 p_4^4 p_5^5+2268000 p_1^2 p_2^3 p_3^3 p_4^5 p_5^5+833490 p_1^2 p_2^2 p_3^4 p_4^5 p_5^5+4484844 p_1 p_2^3 p_3^4 p_4^5 p_5^5+190512 p_1 p_2^2 p_3^4 p_4^6 p_5^5+340200 p_1^2 p_2^2 p_3^3 p_4^5 p_5^6+1029000 p_1 p_2^3 p_3^3 p_4^5 p_5^6+666680 p_1 p_2^2 p_3^4 p_4^5 p_5^6+7056 p_2^2 p_3^4 p_4^6 p_5^6) \textcolor{red}{{z}^{18}}+(137200 p_1^2 p_2^4 p_3^4 p_4^5 p_5^4+411600 p_1^2 p_2^3 p_3^5 p_4^5 p_5^4+2731680 p_1^2 p_2^3 p_3^4 p_4^5 p_5^5+907200 p_1 p_2^3 p_3^5 p_4^5 p_5^5+470400 p_1 p_2^3 p_3^4 p_4^6 p_5^5+537600 p_1^2 p_2^3 p_3^3 p_4^5 p_5^6+308700 p_1^2 p_2^2 p_3^4 p_4^5 p_5^6+1292760 p_1 p_2^3 p_3^4 p_4^5 p_5^6+109760 p_1 p_2^2 p_3^4 p_4^6 p_5^6) \textcolor{red}{{z}^{19}}+(51450 p_1^2 p_2^4 p_3^5 p_4^5 p_5^4+164640 p_1^2 p_2^4 p_3^4 p_4^5 p_5^5+771750 p_1^2 p_2^3 p_3^5 p_4^5 p_5^5+238140 p_1^2 p_2^3 p_3^4 p_4^6 p_5^5+176400 p_1 p_2^3 p_3^5 p_4^6 p_5^5+961632 p_1^2 p_2^3 p_3^4 p_4^5 p_5^6+425250 p_1 p_2^3 p_3^5 p_4^5 p_5^6+35721 p_1^2 p_2^2 p_3^4 p_4^6 p_5^6+277830 p_1 p_2^3 p_3^4 p_4^6 p_5^6+5292 p_1 p_2^2 p_3^4 p_4^6 p_5^7) \textcolor{red}{{z}^{20}}+(109760 p_1^2 p_2^4 p_3^5 p_4^5 p_5^5+158760 p_1^2 p_2^3 p_3^5 p_4^6 p_5^5+87808 p_1^2 p_2^4 p_3^4 p_4^5 p_5^6+411600 p_1^2 p_2^3 p_3^5 p_4^5 p_5^6+202176 p_1^2 p_2^3 p_3^4 p_4^6 p_5^6+195120 p_1 p_2^3 p_3^5 p_4^6 p_5^6+18816 p_1 p_2^3 p_3^4 p_4^6 p_5^7) \textcolor{red}{{z}^{21}}+(24500 p_1^2 p_2^4 p_3^5 p_4^6 p_5^5+74088 p_1^2 p_2^4 p_3^5 p_4^5 p_5^6+19600 p_1^2 p_2^4 p_3^4 p_4^6 p_5^6+209916 p_1^2 p_2^3 p_3^5 p_4^6 p_5^6+21000 p_1 p_2^3 p_3^5 p_4^7 p_5^6+11760 p_1^2 p_2^3 p_3^4 p_4^6 p_5^7+15876 p_1 p_2^3 p_3^5 p_4^6 p_5^7) \textcolor{red}{{z}^{22}}+(49392 p_1^2 p_2^4 p_3^5 p_4^6 p_5^6+27216 p_1^2 p_2^3 p_3^5 p_4^7 p_5^6+17640 p_1^2 p_2^3 p_3^5 p_4^6 p_5^7+4032 p_1 p_2^3 p_3^5 p_4^7 p_5^7) \textcolor{red}{{z}^{23}}+(1225 p_1^2 p_2^4 p_3^6 p_4^6 p_5^6+9450 p_1^2 p_2^4 p_3^5 p_4^7 p_5^6+3920 p_1^2 p_2^4 p_3^5 p_4^6 p_5^7+5670 p_1^2 p_2^3 p_3^5 p_4^7 p_5^7+210 p_1 p_2^3 p_3^5 p_4^7 p_5^8) \textcolor{red}{{z}^{24}}+(700 p_1^2 p_2^4 p_3^6 p_4^7 p_5^6+2240 p_1^2 p_2^4 p_3^5 p_4^7 p_5^7+336 p_1^2 p_2^3 p_3^5 p_4^7 p_5^8) \textcolor{red}{{z}^{25}}+(210 p_1^2 p_2^4 p_3^6 p_4^7 p_5^7+168 p_1^2 p_2^4 p_3^5 p_4^7 p_5^8) \textcolor{red}{{z}^{26}}+28 p_1^2 p_2^4 p_3^6 p_4^7 p_5^8 \textcolor{red}{{z}^{27}}+p_1^2 p_2^4 p_3^6 p_4^8 p_5^8 \textcolor{red}{{z}^{28}}
  \)
  \item[$H_5=$]\(
  1+15 p_5 \textcolor{red}{z}+105 p_4 p_5 \textcolor{red}{{z}^2}+(280 p_3 p_4 p_5+175 p_4 p_5^2) \textcolor{red}{{z}^3}+(315 p_2 p_3 p_4 p_5+1050 p_3 p_4 p_5^2) \textcolor{red}{{z}^4}+(126 p_1 p_2 p_3 p_4 p_5+1701 p_2 p_3 p_4 p_5^2+1176 p_3 p_4^2 p_5^2) \textcolor{red}{{z}^5}+(840 p_1 p_2 p_3 p_4 p_5^2+3675 p_2 p_3 p_4^2 p_5^2+490 p_3 p_4^2 p_5^3) \textcolor{red}{{z}^6}+(2430 p_1 p_2 p_3 p_4^2 p_5^2+1800 p_2 p_3^2 p_4^2 p_5^2+2205 p_2 p_3 p_4^2 p_5^3) \textcolor{red}{{z}^7}+(2205 p_1 p_2 p_3^2 p_4^2 p_5^2+1800 p_1 p_2 p_3 p_4^2 p_5^3+2430 p_2 p_3^2 p_4^2 p_5^3) \textcolor{red}{{z}^8}+(490 p_1 p_2^2 p_3^2 p_4^2 p_5^2+3675 p_1 p_2 p_3^2 p_4^2 p_5^3+840 p_2 p_3^2 p_4^3 p_5^3) \textcolor{red}{{z}^9}+(1176 p_1 p_2^2 p_3^2 p_4^2 p_5^3+1701 p_1 p_2 p_3^2 p_4^3 p_5^3+126 p_2 p_3^2 p_4^3 p_5^4) \textcolor{red}{{z}^{10}}+(1050 p_1 p_2^2 p_3^2 p_4^3 p_5^3+315 p_1 p_2 p_3^2 p_4^3 p_5^4) \textcolor{red}{{z}^{11}}+(175 p_1 p_2^2 p_3^3 p_4^3 p_5^3+280 p_1 p_2^2 p_3^2 p_4^3 p_5^4) \textcolor{red}{{z}^{12}}+105 p_1 p_2^2 p_3^3 p_4^3 p_5^4 \textcolor{red}{{z}^{13}}+15 p_1 p_2^2 p_3^3 p_4^4 p_5^4 \textcolor{red}{{z}^{14}}+p_1 p_2^2 p_3^3 p_4^4 p_5^5 \textcolor{red}{{z}^{15}}
  \)
  \end{itemize}

  \bigskip 
  \noindent { \bf $C_5$-case. }
  For the Lie algebra $C_5 \cong sp(5)$ we obtain
  \begin{itemize}
  \item[$H_1=$]\(
  1+9 p_1 \textcolor{red}{z}+36 p_1 p_2 \textcolor{red}{{z}^2}+84 p_1 p_2 p_3 \textcolor{red}{{z}^3}+126 p_1 p_2 p_3 p_4 \textcolor{red}{{z}^4}+126 p_1 p_2 p_3 p_4 p_5 \textcolor{red}{{z}^5}+84 p_1 p_2 p_3 p_4^2 p_5 \textcolor{red}{{z}^6}+36 p_1 p_2 p_3^2 p_4^2 p_5 \textcolor{red}{{z}^7}+9 p_1 p_2^2 p_3^2 p_4^2 p_5 \textcolor{red}{{z}^8}+p_1^2 p_2^2 p_3^2 p_4^2 p_5 \textcolor{red}{{z}^9}
  \)
  \item[$H_2=$]\(
  1+16 p_2 \textcolor{red}{z}+(36 p_1 p_2+84 p_2 p_3) \textcolor{red}{{z}^2}+(336 p_1 p_2 p_3+224 p_2 p_3 p_4) \textcolor{red}{{z}^3}+(336 p_1 p_2^2 p_3+1134 p_1 p_2 p_3 p_4+350 p_2 p_3 p_4 p_5) \textcolor{red}{{z}^4}+(2016 p_1 p_2^2 p_3 p_4+2016 p_1 p_2 p_3 p_4 p_5+336 p_2 p_3 p_4^2 p_5) \textcolor{red}{{z}^5}+(1176 p_1 p_2^2 p_3^2 p_4+4536 p_1 p_2^2 p_3 p_4 p_5+2100 p_1 p_2 p_3 p_4^2 p_5+196 p_2 p_3^2 p_4^2 p_5) \textcolor{red}{{z}^6}+(4704 p_1 p_2^2 p_3^2 p_4 p_5+5376 p_1 p_2^2 p_3 p_4^2 p_5+1296 p_1 p_2 p_3^2 p_4^2 p_5+64 p_2^2 p_3^2 p_4^2 p_5) \textcolor{red}{{z}^7}+12870 p_1 p_2^2 p_3^2 p_4^2 p_5 \textcolor{red}{{z}^8}+(64 p_1^2 p_2^2 p_3^2 p_4^2 p_5+1296 p_1 p_2^3 p_3^2 p_4^2 p_5+5376 p_1 p_2^2 p_3^3 p_4^2 p_5+4704 p_1 p_2^2 p_3^2 p_4^3 p_5) \textcolor{red}{{z}^9}+(196 p_1^2 p_2^3 p_3^2 p_4^2 p_5+2100 p_1 p_2^3 p_3^3 p_4^2 p_5+4536 p_1 p_2^2 p_3^3 p_4^3 p_5+1176 p_1 p_2^2 p_3^2 p_4^3 p_5^2) \textcolor{red}{{z}^{10}}+(336 p_1^2 p_2^3 p_3^3 p_4^2 p_5+2016 p_1 p_2^3 p_3^3 p_4^3 p_5+2016 p_1 p_2^2 p_3^3 p_4^3 p_5^2) \textcolor{red}{{z}^{11}}+(350 p_1^2 p_2^3 p_3^3 p_4^3 p_5+1134 p_1 p_2^3 p_3^3 p_4^3 p_5^2+336 p_1 p_2^2 p_3^3 p_4^4 p_5^2) \textcolor{red}{{z}^{12}}+(224 p_1^2 p_2^3 p_3^3 p_4^3 p_5^2+336 p_1 p_2^3 p_3^3 p_4^4 p_5^2) \textcolor{red}{{z}^{13}}+(84 p_1^2 p_2^3 p_3^3 p_4^4 p_5^2+36 p_1 p_2^3 p_3^4 p_4^4 p_5^2) \textcolor{red}{{z}^{14}}+16 p_1^2 p_2^3 p_3^4 p_4^4 p_5^2 \textcolor{red}{{z}^{15}}+p_1^2 p_2^4 p_3^4 p_4^4 p_5^2 \textcolor{red}{{z}^{16}}
  \)
  \item[$H_3=$]\(
  1+21 p_3 \textcolor{red}{z}+(84 p_2 p_3+126 p_3 p_4) \textcolor{red}{{z}^2}+(84 p_1 p_2 p_3+896 p_2 p_3 p_4+350 p_3 p_4 p_5) \textcolor{red}{{z}^3}+(1134 p_1 p_2 p_3 p_4+1176 p_2 p_3^2 p_4+3150 p_2 p_3 p_4 p_5+525 p_3 p_4^2 p_5) \textcolor{red}{{z}^4}+(2646 p_1 p_2 p_3^2 p_4+4536 p_1 p_2 p_3 p_4 p_5+7350 p_2 p_3^2 p_4 p_5+5376 p_2 p_3 p_4^2 p_5+441 p_3^2 p_4^2 p_5) \textcolor{red}{{z}^5}+(1176 p_1 p_2^2 p_3^2 p_4+18816 p_1 p_2 p_3^2 p_4 p_5+8400 p_1 p_2 p_3 p_4^2 p_5+25872 p_2 p_3^2 p_4^2 p_5) \textcolor{red}{{z}^6}+(10584 p_1 p_2^2 p_3^2 p_4 p_5+68112 p_1 p_2 p_3^2 p_4^2 p_5+2304 p_2^2 p_3^2 p_4^2 p_5+16464 p_2 p_3^3 p_4^2 p_5+18816 p_2 p_3^2 p_4^3 p_5) \textcolor{red}{{z}^7}+(48510 p_1 p_2^2 p_3^2 p_4^2 p_5+48384 p_1 p_2 p_3^3 p_4^2 p_5+8400 p_2^2 p_3^3 p_4^2 p_5+66150 p_1 p_2 p_3^2 p_4^3 p_5+24696 p_2 p_3^3 p_4^3 p_5+7350 p_2 p_3^2 p_4^3 p_5^2) \textcolor{red}{{z}^8}+(784 p_1^2 p_2^2 p_3^2 p_4^2 p_5+65142 p_1 p_2^2 p_3^3 p_4^2 p_5+75264 p_1 p_2^2 p_3^2 p_4^3 p_5+91854 p_1 p_2 p_3^3 p_4^3 p_5+14336 p_2^2 p_3^3 p_4^3 p_5+29400 p_1 p_2 p_3^2 p_4^3 p_5^2+17150 p_2 p_3^3 p_4^3 p_5^2) \textcolor{red}{{z}^9}+(5376 p_1^2 p_2^2 p_3^3 p_4^2 p_5+18900 p_1 p_2^3 p_3^3 p_4^2 p_5+196812 p_1 p_2^2 p_3^3 p_4^3 p_5+42336 p_1 p_2^2 p_3^2 p_4^3 p_5^2+72576 p_1 p_2 p_3^3 p_4^3 p_5^2+12600 p_2^2 p_3^3 p_4^3 p_5^2+4116 p_2 p_3^3 p_4^4 p_5^2) \textcolor{red}{{z}^{10}}+(4116 p_1^2 p_2^3 p_3^3 p_4^2 p_5+12600 p_1^2 p_2^2 p_3^3 p_4^3 p_5+72576 p_1 p_2^3 p_3^3 p_4^3 p_5+42336 p_1 p_2^2 p_3^4 p_4^3 p_5+196812 p_1 p_2^2 p_3^3 p_4^3 p_5^2+18900 p_1 p_2 p_3^3 p_4^4 p_5^2+5376 p_2^2 p_3^3 p_4^4 p_5^2) \textcolor{red}{{z}^{11}}+(17150 p_1^2 p_2^3 p_3^3 p_4^3 p_5+29400 p_1 p_2^3 p_3^4 p_4^3 p_5+14336 p_1^2 p_2^2 p_3^3 p_4^3 p_5^2+91854 p_1 p_2^3 p_3^3 p_4^3 p_5^2+75264 p_1 p_2^2 p_3^4 p_4^3 p_5^2+65142 p_1 p_2^2 p_3^3 p_4^4 p_5^2+784 p_2^2 p_3^4 p_4^4 p_5^2) \textcolor{red}{{z}^{12}}+(7350 p_1^2 p_2^3 p_3^4 p_4^3 p_5+24696 p_1^2 p_2^3 p_3^3 p_4^3 p_5^2+66150 p_1 p_2^3 p_3^4 p_4^3 p_5^2+8400 p_1^2 p_2^2 p_3^3 p_4^4 p_5^2+48384 p_1 p_2^3 p_3^3 p_4^4 p_5^2+48510 p_1 p_2^2 p_3^4 p_4^4 p_5^2) \textcolor{red}{{z}^{13}}+(18816 p_1^2 p_2^3 p_3^4 p_4^3 p_5^2+16464 p_1^2 p_2^3 p_3^3 p_4^4 p_5^2+2304 p_1^2 p_2^2 p_3^4 p_4^4 p_5^2+68112 p_1 p_2^3 p_3^4 p_4^4 p_5^2+10584 p_1 p_2^2 p_3^4 p_4^5 p_5^2) \textcolor{red}{{z}^{14}}+(25872 p_1^2 p_2^3 p_3^4 p_4^4 p_5^2+8400 p_1 p_2^3 p_3^5 p_4^4 p_5^2+18816 p_1 p_2^3 p_3^4 p_4^5 p_5^2+1176 p_1 p_2^2 p_3^4 p_4^5 p_5^3) \textcolor{red}{{z}^{15}}+(441 p_1^2 p_2^4 p_3^4 p_4^4 p_5^2+5376 p_1^2 p_2^3 p_3^5 p_4^4 p_5^2+7350 p_1^2 p_2^3 p_3^4 p_4^5 p_5^2+4536 p_1 p_2^3 p_3^5 p_4^5 p_5^2+2646 p_1 p_2^3 p_3^4 p_4^5 p_5^3) \textcolor{red}{{z}^{16}}+(525 p_1^2 p_2^4 p_3^5 p_4^4 p_5^2+3150 p_1^2 p_2^3 p_3^5 p_4^5 p_5^2+1176 p_1^2 p_2^3 p_3^4 p_4^5 p_5^3+1134 p_1 p_2^3 p_3^5 p_4^5 p_5^3) \textcolor{red}{{z}^{17}}+(350 p_1^2 p_2^4 p_3^5 p_4^5 p_5^2+896 p_1^2 p_2^3 p_3^5 p_4^5 p_5^3+84 p_1 p_2^3 p_3^5 p_4^6 p_5^3) \textcolor{red}{{z}^{18}}+(126 p_1^2 p_2^4 p_3^5 p_4^5 p_5^3+84 p_1^2 p_2^3 p_3^5 p_4^6 p_5^3) \textcolor{red}{{z}^{19}}+21 p_1^2 p_2^4 p_3^5 p_4^6 p_5^3 \textcolor{red}{{z}^{20}}+p_1^2 p_2^4 p_3^6 p_4^6 p_5^3 \textcolor{red}{{z}^{21}}
  \)
  \item[$H_4=$]\(
  1+24 p_4 \textcolor{red}{z}+(126 p_3 p_4+150 p_4 p_5) \textcolor{red}{{z}^2}+(224 p_2 p_3 p_4+1400 p_3 p_4 p_5+400 p_4^2 p_5) \textcolor{red}{{z}^3}+(126 p_1 p_2 p_3 p_4+3150 p_2 p_3 p_4 p_5+7350 p_3 p_4^2 p_5) \textcolor{red}{{z}^4}+(2016 p_1 p_2 p_3 p_4 p_5+20832 p_2 p_3 p_4^2 p_5+7056 p_3^2 p_4^2 p_5+12600 p_3 p_4^3 p_5) \textcolor{red}{{z}^5}+(15288 p_1 p_2 p_3 p_4^2 p_5+29400 p_2 p_3^2 p_4^2 p_5+57344 p_2 p_3 p_4^3 p_5+23814 p_3^2 p_4^3 p_5+8750 p_3 p_4^3 p_5^2) \textcolor{red}{{z}^6}+(22752 p_1 p_2 p_3^2 p_4^2 p_5+14400 p_2^2 p_3^2 p_4^2 p_5+50400 p_1 p_2 p_3 p_4^3 p_5+178752 p_2 p_3^2 p_4^3 p_5+50400 p_2 p_3 p_4^3 p_5^2+29400 p_3^2 p_4^3 p_5^2) \textcolor{red}{{z}^7}+(16758 p_1 p_2^2 p_3^2 p_4^2 p_5+180900 p_1 p_2 p_3^2 p_4^3 p_5+98304 p_2^2 p_3^2 p_4^3 p_5+98784 p_2 p_3^3 p_4^3 p_5+50400 p_1 p_2 p_3 p_4^3 p_5^2+279300 p_2 p_3^2 p_4^3 p_5^2+11025 p_3^2 p_4^4 p_5^2) \textcolor{red}{{z}^8}+(3136 p_1^2 p_2^2 p_3^2 p_4^2 p_5+143472 p_1 p_2^2 p_3^2 p_4^3 p_5+163296 p_1 p_2 p_3^3 p_4^3 p_5+89600 p_2^2 p_3^3 p_4^3 p_5+321600 p_1 p_2 p_3^2 p_4^3 p_5^2+194400 p_2^2 p_3^2 p_4^3 p_5^2+274400 p_2 p_3^3 p_4^3 p_5^2+117600 p_2 p_3^2 p_4^4 p_5^2) \textcolor{red}{{z}^9}+(29400 p_1^2 p_2^2 p_3^2 p_4^3 p_5+233100 p_1 p_2^2 p_3^3 p_4^3 p_5+322812 p_1 p_2^2 p_3^2 p_4^3 p_5^2+516096 p_1 p_2 p_3^3 p_4^3 p_5^2+315000 p_2^2 p_3^3 p_4^3 p_5^2+142200 p_1 p_2 p_3^2 p_4^4 p_5^2+147456 p_2^2 p_3^2 p_4^4 p_5^2+255192 p_2 p_3^3 p_4^4 p_5^2) \textcolor{red}{{z}^{10}}+(50400 p_1^2 p_2^2 p_3^3 p_4^3 p_5+50400 p_1 p_2^3 p_3^3 p_4^3 p_5+75264 p_1^2 p_2^2 p_3^2 p_4^3 p_5^2+932400 p_1 p_2^2 p_3^3 p_4^3 p_5^2+268128 p_1 p_2^2 p_3^2 p_4^4 p_5^2+550368 p_1 p_2 p_3^3 p_4^4 p_5^2+470400 p_2^2 p_3^3 p_4^4 p_5^2+98784 p_2 p_3^3 p_4^5 p_5^2) \textcolor{red}{{z}^{11}}+(17150 p_1^2 p_2^3 p_3^3 p_4^3 p_5+229376 p_1^2 p_2^2 p_3^3 p_4^3 p_5^2+255150 p_1 p_2^3 p_3^3 p_4^3 p_5^2+78400 p_1^2 p_2^2 p_3^2 p_4^4 p_5^2+1544004 p_1 p_2^2 p_3^3 p_4^4 p_5^2+78400 p_2^2 p_3^4 p_4^4 p_5^2+255150 p_1 p_2 p_3^3 p_4^5 p_5^2+229376 p_2^2 p_3^3 p_4^5 p_5^2+17150 p_2 p_3^3 p_4^5 p_5^3) \textcolor{red}{{z}^{12}}+(98784 p_1^2 p_2^3 p_3^3 p_4^3 p_5^2+470400 p_1^2 p_2^2 p_3^3 p_4^4 p_5^2+550368 p_1 p_2^3 p_3^3 p_4^4 p_5^2+268128 p_1 p_2^2 p_3^4 p_4^4 p_5^2+932400 p_1 p_2^2 p_3^3 p_4^5 p_5^2+75264 p_2^2 p_3^4 p_4^5 p_5^2+50400 p_1 p_2 p_3^3 p_4^5 p_5^3+50400 p_2^2 p_3^3 p_4^5 p_5^3) \textcolor{red}{{z}^{13}}+(255192 p_1^2 p_2^3 p_3^3 p_4^4 p_5^2+147456 p_1^2 p_2^2 p_3^4 p_4^4 p_5^2+142200 p_1 p_2^3 p_3^4 p_4^4 p_5^2+315000 p_1^2 p_2^2 p_3^3 p_4^5 p_5^2+516096 p_1 p_2^3 p_3^3 p_4^5 p_5^2+322812 p_1 p_2^2 p_3^4 p_4^5 p_5^2+233100 p_1 p_2^2 p_3^3 p_4^5 p_5^3+29400 p_2^2 p_3^4 p_4^5 p_5^3) \textcolor{red}{{z}^{14}}+(117600 p_1^2 p_2^3 p_3^4 p_4^4 p_5^2+274400 p_1^2 p_2^3 p_3^3 p_4^5 p_5^2+194400 p_1^2 p_2^2 p_3^4 p_4^5 p_5^2+321600 p_1 p_2^3 p_3^4 p_4^5 p_5^2+89600 p_1^2 p_2^2 p_3^3 p_4^5 p_5^3+163296 p_1 p_2^3 p_3^3 p_4^5 p_5^3+143472 p_1 p_2^2 p_3^4 p_4^5 p_5^3+3136 p_2^2 p_3^4 p_4^6 p_5^3) \textcolor{red}{{z}^{15}}+(11025 p_1^2 p_2^4 p_3^4 p_4^4 p_5^2+279300 p_1^2 p_2^3 p_3^4 p_4^5 p_5^2+50400 p_1 p_2^3 p_3^5 p_4^5 p_5^2+98784 p_1^2 p_2^3 p_3^3 p_4^5 p_5^3+98304 p_1^2 p_2^2 p_3^4 p_4^5 p_5^3+180900 p_1 p_2^3 p_3^4 p_4^5 p_5^3+16758 p_1 p_2^2 p_3^4 p_4^6 p_5^3) \textcolor{red}{{z}^{16}}+(29400 p_1^2 p_2^4 p_3^4 p_4^5 p_5^2+50400 p_1^2 p_2^3 p_3^5 p_4^5 p_5^2+178752 p_1^2 p_2^3 p_3^4 p_4^5 p_5^3+50400 p_1 p_2^3 p_3^5 p_4^5 p_5^3+14400 p_1^2 p_2^2 p_3^4 p_4^6 p_5^3+22752 p_1 p_2^3 p_3^4 p_4^6 p_5^3) \textcolor{red}{{z}^{17}}+(8750 p_1^2 p_2^4 p_3^5 p_4^5 p_5^2+23814 p_1^2 p_2^4 p_3^4 p_4^5 p_5^3+57344 p_1^2 p_2^3 p_3^5 p_4^5 p_5^3+29400 p_1^2 p_2^3 p_3^4 p_4^6 p_5^3+15288 p_1 p_2^3 p_3^5 p_4^6 p_5^3) \textcolor{red}{{z}^{18}}+(12600 p_1^2 p_2^4 p_3^5 p_4^5 p_5^3+7056 p_1^2 p_2^4 p_3^4 p_4^6 p_5^3+20832 p_1^2 p_2^3 p_3^5 p_4^6 p_5^3+2016 p_1 p_2^3 p_3^5 p_4^7 p_5^3) \textcolor{red}{{z}^{19}}+(7350 p_1^2 p_2^4 p_3^5 p_4^6 p_5^3+3150 p_1^2 p_2^3 p_3^5 p_4^7 p_5^3+126 p_1 p_2^3 p_3^5 p_4^7 p_5^4) \textcolor{red}{{z}^{20}}+(400 p_1^2 p_2^4 p_3^6 p_4^6 p_5^3+1400 p_1^2 p_2^4 p_3^5 p_4^7 p_5^3+224 p_1^2 p_2^3 p_3^5 p_4^7 p_5^4) \textcolor{red}{{z}^{21}}+(150 p_1^2 p_2^4 p_3^6 p_4^7 p_5^3+126 p_1^2 p_2^4 p_3^5 p_4^7 p_5^4) \textcolor{red}{{z}^{22}}+24 p_1^2 p_2^4 p_3^6 p_4^7 p_5^4 \textcolor{red}{{z}^{23}}+p_1^2 p_2^4 p_3^6 p_4^8 p_5^4 \textcolor{red}{{z}^{24}}
  \)
  \item[$H_5=$]\(
  1+25 p_5 \textcolor{red}{z}+300 p_4 p_5 \textcolor{red}{{z}^2}+(700 p_3 p_4 p_5+1600 p_4^2 p_5) \textcolor{red}{{z}^3}+(700 p_2 p_3 p_4 p_5+9450 p_3 p_4^2 p_5+2500 p_4^2 p_5^2) \textcolor{red}{{z}^4}+(252 p_1 p_2 p_3 p_4 p_5+10752 p_2 p_3 p_4^2 p_5+15876 p_3^2 p_4^2 p_5+26250 p_3 p_4^2 p_5^2) \textcolor{red}{{z}^5}+(4200 p_1 p_2 p_3 p_4^2 p_5+39200 p_2 p_3^2 p_4^2 p_5+37800 p_2 p_3 p_4^2 p_5^2+78400 p_3^2 p_4^2 p_5^2+17500 p_3 p_4^3 p_5^2) \textcolor{red}{{z}^6}+(16200 p_1 p_2 p_3^2 p_4^2 p_5+25600 p_2^2 p_3^2 p_4^2 p_5+16800 p_1 p_2 p_3 p_4^2 p_5^2+245000 p_2 p_3^2 p_4^2 p_5^2+44800 p_2 p_3 p_4^3 p_5^2+132300 p_3^2 p_4^3 p_5^2) \textcolor{red}{{z}^7}+(22050 p_1 p_2^2 p_3^2 p_4^2 p_5+115200 p_1 p_2 p_3^2 p_4^2 p_5^2+202500 p_2^2 p_3^2 p_4^2 p_5^2+25200 p_1 p_2 p_3 p_4^3 p_5^2+617400 p_2 p_3^2 p_4^3 p_5^2+99225 p_3^2 p_4^4 p_5^2) \textcolor{red}{{z}^8}+(4900 p_1^2 p_2^2 p_3^2 p_4^2 p_5+198450 p_1 p_2^2 p_3^2 p_4^2 p_5^2+353400 p_1 p_2 p_3^2 p_4^3 p_5^2+691200 p_2^2 p_3^2 p_4^3 p_5^2+137200 p_2 p_3^3 p_4^3 p_5^2+627200 p_2 p_3^2 p_4^4 p_5^2+30625 p_3^2 p_4^4 p_5^3) \textcolor{red}{{z}^9}+(50176 p_1^2 p_2^2 p_3^2 p_4^2 p_5^2+798504 p_1 p_2^2 p_3^2 p_4^3 p_5^2+145152 p_1 p_2 p_3^3 p_4^3 p_5^2+280000 p_2^2 p_3^3 p_4^3 p_5^2+405000 p_1 p_2 p_3^2 p_4^4 p_5^2+1048576 p_2^2 p_3^2 p_4^4 p_5^2+296352 p_2 p_3^3 p_4^4 p_5^2+245000 p_2 p_3^2 p_4^4 p_5^3) \textcolor{red}{{z}^{10}}+(235200 p_1^2 p_2^2 p_3^2 p_4^3 p_5^2+491400 p_1 p_2^2 p_3^3 p_4^3 p_5^2+1411200 p_1 p_2^2 p_3^2 p_4^4 p_5^2+340200 p_1 p_2 p_3^3 p_4^4 p_5^2+1075200 p_2^2 p_3^3 p_4^4 p_5^2+180000 p_1 p_2 p_3^2 p_4^4 p_5^3+518400 p_2^2 p_3^2 p_4^4 p_5^3+205800 p_2 p_3^3 p_4^4 p_5^3) \textcolor{red}{{z}^{11}}+(179200 p_1^2 p_2^2 p_3^3 p_4^3 p_5^2+56700 p_1 p_2^3 p_3^3 p_4^3 p_5^2+490000 p_1^2 p_2^2 p_3^2 p_4^4 p_5^2+2118900 p_1 p_2^2 p_3^3 p_4^4 p_5^2+313600 p_2^2 p_3^4 p_4^4 p_5^2+793800 p_1 p_2^2 p_3^2 p_4^4 p_5^3+268800 p_1 p_2 p_3^3 p_4^4 p_5^3+945000 p_2^2 p_3^3 p_4^4 p_5^3+34300 p_2 p_3^3 p_4^5 p_5^3) \textcolor{red}{{z}^{12}}+(34300 p_1^2 p_2^3 p_3^3 p_4^3 p_5^2+945000 p_1^2 p_2^2 p_3^3 p_4^4 p_5^2+268800 p_1 p_2^3 p_3^3 p_4^4 p_5^2+793800 p_1 p_2^2 p_3^4 p_4^4 p_5^2+313600 p_1^2 p_2^2 p_3^2 p_4^4 p_5^3+2118900 p_1 p_2^2 p_3^3 p_4^4 p_5^3+490000 p_2^2 p_3^4 p_4^4 p_5^3+56700 p_1 p_2 p_3^3 p_4^5 p_5^3+179200 p_2^2 p_3^3 p_4^5 p_5^3) \textcolor{red}{{z}^{13}}+(205800 p_1^2 p_2^3 p_3^3 p_4^4 p_5^2+518400 p_1^2 p_2^2 p_3^4 p_4^4 p_5^2+180000 p_1 p_2^3 p_3^4 p_4^4 p_5^2+1075200 p_1^2 p_2^2 p_3^3 p_4^4 p_5^3+340200 p_1 p_2^3 p_3^3 p_4^4 p_5^3+1411200 p_1 p_2^2 p_3^4 p_4^4 p_5^3+491400 p_1 p_2^2 p_3^3 p_4^5 p_5^3+235200 p_2^2 p_3^4 p_4^5 p_5^3) \textcolor{red}{{z}^{14}}+(245000 p_1^2 p_2^3 p_3^4 p_4^4 p_5^2+296352 p_1^2 p_2^3 p_3^3 p_4^4 p_5^3+1048576 p_1^2 p_2^2 p_3^4 p_4^4 p_5^3+405000 p_1 p_2^3 p_3^4 p_4^4 p_5^3+280000 p_1^2 p_2^2 p_3^3 p_4^5 p_5^3+145152 p_1 p_2^3 p_3^3 p_4^5 p_5^3+798504 p_1 p_2^2 p_3^4 p_4^5 p_5^3+50176 p_2^2 p_3^4 p_4^6 p_5^3) \textcolor{red}{{z}^{15}}+(30625 p_1^2 p_2^4 p_3^4 p_4^4 p_5^2+627200 p_1^2 p_2^3 p_3^4 p_4^4 p_5^3+137200 p_1^2 p_2^3 p_3^3 p_4^5 p_5^3+691200 p_1^2 p_2^2 p_3^4 p_4^5 p_5^3+353400 p_1 p_2^3 p_3^4 p_4^5 p_5^3+198450 p_1 p_2^2 p_3^4 p_4^6 p_5^3+4900 p_2^2 p_3^4 p_4^6 p_5^4) \textcolor{red}{{z}^{16}}+(99225 p_1^2 p_2^4 p_3^4 p_4^4 p_5^3+617400 p_1^2 p_2^3 p_3^4 p_4^5 p_5^3+25200 p_1 p_2^3 p_3^5 p_4^5 p_5^3+202500 p_1^2 p_2^2 p_3^4 p_4^6 p_5^3+115200 p_1 p_2^3 p_3^4 p_4^6 p_5^3+22050 p_1 p_2^2 p_3^4 p_4^6 p_5^4) \textcolor{red}{{z}^{17}}+(132300 p_1^2 p_2^4 p_3^4 p_4^5 p_5^3+44800 p_1^2 p_2^3 p_3^5 p_4^5 p_5^3+245000 p_1^2 p_2^3 p_3^4 p_4^6 p_5^3+16800 p_1 p_2^3 p_3^5 p_4^6 p_5^3+25600 p_1^2 p_2^2 p_3^4 p_4^6 p_5^4+16200 p_1 p_2^3 p_3^4 p_4^6 p_5^4) \textcolor{red}{{z}^{18}}+(17500 p_1^2 p_2^4 p_3^5 p_4^5 p_5^3+78400 p_1^2 p_2^4 p_3^4 p_4^6 p_5^3+37800 p_1^2 p_2^3 p_3^5 p_4^6 p_5^3+39200 p_1^2 p_2^3 p_3^4 p_4^6 p_5^4+4200 p_1 p_2^3 p_3^5 p_4^6 p_5^4) \textcolor{red}{{z}^{19}}+(26250 p_1^2 p_2^4 p_3^5 p_4^6 p_5^3+15876 p_1^2 p_2^4 p_3^4 p_4^6 p_5^4+10752 p_1^2 p_2^3 p_3^5 p_4^6 p_5^4+252 p_1 p_2^3 p_3^5 p_4^7 p_5^4) \textcolor{red}{{z}^{20}}+(2500 p_1^2 p_2^4 p_3^6 p_4^6 p_5^3+9450 p_1^2 p_2^4 p_3^5 p_4^6 p_5^4+700 p_1^2 p_2^3 p_3^5 p_4^7 p_5^4) \textcolor{red}{{z}^{21}}+(1600 p_1^2 p_2^4 p_3^6 p_4^6 p_5^4+700 p_1^2 p_2^4 p_3^5 p_4^7 p_5^4) \textcolor{red}{{z}^{22}}+300 p_1^2 p_2^4 p_3^6 p_4^7 p_5^4 \textcolor{red}{{z}^{23}}+25 p_1^2 p_2^4 p_3^6 p_4^8 p_5^4 \textcolor{red}{{z}^{24}}+p_1^2 p_2^4 p_3^6 p_4^8 p_5^5 \textcolor{red}{{z}^{25}}
  \)
  \end{itemize}

  \bigskip
  \noindent { \bf $D_5$-case. }
   In case of the Lie algebra $D_5 \cong so(10)$ we get the following  polynomials
  \begin{itemize}
  \item[$H_1=$]\(
  1+8 p_1 \textcolor{red}{z}+28 p_1 p_2 \textcolor{red}{{z}^2}+56 p_1 p_2 p_3 \textcolor{red}{{z}^3}+(35 p_1 p_2 p_3 p_4+35 p_1 p_2 p_3 p_5) \textcolor{red}{{z}^4}+56 p_1 p_2 p_3 p_4 p_5 \textcolor{red}{{z}^5}+28 p_1 p_2 p_3^2 p_4 p_5 \textcolor{red}{{z}^6}+8 p_1 p_2^2 p_3^2 p_4 p_5 \textcolor{red}{{z}^7}+p_1^2 p_2^2 p_3^2 p_4 p_5 \textcolor{red}{{z}^8}
  \)
  \item[$H_2=$]\(
  1+14 p_2 \textcolor{red}{z}+(28 p_1 p_2+63 p_2 p_3) \textcolor{red}{{z}^2}+(224 p_1 p_2 p_3+70 p_2 p_3 p_4+70 p_2 p_3 p_5) \textcolor{red}{{z}^3}+(196 p_1 p_2^2 p_3+315 p_1 p_2 p_3 p_4+315 p_1 p_2 p_3 p_5+175 p_2 p_3 p_4 p_5) \textcolor{red}{{z}^4}+(490 p_1 p_2^2 p_3 p_4+490 p_1 p_2^2 p_3 p_5+896 p_1 p_2 p_3 p_4 p_5+126 p_2 p_3^2 p_4 p_5) \textcolor{red}{{z}^5}+(245 p_1 p_2^2 p_3^2 p_4+245 p_1 p_2^2 p_3^2 p_5+1764 p_1 p_2^2 p_3 p_4 p_5+700 p_1 p_2 p_3^2 p_4 p_5+49 p_2^2 p_3^2 p_4 p_5) \textcolor{red}{{z}^6}+3432 p_1 p_2^2 p_3^2 p_4 p_5 \textcolor{red}{{z}^7}+(49 p_1^2 p_2^2 p_3^2 p_4 p_5+700 p_1 p_2^3 p_3^2 p_4 p_5+1764 p_1 p_2^2 p_3^3 p_4 p_5+245 p_1 p_2^2 p_3^2 p_4^2 p_5+245 p_1 p_2^2 p_3^2 p_4 p_5^2) \textcolor{red}{{z}^8}+(126 p_1^2 p_2^3 p_3^2 p_4 p_5+896 p_1 p_2^3 p_3^3 p_4 p_5+490 p_1 p_2^2 p_3^3 p_4^2 p_5+490 p_1 p_2^2 p_3^3 p_4 p_5^2) \textcolor{red}{{z}^9}+(175 p_1^2 p_2^3 p_3^3 p_4 p_5+315 p_1 p_2^3 p_3^3 p_4^2 p_5+315 p_1 p_2^3 p_3^3 p_4 p_5^2+196 p_1 p_2^2 p_3^3 p_4^2 p_5^2) \textcolor{red}{{z}^{10}}+(70 p_1^2 p_2^3 p_3^3 p_4^2 p_5+70 p_1^2 p_2^3 p_3^3 p_4 p_5^2+224 p_1 p_2^3 p_3^3 p_4^2 p_5^2) \textcolor{red}{{z}^{11}}+(63 p_1^2 p_2^3 p_3^3 p_4^2 p_5^2+28 p_1 p_2^3 p_3^4 p_4^2 p_5^2) \textcolor{red}{{z}^{12}}+14 p_1^2 p_2^3 p_3^4 p_4^2 p_5^2 \textcolor{red}{{z}^{13}}+p_1^2 p_2^4 p_3^4 p_4^2 p_5^2 \textcolor{red}{{z}^{14}}
  \) %
  \item[$H_3=$]\(
  1+18 p_3 \textcolor{red}{z}+(63 p_2 p_3+45 p_3 p_4+45 p_3 p_5) \textcolor{red}{{z}^2}+(56 p_1 p_2 p_3+280 p_2 p_3 p_4+280 p_2 p_3 p_5+200 p_3 p_4 p_5) \textcolor{red}{{z}^3}+(315 p_1 p_2 p_3 p_4+315 p_2 p_3^2 p_4+315 p_1 p_2 p_3 p_5+315 p_2 p_3^2 p_5+1575 p_2 p_3 p_4 p_5+225 p_3^2 p_4 p_5) \textcolor{red}{{z}^4}+(630 p_1 p_2 p_3^2 p_4+630 p_1 p_2 p_3^2 p_5+2016 p_1 p_2 p_3 p_4 p_5+5292 p_2 p_3^2 p_4 p_5) \textcolor{red}{{z}^5}+(245 p_1 p_2^2 p_3^2 p_4+245 p_1 p_2^2 p_3^2 p_5+9996 p_1 p_2 p_3^2 p_4 p_5+1225 p_2^2 p_3^2 p_4 p_5+5103 p_2 p_3^3 p_4 p_5+875 p_2 p_3^2 p_4^2 p_5+875 p_2 p_3^2 p_4 p_5^2) \textcolor{red}{{z}^6}+(5616 p_1 p_2^2 p_3^2 p_4 p_5+12600 p_1 p_2 p_3^3 p_4 p_5+3528 p_2^2 p_3^3 p_4 p_5+2520 p_1 p_2 p_3^2 p_4^2 p_5+2520 p_2 p_3^3 p_4^2 p_5+2520 p_1 p_2 p_3^2 p_4 p_5^2+2520 p_2 p_3^3 p_4 p_5^2) \textcolor{red}{{z}^7}+(441 p_1^2 p_2^2 p_3^2 p_4 p_5+17172 p_1 p_2^2 p_3^3 p_4 p_5+2205 p_1 p_2^2 p_3^2 p_4^2 p_5+7875 p_1 p_2 p_3^3 p_4^2 p_5+2205 p_2^2 p_3^3 p_4^2 p_5+2205 p_1 p_2^2 p_3^2 p_4 p_5^2+7875 p_1 p_2 p_3^3 p_4 p_5^2+2205 p_2^2 p_3^3 p_4 p_5^2+1575 p_2 p_3^3 p_4^2 p_5^2) \textcolor{red}{{z}^8}+(2450 p_1^2 p_2^2 p_3^3 p_4 p_5+5600 p_1 p_2^3 p_3^3 p_4 p_5+16260 p_1 p_2^2 p_3^3 p_4^2 p_5+16260 p_1 p_2^2 p_3^3 p_4 p_5^2+5600 p_1 p_2 p_3^3 p_4^2 p_5^2+2450 p_2^2 p_3^3 p_4^2 p_5^2) \textcolor{red}{{z}^9}+(1575 p_1^2 p_2^3 p_3^3 p_4 p_5+2205 p_1^2 p_2^2 p_3^3 p_4^2 p_5+7875 p_1 p_2^3 p_3^3 p_4^2 p_5+2205 p_1 p_2^2 p_3^4 p_4^2 p_5+2205 p_1^2 p_2^2 p_3^3 p_4 p_5^2+7875 p_1 p_2^3 p_3^3 p_4 p_5^2+2205 p_1 p_2^2 p_3^4 p_4 p_5^2+17172 p_1 p_2^2 p_3^3 p_4^2 p_5^2+441 p_2^2 p_3^4 p_4^2 p_5^2) \textcolor{red}{{z}^{10}}+(2520 p_1^2 p_2^3 p_3^3 p_4^2 p_5+2520 p_1 p_2^3 p_3^4 p_4^2 p_5+2520 p_1^2 p_2^3 p_3^3 p_4 p_5^2+2520 p_1 p_2^3 p_3^4 p_4 p_5^2+3528 p_1^2 p_2^2 p_3^3 p_4^2 p_5^2+12600 p_1 p_2^3 p_3^3 p_4^2 p_5^2+5616 p_1 p_2^2 p_3^4 p_4^2 p_5^2) \textcolor{red}{{z}^{11}}+(875 p_1^2 p_2^3 p_3^4 p_4^2 p_5+875 p_1^2 p_2^3 p_3^4 p_4 p_5^2+5103 p_1^2 p_2^3 p_3^3 p_4^2 p_5^2+1225 p_1^2 p_2^2 p_3^4 p_4^2 p_5^2+9996 p_1 p_2^3 p_3^4 p_4^2 p_5^2+245 p_1 p_2^2 p_3^4 p_4^3 p_5^2+245 p_1 p_2^2 p_3^4 p_4^2 p_5^3) \textcolor{red}{{z}^{12}}+(5292 p_1^2 p_2^3 p_3^4 p_4^2 p_5^2+2016 p_1 p_2^3 p_3^5 p_4^2 p_5^2+630 p_1 p_2^3 p_3^4 p_4^3 p_5^2+630 p_1 p_2^3 p_3^4 p_4^2 p_5^3) \textcolor{red}{{z}^{13}}+(225 p_1^2 p_2^4 p_3^4 p_4^2 p_5^2+1575 p_1^2 p_2^3 p_3^5 p_4^2 p_5^2+315 p_1^2 p_2^3 p_3^4 p_4^3 p_5^2+315 p_1 p_2^3 p_3^5 p_4^3 p_5^2+315 p_1^2 p_2^3 p_3^4 p_4^2 p_5^3+315 p_1 p_2^3 p_3^5 p_4^2 p_5^3) \textcolor{red}{{z}^{14}}+(200 p_1^2 p_2^4 p_3^5 p_4^2 p_5^2+280 p_1^2 p_2^3 p_3^5 p_4^3 p_5^2+280 p_1^2 p_2^3 p_3^5 p_4^2 p_5^3+56 p_1 p_2^3 p_3^5 p_4^3 p_5^3) \textcolor{red}{{z}^{15}}+(45 p_1^2 p_2^4 p_3^5 p_4^3 p_5^2+45 p_1^2 p_2^4 p_3^5 p_4^2 p_5^3+63 p_1^2 p_2^3 p_3^5 p_4^3 p_5^3) \textcolor{red}{{z}^{16}}+18 p_1^2 p_2^4 p_3^5 p_4^3 p_5^3 \textcolor{red}{{z}^{17}}+p_1^2 p_2^4 p_3^6 p_4^3 p_5^3 \textcolor{red}{{z}^{18}}
  \)
  \item[$H_4=$]\(
  1+10 p_4 \textcolor{red}{z}+45 p_3 p_4 \textcolor{red}{{z}^2}+(70 p_2 p_3 p_4+50 p_3 p_4 p_5) \textcolor{red}{{z}^3}+(35 p_1 p_2 p_3 p_4+175 p_2 p_3 p_4 p_5) \textcolor{red}{{z}^4}+(126 p_1 p_2 p_3 p_4 p_5+126 p_2 p_3^2 p_4 p_5) \textcolor{red}{{z}^5}+(175 p_1 p_2 p_3^2 p_4 p_5+35 p_2 p_3^2 p_4^2 p_5) \textcolor{red}{{z}^6}+(50 p_1 p_2^2 p_3^2 p_4 p_5+70 p_1 p_2 p_3^2 p_4^2 p_5) \textcolor{red}{{z}^7}+45 p_1 p_2^2 p_3^2 p_4^2 p_5 \textcolor{red}{{z}^8}+10 p_1 p_2^2 p_3^3 p_4^2 p_5 \textcolor{red}{{z}^9}+p_1 p_2^2 p_3^3 p_4^2 p_5^2 \textcolor{red}{{z}^{10}}
  \)
  \item[$H_5=$]\(
  1+10 p_5 \textcolor{red}{z}+45 p_3 p_5 \textcolor{red}{{z}^2}+(70 p_2 p_3 p_5+50 p_3 p_4 p_5) \textcolor{red}{{z}^3}+(35 p_1 p_2 p_3 p_5+175 p_2 p_3 p_4 p_5) \textcolor{red}{{z}^4}+(126 p_1 p_2 p_3 p_4 p_5+126 p_2 p_3^2 p_4 p_5) \textcolor{red}{{z}^5}+(175 p_1 p_2 p_3^2 p_4 p_5+35 p_2 p_3^2 p_4 p_5^2) \textcolor{red}{{z}^6}+(50 p_1 p_2^2 p_3^2 p_4 p_5+70 p_1 p_2 p_3^2 p_4 p_5^2) \textcolor{red}{{z}^7}+45 p_1 p_2^2 p_3^2 p_4 p_5^2 \textcolor{red}{{z}^8}+10 p_1 p_2^2 p_3^3 p_4 p_5^2 \textcolor{red}{{z}^9}+p_1 p_2^2 p_3^3 p_4^2 p_5^2 \textcolor{red}{{z}^{10}}
  \)
  \end{itemize}
 
 
 \renewcommand{\thesubsection}{\Alph{subsection}}
 
 \subsection{Polynomials for $E_6$}
 
 In this subsection we present polynomials corresponding to the Lie algebra $E_6$ \cite{BolIvas-17}.

 \vspace{1em}
 
 \newcounter{mathematicapage}
 \tolerance7000
 \small
 \begin{itemize}
 \item[$H_1=$]\(p_1^2 p_2^3 p_3^4 p_4^3 p_5^2 p_6^2 \textcolor{red}{z^{16}}+16 p_1^2 p_2^3 p_3^4 p_4^3 p_5 p_6^2 \textcolor{red}{z^{15}}+120 p_1^2 p_2^3 p_3^4 p_4^2 p_5 p_6^2
 \textcolor{red}{z^{14}}+560 p_1^2 p_2^3 p_3^3 p_4^2 p_5 p_6^2 \textcolor{red}{z^{13}}+(1050 p_1^2 p_2^2 p_4^2 p_5 p_6^2 p_3^3+770 p_1^2 p_2^3 p_4^2 p_5 p_6 p_3^3) \textcolor{red}{z^{12}}+(672
 p_1 p_2^2 p_4^2 p_5 p_6^2 p_3^3+3696 p_1^2 p_2^2 p_4^2 p_5 p_6 p_3^3) \textcolor{red}{z^{11}}+(3696 p_1 p_2^2 p_4^2 p_5 p_6 p_3^3+4312 p_1^2 p_2^2 p_4^2
 p_5 p_6 p_3^2) \textcolor{red}{z^{10}}+(8800 p_1 p_2^2 p_4^2 p_5 p_6 p_3^2+2640 p_1^2 p_2^2 p_4 p_5 p_6 p_3^2) \textcolor{red}{z^9}+(660 p_1^2 p_2^2 p_4 p_6
 p_3^2+4125 p_1 p_2 p_4^2 p_5 p_6 p_3^2+8085 p_1 p_2^2 p_4 p_5 p_6 p_3^2) \textcolor{red}{z^8}+(2640 p_1 p_2^2 p_4 p_6 p_3^2+8800 p_1 p_2 p_4 p_5 p_6 p_3^2)
 \textcolor{red}{z^7}+(4312 p_1 p_2 p_4 p_6 p_3^2+3696 p_1 p_2 p_4 p_5 p_6 p_3) \textcolor{red}{z^6}+(672 p_1 p_2 p_3 p_4 p_5+3696 p_1 p_2 p_3 p_4 p_6) \textcolor{red}{z^5}+(1050
 p_1 p_2 p_3 p_4+770 p_1 p_2 p_3 p_6) \textcolor{red}{z^4}+560 p_1 p_2 p_3 \textcolor{red}{z^3}+120 p_1 p_2 \textcolor{red}{z^2}+16 p_1 \textcolor{red}{z}+1\)
 
 \vspace{1em}
\item[$H_2=$]\(p_1^3 p_2^6 p_3^8 p_4^6 p_5^3 p_6^4 \textcolor{red}{z^{30}}+30 p_1^3 p_2^6 p_3^8 p_4^5 p_5^3 p_6^4 \textcolor{red}{z^{29}}+(120 p_1^3 p_2^6 p_4^5 p_5^2
 p_6^4 p_3^8+315 p_1^3 p_2^6 p_4^5 p_5^3 p_6^4 p_3^7) \textcolor{red}{z^{28}}+(1050 p_1^3 p_2^5 p_4^5 p_5^3 p_6^4 p_3^7+2240 p_1^3 p_2^6 p_4^5 p_5^2 p_6^4
 p_3^7+770 p_1^3 p_2^6 p_4^5 p_5^3 p_6^3 p_3^7) \textcolor{red}{z^{27}}+(1050 p_1^2 p_2^5 p_4^5 p_5^3 p_6^4 p_3^7+9450 p_1^3 p_2^5 p_4^5 p_5^2 p_6^4 p_3^7+4200
 p_1^3 p_2^6 p_4^4 p_5^2 p_6^4 p_3^7+5775 p_1^3 p_2^5 p_4^5 p_5^3 p_6^3 p_3^7+6930 p_1^3 p_2^6 p_4^5 p_5^2 p_6^3 p_3^7) \textcolor{red}{z^{26}}+(10752 p_1^2
 p_2^5 p_4^5 p_5^2 p_6^4 p_3^7+31500 p_1^3 p_2^5 p_4^4 p_5^2 p_6^4 p_3^7+8316 p_1^2 p_2^5 p_4^5 p_5^3 p_6^3 p_3^7+59136 p_1^3 p_2^5 p_4^5 p_5^2 p_6^3
 p_3^7+23100 p_1^3 p_2^6 p_4^4 p_5^2 p_6^3 p_3^7+9702 p_1^3 p_2^5 p_4^5 p_5^3 p_6^3 p_3^6) \textcolor{red}{z^{25}}+(45360 p_1^2 p_2^5 p_4^4 p_5^2 p_6^4
 p_3^7+92400 p_1^2 p_2^5 p_4^5 p_5^2 p_6^3 p_3^7+249480 p_1^3 p_2^5 p_4^4 p_5^2 p_6^3 p_3^7+36750 p_1^3 p_2^5 p_4^4 p_5^2 p_6^4 p_3^6+26950 p_1^2
 p_2^5 p_4^5 p_5^3 p_6^3 p_3^6+8085 p_1^3 p_2^5 p_4^4 p_5^3 p_6^3 p_3^6+107800 p_1^3 p_2^5 p_4^5 p_5^2 p_6^3 p_3^6+26950 p_1^3 p_2^6 p_4^4 p_5^2 p_6^3
 p_3^6) \textcolor{red}{z^{24}}+(443520 p_1^2 p_2^5 p_4^4 p_5^2 p_6^3 p_3^7+94080 p_1^2 p_2^5 p_4^4 p_5^2 p_6^4 p_3^6+16500 p_1^2 p_2^4 p_4^5 p_5^3 p_6^3
 p_3^6+32340 p_1^2 p_2^5 p_4^4 p_5^3 p_6^3 p_3^6+316800 p_1^2 p_2^5 p_4^5 p_5^2 p_6^3 p_3^6+1132560 p_1^3 p_2^5 p_4^4 p_5^2 p_6^3 p_3^6) \textcolor{red}{z^{23}}+(44100
 p_1^2 p_2^4 p_4^4 p_5^2 p_6^4 p_3^6+44550 p_1^2 p_2^4 p_4^4 p_5^3 p_6^3 p_3^6+202125 p_1^2 p_2^4 p_4^5 p_5^2 p_6^3 p_3^6+3256110 p_1^2 p_2^5 p_4^4
 p_5^2 p_6^3 p_3^6+242550 p_1^3 p_2^4 p_4^4 p_5^2 p_6^3 p_3^6+495000 p_1^3 p_2^5 p_4^3 p_5^2 p_6^3 p_3^6+32340 p_1^3 p_2^5 p_4^4 p_5 p_6^3 p_3^6+177870
 p_1^3 p_2^5 p_4^4 p_5^2 p_6^2 p_3^6+1358280 p_1^3 p_2^5 p_4^4 p_5^2 p_6^3 p_3^5) \textcolor{red}{z^{22}}+(2674100 p_1^2 p_2^4 p_4^4 p_5^2 p_6^3 p_3^6+2182950
 p_1^2 p_2^5 p_4^3 p_5^2 p_6^3 p_3^6+168960 p_1^2 p_2^5 p_4^4 p_5 p_6^3 p_3^6+178200 p_1^3 p_2^5 p_4^3 p_5 p_6^3 p_3^6+711480 p_1^2 p_2^5 p_4^4 p_5^2
 p_6^2 p_3^6+23100 p_1^2 p_2^4 p_4^4 p_5^3 p_6^3 p_3^5+4928000 p_1^2 p_2^5 p_4^4 p_5^2 p_6^3 p_3^5+1131900 p_1^3 p_2^4 p_4^4 p_5^2 p_6^3 p_3^5+1478400
 p_1^3 p_2^5 p_4^3 p_5^2 p_6^3 p_3^5+830060 p_1^3 p_2^5 p_4^4 p_5^2 p_6^2 p_3^5) \textcolor{red}{z^{21}}+(155232 p_1 p_2^4 p_4^4 p_5^2 p_6^3 p_3^6+3234000
 p_1^2 p_2^4 p_4^3 p_5^2 p_6^3 p_3^6+349272 p_1^2 p_2^4 p_4^4 p_5 p_6^3 p_3^6+970200 p_1^2 p_2^5 p_4^3 p_5 p_6^3 p_3^6+853776 p_1^2 p_2^4 p_4^4 p_5^2
 p_6^2 p_3^6+9315306 p_1^2 p_2^4 p_4^4 p_5^2 p_6^3 p_3^5+7074375 p_1^2 p_2^5 p_4^3 p_5^2 p_6^3 p_3^5+1559250 p_1^3 p_2^4 p_4^3 p_5^2 p_6^3 p_3^5+577500
 p_1^3 p_2^5 p_4^3 p_5 p_6^3 p_3^5+5082 p_1^2 p_2^4 p_4^4 p_5^3 p_6^2 p_3^5+3811500 p_1^2 p_2^5 p_4^4 p_5^2 p_6^2 p_3^5+996072 p_1^3 p_2^4 p_4^4 p_5^2
 p_6^2 p_3^5+1143450 p_1^3 p_2^5 p_4^3 p_5^2 p_6^2 p_3^5) \textcolor{red}{z^{20}}+(2069760 p_1^2 p_2^4 p_4^3 p_5 p_6^3 p_3^6+1478400 p_1 p_2^4 p_4^4 p_5^2
 p_6^3 p_3^5+4331250 p_1^2 p_2^3 p_4^4 p_5^2 p_6^3 p_3^5+21801780 p_1^2 p_2^4 p_4^3 p_5^2 p_6^3 p_3^5+369600 p_1^2 p_2^4 p_4^4 p_5 p_6^3 p_3^5+3326400
 p_1^2 p_2^5 p_4^3 p_5 p_6^3 p_3^5+693000 p_1^3 p_2^4 p_4^3 p_5 p_6^3 p_3^5+11384100 p_1^2 p_2^4 p_4^4 p_5^2 p_6^2 p_3^5+6225450 p_1^2 p_2^5 p_4^3
 p_5^2 p_6^2 p_3^5+2439360 p_1^3 p_2^4 p_4^3 p_5^2 p_6^2 p_3^5+508200 p_1^3 p_2^5 p_4^3 p_5 p_6^2 p_3^5) \textcolor{red}{z^{19}}+(1559250 p_1 p_2^3 p_4^4
 p_5^2 p_6^3 p_3^5+3056130 p_1 p_2^4 p_4^3 p_5^2 p_6^3 p_3^5+14437500 p_1^2 p_2^3 p_4^3 p_5^2 p_6^3 p_3^5+14314300 p_1^2 p_2^4 p_4^3 p_5 p_6^3 p_3^5+2032800
 p_1 p_2^4 p_4^4 p_5^2 p_6^2 p_3^5+8575875 p_1^2 p_2^3 p_4^4 p_5^2 p_6^2 p_3^5+28420210 p_1^2 p_2^4 p_4^3 p_5^2 p_6^2 p_3^5+127050 p_1^2 p_2^4 p_4^4
 p_5 p_6^2 p_3^5+3176250 p_1^2 p_2^5 p_4^3 p_5 p_6^2 p_3^5+1372140 p_1^3 p_2^4 p_4^3 p_5 p_6^2 p_3^5+6338640 p_1^2 p_2^4 p_4^3 p_5^2 p_6^3 p_3^4+2371600
 p_1^2 p_2^4 p_4^4 p_5^2 p_6^2 p_3^4+711480 p_1^3 p_2^4 p_4^3 p_5^2 p_6^2 p_3^4) \textcolor{red}{z^{18}}+(5913600 p_1 p_2^3 p_4^3 p_5^2 p_6^3 p_3^5+1774080
 p_1 p_2^4 p_4^3 p_5 p_6^3 p_3^5+10187100 p_1^2 p_2^3 p_4^3 p_5 p_6^3 p_3^5+577500 p_1^2 p_2^4 p_4^2 p_5 p_6^3 p_3^5+3811500 p_1 p_2^3 p_4^4 p_5^2
 p_6^2 p_3^5+7470540 p_1 p_2^4 p_4^3 p_5^2 p_6^2 p_3^5+32524800 p_1^2 p_2^3 p_4^3 p_5^2 p_6^2 p_3^5+18705960 p_1^2 p_2^4 p_4^3 p_5 p_6^2 p_3^5+8731800
 p_1^2 p_2^3 p_4^3 p_5^2 p_6^3 p_3^4+8279040 p_1^2 p_2^4 p_4^3 p_5 p_6^3 p_3^4+4446750 p_1^2 p_2^3 p_4^4 p_5^2 p_6^2 p_3^4+16625700 p_1^2 p_2^4 p_4^3
 p_5^2 p_6^2 p_3^4+711480 p_1^3 p_2^4 p_4^3 p_5 p_6^2 p_3^4) \textcolor{red}{z^{17}}+(4527600 p_1 p_2^3 p_4^3 p_5 p_6^3 p_3^5+18295200 p_1 p_2^3 p_4^3 p_5^2
 p_6^2 p_3^5+5488560 p_1 p_2^4 p_4^3 p_5 p_6^2 p_3^5+24901800 p_1^2 p_2^3 p_4^3 p_5 p_6^2 p_3^5+508200 p_1^2 p_2^4 p_4^2 p_5 p_6^2 p_3^5+3880800 p_1
 p_2^3 p_4^3 p_5^2 p_6^3 p_3^4+11884950 p_1^2 p_2^3 p_4^3 p_5 p_6^3 p_3^4+2182950 p_1^2 p_2^4 p_4^2 p_5 p_6^3 p_3^4+2268750 p_1 p_2^3 p_4^4 p_5^2
 p_6^2 p_3^4+4446750 p_1 p_2^4 p_4^3 p_5^2 p_6^2 p_3^4+45530550 p_1^2 p_2^3 p_4^3 p_5^2 p_6^2 p_3^4+1334025 p_1^2 p_2^4 p_4^2 p_5^2 p_6^2 p_3^4+18478980
 p_1^2 p_2^4 p_4^3 p_5 p_6^2 p_3^4+108900 p_1^3 p_2^4 p_4^2 p_5 p_6^2 p_3^4+1584660 p_1^2 p_2^4 p_4^3 p_5^2 p_6 p_3^4) \textcolor{red}{z^{16}}+(15937152
 p_1 p_2^3 p_4^3 p_5 p_6^2 p_3^5+5588352 p_1 p_2^3 p_4^3 p_5 p_6^3 p_3^4+3234000 p_1^2 p_2^3 p_4^2 p_5 p_6^3 p_3^4+34036496 p_1 p_2^3 p_4^3 p_5^2
 p_6^2 p_3^4+5808000 p_1^2 p_2^3 p_4^2 p_5^2 p_6^2 p_3^4+5808000 p_1 p_2^4 p_4^3 p_5 p_6^2 p_3^4+53742416 p_1^2 p_2^3 p_4^3 p_5 p_6^2 p_3^4+6203600
 p_1^2 p_2^4 p_4^2 p_5 p_6^2 p_3^4+5588352 p_1^2 p_2^3 p_4^3 p_5^2 p_6 p_3^4+3234000 p_1^2 p_2^4 p_4^3 p_5 p_6 p_3^4+15937152 p_1^2 p_2^3 p_4^3 p_5^2
 p_6^2 p_3^3) \textcolor{red}{z^{15}}+(108900 p_1^2 p_3^4 p_4^2 p_6^2 p_2^4+1334025 p_1 p_3^4 p_4^2 p_5 p_6^2 p_2^4+508200 p_1^2 p_3^3 p_4^2 p_5 p_6^2 p_2^4+2182950
 p_1^2 p_3^4 p_4^2 p_5 p_6 p_2^4+1584660 p_1 p_3^4 p_4^2 p_5 p_6^3 p_2^3+18295200 p_1 p_3^3 p_4^3 p_5^2 p_6^2 p_2^3+4446750 p_1 p_3^4 p_4^2 p_5^2
 p_6^2 p_2^3+5488560 p_1^2 p_3^3 p_4^2 p_5^2 p_6^2 p_2^3+45530550 p_1 p_3^4 p_4^3 p_5 p_6^2 p_2^3+24901800 p_1^2 p_3^3 p_4^3 p_5 p_6^2 p_2^3+18478980
 p_1^2 p_3^4 p_4^2 p_5 p_6^2 p_2^3+3880800 p_1 p_3^4 p_4^3 p_5^2 p_6 p_2^3+4527600 p_1^2 p_3^3 p_4^3 p_5^2 p_6 p_2^3+11884950 p_1^2 p_3^4 p_4^3 p_5
 p_6 p_2^3+2268750 p_1 p_3^4 p_4^3 p_5^2 p_6^2 p_2^2) \textcolor{red}{z^{14}}+(577500 p_1^2 p_3^3 p_4^2 p_5 p_6 p_2^4+711480 p_1^2 p_3^4 p_4^2 p_6^2 p_2^3+7470540
 p_1 p_3^3 p_4^2 p_5^2 p_6^2 p_2^3+32524800 p_1 p_3^3 p_4^3 p_5 p_6^2 p_2^3+16625700 p_1 p_3^4 p_4^2 p_5 p_6^2 p_2^3+18705960 p_1^2 p_3^3 p_4^2 p_5
 p_6^2 p_2^3+5913600 p_1 p_3^3 p_4^3 p_5^2 p_6 p_2^3+1774080 p_1^2 p_3^3 p_4^2 p_5^2 p_6 p_2^3+8731800 p_1 p_3^4 p_4^3 p_5 p_6 p_2^3+10187100 p_1^2
 p_3^3 p_4^3 p_5 p_6 p_2^3+8279040 p_1^2 p_3^4 p_4^2 p_5 p_6 p_2^3+3811500 p_1 p_3^3 p_4^3 p_5^2 p_6^2 p_2^2+4446750 p_1 p_3^4 p_4^3 p_5 p_6^2 p_2^2)
 \textcolor{red}{z^{13}}+(711480 p_1 p_2^3 p_4^2 p_6^2 p_3^4+2371600 p_1 p_2^2 p_4^2 p_5 p_6^2 p_3^4+6338640 p_1 p_2^3 p_4^2 p_5 p_6 p_3^4+1372140 p_1^2 p_2^3
 p_4^2 p_6^2 p_3^3+2032800 p_1 p_2^2 p_4^2 p_5^2 p_6^2 p_3^3+8575875 p_1 p_2^2 p_4^3 p_5 p_6^2 p_3^3+28420210 p_1 p_2^3 p_4^2 p_5 p_6^2 p_3^3+127050
 p_1^2 p_2^2 p_4^2 p_5 p_6^2 p_3^3+3176250 p_1^2 p_2^3 p_4 p_5 p_6^2 p_3^3+1559250 p_1 p_2^2 p_4^3 p_5^2 p_6 p_3^3+3056130 p_1 p_2^3 p_4^2 p_5^2 p_6
 p_3^3+14437500 p_1 p_2^3 p_4^3 p_5 p_6 p_3^3+14314300 p_1^2 p_2^3 p_4^2 p_5 p_6 p_3^3) \textcolor{red}{z^{12}}+(2439360 p_1 p_3^3 p_4^2 p_6^2 p_2^3+508200
 p_1^2 p_3^3 p_4 p_6^2 p_2^3+6225450 p_1 p_3^3 p_4 p_5 p_6^2 p_2^3+693000 p_1^2 p_3^3 p_4^2 p_6 p_2^3+21801780 p_1 p_3^3 p_4^2 p_5 p_6 p_2^3+2069760
 p_1^2 p_3^2 p_4^2 p_5 p_6 p_2^3+3326400 p_1^2 p_3^3 p_4 p_5 p_6 p_2^3+11384100 p_1 p_3^3 p_4^2 p_5 p_6^2 p_2^2+1478400 p_1 p_3^3 p_4^2 p_5^2 p_6
 p_2^2+4331250 p_1 p_3^3 p_4^3 p_5 p_6 p_2^2+369600 p_1^2 p_3^3 p_4^2 p_5 p_6 p_2^2) \textcolor{red}{z^{11}}+(1143450 p_1 p_3^3 p_4 p_6^2 p_2^3+1559250
 p_1 p_3^3 p_4^2 p_6 p_2^3+577500 p_1^2 p_3^3 p_4 p_6 p_2^3+3234000 p_1 p_3^2 p_4^2 p_5 p_6 p_2^3+7074375 p_1 p_3^3 p_4 p_5 p_6 p_2^3+970200 p_1^2
 p_3^2 p_4 p_5 p_6 p_2^3+996072 p_1 p_3^3 p_4^2 p_6^2 p_2^2+5082 p_3^3 p_4^2 p_5 p_6^2 p_2^2+853776 p_1 p_3^2 p_4^2 p_5 p_6^2 p_2^2+3811500 p_1 p_3^3
 p_4 p_5 p_6^2 p_2^2+155232 p_1 p_3^2 p_4^2 p_5^2 p_6 p_2^2+9315306 p_1 p_3^3 p_4^2 p_5 p_6 p_2^2+349272 p_1^2 p_3^2 p_4^2 p_5 p_6 p_2^2) \textcolor{red}{z^{10}}+(1478400
 p_1 p_3^3 p_4 p_6 p_2^3+178200 p_1^2 p_3^2 p_4 p_6 p_2^3+2182950 p_1 p_3^2 p_4 p_5 p_6 p_2^3+830060 p_1 p_3^3 p_4 p_6^2 p_2^2+711480 p_1 p_3^2 p_4
 p_5 p_6^2 p_2^2+1131900 p_1 p_3^3 p_4^2 p_6 p_2^2+23100 p_3^3 p_4^2 p_5 p_6 p_2^2+2674100 p_1 p_3^2 p_4^2 p_5 p_6 p_2^2+4928000 p_1 p_3^3 p_4 p_5
 p_6 p_2^2+168960 p_1^2 p_3^2 p_4 p_5 p_6 p_2^2) \textcolor{red}{z^9}+(495000 p_1 p_3^2 p_4 p_6 p_2^3+177870 p_1 p_3^2 p_4 p_6^2 p_2^2+44100 p_1 p_3^2 p_4^2
 p_5 p_2^2+242550 p_1 p_3^2 p_4^2 p_6 p_2^2+1358280 p_1 p_3^3 p_4 p_6 p_2^2+32340 p_1^2 p_3^2 p_4 p_6 p_2^2+44550 p_3^2 p_4^2 p_5 p_6 p_2^2+3256110
 p_1 p_3^2 p_4 p_5 p_6 p_2^2+202125 p_1 p_3^2 p_4^2 p_5 p_6 p_2) \textcolor{red}{z^8}+(94080 p_1 p_3^2 p_4 p_5 p_2^2+1132560 p_1 p_3^2 p_4 p_6 p_2^2+32340
 p_3^2 p_4 p_5 p_6 p_2^2+443520 p_1 p_3 p_4 p_5 p_6 p_2^2+16500 p_3^2 p_4^2 p_5 p_6 p_2+316800 p_1 p_3^2 p_4 p_5 p_6 p_2) \textcolor{red}{z^7}+(36750 p_1
 p_3^2 p_4 p_2^2+45360 p_1 p_3 p_4 p_5 p_2^2+26950 p_1 p_3^2 p_6 p_2^2+8085 p_3^2 p_4 p_6 p_2^2+249480 p_1 p_3 p_4 p_6 p_2^2+107800 p_1 p_3^2 p_4
 p_6 p_2+26950 p_3^2 p_4 p_5 p_6 p_2+92400 p_1 p_3 p_4 p_5 p_6 p_2) \textcolor{red}{z^6}+(31500 p_1 p_3 p_4 p_2^2+23100 p_1 p_3 p_6 p_2^2+10752 p_1 p_3
 p_4 p_5 p_2+9702 p_3^2 p_4 p_6 p_2+59136 p_1 p_3 p_4 p_6 p_2+8316 p_3 p_4 p_5 p_6 p_2) \textcolor{red}{z^5}+(4200 p_1 p_3 p_2^2+9450 p_1 p_3 p_4 p_2+1050
 p_3 p_4 p_5 p_2+6930 p_1 p_3 p_6 p_2+5775 p_3 p_4 p_6 p_2) \textcolor{red}{z^4}+(2240 p_1 p_2 p_3+1050 p_2 p_4 p_3+770 p_2 p_6 p_3) \textcolor{red}{z^3}+(120
 p_1 p_2+315 p_3 p_2) \textcolor{red}{z^2}+30 p_2 \textcolor{red}{z}+1\)
 
 \vspace{1em}
 \item[$H_3=$]\(p_1^4 p_2^8 p_3^{12} p_4^8 p_5^4 p_6^6 \textcolor{red}{z^{42}}+42 p_1^4 p_2^8 p_3^{11} p_4^8 p_5^4 p_6^6 \textcolor{red}{z^{41}}+(315 p_1^4 p_2^7 p_4^8
 p_5^4 p_6^6 p_3^{11}+315 p_1^4 p_2^8 p_4^7 p_5^4 p_6^6 p_3^{11}+231 p_1^4 p_2^8 p_4^8 p_5^4 p_6^5 p_3^{11}) \textcolor{red}{z^{40}}+(560 p_1^3 p_2^7 p_4^8
 p_5^4 p_6^6 p_3^{11}+4200 p_1^4 p_2^7 p_4^7 p_5^4 p_6^6 p_3^{11}+560 p_1^4 p_2^8 p_4^7 p_5^3 p_6^6 p_3^{11}+3080 p_1^4 p_2^7 p_4^8 p_5^4 p_6^5 p_3^{11}+3080
 p_1^4 p_2^8 p_4^7 p_5^4 p_6^5 p_3^{11}) \textcolor{red}{z^{39}}+(9450 p_1^3 p_2^7 p_4^7 p_5^4 p_6^6 p_3^{11}+9450 p_1^4 p_2^7 p_4^7 p_5^3 p_6^6 p_3^{11}+6930
 p_1^3 p_2^7 p_4^8 p_5^4 p_6^5 p_3^{11}+51975 p_1^4 p_2^7 p_4^7 p_5^4 p_6^5 p_3^{11}+6930 p_1^4 p_2^8 p_4^7 p_5^3 p_6^5 p_3^{11}+11025 p_1^4 p_2^7
 p_4^7 p_5^4 p_6^6 p_3^{10}+8085 p_1^4 p_2^7 p_4^8 p_5^4 p_6^5 p_3^{10}+8085 p_1^4 p_2^8 p_4^7 p_5^4 p_6^5 p_3^{10}) \textcolor{red}{z^{38}}+(24192 p_1^3
 p_2^7 p_4^7 p_5^3 p_6^6 p_3^{11}+133056 p_1^3 p_2^7 p_4^7 p_5^4 p_6^5 p_3^{11}+133056 p_1^4 p_2^7 p_4^7 p_5^3 p_6^5 p_3^{11}+44100 p_1^3 p_2^7 p_4^7
 p_5^4 p_6^6 p_3^{10}+44100 p_1^4 p_2^7 p_4^7 p_5^3 p_6^6 p_3^{10}+32340 p_1^3 p_2^7 p_4^8 p_5^4 p_6^5 p_3^{10}+407484 p_1^4 p_2^7 p_4^7 p_5^4 p_6^5
 p_3^{10}+32340 p_1^4 p_2^8 p_4^7 p_5^3 p_6^5 p_3^{10}) \textcolor{red}{z^{37}}+(369600 p_1^3 p_2^7 p_4^7 p_5^3 p_6^5 p_3^{11}+36750 p_1^3 p_2^6 p_4^7 p_5^4
 p_6^6 p_3^{10}+200704 p_1^3 p_2^7 p_4^7 p_5^3 p_6^6 p_3^{10}+36750 p_1^4 p_2^7 p_4^6 p_5^3 p_6^6 p_3^{10}+26950 p_1^3 p_2^6 p_4^8 p_5^4 p_6^5 p_3^{10}+1539384
 p_1^3 p_2^7 p_4^7 p_5^4 p_6^5 p_3^{10}+202125 p_1^4 p_2^6 p_4^7 p_5^4 p_6^5 p_3^{10}+202125 p_1^4 p_2^7 p_4^6 p_5^4 p_6^5 p_3^{10}+1539384 p_1^4
 p_2^7 p_4^7 p_5^3 p_6^5 p_3^{10}+26950 p_1^4 p_2^8 p_4^6 p_5^3 p_6^5 p_3^{10}+148225 p_1^4 p_2^7 p_4^7 p_5^4 p_6^4 p_3^{10}+916839 p_1^4 p_2^7 p_4^7
 p_5^4 p_6^5 p_3^9) \textcolor{red}{z^{36}}+(211680 p_1^3 p_2^6 p_4^7 p_5^3 p_6^6 p_3^{10}+211680 p_1^3 p_2^7 p_4^6 p_5^3 p_6^6 p_3^{10}+1853280 p_1^3 p_2^6
 p_4^7 p_5^4 p_6^5 p_3^{10}+1164240 p_1^3 p_2^7 p_4^6 p_5^4 p_6^5 p_3^{10}+6044544 p_1^3 p_2^7 p_4^7 p_5^3 p_6^5 p_3^{10}+1164240 p_1^4 p_2^6 p_4^7
 p_5^3 p_6^5 p_3^{10}+1853280 p_1^4 p_2^7 p_4^6 p_5^3 p_6^5 p_3^{10}+853776 p_1^3 p_2^7 p_4^7 p_5^4 p_6^4 p_3^{10}+853776 p_1^4 p_2^7 p_4^7 p_5^3
 p_6^4 p_3^{10}+4527600 p_1^3 p_2^7 p_4^7 p_5^4 p_6^5 p_3^9+1358280 p_1^4 p_2^6 p_4^7 p_5^4 p_6^5 p_3^9+1358280 p_1^4 p_2^7 p_4^6 p_5^4 p_6^5 p_3^9+4527600
 p_1^4 p_2^7 p_4^7 p_5^3 p_6^5 p_3^9+996072 p_1^4 p_2^7 p_4^7 p_5^4 p_6^4 p_3^9) \textcolor{red}{z^{35}}+(396900 p_1^3 p_2^6 p_4^6 p_5^3 p_6^6 p_3^{10}+291060
 p_1^2 p_2^6 p_4^7 p_5^4 p_6^5 p_3^{10}+2182950 p_1^3 p_2^6 p_4^6 p_5^4 p_6^5 p_3^{10}+9168390 p_1^3 p_2^6 p_4^7 p_5^3 p_6^5 p_3^{10}+9168390 p_1^3
 p_2^7 p_4^6 p_5^3 p_6^5 p_3^{10}+2182950 p_1^4 p_2^6 p_4^6 p_5^3 p_6^5 p_3^{10}+291060 p_1^4 p_2^7 p_4^6 p_5^2 p_6^5 p_3^{10}+1600830 p_1^3 p_2^6
 p_4^7 p_5^4 p_6^4 p_3^{10}+5336100 p_1^3 p_2^7 p_4^7 p_5^3 p_6^4 p_3^{10}+1600830 p_1^4 p_2^7 p_4^6 p_5^3 p_6^4 p_3^{10}+13222440 p_1^3 p_2^6 p_4^7
 p_5^4 p_6^5 p_3^9+8489250 p_1^3 p_2^7 p_4^6 p_5^4 p_6^5 p_3^9+2546775 p_1^4 p_2^6 p_4^6 p_5^4 p_6^5 p_3^9+23654400 p_1^3 p_2^7 p_4^7 p_5^3 p_6^5
 p_3^9+8489250 p_1^4 p_2^6 p_4^7 p_5^3 p_6^5 p_3^9+13222440 p_1^4 p_2^7 p_4^6 p_5^3 p_6^5 p_3^9+6225450 p_1^3 p_2^7 p_4^7 p_5^4 p_6^4 p_3^9+1867635
 p_1^4 p_2^6 p_4^7 p_5^4 p_6^4 p_3^9+1867635 p_1^4 p_2^7 p_4^6 p_5^4 p_6^4 p_3^9+6225450 p_1^4 p_2^7 p_4^7 p_5^3 p_6^4 p_3^9) \textcolor{red}{z^{34}}+(2069760
 p_1^2 p_2^6 p_4^7 p_5^3 p_6^5 p_3^{10}+24147200 p_1^3 p_2^6 p_4^6 p_5^3 p_6^5 p_3^{10}+2069760 p_1^3 p_2^7 p_4^6 p_5^2 p_6^5 p_3^{10}+11383680 p_1^3
 p_2^6 p_4^7 p_5^3 p_6^4 p_3^{10}+11383680 p_1^3 p_2^7 p_4^6 p_5^3 p_6^4 p_3^{10}+205800 p_1^3 p_2^6 p_4^6 p_5^3 p_6^6 p_3^9+3773000 p_1^2 p_2^6 p_4^7
 p_5^4 p_6^5 p_3^9+9240000 p_1^3 p_2^5 p_4^7 p_5^4 p_6^5 p_3^9+37560600 p_1^3 p_2^6 p_4^6 p_5^4 p_6^5 p_3^9+82222140 p_1^3 p_2^6 p_4^7 p_5^3 p_6^5
 p_3^9+82222140 p_1^3 p_2^7 p_4^6 p_5^3 p_6^5 p_3^9+37560600 p_1^4 p_2^6 p_4^6 p_5^3 p_6^5 p_3^9+9240000 p_1^4 p_2^7 p_4^5 p_5^3 p_6^5 p_3^9+3773000
 p_1^4 p_2^7 p_4^6 p_5^2 p_6^5 p_3^9+27544440 p_1^3 p_2^6 p_4^7 p_5^4 p_6^4 p_3^9+13280960 p_1^3 p_2^7 p_4^6 p_5^4 p_6^4 p_3^9+6225450 p_1^4 p_2^6
 p_4^6 p_5^4 p_6^4 p_3^9+41164200 p_1^3 p_2^7 p_4^7 p_5^3 p_6^4 p_3^9+13280960 p_1^4 p_2^6 p_4^7 p_5^3 p_6^4 p_3^9+27544440 p_1^4 p_2^7 p_4^6 p_5^3
 p_6^4 p_3^9) \textcolor{red}{z^{33}}+(5588352 p_1^2 p_2^6 p_4^6 p_5^3 p_6^5 p_3^{10}+5588352 p_1^3 p_2^6 p_4^6 p_5^2 p_6^5 p_3^{10}+30735936 p_1^3 p_2^6
 p_4^6 p_5^3 p_6^4 p_3^{10}+5197500 p_1^2 p_2^5 p_4^7 p_5^4 p_6^5 p_3^9+10187100 p_1^2 p_2^6 p_4^6 p_5^4 p_6^5 p_3^9+38981250 p_1^3 p_2^5 p_4^6 p_5^4
 p_6^5 p_3^9+28385280 p_1^2 p_2^6 p_4^7 p_5^3 p_6^5 p_3^9+63669375 p_1^3 p_2^5 p_4^7 p_5^3 p_6^5 p_3^9+440527626 p_1^3 p_2^6 p_4^6 p_5^3 p_6^5 p_3^9+63669375
 p_1^3 p_2^7 p_4^5 p_5^3 p_6^5 p_3^9+38981250 p_1^4 p_2^6 p_4^5 p_5^3 p_6^5 p_3^9+28385280 p_1^3 p_2^7 p_4^6 p_5^2 p_6^5 p_3^9+10187100 p_1^4 p_2^6
 p_4^6 p_5^2 p_6^5 p_3^9+5197500 p_1^4 p_2^7 p_4^5 p_5^2 p_6^5 p_3^9+7470540 p_1^2 p_2^6 p_4^7 p_5^4 p_6^4 p_3^9+28586250 p_1^3 p_2^5 p_4^7 p_5^4
 p_6^4 p_3^9+87268104 p_1^3 p_2^6 p_4^6 p_5^4 p_6^4 p_3^9+202848030 p_1^3 p_2^6 p_4^7 p_5^3 p_6^4 p_3^9+202848030 p_1^3 p_2^7 p_4^6 p_5^3 p_6^4 p_3^9+87268104
 p_1^4 p_2^6 p_4^6 p_5^3 p_6^4 p_3^9+28586250 p_1^4 p_2^7 p_4^5 p_5^3 p_6^4 p_3^9+7470540 p_1^4 p_2^7 p_4^6 p_5^2 p_6^4 p_3^9+11884950 p_1^3 p_2^6
 p_4^6 p_5^4 p_6^5 p_3^8+11884950 p_1^4 p_2^6 p_4^6 p_5^3 p_6^5 p_3^8+8715630 p_1^3 p_2^6 p_4^7 p_5^4 p_6^4 p_3^8+2614689 p_1^4 p_2^6 p_4^6 p_5^4
 p_6^4 p_3^8+8715630 p_1^4 p_2^7 p_4^6 p_5^3 p_6^4 p_3^8) \textcolor{red}{z^{32}}+(24948000 p_1^2 p_2^5 p_4^6 p_5^4 p_6^5 p_3^9+40748400 p_1^2 p_2^5 p_4^7
 p_5^3 p_6^5 p_3^9+177031008 p_1^2 p_2^6 p_4^6 p_5^3 p_6^5 p_3^9+467082000 p_1^3 p_2^5 p_4^6 p_5^3 p_6^5 p_3^9+467082000 p_1^3 p_2^6 p_4^5 p_5^3 p_6^5
 p_3^9+177031008 p_1^3 p_2^6 p_4^6 p_5^2 p_6^5 p_3^9+40748400 p_1^3 p_2^7 p_4^5 p_5^2 p_6^5 p_3^9+24948000 p_1^4 p_2^6 p_4^5 p_5^2 p_6^5 p_3^9+18295200
 p_1^2 p_2^5 p_4^7 p_5^4 p_6^4 p_3^9+35858592 p_1^2 p_2^6 p_4^6 p_5^4 p_6^4 p_3^9+137214000 p_1^3 p_2^5 p_4^6 p_5^4 p_6^4 p_3^9+60984000 p_1^2 p_2^6
 p_4^7 p_5^3 p_6^4 p_3^9+224116200 p_1^3 p_2^5 p_4^7 p_5^3 p_6^4 p_3^9+1339753968 p_1^3 p_2^6 p_4^6 p_5^3 p_6^4 p_3^9+224116200 p_1^3 p_2^7 p_4^5
 p_5^3 p_6^4 p_3^9+137214000 p_1^4 p_2^6 p_4^5 p_5^3 p_6^4 p_3^9+60984000 p_1^3 p_2^7 p_4^6 p_5^2 p_6^4 p_3^9+35858592 p_1^4 p_2^6 p_4^6 p_5^2 p_6^4
 p_3^9+18295200 p_1^4 p_2^7 p_4^5 p_5^2 p_6^4 p_3^9+29106000 p_1^3 p_2^5 p_4^6 p_5^4 p_6^5 p_3^8+191866752 p_1^3 p_2^6 p_4^6 p_5^3 p_6^5 p_3^8+29106000
 p_1^4 p_2^6 p_4^5 p_5^3 p_6^5 p_3^8+21344400 p_1^3 p_2^5 p_4^7 p_5^4 p_6^4 p_3^8+66594528 p_1^3 p_2^6 p_4^6 p_5^4 p_6^4 p_3^8+71148000 p_1^3 p_2^6
 p_4^7 p_5^3 p_6^4 p_3^8+71148000 p_1^3 p_2^7 p_4^6 p_5^3 p_6^4 p_3^8+66594528 p_1^4 p_2^6 p_4^6 p_5^3 p_6^4 p_3^8+21344400 p_1^4 p_2^7 p_4^5 p_5^3
 p_6^4 p_3^8) \textcolor{red}{z^{31}}+(353089660 p_1^2 p_2^5 p_4^6 p_5^3 p_6^5 p_3^9+247546530 p_1^2 p_2^6 p_4^5 p_5^3 p_6^5 p_3^9+707437500 p_1^3 p_2^5
 p_4^5 p_5^3 p_6^5 p_3^9+60555264 p_1^2 p_2^6 p_4^6 p_5^2 p_6^5 p_3^9+247546530 p_1^3 p_2^5 p_4^6 p_5^2 p_6^5 p_3^9+353089660 p_1^3 p_2^6 p_4^5 p_5^2
 p_6^5 p_3^9+111143340 p_1^2 p_2^5 p_4^6 p_5^4 p_6^4 p_3^9+155636250 p_1^2 p_2^5 p_4^7 p_5^3 p_6^4 p_3^9+542666124 p_1^2 p_2^6 p_4^6 p_5^3 p_6^4 p_3^9+1941993130
 p_1^3 p_2^5 p_4^6 p_5^3 p_6^4 p_3^9+1941993130 p_1^3 p_2^6 p_4^5 p_5^3 p_6^4 p_3^9+542666124 p_1^3 p_2^6 p_4^6 p_5^2 p_6^4 p_3^9+155636250 p_1^3
 p_2^7 p_4^5 p_5^2 p_6^4 p_3^9+111143340 p_1^4 p_2^6 p_4^5 p_5^2 p_6^4 p_3^9+5478396 p_1^3 p_2^6 p_4^6 p_5^3 p_6^3 p_3^9+20212500 p_1^2 p_2^5 p_4^6
 p_5^4 p_6^5 p_3^8+110387200 p_1^2 p_2^6 p_4^6 p_5^3 p_6^5 p_3^8+388031490 p_1^3 p_2^5 p_4^6 p_5^3 p_6^5 p_3^8+388031490 p_1^3 p_2^6 p_4^5 p_5^3 p_6^5
 p_3^8+110387200 p_1^3 p_2^6 p_4^6 p_5^2 p_6^5 p_3^8+20212500 p_1^4 p_2^6 p_4^5 p_5^2 p_6^5 p_3^8+14822500 p_1^2 p_2^5 p_4^7 p_5^4 p_6^4 p_3^8+29052100
 p_1^2 p_2^6 p_4^6 p_5^4 p_6^4 p_3^8+253998360 p_1^3 p_2^5 p_4^6 p_5^4 p_6^4 p_3^8+8715630 p_1^3 p_2^6 p_4^5 p_5^4 p_6^4 p_3^8+181575625 p_1^3 p_2^5
 p_4^7 p_5^3 p_6^4 p_3^8+1554121926 p_1^3 p_2^6 p_4^6 p_5^3 p_6^4 p_3^8+8715630 p_1^4 p_2^5 p_4^6 p_5^3 p_6^4 p_3^8+181575625 p_1^3 p_2^7 p_4^5 p_5^3
 p_6^4 p_3^8+253998360 p_1^4 p_2^6 p_4^5 p_5^3 p_6^4 p_3^8+29052100 p_1^4 p_2^6 p_4^6 p_5^2 p_6^4 p_3^8+14822500 p_1^4 p_2^7 p_4^5 p_5^2 p_6^4 p_3^8+6391462
 p_1^3 p_2^6 p_4^6 p_5^4 p_6^3 p_3^8+6391462 p_1^4 p_2^6 p_4^6 p_5^3 p_6^3 p_3^8) \textcolor{red}{z^{30}}+(651974400 p_1^2 p_2^5 p_4^5 p_5^3 p_6^5 p_3^9+195592320
 p_1^2 p_2^5 p_4^6 p_5^2 p_6^5 p_3^9+195592320 p_1^2 p_2^6 p_4^5 p_5^2 p_6^5 p_3^9+651974400 p_1^3 p_2^5 p_4^5 p_5^2 p_6^5 p_3^9+1644128640 p_1^2
 p_2^5 p_4^6 p_5^3 p_6^4 p_3^9+1075757760 p_1^2 p_2^6 p_4^5 p_5^3 p_6^4 p_3^9+3585859200 p_1^3 p_2^5 p_4^5 p_5^3 p_6^4 p_3^9+292723200 p_1^2 p_2^6
 p_4^6 p_5^2 p_6^4 p_3^9+1075757760 p_1^3 p_2^5 p_4^6 p_5^2 p_6^4 p_3^9+1644128640 p_1^3 p_2^6 p_4^5 p_5^2 p_6^4 p_3^9+413887320 p_1^2 p_2^5 p_4^6
 p_5^3 p_6^5 p_3^8+356548500 p_1^2 p_2^6 p_4^5 p_5^3 p_6^5 p_3^8+1189465200 p_1^3 p_2^5 p_4^5 p_5^3 p_6^5 p_3^8+356548500 p_1^3 p_2^5 p_4^6 p_5^2
 p_6^5 p_3^8+413887320 p_1^3 p_2^6 p_4^5 p_5^2 p_6^5 p_3^8+268939440 p_1^2 p_2^5 p_4^6 p_5^4 p_6^4 p_3^8+48024900 p_1^3 p_2^5 p_4^5 p_5^4 p_6^4 p_3^8+133402500
 p_1^2 p_2^5 p_4^7 p_5^3 p_6^4 p_3^8+672348600 p_1^2 p_2^6 p_4^6 p_5^3 p_6^4 p_3^8+4320547560 p_1^3 p_2^5 p_4^6 p_5^3 p_6^4 p_3^8+4320547560 p_1^3
 p_2^6 p_4^5 p_5^3 p_6^4 p_3^8+48024900 p_1^4 p_2^5 p_4^5 p_5^3 p_6^4 p_3^8+672348600 p_1^3 p_2^6 p_4^6 p_5^2 p_6^4 p_3^8+133402500 p_1^3 p_2^7 p_4^5
 p_5^2 p_6^4 p_3^8+268939440 p_1^4 p_2^6 p_4^5 p_5^2 p_6^4 p_3^8+35218260 p_1^3 p_2^5 p_4^6 p_5^4 p_6^3 p_3^8+180457200 p_1^3 p_2^6 p_4^6 p_5^3 p_6^3
 p_3^8+35218260 p_1^4 p_2^6 p_4^5 p_5^3 p_6^3 p_3^8+119528640 p_1^3 p_2^5 p_4^6 p_5^4 p_6^4 p_3^7+398428800 p_1^3 p_2^6 p_4^6 p_5^3 p_6^4 p_3^7+119528640
 p_1^4 p_2^6 p_4^5 p_5^3 p_6^4 p_3^7) \textcolor{red}{z^{29}}+(651974400 p_1^2 p_2^5 p_4^5 p_5^2 p_6^5 p_3^9+3585859200 p_1^2 p_2^5 p_4^5 p_5^3 p_6^4 p_3^9+1075757760
 p_1^2 p_2^5 p_4^6 p_5^2 p_6^4 p_3^9+1075757760 p_1^2 p_2^6 p_4^5 p_5^2 p_6^4 p_3^9+3585859200 p_1^3 p_2^5 p_4^5 p_5^2 p_6^4 p_3^9+54573750 p_1^2
 p_2^4 p_4^6 p_5^3 p_6^5 p_3^8+1671169500 p_1^2 p_2^5 p_4^5 p_5^3 p_6^5 p_3^8+298045440 p_1^2 p_2^5 p_4^6 p_5^2 p_6^5 p_3^8+298045440 p_1^2 p_2^6
 p_4^5 p_5^2 p_6^5 p_3^8+1671169500 p_1^3 p_2^5 p_4^5 p_5^2 p_6^5 p_3^8+54573750 p_1^3 p_2^6 p_4^4 p_5^2 p_6^5 p_3^8+24502500 p_1^2 p_2^4 p_4^6 p_5^4
 p_6^4 p_3^8+48024900 p_1^2 p_2^5 p_4^5 p_5^4 p_6^4 p_3^8+4609356210 p_1^2 p_2^5 p_4^6 p_5^3 p_6^4 p_3^8+300155625 p_1^3 p_2^4 p_4^6 p_5^3 p_6^4 p_3^8+3054383640
 p_1^2 p_2^6 p_4^5 p_5^3 p_6^4 p_3^8+14283282150 p_1^3 p_2^5 p_4^5 p_5^3 p_6^4 p_3^8+300155625 p_1^3 p_2^6 p_4^4 p_5^3 p_6^4 p_3^8+446054400 p_1^2
 p_2^6 p_4^6 p_5^2 p_6^4 p_3^8+3054383640 p_1^3 p_2^5 p_4^6 p_5^2 p_6^4 p_3^8+4609356210 p_1^3 p_2^6 p_4^5 p_5^2 p_6^4 p_3^8+48024900 p_1^4 p_2^5
 p_4^5 p_5^2 p_6^4 p_3^8+24502500 p_1^4 p_2^6 p_4^4 p_5^2 p_6^4 p_3^8+35218260 p_1^2 p_2^5 p_4^6 p_5^4 p_6^3 p_3^8+117394200 p_1^2 p_2^6 p_4^6 p_5^3
 p_6^3 p_3^8+607296690 p_1^3 p_2^5 p_4^6 p_5^3 p_6^3 p_3^8+607296690 p_1^3 p_2^6 p_4^5 p_5^3 p_6^3 p_3^8+117394200 p_1^3 p_2^6 p_4^6 p_5^2 p_6^3 p_3^8+35218260
 p_1^4 p_2^6 p_4^5 p_5^2 p_6^3 p_3^8+499167900 p_1^3 p_2^5 p_4^5 p_5^3 p_6^5 p_3^7+186763500 p_1^2 p_2^5 p_4^6 p_5^4 p_6^4 p_3^7+56029050 p_1^3 p_2^5
 p_4^5 p_5^4 p_6^4 p_3^7+2655776970 p_1^3 p_2^5 p_4^6 p_5^3 p_6^4 p_3^7+2655776970 p_1^3 p_2^6 p_4^5 p_5^3 p_6^4 p_3^7+56029050 p_1^4 p_2^5 p_4^5
 p_5^3 p_6^4 p_3^7+186763500 p_1^4 p_2^6 p_4^5 p_5^2 p_6^4 p_3^7+41087970 p_1^3 p_2^5 p_4^6 p_5^4 p_6^3 p_3^7+136959900 p_1^3 p_2^6 p_4^6 p_5^3 p_6^3
 p_3^7+41087970 p_1^4 p_2^6 p_4^5 p_5^3 p_6^3 p_3^7) \textcolor{red}{z^{28}}+(4079910912 p_1^2 p_2^5 p_4^5 p_5^2 p_6^4 p_3^9+323400000 p_1^2 p_2^4 p_4^5
 p_5^3 p_6^5 p_3^8+2253071744 p_1^2 p_2^5 p_4^5 p_5^2 p_6^5 p_3^8+323400000 p_1^3 p_2^5 p_4^4 p_5^2 p_6^5 p_3^8+11383680 p_1 p_2^5 p_4^6 p_5^3 p_6^4
 p_3^8+830060000 p_1^2 p_2^4 p_4^6 p_5^3 p_6^4 p_3^8+18476731056 p_1^2 p_2^5 p_4^5 p_5^3 p_6^4 p_3^8+1778700000 p_1^3 p_2^4 p_4^5 p_5^3 p_6^4 p_3^8+1778700000
 p_1^3 p_2^5 p_4^4 p_5^3 p_6^4 p_3^8+4063973760 p_1^2 p_2^5 p_4^6 p_5^2 p_6^4 p_3^8+4063973760 p_1^2 p_2^6 p_4^5 p_5^2 p_6^4 p_3^8+18476731056 p_1^3
 p_2^5 p_4^5 p_5^2 p_6^4 p_3^8+830060000 p_1^3 p_2^6 p_4^4 p_5^2 p_6^4 p_3^8+11383680 p_1^3 p_2^6 p_4^5 p_5 p_6^4 p_3^8+871627680 p_1^2 p_2^5 p_4^6
 p_5^3 p_6^3 p_3^8+766975440 p_1^2 p_2^6 p_4^5 p_5^3 p_6^3 p_3^8+2414513024 p_1^3 p_2^5 p_4^5 p_5^3 p_6^3 p_3^8+766975440 p_1^3 p_2^5 p_4^6 p_5^2
 p_6^3 p_3^8+871627680 p_1^3 p_2^6 p_4^5 p_5^2 p_6^3 p_3^8+887409600 p_1^2 p_2^5 p_4^5 p_5^3 p_6^5 p_3^7+887409600 p_1^3 p_2^5 p_4^5 p_5^2 p_6^5 p_3^7+50820000
 p_1^2 p_2^4 p_4^6 p_5^4 p_6^4 p_3^7+99607200 p_1^2 p_2^5 p_4^5 p_5^4 p_6^4 p_3^7+3252073440 p_1^2 p_2^5 p_4^6 p_5^3 p_6^4 p_3^7+622545000 p_1^3 p_2^4
 p_4^6 p_5^3 p_6^4 p_3^7+2075150000 p_1^2 p_2^6 p_4^5 p_5^3 p_6^4 p_3^7+19480302576 p_1^3 p_2^5 p_4^5 p_5^3 p_6^4 p_3^7+622545000 p_1^3 p_2^6 p_4^4
 p_5^3 p_6^4 p_3^7+2075150000 p_1^3 p_2^5 p_4^6 p_5^2 p_6^4 p_3^7+3252073440 p_1^3 p_2^6 p_4^5 p_5^2 p_6^4 p_3^7+99607200 p_1^4 p_2^5 p_4^5 p_5^2
 p_6^4 p_3^7+50820000 p_1^4 p_2^6 p_4^4 p_5^2 p_6^4 p_3^7+73045280 p_1^2 p_2^5 p_4^6 p_5^4 p_6^3 p_3^7+21913584 p_1^3 p_2^5 p_4^5 p_5^4 p_6^3 p_3^7+1016898960
 p_1^3 p_2^5 p_4^6 p_5^3 p_6^3 p_3^7+1016898960 p_1^3 p_2^6 p_4^5 p_5^3 p_6^3 p_3^7+21913584 p_1^4 p_2^5 p_4^5 p_5^3 p_6^3 p_3^7+73045280 p_1^4 p_2^6
 p_4^5 p_5^2 p_6^3 p_3^7) \textcolor{red}{z^{27}}+(356548500 p_1^2 p_2^4 p_4^5 p_5^2 p_6^5 p_3^8+356548500 p_1^2 p_2^5 p_4^4 p_5^2 p_6^5 p_3^8+48024900
 p_1 p_2^4 p_4^6 p_5^3 p_6^4 p_3^8+94128804 p_1 p_2^5 p_4^5 p_5^3 p_6^4 p_3^8+5042614500 p_1^2 p_2^4 p_4^5 p_5^3 p_6^4 p_3^8+1961016750 p_1^2 p_2^5
 p_4^4 p_5^3 p_6^4 p_3^8+588305025 p_1^2 p_2^4 p_4^6 p_5^2 p_6^4 p_3^8+27835512516 p_1^2 p_2^5 p_4^5 p_5^2 p_6^4 p_3^8+1961016750 p_1^3 p_2^4 p_4^5
 p_5^2 p_6^4 p_3^8+588305025 p_1^2 p_2^6 p_4^4 p_5^2 p_6^4 p_3^8+5042614500 p_1^3 p_2^5 p_4^4 p_5^2 p_6^4 p_3^8+94128804 p_1^3 p_2^5 p_4^5 p_5 p_6^4
 p_3^8+48024900 p_1^3 p_2^6 p_4^4 p_5 p_6^4 p_3^8+264136950 p_1^2 p_2^4 p_4^6 p_5^3 p_6^3 p_3^8+4790284884 p_1^2 p_2^5 p_4^5 p_5^3 p_6^3 p_3^8+880456500
 p_1^2 p_2^5 p_4^6 p_5^2 p_6^3 p_3^8+880456500 p_1^2 p_2^6 p_4^5 p_5^2 p_6^3 p_3^8+4790284884 p_1^3 p_2^5 p_4^5 p_5^2 p_6^3 p_3^8+264136950 p_1^3
 p_2^6 p_4^4 p_5^2 p_6^3 p_3^8+305613000 p_1^2 p_2^4 p_4^5 p_5^3 p_6^5 p_3^7+1669054464 p_1^2 p_2^5 p_4^5 p_5^2 p_6^5 p_3^7+305613000 p_1^3 p_2^5
 p_4^4 p_5^2 p_6^5 p_3^7+34303500 p_1^2 p_2^4 p_4^5 p_5^4 p_6^4 p_3^7+1413304200 p_1^2 p_2^4 p_4^6 p_5^3 p_6^4 p_3^7+30824064054 p_1^2 p_2^5 p_4^5
 p_5^3 p_6^4 p_3^7+6565308750 p_1^3 p_2^4 p_4^5 p_5^3 p_6^4 p_3^7+6565308750 p_1^3 p_2^5 p_4^4 p_5^3 p_6^4 p_3^7+3902976000 p_1^2 p_2^5 p_4^6 p_5^2
 p_6^4 p_3^7+3902976000 p_1^2 p_2^6 p_4^5 p_5^2 p_6^4 p_3^7+30824064054 p_1^3 p_2^5 p_4^5 p_5^2 p_6^4 p_3^7+1413304200 p_1^3 p_2^6 p_4^4 p_5^2 p_6^4
 p_3^7+34303500 p_1^4 p_2^5 p_4^4 p_5^2 p_6^4 p_3^7+25155900 p_1^2 p_2^4 p_4^6 p_5^4 p_6^3 p_3^7+49305564 p_1^2 p_2^5 p_4^5 p_5^4 p_6^3 p_3^7+1445944500
 p_1^2 p_2^5 p_4^6 p_5^3 p_6^3 p_3^7+308159775 p_1^3 p_2^4 p_4^6 p_5^3 p_6^3 p_3^7+1027199250 p_1^2 p_2^6 p_4^5 p_5^3 p_6^3 p_3^7+8951787306 p_1^3
 p_2^5 p_4^5 p_5^3 p_6^3 p_3^7+308159775 p_1^3 p_2^6 p_4^4 p_5^3 p_6^3 p_3^7+1027199250 p_1^3 p_2^5 p_4^6 p_5^2 p_6^3 p_3^7+1445944500 p_1^3 p_2^6
 p_4^5 p_5^2 p_6^3 p_3^7+49305564 p_1^4 p_2^5 p_4^5 p_5^2 p_6^3 p_3^7+25155900 p_1^4 p_2^6 p_4^4 p_5^2 p_6^3 p_3^7+8199664704 p_1^3 p_2^5 p_4^5 p_5^3
 p_6^4 p_3^6) \textcolor{red}{z^{26}}+(409812480 p_1 p_2^4 p_4^5 p_5^3 p_6^4 p_3^8+122943744 p_1 p_2^5 p_4^5 p_5^2 p_6^4 p_3^8+7991343360 p_1^2 p_2^4 p_4^5
 p_5^2 p_6^4 p_3^8+7991343360 p_1^2 p_2^5 p_4^4 p_5^2 p_6^4 p_3^8+122943744 p_1^2 p_2^5 p_4^5 p_5 p_6^4 p_3^8+409812480 p_1^3 p_2^5 p_4^4 p_5 p_6^4
 p_3^8+2253968640 p_1^2 p_2^4 p_4^5 p_5^3 p_6^3 p_3^8+8611029504 p_1^2 p_2^5 p_4^5 p_5^2 p_6^3 p_3^8+2253968640 p_1^3 p_2^5 p_4^4 p_5^2 p_6^3 p_3^8+599001480
 p_1^2 p_2^4 p_4^5 p_5^2 p_6^5 p_3^7+599001480 p_1^2 p_2^5 p_4^4 p_5^2 p_6^5 p_3^7+114345000 p_1 p_2^4 p_4^6 p_5^3 p_6^4 p_3^7+224116200 p_1 p_2^5
 p_4^5 p_5^3 p_6^4 p_3^7+19609868880 p_1^2 p_2^4 p_4^5 p_5^3 p_6^4 p_3^7+9158882040 p_1^2 p_2^5 p_4^4 p_5^3 p_6^4 p_3^7+2858625000 p_1^3 p_2^4 p_4^4
 p_5^3 p_6^4 p_3^7+1400726250 p_1^2 p_2^4 p_4^6 p_5^2 p_6^4 p_3^7+59798117736 p_1^2 p_2^5 p_4^5 p_5^2 p_6^4 p_3^7+9158882040 p_1^3 p_2^4 p_4^5 p_5^2
 p_6^4 p_3^7+1400726250 p_1^2 p_2^6 p_4^4 p_5^2 p_6^4 p_3^7+19609868880 p_1^3 p_2^5 p_4^4 p_5^2 p_6^4 p_3^7+224116200 p_1^3 p_2^5 p_4^5 p_5 p_6^4
 p_3^7+114345000 p_1^3 p_2^6 p_4^4 p_5 p_6^4 p_3^7+30187080 p_1^2 p_2^4 p_4^5 p_5^4 p_6^3 p_3^7+966735000 p_1^2 p_2^4 p_4^6 p_5^3 p_6^3 p_3^7+17946116496
 p_1^2 p_2^5 p_4^5 p_5^3 p_6^3 p_3^7+3421632060 p_1^3 p_2^4 p_4^5 p_5^3 p_6^3 p_3^7+3421632060 p_1^3 p_2^5 p_4^4 p_5^3 p_6^3 p_3^7+2096325000 p_1^2
 p_2^5 p_4^6 p_5^2 p_6^3 p_3^7+2096325000 p_1^2 p_2^6 p_4^5 p_5^2 p_6^3 p_3^7+17946116496 p_1^3 p_2^5 p_4^5 p_5^2 p_6^3 p_3^7+966735000 p_1^3 p_2^6
 p_4^4 p_5^2 p_6^3 p_3^7+30187080 p_1^4 p_2^5 p_4^4 p_5^2 p_6^3 p_3^7+439267752 p_1^3 p_2^5 p_4^5 p_5^3 p_6^2 p_3^7+16734009600 p_1^2 p_2^5 p_4^5
 p_5^3 p_6^4 p_3^6+5020202880 p_1^3 p_2^4 p_4^5 p_5^3 p_6^4 p_3^6+5020202880 p_1^3 p_2^5 p_4^4 p_5^3 p_6^4 p_3^6+16734009600 p_1^3 p_2^5 p_4^5 p_5^2
 p_6^4 p_3^6+6754454784 p_1^3 p_2^5 p_4^5 p_5^3 p_6^3 p_3^6) \textcolor{red}{z^{25}}+(557800320 p_1 p_2^4 p_4^5 p_5^2 p_6^4 p_3^8+1859334400 p_1^2 p_2^4
 p_4^4 p_5^2 p_6^4 p_3^8+557800320 p_1^2 p_2^5 p_4^4 p_5 p_6^4 p_3^8+3067901760 p_1^2 p_2^4 p_4^5 p_5^2 p_6^3 p_3^8+3067901760 p_1^2 p_2^5 p_4^4 p_5^2
 p_6^3 p_3^8+221852400 p_1^2 p_2^4 p_4^4 p_5^2 p_6^5 p_3^7+2212838320 p_1 p_2^4 p_4^5 p_5^3 p_6^4 p_3^7+1985156250 p_1^2 p_2^3 p_4^5 p_5^3 p_6^4 p_3^7+6097637700
 p_1^2 p_2^4 p_4^4 p_5^3 p_6^4 p_3^7+520396800 p_1 p_2^5 p_4^5 p_5^2 p_6^4 p_3^7+40830215370 p_1^2 p_2^4 p_4^5 p_5^2 p_6^4 p_3^7+40830215370 p_1^2
 p_2^5 p_4^4 p_5^2 p_6^4 p_3^7+6097637700 p_1^3 p_2^4 p_4^4 p_5^2 p_6^4 p_3^7+1985156250 p_1^3 p_2^5 p_4^3 p_5^2 p_6^4 p_3^7+520396800 p_1^2 p_2^5
 p_4^5 p_5 p_6^4 p_3^7+2212838320 p_1^3 p_2^5 p_4^4 p_5 p_6^4 p_3^7+69877500 p_1 p_2^4 p_4^6 p_5^3 p_6^3 p_3^7+136959900 p_1 p_2^5 p_4^5 p_5^3 p_6^3
 p_3^7+16980581310 p_1^2 p_2^4 p_4^5 p_5^3 p_6^3 p_3^7+5281747240 p_1^2 p_2^5 p_4^4 p_5^3 p_6^3 p_3^7+2862182400 p_1^3 p_2^4 p_4^4 p_5^3 p_6^3 p_3^7+855999375
 p_1^2 p_2^4 p_4^6 p_5^2 p_6^3 p_3^7+41763159900 p_1^2 p_2^5 p_4^5 p_5^2 p_6^3 p_3^7+5281747240 p_1^3 p_2^4 p_4^5 p_5^2 p_6^3 p_3^7+855999375 p_1^2
 p_2^6 p_4^4 p_5^2 p_6^3 p_3^7+16980581310 p_1^3 p_2^5 p_4^4 p_5^2 p_6^3 p_3^7+136959900 p_1^3 p_2^5 p_4^5 p_5 p_6^3 p_3^7+69877500 p_1^3 p_2^6 p_4^4
 p_5 p_6^3 p_3^7+1220188200 p_1^2 p_2^5 p_4^5 p_5^3 p_6^2 p_3^7+1220188200 p_1^3 p_2^5 p_4^5 p_5^2 p_6^2 p_3^7+18993314340 p_1^2 p_2^4 p_4^5 p_5^3
 p_6^4 p_3^6+10676646750 p_1^2 p_2^5 p_4^4 p_5^3 p_6^4 p_3^6+3890016900 p_1^3 p_2^4 p_4^4 p_5^3 p_6^4 p_3^6+38856294400 p_1^2 p_2^5 p_4^5 p_5^2 p_6^4
 p_3^6+10676646750 p_1^3 p_2^4 p_4^5 p_5^2 p_6^4 p_3^6+18993314340 p_1^3 p_2^5 p_4^4 p_5^2 p_6^4 p_3^6+3913140 p_1^2 p_2^4 p_4^5 p_5^4 p_6^3 p_3^6+81523750
 p_1^2 p_2^4 p_4^6 p_5^3 p_6^3 p_3^6+16798204500 p_1^2 p_2^5 p_4^5 p_5^3 p_6^3 p_3^6+5039461350 p_1^3 p_2^4 p_4^5 p_5^3 p_6^3 p_3^6+5039461350 p_1^3
 p_2^5 p_4^4 p_5^3 p_6^3 p_3^6+16798204500 p_1^3 p_2^5 p_4^5 p_5^2 p_6^3 p_3^6+81523750 p_1^3 p_2^6 p_4^4 p_5^2 p_6^3 p_3^6+3913140 p_1^4 p_2^5 p_4^4
 p_5^2 p_6^3 p_3^6+1423552900 p_1^3 p_2^5 p_4^5 p_5^3 p_6^2 p_3^6) \textcolor{red}{z^{24}}+(457380000 p_1 p_2^3 p_4^5 p_5^3 p_6^4 p_3^7+896464800 p_1 p_2^4
 p_4^4 p_5^3 p_6^4 p_3^7+4201797600 p_1 p_2^4 p_4^5 p_5^2 p_6^4 p_3^7+5602905000 p_1^2 p_2^3 p_4^5 p_5^2 p_6^4 p_3^7+268939440 p_1 p_2^5 p_4^4 p_5^2
 p_6^4 p_3^7+32647658400 p_1^2 p_2^4 p_4^4 p_5^2 p_6^4 p_3^7+5602905000 p_1^2 p_2^5 p_4^3 p_5^2 p_6^4 p_3^7+268939440 p_1^2 p_2^4 p_4^5 p_5 p_6^4
 p_3^7+4201797600 p_1^2 p_2^5 p_4^4 p_5 p_6^4 p_3^7+896464800 p_1^3 p_2^4 p_4^4 p_5 p_6^4 p_3^7+457380000 p_1^3 p_2^5 p_4^3 p_5 p_6^4 p_3^7+1537683840
 p_1 p_2^4 p_4^5 p_5^3 p_6^3 p_3^7+2515590000 p_1^2 p_2^3 p_4^5 p_5^3 p_6^3 p_3^7+6940533600 p_1^2 p_2^4 p_4^4 p_5^3 p_6^3 p_3^7+402494400 p_1 p_2^5
 p_4^5 p_5^2 p_6^3 p_3^7+38328942960 p_1^2 p_2^4 p_4^5 p_5^2 p_6^3 p_3^7+38328942960 p_1^2 p_2^5 p_4^4 p_5^2 p_6^3 p_3^7+6940533600 p_1^3 p_2^4 p_4^4
 p_5^2 p_6^3 p_3^7+2515590000 p_1^3 p_2^5 p_4^3 p_5^2 p_6^3 p_3^7+402494400 p_1^2 p_2^5 p_4^5 p_5 p_6^3 p_3^7+1537683840 p_1^3 p_2^5 p_4^4 p_5 p_6^3
 p_3^7+1075757760 p_1^2 p_2^4 p_4^5 p_5^3 p_6^2 p_3^7+3585859200 p_1^2 p_2^5 p_4^5 p_5^2 p_6^2 p_3^7+1075757760 p_1^3 p_2^5 p_4^4 p_5^2 p_6^2 p_3^7+2561328000
 p_1 p_2^4 p_4^5 p_5^3 p_6^4 p_3^6+4802490000 p_1^2 p_2^3 p_4^5 p_5^3 p_6^4 p_3^6+14386125600 p_1^2 p_2^4 p_4^4 p_5^3 p_6^4 p_3^6+48870138240 p_1^2
 p_2^4 p_4^5 p_5^2 p_6^4 p_3^6+48870138240 p_1^2 p_2^5 p_4^4 p_5^2 p_6^4 p_3^6+14386125600 p_1^3 p_2^4 p_4^4 p_5^2 p_6^4 p_3^6+4802490000 p_1^3 p_2^5
 p_4^3 p_5^2 p_6^4 p_3^6+2561328000 p_1^3 p_2^5 p_4^4 p_5 p_6^4 p_3^6+28131531120 p_1^2 p_2^4 p_4^5 p_5^3 p_6^3 p_3^6+12664602720 p_1^2 p_2^5 p_4^4
 p_5^3 p_6^3 p_3^6+6411081600 p_1^3 p_2^4 p_4^4 p_5^3 p_6^3 p_3^6+45964195200 p_1^2 p_2^5 p_4^5 p_5^2 p_6^3 p_3^6+12664602720 p_1^3 p_2^4 p_4^5 p_5^2
 p_6^3 p_3^6+28131531120 p_1^3 p_2^5 p_4^4 p_5^2 p_6^3 p_3^6+4183502400 p_1^2 p_2^5 p_4^5 p_5^3 p_6^2 p_3^6+1255050720 p_1^3 p_2^4 p_4^5 p_5^3 p_6^2
 p_3^6+1255050720 p_1^3 p_2^5 p_4^4 p_5^3 p_6^2 p_3^6+4183502400 p_1^3 p_2^5 p_4^5 p_5^2 p_6^2 p_3^6) \textcolor{red}{z^{23}}+(1400726250 p_1 p_2^3 p_4^5
 p_5^2 p_6^4 p_3^7+3186932364 p_1 p_2^4 p_4^4 p_5^2 p_6^4 p_3^7+4669087500 p_1^2 p_2^3 p_4^4 p_5^2 p_6^4 p_3^7+4669087500 p_1^2 p_2^4 p_4^3 p_5^2
 p_6^4 p_3^7+3186932364 p_1^2 p_2^4 p_4^4 p_5 p_6^4 p_3^7+1400726250 p_1^2 p_2^5 p_4^3 p_5 p_6^4 p_3^7+628897500 p_1 p_2^3 p_4^5 p_5^3 p_6^3 p_3^7+1232639100
 p_1 p_2^4 p_4^4 p_5^3 p_6^3 p_3^7+3421632060 p_1 p_2^4 p_4^5 p_5^2 p_6^3 p_3^7+7703994375 p_1^2 p_2^3 p_4^5 p_5^2 p_6^3 p_3^7+369791730 p_1 p_2^5
 p_4^4 p_5^2 p_6^3 p_3^7+38630489436 p_1^2 p_2^4 p_4^4 p_5^2 p_6^3 p_3^7+7703994375 p_1^2 p_2^5 p_4^3 p_5^2 p_6^3 p_3^7+369791730 p_1^2 p_2^4 p_4^5
 p_5 p_6^3 p_3^7+3421632060 p_1^2 p_2^5 p_4^4 p_5 p_6^3 p_3^7+1232639100 p_1^3 p_2^4 p_4^4 p_5 p_6^3 p_3^7+628897500 p_1^3 p_2^5 p_4^3 p_5 p_6^3 p_3^7+3294508140
 p_1^2 p_2^4 p_4^5 p_5^2 p_6^2 p_3^7+3294508140 p_1^2 p_2^5 p_4^4 p_5^2 p_6^2 p_3^7+1200622500 p_1 p_2^3 p_4^5 p_5^3 p_6^4 p_3^6+2353220100 p_1 p_2^4
 p_4^4 p_5^3 p_6^4 p_3^6+4002075000 p_1^2 p_2^3 p_4^4 p_5^3 p_6^4 p_3^6+6556999680 p_1 p_2^4 p_4^5 p_5^2 p_6^4 p_3^6+14707625625 p_1^2 p_2^3 p_4^5
 p_5^2 p_6^4 p_3^6+76881768516 p_1^2 p_2^4 p_4^4 p_5^2 p_6^4 p_3^6+14707625625 p_1^2 p_2^5 p_4^3 p_5^2 p_6^4 p_3^6+4002075000 p_1^3 p_2^4 p_4^3 p_5^2
 p_6^4 p_3^6+6556999680 p_1^2 p_2^5 p_4^4 p_5 p_6^4 p_3^6+2353220100 p_1^3 p_2^4 p_4^4 p_5 p_6^4 p_3^6+1200622500 p_1^3 p_2^5 p_4^3 p_5 p_6^4 p_3^6+3447314640
 p_1 p_2^4 p_4^5 p_5^3 p_6^3 p_3^6+10017315000 p_1^2 p_2^3 p_4^5 p_5^3 p_6^3 p_3^6+27874845306 p_1^2 p_2^4 p_4^4 p_5^3 p_6^3 p_3^6+80030488230 p_1^2
 p_2^4 p_4^5 p_5^2 p_6^3 p_3^6+80030488230 p_1^2 p_2^5 p_4^4 p_5^2 p_6^3 p_3^6+27874845306 p_1^3 p_2^4 p_4^4 p_5^2 p_6^3 p_3^6+10017315000 p_1^3 p_2^5
 p_4^3 p_5^2 p_6^3 p_3^6+3447314640 p_1^3 p_2^5 p_4^4 p_5 p_6^3 p_3^6+8519617260 p_1^2 p_2^4 p_4^5 p_5^3 p_6^2 p_3^6+3843592830 p_1^2 p_2^5 p_4^4
 p_5^3 p_6^2 p_3^6+1967099904 p_1^3 p_2^4 p_4^4 p_5^3 p_6^2 p_3^6+13340250000 p_1^2 p_2^5 p_4^5 p_5^2 p_6^2 p_3^6+3843592830 p_1^3 p_2^4 p_4^5 p_5^2
 p_6^2 p_3^6+8519617260 p_1^3 p_2^5 p_4^4 p_5^2 p_6^2 p_3^6+2629630080 p_1^2 p_2^4 p_4^5 p_5^3 p_6^3 p_3^5+788889024 p_1^3 p_2^4 p_4^4 p_5^3 p_6^3
 p_3^5+2629630080 p_1^3 p_2^5 p_4^4 p_5^2 p_6^3 p_3^5) \textcolor{red}{z^{22}}+(1328096000 p_1 p_2^3 p_4^4 p_5^2 p_6^4 p_3^7+398428800 p_1 p_2^4 p_4^4 p_5
 p_6^4 p_3^7+1328096000 p_1^2 p_2^4 p_4^3 p_5 p_6^4 p_3^7+2191358400 p_1 p_2^3 p_4^5 p_5^2 p_6^3 p_3^7+4881115008 p_1 p_2^4 p_4^4 p_5^2 p_6^3 p_3^7+7304528000
 p_1^2 p_2^3 p_4^4 p_5^2 p_6^3 p_3^7+7304528000 p_1^2 p_2^4 p_4^3 p_5^2 p_6^3 p_3^7+4881115008 p_1^2 p_2^4 p_4^4 p_5 p_6^3 p_3^7+2191358400 p_1^2
 p_2^5 p_4^3 p_5 p_6^3 p_3^7+3123681792 p_1^2 p_2^4 p_4^4 p_5^2 p_6^2 p_3^7+1138368000 p_1 p_2^3 p_4^4 p_5^3 p_6^4 p_3^6+3890016900 p_1 p_2^3 p_4^5
 p_5^2 p_6^4 p_3^6+10875161528 p_1 p_2^4 p_4^4 p_5^2 p_6^4 p_3^6+28921662000 p_1^2 p_2^3 p_4^4 p_5^2 p_6^4 p_3^6+28921662000 p_1^2 p_2^4 p_4^3 p_5^2
 p_6^4 p_3^6+10875161528 p_1^2 p_2^4 p_4^4 p_5 p_6^4 p_3^6+3890016900 p_1^2 p_2^5 p_4^3 p_5 p_6^4 p_3^6+1138368000 p_1^3 p_2^4 p_4^3 p_5 p_6^4 p_3^6+2752521200
 p_1 p_2^3 p_4^5 p_5^3 p_6^3 p_3^6+5394941552 p_1 p_2^4 p_4^4 p_5^3 p_6^3 p_3^6+10284615000 p_1^2 p_2^3 p_4^4 p_5^3 p_6^3 p_3^6+97828500 p_1^2 p_2^4
 p_4^3 p_5^3 p_6^3 p_3^6+10284615000 p_1 p_2^4 p_4^5 p_5^2 p_6^3 p_3^6+33718384700 p_1^2 p_2^3 p_4^5 p_5^2 p_6^3 p_3^6+97828500 p_1 p_2^5 p_4^4 p_5^2
 p_6^3 p_3^6+163830961008 p_1^2 p_2^4 p_4^4 p_5^2 p_6^3 p_3^6+97828500 p_1^3 p_2^3 p_4^4 p_5^2 p_6^3 p_3^6+33718384700 p_1^2 p_2^5 p_4^3 p_5^2 p_6^3
 p_3^6+10284615000 p_1^3 p_2^4 p_4^3 p_5^2 p_6^3 p_3^6+97828500 p_1^2 p_2^4 p_4^5 p_5 p_6^3 p_3^6+10284615000 p_1^2 p_2^5 p_4^4 p_5 p_6^3 p_3^6+5394941552
 p_1^3 p_2^4 p_4^4 p_5 p_6^3 p_3^6+2752521200 p_1^3 p_2^5 p_4^3 p_5 p_6^3 p_3^6+1138368000 p_1 p_2^4 p_4^5 p_5^3 p_6^2 p_3^6+3890016900 p_1^2 p_2^3
 p_4^5 p_5^3 p_6^2 p_3^6+10875161528 p_1^2 p_2^4 p_4^4 p_5^3 p_6^2 p_3^6+28921662000 p_1^2 p_2^4 p_4^5 p_5^2 p_6^2 p_3^6+28921662000 p_1^2 p_2^5 p_4^4
 p_5^2 p_6^2 p_3^6+10875161528 p_1^3 p_2^4 p_4^4 p_5^2 p_6^2 p_3^6+3890016900 p_1^3 p_2^5 p_4^3 p_5^2 p_6^2 p_3^6+1138368000 p_1^3 p_2^5 p_4^4 p_5
 p_6^2 p_3^6+3123681792 p_1^2 p_2^4 p_4^4 p_5^2 p_6^4 p_3^5+2191358400 p_1^2 p_2^3 p_4^5 p_5^3 p_6^3 p_3^5+4881115008 p_1^2 p_2^4 p_4^4 p_5^3 p_6^3
 p_3^5+7304528000 p_1^2 p_2^4 p_4^5 p_5^2 p_6^3 p_3^5+7304528000 p_1^2 p_2^5 p_4^4 p_5^2 p_6^3 p_3^5+4881115008 p_1^3 p_2^4 p_4^4 p_5^2 p_6^3 p_3^5+2191358400
 p_1^3 p_2^5 p_4^3 p_5^2 p_6^3 p_3^5+1328096000 p_1^2 p_2^4 p_4^5 p_5^3 p_6^2 p_3^5+398428800 p_1^3 p_2^4 p_4^4 p_5^3 p_6^2 p_3^5+1328096000 p_1^3
 p_2^5 p_4^4 p_5^2 p_6^2 p_3^5) \textcolor{red}{z^{21}}+(2629630080 p_1 p_2^3 p_4^4 p_5^2 p_6^3 p_3^7+788889024 p_1 p_2^4 p_4^4 p_5 p_6^3 p_3^7+2629630080
 p_1^2 p_2^4 p_4^3 p_5 p_6^3 p_3^7+8519617260 p_1 p_2^3 p_4^4 p_5^2 p_6^4 p_3^6+3843592830 p_1 p_2^4 p_4^3 p_5^2 p_6^4 p_3^6+13340250000 p_1^2 p_2^3
 p_4^3 p_5^2 p_6^4 p_3^6+1967099904 p_1 p_2^4 p_4^4 p_5 p_6^4 p_3^6+3843592830 p_1^2 p_2^3 p_4^4 p_5 p_6^4 p_3^6+8519617260 p_1^2 p_2^4 p_4^3 p_5
 p_6^4 p_3^6+3447314640 p_1 p_2^3 p_4^4 p_5^3 p_6^3 p_3^6+10017315000 p_1 p_2^3 p_4^5 p_5^2 p_6^3 p_3^6+27874845306 p_1 p_2^4 p_4^4 p_5^2 p_6^3 p_3^6+80030488230
 p_1^2 p_2^3 p_4^4 p_5^2 p_6^3 p_3^6+80030488230 p_1^2 p_2^4 p_4^3 p_5^2 p_6^3 p_3^6+27874845306 p_1^2 p_2^4 p_4^4 p_5 p_6^3 p_3^6+10017315000 p_1^2
 p_2^5 p_4^3 p_5 p_6^3 p_3^6+3447314640 p_1^3 p_2^4 p_4^3 p_5 p_6^3 p_3^6+1200622500 p_1 p_2^3 p_4^5 p_5^3 p_6^2 p_3^6+2353220100 p_1 p_2^4 p_4^4
 p_5^3 p_6^2 p_3^6+6556999680 p_1^2 p_2^3 p_4^4 p_5^3 p_6^2 p_3^6+4002075000 p_1 p_2^4 p_4^5 p_5^2 p_6^2 p_3^6+14707625625 p_1^2 p_2^3 p_4^5 p_5^2
 p_6^2 p_3^6+76881768516 p_1^2 p_2^4 p_4^4 p_5^2 p_6^2 p_3^6+14707625625 p_1^2 p_2^5 p_4^3 p_5^2 p_6^2 p_3^6+6556999680 p_1^3 p_2^4 p_4^3 p_5^2 p_6^2
 p_3^6+4002075000 p_1^2 p_2^5 p_4^4 p_5 p_6^2 p_3^6+2353220100 p_1^3 p_2^4 p_4^4 p_5 p_6^2 p_3^6+1200622500 p_1^3 p_2^5 p_4^3 p_5 p_6^2 p_3^6+3294508140
 p_1^2 p_2^3 p_4^4 p_5^2 p_6^4 p_3^5+3294508140 p_1^2 p_2^4 p_4^3 p_5^2 p_6^4 p_3^5+628897500 p_1 p_2^3 p_4^5 p_5^3 p_6^3 p_3^5+1232639100 p_1 p_2^4
 p_4^4 p_5^3 p_6^3 p_3^5+3421632060 p_1^2 p_2^3 p_4^4 p_5^3 p_6^3 p_3^5+369791730 p_1^2 p_2^4 p_4^3 p_5^3 p_6^3 p_3^5+7703994375 p_1^2 p_2^3 p_4^5
 p_5^2 p_6^3 p_3^5+38630489436 p_1^2 p_2^4 p_4^4 p_5^2 p_6^3 p_3^5+369791730 p_1^3 p_2^3 p_4^4 p_5^2 p_6^3 p_3^5+7703994375 p_1^2 p_2^5 p_4^3 p_5^2
 p_6^3 p_3^5+3421632060 p_1^3 p_2^4 p_4^3 p_5^2 p_6^3 p_3^5+1232639100 p_1^3 p_2^4 p_4^4 p_5 p_6^3 p_3^5+628897500 p_1^3 p_2^5 p_4^3 p_5 p_6^3 p_3^5+1400726250
 p_1^2 p_2^3 p_4^5 p_5^3 p_6^2 p_3^5+3186932364 p_1^2 p_2^4 p_4^4 p_5^3 p_6^2 p_3^5+4669087500 p_1^2 p_2^4 p_4^5 p_5^2 p_6^2 p_3^5+4669087500 p_1^2
 p_2^5 p_4^4 p_5^2 p_6^2 p_3^5+3186932364 p_1^3 p_2^4 p_4^4 p_5^2 p_6^2 p_3^5+1400726250 p_1^3 p_2^5 p_4^3 p_5^2 p_6^2 p_3^5) \textcolor{red}{z^{20}}+(4183502400
 p_1 p_2^3 p_4^3 p_5^2 p_6^4 p_3^6+1255050720 p_1 p_2^3 p_4^4 p_5 p_6^4 p_3^6+1255050720 p_1 p_2^4 p_4^3 p_5 p_6^4 p_3^6+4183502400 p_1^2 p_2^3 p_4^3
 p_5 p_6^4 p_3^6+28131531120 p_1 p_2^3 p_4^4 p_5^2 p_6^3 p_3^6+12664602720 p_1 p_2^4 p_4^3 p_5^2 p_6^3 p_3^6+45964195200 p_1^2 p_2^3 p_4^3 p_5^2 p_6^3
 p_3^6+6411081600 p_1 p_2^4 p_4^4 p_5 p_6^3 p_3^6+12664602720 p_1^2 p_2^3 p_4^4 p_5 p_6^3 p_3^6+28131531120 p_1^2 p_2^4 p_4^3 p_5 p_6^3 p_3^6+2561328000
 p_1 p_2^3 p_4^4 p_5^3 p_6^2 p_3^6+4802490000 p_1 p_2^3 p_4^5 p_5^2 p_6^2 p_3^6+14386125600 p_1 p_2^4 p_4^4 p_5^2 p_6^2 p_3^6+48870138240 p_1^2 p_2^3
 p_4^4 p_5^2 p_6^2 p_3^6+48870138240 p_1^2 p_2^4 p_4^3 p_5^2 p_6^2 p_3^6+14386125600 p_1^2 p_2^4 p_4^4 p_5 p_6^2 p_3^6+4802490000 p_1^2 p_2^5 p_4^3
 p_5 p_6^2 p_3^6+2561328000 p_1^3 p_2^4 p_4^3 p_5 p_6^2 p_3^6+1075757760 p_1 p_2^3 p_4^4 p_5^2 p_6^4 p_3^5+3585859200 p_1^2 p_2^3 p_4^3 p_5^2 p_6^4
 p_3^5+1075757760 p_1^2 p_2^4 p_4^3 p_5 p_6^4 p_3^5+1537683840 p_1 p_2^3 p_4^4 p_5^3 p_6^3 p_3^5+402494400 p_1^2 p_2^3 p_4^3 p_5^3 p_6^3 p_3^5+2515590000
 p_1 p_2^3 p_4^5 p_5^2 p_6^3 p_3^5+6940533600 p_1 p_2^4 p_4^4 p_5^2 p_6^3 p_3^5+38328942960 p_1^2 p_2^3 p_4^4 p_5^2 p_6^3 p_3^5+38328942960 p_1^2
 p_2^4 p_4^3 p_5^2 p_6^3 p_3^5+402494400 p_1^3 p_2^3 p_4^3 p_5^2 p_6^3 p_3^5+6940533600 p_1^2 p_2^4 p_4^4 p_5 p_6^3 p_3^5+2515590000 p_1^2 p_2^5 p_4^3
 p_5 p_6^3 p_3^5+1537683840 p_1^3 p_2^4 p_4^3 p_5 p_6^3 p_3^5+457380000 p_1 p_2^3 p_4^5 p_5^3 p_6^2 p_3^5+896464800 p_1 p_2^4 p_4^4 p_5^3 p_6^2 p_3^5+4201797600
 p_1^2 p_2^3 p_4^4 p_5^3 p_6^2 p_3^5+268939440 p_1^2 p_2^4 p_4^3 p_5^3 p_6^2 p_3^5+5602905000 p_1^2 p_2^3 p_4^5 p_5^2 p_6^2 p_3^5+32647658400 p_1^2
 p_2^4 p_4^4 p_5^2 p_6^2 p_3^5+268939440 p_1^3 p_2^3 p_4^4 p_5^2 p_6^2 p_3^5+5602905000 p_1^2 p_2^5 p_4^3 p_5^2 p_6^2 p_3^5+4201797600 p_1^3 p_2^4
 p_4^3 p_5^2 p_6^2 p_3^5+896464800 p_1^3 p_2^4 p_4^4 p_5 p_6^2 p_3^5+457380000 p_1^3 p_2^5 p_4^3 p_5 p_6^2 p_3^5) \textcolor{red}{z^{19}}+(1423552900 p_1
 p_2^3 p_4^3 p_5 p_6^4 p_3^6+3913140 p_1^2 p_2^4 p_4^3 p_6^3 p_3^6+3913140 p_2^3 p_4^4 p_5^2 p_6^3 p_3^6+81523750 p_1 p_2^2 p_4^4 p_5^2 p_6^3 p_3^6+16798204500
 p_1 p_2^3 p_4^3 p_5^2 p_6^3 p_3^6+5039461350 p_1 p_2^3 p_4^4 p_5 p_6^3 p_3^6+5039461350 p_1 p_2^4 p_4^3 p_5 p_6^3 p_3^6+16798204500 p_1^2 p_2^3 p_4^3
 p_5 p_6^3 p_3^6+81523750 p_1^2 p_2^4 p_4^2 p_5 p_6^3 p_3^6+18993314340 p_1 p_2^3 p_4^4 p_5^2 p_6^2 p_3^6+10676646750 p_1 p_2^4 p_4^3 p_5^2 p_6^2
 p_3^6+38856294400 p_1^2 p_2^3 p_4^3 p_5^2 p_6^2 p_3^6+3890016900 p_1 p_2^4 p_4^4 p_5 p_6^2 p_3^6+10676646750 p_1^2 p_2^3 p_4^4 p_5 p_6^2 p_3^6+18993314340
 p_1^2 p_2^4 p_4^3 p_5 p_6^2 p_3^6+1220188200 p_1 p_2^3 p_4^3 p_5^2 p_6^4 p_3^5+1220188200 p_1^2 p_2^3 p_4^3 p_5 p_6^4 p_3^5+69877500 p_1 p_2^2 p_4^4
 p_5^3 p_6^3 p_3^5+136959900 p_1 p_2^3 p_4^3 p_5^3 p_6^3 p_3^5+16980581310 p_1 p_2^3 p_4^4 p_5^2 p_6^3 p_3^5+855999375 p_1^2 p_2^2 p_4^4 p_5^2 p_6^3
 p_3^5+5281747240 p_1 p_2^4 p_4^3 p_5^2 p_6^3 p_3^5+41763159900 p_1^2 p_2^3 p_4^3 p_5^2 p_6^3 p_3^5+855999375 p_1^2 p_2^4 p_4^2 p_5^2 p_6^3 p_3^5+2862182400
 p_1 p_2^4 p_4^4 p_5 p_6^3 p_3^5+5281747240 p_1^2 p_2^3 p_4^4 p_5 p_6^3 p_3^5+16980581310 p_1^2 p_2^4 p_4^3 p_5 p_6^3 p_3^5+136959900 p_1^3 p_2^3
 p_4^3 p_5 p_6^3 p_3^5+69877500 p_1^3 p_2^4 p_4^2 p_5 p_6^3 p_3^5+2212838320 p_1 p_2^3 p_4^4 p_5^3 p_6^2 p_3^5+520396800 p_1^2 p_2^3 p_4^3 p_5^3 p_6^2
 p_3^5+1985156250 p_1 p_2^3 p_4^5 p_5^2 p_6^2 p_3^5+6097637700 p_1 p_2^4 p_4^4 p_5^2 p_6^2 p_3^5+40830215370 p_1^2 p_2^3 p_4^4 p_5^2 p_6^2 p_3^5+40830215370
 p_1^2 p_2^4 p_4^3 p_5^2 p_6^2 p_3^5+520396800 p_1^3 p_2^3 p_4^3 p_5^2 p_6^2 p_3^5+6097637700 p_1^2 p_2^4 p_4^4 p_5 p_6^2 p_3^5+1985156250 p_1^2 p_2^5
 p_4^3 p_5 p_6^2 p_3^5+2212838320 p_1^3 p_2^4 p_4^3 p_5 p_6^2 p_3^5+221852400 p_1^2 p_2^4 p_4^4 p_5^2 p_6 p_3^5+3067901760 p_1^2 p_2^3 p_4^4 p_5^2
 p_6^3 p_3^4+3067901760 p_1^2 p_2^4 p_4^3 p_5^2 p_6^3 p_3^4+557800320 p_1^2 p_2^3 p_4^4 p_5^3 p_6^2 p_3^4+1859334400 p_1^2 p_2^4 p_4^4 p_5^2 p_6^2
 p_3^4+557800320 p_1^3 p_2^4 p_4^3 p_5^2 p_6^2 p_3^4) \textcolor{red}{z^{18}}+(6754454784 p_1 p_2^3 p_4^3 p_5 p_6^3 p_3^6+16734009600 p_1 p_2^3 p_4^3 p_5^2
 p_6^2 p_3^6+5020202880 p_1 p_2^3 p_4^4 p_5 p_6^2 p_3^6+5020202880 p_1 p_2^4 p_4^3 p_5 p_6^2 p_3^6+16734009600 p_1^2 p_2^3 p_4^3 p_5 p_6^2 p_3^6+439267752
 p_1 p_2^3 p_4^3 p_5 p_6^4 p_3^5+30187080 p_1^2 p_2^4 p_4^3 p_6^3 p_3^5+30187080 p_2^3 p_4^4 p_5^2 p_6^3 p_3^5+966735000 p_1 p_2^2 p_4^4 p_5^2 p_6^3
 p_3^5+17946116496 p_1 p_2^3 p_4^3 p_5^2 p_6^3 p_3^5+2096325000 p_1^2 p_2^2 p_4^3 p_5^2 p_6^3 p_3^5+2096325000 p_1^2 p_2^3 p_4^2 p_5^2 p_6^3 p_3^5+3421632060
 p_1 p_2^3 p_4^4 p_5 p_6^3 p_3^5+3421632060 p_1 p_2^4 p_4^3 p_5 p_6^3 p_3^5+17946116496 p_1^2 p_2^3 p_4^3 p_5 p_6^3 p_3^5+966735000 p_1^2 p_2^4 p_4^2
 p_5 p_6^3 p_3^5+114345000 p_1 p_2^2 p_4^4 p_5^3 p_6^2 p_3^5+224116200 p_1 p_2^3 p_4^3 p_5^3 p_6^2 p_3^5+19609868880 p_1 p_2^3 p_4^4 p_5^2 p_6^2 p_3^5+1400726250
 p_1^2 p_2^2 p_4^4 p_5^2 p_6^2 p_3^5+9158882040 p_1 p_2^4 p_4^3 p_5^2 p_6^2 p_3^5+59798117736 p_1^2 p_2^3 p_4^3 p_5^2 p_6^2 p_3^5+1400726250 p_1^2
 p_2^4 p_4^2 p_5^2 p_6^2 p_3^5+2858625000 p_1 p_2^4 p_4^4 p_5 p_6^2 p_3^5+9158882040 p_1^2 p_2^3 p_4^4 p_5 p_6^2 p_3^5+19609868880 p_1^2 p_2^4 p_4^3
 p_5 p_6^2 p_3^5+224116200 p_1^3 p_2^3 p_4^3 p_5 p_6^2 p_3^5+114345000 p_1^3 p_2^4 p_4^2 p_5 p_6^2 p_3^5+599001480 p_1^2 p_2^3 p_4^4 p_5^2 p_6 p_3^5+599001480
 p_1^2 p_2^4 p_4^3 p_5^2 p_6 p_3^5+2253968640 p_1 p_2^3 p_4^4 p_5^2 p_6^3 p_3^4+8611029504 p_1^2 p_2^3 p_4^3 p_5^2 p_6^3 p_3^4+2253968640 p_1^2 p_2^4
 p_4^3 p_5 p_6^3 p_3^4+409812480 p_1 p_2^3 p_4^4 p_5^3 p_6^2 p_3^4+122943744 p_1^2 p_2^3 p_4^3 p_5^3 p_6^2 p_3^4+7991343360 p_1^2 p_2^3 p_4^4 p_5^2
 p_6^2 p_3^4+7991343360 p_1^2 p_2^4 p_4^3 p_5^2 p_6^2 p_3^4+122943744 p_1^3 p_2^3 p_4^3 p_5^2 p_6^2 p_3^4+409812480 p_1^3 p_2^4 p_4^3 p_5 p_6^2 p_3^4)
 \textcolor{red}{z^{17}}+(8199664704 p_1 p_2^3 p_4^3 p_5 p_6^2 p_3^6+49305564 p_1^2 p_2^3 p_4^3 p_6^3 p_3^5+25155900 p_1^2 p_2^4 p_4^2 p_6^3 p_3^5+25155900 p_2^2
 p_4^4 p_5^2 p_6^3 p_3^5+49305564 p_2^3 p_4^3 p_5^2 p_6^3 p_3^5+1445944500 p_1 p_2^2 p_4^3 p_5^2 p_6^3 p_3^5+1027199250 p_1 p_2^3 p_4^2 p_5^2 p_6^3
 p_3^5+308159775 p_1 p_2^2 p_4^4 p_5 p_6^3 p_3^5+8951787306 p_1 p_2^3 p_4^3 p_5 p_6^3 p_3^5+1027199250 p_1^2 p_2^2 p_4^3 p_5 p_6^3 p_3^5+308159775
 p_1 p_2^4 p_4^2 p_5 p_6^3 p_3^5+1445944500 p_1^2 p_2^3 p_4^2 p_5 p_6^3 p_3^5+34303500 p_1^2 p_2^4 p_4^3 p_6^2 p_3^5+34303500 p_2^3 p_4^4 p_5^2 p_6^2
 p_3^5+1413304200 p_1 p_2^2 p_4^4 p_5^2 p_6^2 p_3^5+30824064054 p_1 p_2^3 p_4^3 p_5^2 p_6^2 p_3^5+3902976000 p_1^2 p_2^2 p_4^3 p_5^2 p_6^2 p_3^5+3902976000
 p_1^2 p_2^3 p_4^2 p_5^2 p_6^2 p_3^5+6565308750 p_1 p_2^3 p_4^4 p_5 p_6^2 p_3^5+6565308750 p_1 p_2^4 p_4^3 p_5 p_6^2 p_3^5+30824064054 p_1^2 p_2^3
 p_4^3 p_5 p_6^2 p_3^5+1413304200 p_1^2 p_2^4 p_4^2 p_5 p_6^2 p_3^5+305613000 p_1 p_2^3 p_4^4 p_5^2 p_6 p_3^5+1669054464 p_1^2 p_2^3 p_4^3 p_5^2 p_6
 p_3^5+305613000 p_1^2 p_2^4 p_4^3 p_5 p_6 p_3^5+264136950 p_1 p_2^2 p_4^4 p_5^2 p_6^3 p_3^4+4790284884 p_1 p_2^3 p_4^3 p_5^2 p_6^3 p_3^4+880456500
 p_1^2 p_2^2 p_4^3 p_5^2 p_6^3 p_3^4+880456500 p_1^2 p_2^3 p_4^2 p_5^2 p_6^3 p_3^4+4790284884 p_1^2 p_2^3 p_4^3 p_5 p_6^3 p_3^4+264136950 p_1^2 p_2^4
 p_4^2 p_5 p_6^3 p_3^4+48024900 p_1 p_2^2 p_4^4 p_5^3 p_6^2 p_3^4+94128804 p_1 p_2^3 p_4^3 p_5^3 p_6^2 p_3^4+5042614500 p_1 p_2^3 p_4^4 p_5^2 p_6^2
 p_3^4+588305025 p_1^2 p_2^2 p_4^4 p_5^2 p_6^2 p_3^4+1961016750 p_1 p_2^4 p_4^3 p_5^2 p_6^2 p_3^4+27835512516 p_1^2 p_2^3 p_4^3 p_5^2 p_6^2 p_3^4+588305025
 p_1^2 p_2^4 p_4^2 p_5^2 p_6^2 p_3^4+1961016750 p_1^2 p_2^3 p_4^4 p_5 p_6^2 p_3^4+5042614500 p_1^2 p_2^4 p_4^3 p_5 p_6^2 p_3^4+94128804 p_1^3 p_2^3
 p_4^3 p_5 p_6^2 p_3^4+48024900 p_1^3 p_2^4 p_4^2 p_5 p_6^2 p_3^4+356548500 p_1^2 p_2^3 p_4^4 p_5^2 p_6 p_3^4+356548500 p_1^2 p_2^4 p_4^3 p_5^2 p_6
 p_3^4) \textcolor{red}{z^{16}}+(21913584 p_1 p_2^3 p_4^3 p_6^3 p_3^5+73045280 p_1^2 p_2^3 p_4^2 p_6^3 p_3^5+73045280 p_2^2 p_4^3 p_5^2 p_6^3 p_3^5+21913584
 p_2^3 p_4^3 p_5 p_6^3 p_3^5+1016898960 p_1 p_2^2 p_4^3 p_5 p_6^3 p_3^5+1016898960 p_1 p_2^3 p_4^2 p_5 p_6^3 p_3^5+99607200 p_1^2 p_2^3 p_4^3 p_6^2
 p_3^5+50820000 p_1^2 p_2^4 p_4^2 p_6^2 p_3^5+50820000 p_2^2 p_4^4 p_5^2 p_6^2 p_3^5+99607200 p_2^3 p_4^3 p_5^2 p_6^2 p_3^5+3252073440 p_1 p_2^2 p_4^3
 p_5^2 p_6^2 p_3^5+2075150000 p_1 p_2^3 p_4^2 p_5^2 p_6^2 p_3^5+622545000 p_1 p_2^2 p_4^4 p_5 p_6^2 p_3^5+19480302576 p_1 p_2^3 p_4^3 p_5 p_6^2 p_3^5+2075150000
 p_1^2 p_2^2 p_4^3 p_5 p_6^2 p_3^5+622545000 p_1 p_2^4 p_4^2 p_5 p_6^2 p_3^5+3252073440 p_1^2 p_2^3 p_4^2 p_5 p_6^2 p_3^5+887409600 p_1 p_2^3 p_4^3
 p_5^2 p_6 p_3^5+887409600 p_1^2 p_2^3 p_4^3 p_5 p_6 p_3^5+871627680 p_1 p_2^2 p_4^3 p_5^2 p_6^3 p_3^4+766975440 p_1 p_2^3 p_4^2 p_5^2 p_6^3 p_3^4+2414513024
 p_1 p_2^3 p_4^3 p_5 p_6^3 p_3^4+766975440 p_1^2 p_2^2 p_4^3 p_5 p_6^3 p_3^4+871627680 p_1^2 p_2^3 p_4^2 p_5 p_6^3 p_3^4+11383680 p_1 p_2^2 p_4^3
 p_5^3 p_6^2 p_3^4+830060000 p_1 p_2^2 p_4^4 p_5^2 p_6^2 p_3^4+18476731056 p_1 p_2^3 p_4^3 p_5^2 p_6^2 p_3^4+4063973760 p_1^2 p_2^2 p_4^3 p_5^2 p_6^2
 p_3^4+4063973760 p_1^2 p_2^3 p_4^2 p_5^2 p_6^2 p_3^4+1778700000 p_1 p_2^3 p_4^4 p_5 p_6^2 p_3^4+1778700000 p_1 p_2^4 p_4^3 p_5 p_6^2 p_3^4+18476731056
 p_1^2 p_2^3 p_4^3 p_5 p_6^2 p_3^4+830060000 p_1^2 p_2^4 p_4^2 p_5 p_6^2 p_3^4+11383680 p_1^3 p_2^3 p_4^2 p_5 p_6^2 p_3^4+323400000 p_1 p_2^3 p_4^4
 p_5^2 p_6 p_3^4+2253071744 p_1^2 p_2^3 p_4^3 p_5^2 p_6 p_3^4+323400000 p_1^2 p_2^4 p_4^3 p_5 p_6 p_3^4+4079910912 p_1^2 p_2^3 p_4^3 p_5^2 p_6^2 p_3^3)
 \textcolor{red}{z^{15}}+(41087970 p_1 p_2^3 p_4^2 p_6^3 p_3^5+41087970 p_2^2 p_4^3 p_5 p_6^3 p_3^5+136959900 p_1 p_2^2 p_4^2 p_5 p_6^3 p_3^5+56029050 p_1 p_2^3
 p_4^3 p_6^2 p_3^5+186763500 p_1^2 p_2^3 p_4^2 p_6^2 p_3^5+186763500 p_2^2 p_4^3 p_5^2 p_6^2 p_3^5+56029050 p_2^3 p_4^3 p_5 p_6^2 p_3^5+2655776970
 p_1 p_2^2 p_4^3 p_5 p_6^2 p_3^5+2655776970 p_1 p_2^3 p_4^2 p_5 p_6^2 p_3^5+499167900 p_1 p_2^3 p_4^3 p_5 p_6 p_3^5+35218260 p_1^2 p_2^3 p_4^2 p_6^3
 p_3^4+35218260 p_2^2 p_4^3 p_5^2 p_6^3 p_3^4+117394200 p_1 p_2^2 p_4^2 p_5^2 p_6^3 p_3^4+607296690 p_1 p_2^2 p_4^3 p_5 p_6^3 p_3^4+607296690 p_1
 p_2^3 p_4^2 p_5 p_6^3 p_3^4+117394200 p_1^2 p_2^2 p_4^2 p_5 p_6^3 p_3^4+48024900 p_1^2 p_2^3 p_4^3 p_6^2 p_3^4+24502500 p_1^2 p_2^4 p_4^2 p_6^2 p_3^4+24502500
 p_2^2 p_4^4 p_5^2 p_6^2 p_3^4+48024900 p_2^3 p_4^3 p_5^2 p_6^2 p_3^4+4609356210 p_1 p_2^2 p_4^3 p_5^2 p_6^2 p_3^4+3054383640 p_1 p_2^3 p_4^2 p_5^2
 p_6^2 p_3^4+446054400 p_1^2 p_2^2 p_4^2 p_5^2 p_6^2 p_3^4+300155625 p_1 p_2^2 p_4^4 p_5 p_6^2 p_3^4+14283282150 p_1 p_2^3 p_4^3 p_5 p_6^2 p_3^4+3054383640
 p_1^2 p_2^2 p_4^3 p_5 p_6^2 p_3^4+300155625 p_1 p_2^4 p_4^2 p_5 p_6^2 p_3^4+4609356210 p_1^2 p_2^3 p_4^2 p_5 p_6^2 p_3^4+54573750 p_1 p_2^2 p_4^4
 p_5^2 p_6 p_3^4+1671169500 p_1 p_2^3 p_4^3 p_5^2 p_6 p_3^4+298045440 p_1^2 p_2^2 p_4^3 p_5^2 p_6 p_3^4+298045440 p_1^2 p_2^3 p_4^2 p_5^2 p_6 p_3^4+1671169500
 p_1^2 p_2^3 p_4^3 p_5 p_6 p_3^4+54573750 p_1^2 p_2^4 p_4^2 p_5 p_6 p_3^4+3585859200 p_1 p_2^3 p_4^3 p_5^2 p_6^2 p_3^3+1075757760 p_1^2 p_2^2 p_4^3
 p_5^2 p_6^2 p_3^3+1075757760 p_1^2 p_2^3 p_4^2 p_5^2 p_6^2 p_3^3+3585859200 p_1^2 p_2^3 p_4^3 p_5 p_6^2 p_3^3+651974400 p_1^2 p_2^3 p_4^3 p_5^2 p_6
 p_3^3) \textcolor{red}{z^{14}}+(119528640 p_1 p_2^3 p_4^2 p_6^2 p_3^5+119528640 p_2^2 p_4^3 p_5 p_6^2 p_3^5+398428800 p_1 p_2^2 p_4^2 p_5 p_6^2 p_3^5+35218260
 p_1 p_2^3 p_4^2 p_6^3 p_3^4+35218260 p_2^2 p_4^3 p_5 p_6^3 p_3^4+180457200 p_1 p_2^2 p_4^2 p_5 p_6^3 p_3^4+48024900 p_1 p_2^3 p_4^3 p_6^2 p_3^4+268939440
 p_1^2 p_2^3 p_4^2 p_6^2 p_3^4+268939440 p_2^2 p_4^3 p_5^2 p_6^2 p_3^4+133402500 p_1 p_2 p_4^3 p_5^2 p_6^2 p_3^4+672348600 p_1 p_2^2 p_4^2 p_5^2 p_6^2
 p_3^4+48024900 p_2^3 p_4^3 p_5 p_6^2 p_3^4+4320547560 p_1 p_2^2 p_4^3 p_5 p_6^2 p_3^4+4320547560 p_1 p_2^3 p_4^2 p_5 p_6^2 p_3^4+672348600 p_1^2
 p_2^2 p_4^2 p_5 p_6^2 p_3^4+133402500 p_1^2 p_2^3 p_4 p_5 p_6^2 p_3^4+413887320 p_1 p_2^2 p_4^3 p_5^2 p_6 p_3^4+356548500 p_1 p_2^3 p_4^2 p_5^2 p_6
 p_3^4+1189465200 p_1 p_2^3 p_4^3 p_5 p_6 p_3^4+356548500 p_1^2 p_2^2 p_4^3 p_5 p_6 p_3^4+413887320 p_1^2 p_2^3 p_4^2 p_5 p_6 p_3^4+1644128640 p_1
 p_2^2 p_4^3 p_5^2 p_6^2 p_3^3+1075757760 p_1 p_2^3 p_4^2 p_5^2 p_6^2 p_3^3+292723200 p_1^2 p_2^2 p_4^2 p_5^2 p_6^2 p_3^3+3585859200 p_1 p_2^3 p_4^3
 p_5 p_6^2 p_3^3+1075757760 p_1^2 p_2^2 p_4^3 p_5 p_6^2 p_3^3+1644128640 p_1^2 p_2^3 p_4^2 p_5 p_6^2 p_3^3+651974400 p_1 p_2^3 p_4^3 p_5^2 p_6 p_3^3+195592320
 p_1^2 p_2^2 p_4^3 p_5^2 p_6 p_3^3+195592320 p_1^2 p_2^3 p_4^2 p_5^2 p_6 p_3^3+651974400 p_1^2 p_2^3 p_4^3 p_5 p_6 p_3^3) \textcolor{red}{z^{13}}+(6391462
 p_1 p_2^2 p_4^2 p_6^3 p_3^4+6391462 p_2^2 p_4^2 p_5 p_6^3 p_3^4+8715630 p_1 p_2^2 p_4^3 p_6^2 p_3^4+253998360 p_1 p_2^3 p_4^2 p_6^2 p_3^4+29052100
 p_1^2 p_2^2 p_4^2 p_6^2 p_3^4+14822500 p_2 p_4^3 p_5^2 p_6^2 p_3^4+29052100 p_2^2 p_4^2 p_5^2 p_6^2 p_3^4+14822500 p_1^2 p_2^3 p_4 p_6^2 p_3^4+253998360
 p_2^2 p_4^3 p_5 p_6^2 p_3^4+181575625 p_1 p_2 p_4^3 p_5 p_6^2 p_3^4+8715630 p_2^3 p_4^2 p_5 p_6^2 p_3^4+1554121926 p_1 p_2^2 p_4^2 p_5 p_6^2 p_3^4+181575625
 p_1 p_2^3 p_4 p_5 p_6^2 p_3^4+20212500 p_1^2 p_2^3 p_4^2 p_6 p_3^4+20212500 p_2^2 p_4^3 p_5^2 p_6 p_3^4+110387200 p_1 p_2^2 p_4^2 p_5^2 p_6 p_3^4+388031490
 p_1 p_2^2 p_4^3 p_5 p_6 p_3^4+388031490 p_1 p_2^3 p_4^2 p_5 p_6 p_3^4+110387200 p_1^2 p_2^2 p_4^2 p_5 p_6 p_3^4+5478396 p_1 p_2^2 p_4^2 p_5 p_6^3
 p_3^3+111143340 p_1^2 p_2^3 p_4^2 p_6^2 p_3^3+111143340 p_2^2 p_4^3 p_5^2 p_6^2 p_3^3+155636250 p_1 p_2 p_4^3 p_5^2 p_6^2 p_3^3+542666124 p_1 p_2^2
 p_4^2 p_5^2 p_6^2 p_3^3+1941993130 p_1 p_2^2 p_4^3 p_5 p_6^2 p_3^3+1941993130 p_1 p_2^3 p_4^2 p_5 p_6^2 p_3^3+542666124 p_1^2 p_2^2 p_4^2 p_5 p_6^2
 p_3^3+155636250 p_1^2 p_2^3 p_4 p_5 p_6^2 p_3^3+353089660 p_1 p_2^2 p_4^3 p_5^2 p_6 p_3^3+247546530 p_1 p_2^3 p_4^2 p_5^2 p_6 p_3^3+60555264 p_1^2
 p_2^2 p_4^2 p_5^2 p_6 p_3^3+707437500 p_1 p_2^3 p_4^3 p_5 p_6 p_3^3+247546530 p_1^2 p_2^2 p_4^3 p_5 p_6 p_3^3+353089660 p_1^2 p_2^3 p_4^2 p_5 p_6
 p_3^3) \textcolor{red}{z^{12}}+(66594528 p_1 p_2^2 p_4^2 p_6^2 p_3^4+21344400 p_1 p_2^3 p_4 p_6^2 p_3^4+21344400 p_2 p_4^3 p_5 p_6^2 p_3^4+66594528 p_2^2
 p_4^2 p_5 p_6^2 p_3^4+71148000 p_1 p_2 p_4^2 p_5 p_6^2 p_3^4+71148000 p_1 p_2^2 p_4 p_5 p_6^2 p_3^4+29106000 p_1 p_2^3 p_4^2 p_6 p_3^4+29106000 p_2^2
 p_4^3 p_5 p_6 p_3^4+191866752 p_1 p_2^2 p_4^2 p_5 p_6 p_3^4+137214000 p_1 p_2^3 p_4^2 p_6^2 p_3^3+35858592 p_1^2 p_2^2 p_4^2 p_6^2 p_3^3+18295200
 p_2 p_4^3 p_5^2 p_6^2 p_3^3+35858592 p_2^2 p_4^2 p_5^2 p_6^2 p_3^3+60984000 p_1 p_2 p_4^2 p_5^2 p_6^2 p_3^3+18295200 p_1^2 p_2^3 p_4 p_6^2 p_3^3+137214000
 p_2^2 p_4^3 p_5 p_6^2 p_3^3+224116200 p_1 p_2 p_4^3 p_5 p_6^2 p_3^3+1339753968 p_1 p_2^2 p_4^2 p_5 p_6^2 p_3^3+224116200 p_1 p_2^3 p_4 p_5 p_6^2
 p_3^3+60984000 p_1^2 p_2^2 p_4 p_5 p_6^2 p_3^3+24948000 p_1^2 p_2^3 p_4^2 p_6 p_3^3+24948000 p_2^2 p_4^3 p_5^2 p_6 p_3^3+40748400 p_1 p_2 p_4^3 p_5^2
 p_6 p_3^3+177031008 p_1 p_2^2 p_4^2 p_5^2 p_6 p_3^3+467082000 p_1 p_2^2 p_4^3 p_5 p_6 p_3^3+467082000 p_1 p_2^3 p_4^2 p_5 p_6 p_3^3+177031008 p_1^2
 p_2^2 p_4^2 p_5 p_6 p_3^3+40748400 p_1^2 p_2^3 p_4 p_5 p_6 p_3^3) \textcolor{red}{z^{11}}+(2614689 p_2^2 p_4^2 p_6^2 p_3^4+8715630 p_1 p_2^2 p_4 p_6^2
 p_3^4+8715630 p_2 p_4^2 p_5 p_6^2 p_3^4+11884950 p_1 p_2^2 p_4^2 p_6 p_3^4+11884950 p_2^2 p_4^2 p_5 p_6 p_3^4+87268104 p_1 p_2^2 p_4^2 p_6^2 p_3^3+7470540
 p_2 p_4^2 p_5^2 p_6^2 p_3^3+28586250 p_1 p_2^3 p_4 p_6^2 p_3^3+7470540 p_1^2 p_2^2 p_4 p_6^2 p_3^3+28586250 p_2 p_4^3 p_5 p_6^2 p_3^3+87268104 p_2^2
 p_4^2 p_5 p_6^2 p_3^3+202848030 p_1 p_2 p_4^2 p_5 p_6^2 p_3^3+202848030 p_1 p_2^2 p_4 p_5 p_6^2 p_3^3+38981250 p_1 p_2^3 p_4^2 p_6 p_3^3+10187100
 p_1^2 p_2^2 p_4^2 p_6 p_3^3+5197500 p_2 p_4^3 p_5^2 p_6 p_3^3+10187100 p_2^2 p_4^2 p_5^2 p_6 p_3^3+28385280 p_1 p_2 p_4^2 p_5^2 p_6 p_3^3+5197500
 p_1^2 p_2^3 p_4 p_6 p_3^3+38981250 p_2^2 p_4^3 p_5 p_6 p_3^3+63669375 p_1 p_2 p_4^3 p_5 p_6 p_3^3+440527626 p_1 p_2^2 p_4^2 p_5 p_6 p_3^3+63669375
 p_1 p_2^3 p_4 p_5 p_6 p_3^3+28385280 p_1^2 p_2^2 p_4 p_5 p_6 p_3^3+30735936 p_1 p_2^2 p_4^2 p_5 p_6^2 p_3^2+5588352 p_1 p_2^2 p_4^2 p_5^2 p_6 p_3^2+5588352
 p_1^2 p_2^2 p_4^2 p_5 p_6 p_3^2) \textcolor{red}{z^{10}}+(6225450 p_2^2 p_4^2 p_6^2 p_3^3+13280960 p_1 p_2 p_4^2 p_6^2 p_3^3+27544440 p_1 p_2^2 p_4 p_6^2
 p_3^3+27544440 p_2 p_4^2 p_5 p_6^2 p_3^3+13280960 p_2^2 p_4 p_5 p_6^2 p_3^3+41164200 p_1 p_2 p_4 p_5 p_6^2 p_3^3+205800 p_1 p_2^2 p_4^2 p_5 p_3^3+37560600
 p_1 p_2^2 p_4^2 p_6 p_3^3+3773000 p_2 p_4^2 p_5^2 p_6 p_3^3+9240000 p_1 p_2^3 p_4 p_6 p_3^3+3773000 p_1^2 p_2^2 p_4 p_6 p_3^3+9240000 p_2 p_4^3 p_5
 p_6 p_3^3+37560600 p_2^2 p_4^2 p_5 p_6 p_3^3+82222140 p_1 p_2 p_4^2 p_5 p_6 p_3^3+82222140 p_1 p_2^2 p_4 p_5 p_6 p_3^3+11383680 p_1 p_2 p_4^2 p_5
 p_6^2 p_3^2+11383680 p_1 p_2^2 p_4 p_5 p_6^2 p_3^2+2069760 p_1 p_2 p_4^2 p_5^2 p_6 p_3^2+24147200 p_1 p_2^2 p_4^2 p_5 p_6 p_3^2+2069760 p_1^2 p_2^2
 p_4 p_5 p_6 p_3^2) \textcolor{red}{z^9}+(1867635 p_2 p_4^2 p_6^2 p_3^3+1867635 p_2^2 p_4 p_6^2 p_3^3+6225450 p_1 p_2 p_4 p_6^2 p_3^3+6225450 p_2 p_4 p_5
 p_6^2 p_3^3+2546775 p_2^2 p_4^2 p_6 p_3^3+8489250 p_1 p_2 p_4^2 p_6 p_3^3+13222440 p_1 p_2^2 p_4 p_6 p_3^3+13222440 p_2 p_4^2 p_5 p_6 p_3^3+8489250
 p_2^2 p_4 p_5 p_6 p_3^3+23654400 p_1 p_2 p_4 p_5 p_6 p_3^3+1600830 p_1 p_2^2 p_4 p_6^2 p_3^2+1600830 p_2 p_4^2 p_5 p_6^2 p_3^2+5336100 p_1 p_2 p_4
 p_5 p_6^2 p_3^2+396900 p_1 p_2^2 p_4^2 p_5 p_3^2+2182950 p_1 p_2^2 p_4^2 p_6 p_3^2+291060 p_2 p_4^2 p_5^2 p_6 p_3^2+291060 p_1^2 p_2^2 p_4 p_6 p_3^2+2182950
 p_2^2 p_4^2 p_5 p_6 p_3^2+9168390 p_1 p_2 p_4^2 p_5 p_6 p_3^2+9168390 p_1 p_2^2 p_4 p_5 p_6 p_3^2) \textcolor{red}{z^8}+(996072 p_2 p_4 p_6^2 p_3^3+1358280
 p_2 p_4^2 p_6 p_3^3+1358280 p_2^2 p_4 p_6 p_3^3+4527600 p_1 p_2 p_4 p_6 p_3^3+4527600 p_2 p_4 p_5 p_6 p_3^3+853776 p_1 p_2 p_4 p_6^2 p_3^2+853776
 p_2 p_4 p_5 p_6^2 p_3^2+211680 p_1 p_2 p_4^2 p_5 p_3^2+211680 p_1 p_2^2 p_4 p_5 p_3^2+1164240 p_1 p_2 p_4^2 p_6 p_3^2+1853280 p_1 p_2^2 p_4 p_6 p_3^2+1853280
 p_2 p_4^2 p_5 p_6 p_3^2+1164240 p_2^2 p_4 p_5 p_6 p_3^2+6044544 p_1 p_2 p_4 p_5 p_6 p_3^2) \textcolor{red}{z^7}+(916839 p_2 p_4 p_6 p_3^3+148225 p_2 p_4
 p_6^2 p_3^2+36750 p_1 p_2^2 p_4 p_3^2+36750 p_2 p_4^2 p_5 p_3^2+200704 p_1 p_2 p_4 p_5 p_3^2+26950 p_1 p_2^2 p_6 p_3^2+202125 p_2 p_4^2 p_6 p_3^2+202125
 p_2^2 p_4 p_6 p_3^2+1539384 p_1 p_2 p_4 p_6 p_3^2+26950 p_4^2 p_5 p_6 p_3^2+1539384 p_2 p_4 p_5 p_6 p_3^2+369600 p_1 p_2 p_4 p_5 p_6 p_3) \textcolor{red}{z^6}+(44100
 p_1 p_2 p_4 p_3^2+44100 p_2 p_4 p_5 p_3^2+32340 p_1 p_2 p_6 p_3^2+407484 p_2 p_4 p_6 p_3^2+32340 p_4 p_5 p_6 p_3^2+24192 p_1 p_2 p_4 p_5 p_3+133056
 p_1 p_2 p_4 p_6 p_3+133056 p_2 p_4 p_5 p_6 p_3) \textcolor{red}{z^5}+(11025 p_2 p_4 p_3^2+8085 p_2 p_6 p_3^2+8085 p_4 p_6 p_3^2+9450 p_1 p_2 p_4 p_3+9450
 p_2 p_4 p_5 p_3+6930 p_1 p_2 p_6 p_3+51975 p_2 p_4 p_6 p_3+6930 p_4 p_5 p_6 p_3) \textcolor{red}{z^4}+(560 p_1 p_2 p_3+4200 p_2 p_4 p_3+560 p_4 p_5 p_3+3080
 p_2 p_6 p_3+3080 p_4 p_6 p_3) \textcolor{red}{z^3}+(315 p_2 p_3+315 p_4 p_3+231 p_6 p_3) \textcolor{red}{z^2}+42 p_3 \textcolor{red}{z}+1\)
 
 \vspace{1em}
 \item[$H_4=$]\(p_1^3 p_2^6 p_3^8 p_4^6 p_5^3 p_6^4 \textcolor{red}{z^{30}}+30 p_1^3 p_2^5 p_3^8 p_4^6 p_5^3 p_6^4 \textcolor{red}{z^{29}}+(120 p_1^2 p_2^5 p_4^6 p_5^3
 p_6^4 p_3^8+315 p_1^3 p_2^5 p_4^6 p_5^3 p_6^4 p_3^7) \textcolor{red}{z^{28}}+(2240 p_1^2 p_2^5 p_4^6 p_5^3 p_6^4 p_3^7+1050 p_1^3 p_2^5 p_4^5 p_5^3 p_6^4
 p_3^7+770 p_1^3 p_2^5 p_4^6 p_5^3 p_6^3 p_3^7) \textcolor{red}{z^{27}}+(4200 p_1^2 p_2^4 p_4^6 p_5^3 p_6^4 p_3^7+9450 p_1^2 p_2^5 p_4^5 p_5^3 p_6^4 p_3^7+1050
 p_1^3 p_2^5 p_4^5 p_5^2 p_6^4 p_3^7+6930 p_1^2 p_2^5 p_4^6 p_5^3 p_6^3 p_3^7+5775 p_1^3 p_2^5 p_4^5 p_5^3 p_6^3 p_3^7) \textcolor{red}{z^{26}}+(31500 p_1^2
 p_2^4 p_4^5 p_5^3 p_6^4 p_3^7+10752 p_1^2 p_2^5 p_4^5 p_5^2 p_6^4 p_3^7+23100 p_1^2 p_2^4 p_4^6 p_5^3 p_6^3 p_3^7+59136 p_1^2 p_2^5 p_4^5 p_5^3 p_6^3
 p_3^7+8316 p_1^3 p_2^5 p_4^5 p_5^2 p_6^3 p_3^7+9702 p_1^3 p_2^5 p_4^5 p_5^3 p_6^3 p_3^6) \textcolor{red}{z^{25}}+(45360 p_1^2 p_2^4 p_4^5 p_5^2 p_6^4 p_3^7+249480
 p_1^2 p_2^4 p_4^5 p_5^3 p_6^3 p_3^7+92400 p_1^2 p_2^5 p_4^5 p_5^2 p_6^3 p_3^7+36750 p_1^2 p_2^4 p_4^5 p_5^3 p_6^4 p_3^6+26950 p_1^2 p_2^4 p_4^6 p_5^3
 p_6^3 p_3^6+107800 p_1^2 p_2^5 p_4^5 p_5^3 p_6^3 p_3^6+8085 p_1^3 p_2^4 p_4^5 p_5^3 p_6^3 p_3^6+26950 p_1^3 p_2^5 p_4^5 p_5^2 p_6^3 p_3^6)
 \textcolor{red}{z^{24}}+(443520 p_1^2 p_2^4 p_4^5 p_5^2 p_6^3 p_3^7+94080 p_1^2 p_2^4 p_4^5 p_5^2 p_6^4 p_3^6+1132560 p_1^2 p_2^4 p_4^5 p_5^3 p_6^3 p_3^6+316800
 p_1^2 p_2^5 p_4^5 p_5^2 p_6^3 p_3^6+32340 p_1^3 p_2^4 p_4^5 p_5^2 p_6^3 p_3^6+16500 p_1^3 p_2^5 p_4^4 p_5^2 p_6^3 p_3^6) \textcolor{red}{z^{23}}+(44100
 p_1^2 p_2^4 p_4^4 p_5^2 p_6^4 p_3^6+32340 p_1 p_2^4 p_4^5 p_5^3 p_6^3 p_3^6+495000 p_1^2 p_2^3 p_4^5 p_5^3 p_6^3 p_3^6+242550 p_1^2 p_2^4 p_4^4 p_5^3
 p_6^3 p_3^6+3256110 p_1^2 p_2^4 p_4^5 p_5^2 p_6^3 p_3^6+202125 p_1^2 p_2^5 p_4^4 p_5^2 p_6^3 p_3^6+44550 p_1^3 p_2^4 p_4^4 p_5^2 p_6^3 p_3^6+177870
 p_1^2 p_2^4 p_4^5 p_5^3 p_6^2 p_3^6+1358280 p_1^2 p_2^4 p_4^5 p_5^3 p_6^3 p_3^5) \textcolor{red}{z^{22}}+(178200 p_1 p_2^3 p_4^5 p_5^3 p_6^3 p_3^6+168960
 p_1 p_2^4 p_4^5 p_5^2 p_6^3 p_3^6+2182950 p_1^2 p_2^3 p_4^5 p_5^2 p_6^3 p_3^6+2674100 p_1^2 p_2^4 p_4^4 p_5^2 p_6^3 p_3^6+711480 p_1^2 p_2^4 p_4^5
 p_5^2 p_6^2 p_3^6+1478400 p_1^2 p_2^3 p_4^5 p_5^3 p_6^3 p_3^5+1131900 p_1^2 p_2^4 p_4^4 p_5^3 p_6^3 p_3^5+4928000 p_1^2 p_2^4 p_4^5 p_5^2 p_6^3 p_3^5+23100
 p_1^3 p_2^4 p_4^4 p_5^2 p_6^3 p_3^5+830060 p_1^2 p_2^4 p_4^5 p_5^3 p_6^2 p_3^5) \textcolor{red}{z^{21}}+(970200 p_1 p_2^3 p_4^5 p_5^2 p_6^3 p_3^6+349272
 p_1 p_2^4 p_4^4 p_5^2 p_6^3 p_3^6+3234000 p_1^2 p_2^3 p_4^4 p_5^2 p_6^3 p_3^6+155232 p_1^2 p_2^4 p_4^4 p_5 p_6^3 p_3^6+853776 p_1^2 p_2^4 p_4^4 p_5^2
 p_6^2 p_3^6+577500 p_1 p_2^3 p_4^5 p_5^3 p_6^3 p_3^5+1559250 p_1^2 p_2^3 p_4^4 p_5^3 p_6^3 p_3^5+7074375 p_1^2 p_2^3 p_4^5 p_5^2 p_6^3 p_3^5+9315306
 p_1^2 p_2^4 p_4^4 p_5^2 p_6^3 p_3^5+1143450 p_1^2 p_2^3 p_4^5 p_5^3 p_6^2 p_3^5+996072 p_1^2 p_2^4 p_4^4 p_5^3 p_6^2 p_3^5+3811500 p_1^2 p_2^4 p_4^5
 p_5^2 p_6^2 p_3^5+5082 p_1^3 p_2^4 p_4^4 p_5^2 p_6^2 p_3^5) \textcolor{red}{z^{20}}+(2069760 p_1 p_2^3 p_4^4 p_5^2 p_6^3 p_3^6+693000 p_1 p_2^3 p_4^4 p_5^3
 p_6^3 p_3^5+3326400 p_1 p_2^3 p_4^5 p_5^2 p_6^3 p_3^5+369600 p_1 p_2^4 p_4^4 p_5^2 p_6^3 p_3^5+21801780 p_1^2 p_2^3 p_4^4 p_5^2 p_6^3 p_3^5+4331250
 p_1^2 p_2^4 p_4^3 p_5^2 p_6^3 p_3^5+1478400 p_1^2 p_2^4 p_4^4 p_5 p_6^3 p_3^5+508200 p_1 p_2^3 p_4^5 p_5^3 p_6^2 p_3^5+2439360 p_1^2 p_2^3 p_4^4
 p_5^3 p_6^2 p_3^5+6225450 p_1^2 p_2^3 p_4^5 p_5^2 p_6^2 p_3^5+11384100 p_1^2 p_2^4 p_4^4 p_5^2 p_6^2 p_3^5) \textcolor{red}{z^{19}}+(14314300 p_1 p_2^3
 p_4^4 p_5^2 p_6^3 p_3^5+14437500 p_1^2 p_2^3 p_4^3 p_5^2 p_6^3 p_3^5+3056130 p_1^2 p_2^3 p_4^4 p_5 p_6^3 p_3^5+1559250 p_1^2 p_2^4 p_4^3 p_5 p_6^3
 p_3^5+1372140 p_1 p_2^3 p_4^4 p_5^3 p_6^2 p_3^5+3176250 p_1 p_2^3 p_4^5 p_5^2 p_6^2 p_3^5+127050 p_1 p_2^4 p_4^4 p_5^2 p_6^2 p_3^5+28420210 p_1^2
 p_2^3 p_4^4 p_5^2 p_6^2 p_3^5+8575875 p_1^2 p_2^4 p_4^3 p_5^2 p_6^2 p_3^5+2032800 p_1^2 p_2^4 p_4^4 p_5 p_6^2 p_3^5+6338640 p_1^2 p_2^3 p_4^4 p_5^2
 p_6^3 p_3^4+711480 p_1^2 p_2^3 p_4^4 p_5^3 p_6^2 p_3^4+2371600 p_1^2 p_2^4 p_4^4 p_5^2 p_6^2 p_3^4) \textcolor{red}{z^{18}}+(577500 p_1 p_2^2 p_4^4 p_5^2
 p_6^3 p_3^5+10187100 p_1 p_2^3 p_4^3 p_5^2 p_6^3 p_3^5+1774080 p_1 p_2^3 p_4^4 p_5 p_6^3 p_3^5+5913600 p_1^2 p_2^3 p_4^3 p_5 p_6^3 p_3^5+18705960
 p_1 p_2^3 p_4^4 p_5^2 p_6^2 p_3^5+32524800 p_1^2 p_2^3 p_4^3 p_5^2 p_6^2 p_3^5+7470540 p_1^2 p_2^3 p_4^4 p_5 p_6^2 p_3^5+3811500 p_1^2 p_2^4 p_4^3
 p_5 p_6^2 p_3^5+8279040 p_1 p_2^3 p_4^4 p_5^2 p_6^3 p_3^4+8731800 p_1^2 p_2^3 p_4^3 p_5^2 p_6^3 p_3^4+711480 p_1 p_2^3 p_4^4 p_5^3 p_6^2 p_3^4+16625700
 p_1^2 p_2^3 p_4^4 p_5^2 p_6^2 p_3^4+4446750 p_1^2 p_2^4 p_4^3 p_5^2 p_6^2 p_3^4) \textcolor{red}{z^{17}}+(4527600 p_1 p_2^3 p_4^3 p_5 p_6^3 p_3^5+508200
 p_1 p_2^2 p_4^4 p_5^2 p_6^2 p_3^5+24901800 p_1 p_2^3 p_4^3 p_5^2 p_6^2 p_3^5+5488560 p_1 p_2^3 p_4^4 p_5 p_6^2 p_3^5+18295200 p_1^2 p_2^3 p_4^3 p_5
 p_6^2 p_3^5+2182950 p_1 p_2^2 p_4^4 p_5^2 p_6^3 p_3^4+11884950 p_1 p_2^3 p_4^3 p_5^2 p_6^3 p_3^4+3880800 p_1^2 p_2^3 p_4^3 p_5 p_6^3 p_3^4+108900
 p_1 p_2^2 p_4^4 p_5^3 p_6^2 p_3^4+18478980 p_1 p_2^3 p_4^4 p_5^2 p_6^2 p_3^4+1334025 p_1^2 p_2^2 p_4^4 p_5^2 p_6^2 p_3^4+45530550 p_1^2 p_2^3 p_4^3
 p_5^2 p_6^2 p_3^4+4446750 p_1^2 p_2^3 p_4^4 p_5 p_6^2 p_3^4+2268750 p_1^2 p_2^4 p_4^3 p_5 p_6^2 p_3^4+1584660 p_1^2 p_2^3 p_4^4 p_5^2 p_6 p_3^4)
 \textcolor{red}{z^{16}}+(15937152 p_1 p_2^3 p_4^3 p_5 p_6^2 p_3^5+3234000 p_1 p_2^2 p_4^3 p_5^2 p_6^3 p_3^4+5588352 p_1 p_2^3 p_4^3 p_5 p_6^3 p_3^4+6203600 p_1
 p_2^2 p_4^4 p_5^2 p_6^2 p_3^4+53742416 p_1 p_2^3 p_4^3 p_5^2 p_6^2 p_3^4+5808000 p_1^2 p_2^2 p_4^3 p_5^2 p_6^2 p_3^4+5808000 p_1 p_2^3 p_4^4 p_5
 p_6^2 p_3^4+34036496 p_1^2 p_2^3 p_4^3 p_5 p_6^2 p_3^4+3234000 p_1 p_2^3 p_4^4 p_5^2 p_6 p_3^4+5588352 p_1^2 p_2^3 p_4^3 p_5^2 p_6 p_3^4+15937152
 p_1^2 p_2^3 p_4^3 p_5^2 p_6^2 p_3^3) \textcolor{red}{z^{15}}+(1584660 p_1 p_2^2 p_4^3 p_5 p_6^3 p_3^4+108900 p_2^2 p_4^4 p_5^2 p_6^2 p_3^4+18478980 p_1
 p_2^2 p_4^3 p_5^2 p_6^2 p_3^4+1334025 p_1 p_2^2 p_4^4 p_5 p_6^2 p_3^4+45530550 p_1 p_2^3 p_4^3 p_5 p_6^2 p_3^4+4446750 p_1^2 p_2^2 p_4^3 p_5 p_6^2
 p_3^4+2268750 p_1^2 p_2^3 p_4^2 p_5 p_6^2 p_3^4+2182950 p_1 p_2^2 p_4^4 p_5^2 p_6 p_3^4+11884950 p_1 p_2^3 p_4^3 p_5^2 p_6 p_3^4+3880800 p_1^2 p_2^3
 p_4^3 p_5 p_6 p_3^4+508200 p_1 p_2^2 p_4^4 p_5^2 p_6^2 p_3^3+24901800 p_1 p_2^3 p_4^3 p_5^2 p_6^2 p_3^3+5488560 p_1^2 p_2^2 p_4^3 p_5^2 p_6^2 p_3^3+18295200
 p_1^2 p_2^3 p_4^3 p_5 p_6^2 p_3^3+4527600 p_1^2 p_2^3 p_4^3 p_5^2 p_6 p_3^3) \textcolor{red}{z^{14}}+(711480 p_2^2 p_4^3 p_5^2 p_6^2 p_3^4+16625700 p_1
 p_2^2 p_4^3 p_5 p_6^2 p_3^4+4446750 p_1 p_2^3 p_4^2 p_5 p_6^2 p_3^4+8279040 p_1 p_2^2 p_4^3 p_5^2 p_6 p_3^4+8731800 p_1 p_2^3 p_4^3 p_5 p_6 p_3^4+18705960
 p_1 p_2^2 p_4^3 p_5^2 p_6^2 p_3^3+32524800 p_1 p_2^3 p_4^3 p_5 p_6^2 p_3^3+7470540 p_1^2 p_2^2 p_4^3 p_5 p_6^2 p_3^3+3811500 p_1^2 p_2^3 p_4^2 p_5
 p_6^2 p_3^3+577500 p_1 p_2^2 p_4^4 p_5^2 p_6 p_3^3+10187100 p_1 p_2^3 p_4^3 p_5^2 p_6 p_3^3+1774080 p_1^2 p_2^2 p_4^3 p_5^2 p_6 p_3^3+5913600 p_1^2
 p_2^3 p_4^3 p_5 p_6 p_3^3) \textcolor{red}{z^{13}}+(711480 p_2^2 p_4^3 p_5 p_6^2 p_3^4+2371600 p_1 p_2^2 p_4^2 p_5 p_6^2 p_3^4+6338640 p_1 p_2^2 p_4^3
 p_5 p_6 p_3^4+1372140 p_2^2 p_4^3 p_5^2 p_6^2 p_3^3+3176250 p_1 p_2 p_4^3 p_5^2 p_6^2 p_3^3+127050 p_1 p_2^2 p_4^2 p_5^2 p_6^2 p_3^3+28420210 p_1
 p_2^2 p_4^3 p_5 p_6^2 p_3^3+8575875 p_1 p_2^3 p_4^2 p_5 p_6^2 p_3^3+2032800 p_1^2 p_2^2 p_4^2 p_5 p_6^2 p_3^3+14314300 p_1 p_2^2 p_4^3 p_5^2 p_6
 p_3^3+14437500 p_1 p_2^3 p_4^3 p_5 p_6 p_3^3+3056130 p_1^2 p_2^2 p_4^3 p_5 p_6 p_3^3+1559250 p_1^2 p_2^3 p_4^2 p_5 p_6 p_3^3) \textcolor{red}{z^{12}}+(508200
 p_2 p_4^3 p_5^2 p_6^2 p_3^3+2439360 p_2^2 p_4^3 p_5 p_6^2 p_3^3+6225450 p_1 p_2 p_4^3 p_5 p_6^2 p_3^3+11384100 p_1 p_2^2 p_4^2 p_5 p_6^2 p_3^3+693000
 p_2^2 p_4^3 p_5^2 p_6 p_3^3+3326400 p_1 p_2 p_4^3 p_5^2 p_6 p_3^3+369600 p_1 p_2^2 p_4^2 p_5^2 p_6 p_3^3+21801780 p_1 p_2^2 p_4^3 p_5 p_6 p_3^3+4331250
 p_1 p_2^3 p_4^2 p_5 p_6 p_3^3+1478400 p_1^2 p_2^2 p_4^2 p_5 p_6 p_3^3+2069760 p_1 p_2^2 p_4^3 p_5^2 p_6 p_3^2) \textcolor{red}{z^{11}}+(5082 p_1 p_2^2
 p_4^2 p_6^2 p_3^3+1143450 p_2 p_4^3 p_5 p_6^2 p_3^3+996072 p_2^2 p_4^2 p_5 p_6^2 p_3^3+3811500 p_1 p_2 p_4^2 p_5 p_6^2 p_3^3+577500 p_2 p_4^3 p_5^2
 p_6 p_3^3+1559250 p_2^2 p_4^3 p_5 p_6 p_3^3+7074375 p_1 p_2 p_4^3 p_5 p_6 p_3^3+9315306 p_1 p_2^2 p_4^2 p_5 p_6 p_3^3+853776 p_1 p_2^2 p_4^2 p_5
 p_6^2 p_3^2+970200 p_1 p_2 p_4^3 p_5^2 p_6 p_3^2+349272 p_1 p_2^2 p_4^2 p_5^2 p_6 p_3^2+3234000 p_1 p_2^2 p_4^3 p_5 p_6 p_3^2+155232 p_1^2 p_2^2
 p_4^2 p_5 p_6 p_3^2) \textcolor{red}{z^{10}}+(830060 p_2 p_4^2 p_5 p_6^2 p_3^3+23100 p_1 p_2^2 p_4^2 p_6 p_3^3+1478400 p_2 p_4^3 p_5 p_6 p_3^3+1131900
 p_2^2 p_4^2 p_5 p_6 p_3^3+4928000 p_1 p_2 p_4^2 p_5 p_6 p_3^3+711480 p_1 p_2 p_4^2 p_5 p_6^2 p_3^2+178200 p_2 p_4^3 p_5^2 p_6 p_3^2+168960 p_1 p_2
 p_4^2 p_5^2 p_6 p_3^2+2182950 p_1 p_2 p_4^3 p_5 p_6 p_3^2+2674100 p_1 p_2^2 p_4^2 p_5 p_6 p_3^2) \textcolor{red}{z^9}+(1358280 p_2 p_4^2 p_5 p_6 p_3^3+177870
 p_2 p_4^2 p_5 p_6^2 p_3^2+44100 p_1 p_2^2 p_4^2 p_5 p_3^2+44550 p_1 p_2^2 p_4^2 p_6 p_3^2+32340 p_2 p_4^2 p_5^2 p_6 p_3^2+495000 p_2 p_4^3 p_5 p_6
 p_3^2+242550 p_2^2 p_4^2 p_5 p_6 p_3^2+3256110 p_1 p_2 p_4^2 p_5 p_6 p_3^2+202125 p_1 p_2^2 p_4 p_5 p_6 p_3^2) \textcolor{red}{z^8}+(94080 p_1 p_2 p_4^2
 p_5 p_3^2+32340 p_1 p_2 p_4^2 p_6 p_3^2+16500 p_1 p_2^2 p_4 p_6 p_3^2+1132560 p_2 p_4^2 p_5 p_6 p_3^2+316800 p_1 p_2 p_4 p_5 p_6 p_3^2+443520 p_1
 p_2 p_4^2 p_5 p_6 p_3) \textcolor{red}{z^7}+(36750 p_2 p_4^2 p_5 p_3^2+8085 p_2 p_4^2 p_6 p_3^2+26950 p_1 p_2 p_4 p_6 p_3^2+26950 p_4^2 p_5 p_6 p_3^2+107800
 p_2 p_4 p_5 p_6 p_3^2+45360 p_1 p_2 p_4^2 p_5 p_3+249480 p_2 p_4^2 p_5 p_6 p_3+92400 p_1 p_2 p_4 p_5 p_6 p_3) \textcolor{red}{z^6}+(9702 p_2 p_4 p_6 p_3^2+31500
 p_2 p_4^2 p_5 p_3+10752 p_1 p_2 p_4 p_5 p_3+8316 p_1 p_2 p_4 p_6 p_3+23100 p_4^2 p_5 p_6 p_3+59136 p_2 p_4 p_5 p_6 p_3) \textcolor{red}{z^5}+(4200 p_3
 p_5 p_4^2+1050 p_1 p_2 p_3 p_4+9450 p_2 p_3 p_5 p_4+5775 p_2 p_3 p_6 p_4+6930 p_3 p_5 p_6 p_4) \textcolor{red}{z^4}+(1050 p_2 p_3 p_4+2240 p_3 p_5 p_4+770
 p_3 p_6 p_4) \textcolor{red}{z^3}+(315 p_3 p_4+120 p_5 p_4) \textcolor{red}{z^2}+30 p_4 \textcolor{red}{z}+1\)
 
 \vspace{1em}
 \item[$H_5=$]\(p_1^2 p_2^3 p_3^4 p_4^3 p_5^2 p_6^2 \textcolor{red}{z^{16}}+16 p_1 p_2^3 p_3^4 p_4^3 p_5^2 p_6^2 \textcolor{red}{z^{15}}+120 p_1 p_2^2 p_3^4 p_4^3 p_5^2 p_6^2
 \textcolor{red}{z^{14}}+560 p_1 p_2^2 p_3^3 p_4^3 p_5^2 p_6^2 \textcolor{red}{z^{13}}+(1050 p_1 p_2^2 p_4^2 p_5^2 p_6^2 p_3^3+770 p_1 p_2^2 p_4^3 p_5^2 p_6 p_3^3) \textcolor{red}{z^{12}}+(672
 p_1 p_2^2 p_4^2 p_5 p_6^2 p_3^3+3696 p_1 p_2^2 p_4^2 p_5^2 p_6 p_3^3) \textcolor{red}{z^{11}}+(3696 p_1 p_2^2 p_4^2 p_5 p_6 p_3^3+4312 p_1 p_2^2 p_4^2
 p_5^2 p_6 p_3^2) \textcolor{red}{z^{10}}+(2640 p_1 p_2 p_3^2 p_5^2 p_6 p_4^2+8800 p_1 p_2^2 p_3^2 p_5 p_6 p_4^2) \textcolor{red}{z^9}+(660 p_2 p_4^2 p_5^2 p_6
 p_3^2+8085 p_1 p_2 p_4^2 p_5 p_6 p_3^2+4125 p_1 p_2^2 p_4 p_5 p_6 p_3^2) \textcolor{red}{z^8}+(2640 p_2 p_4^2 p_5 p_6 p_3^2+8800 p_1 p_2 p_4 p_5 p_6 p_3^2)
 \textcolor{red}{z^7}+(4312 p_2 p_4 p_5 p_6 p_3^2+3696 p_1 p_2 p_4 p_5 p_6 p_3) \textcolor{red}{z^6}+(672 p_1 p_2 p_3 p_4 p_5+3696 p_2 p_3 p_4 p_6 p_5) \textcolor{red}{z^5}+(1050
 p_2 p_3 p_4 p_5+770 p_3 p_4 p_6 p_5) \textcolor{red}{z^4}+560 p_3 p_4 p_5 \textcolor{red}{z^3}+120 p_4 p_5 \textcolor{red}{z^2}+16 p_5 \textcolor{red}{z}+1\)
 
 \vspace{1em}
 \item[$H_6=$]\(p_1^2 p_2^4 p_3^6 p_4^4 p_5^2 p_6^4 \textcolor{red}{z^{22}}+22 p_1^2 p_2^4 p_3^6 p_4^4 p_5^2 p_6^3 \textcolor{red}{z^{21}}+231 p_1^2 p_2^4 p_3^5 p_4^4 p_5^2
 p_6^3 \textcolor{red}{z^{20}}+(770 p_1^2 p_2^3 p_4^4 p_5^2 p_6^3 p_3^5+770 p_1^2 p_2^4 p_4^3 p_5^2 p_6^3 p_3^5) \textcolor{red}{z^{19}}+(770 p_1 p_2^3 p_4^4 p_5^2
 p_6^3 p_3^5+5775 p_1^2 p_2^3 p_4^3 p_5^2 p_6^3 p_3^5+770 p_1^2 p_2^4 p_4^3 p_5 p_6^3 p_3^5) \textcolor{red}{z^{18}}+(8316 p_1 p_2^3 p_4^3 p_5^2 p_6^3 p_3^5+8316
 p_1^2 p_2^3 p_4^3 p_5 p_6^3 p_3^5+9702 p_1^2 p_2^3 p_4^3 p_5^2 p_6^3 p_3^4) \textcolor{red}{z^{17}}+(14784 p_1 p_2^3 p_4^3 p_5 p_6^3 p_3^5+26950 p_1 p_2^3
 p_4^3 p_5^2 p_6^3 p_3^4+26950 p_1^2 p_2^3 p_4^3 p_5 p_6^3 p_3^4+5929 p_1^2 p_2^3 p_4^3 p_5^2 p_6^2 p_3^4) \textcolor{red}{z^{16}}+(16500 p_1 p_2^2 p_4^3
 p_5^2 p_6^3 p_3^4+90112 p_1 p_2^3 p_4^3 p_5 p_6^3 p_3^4+16500 p_1^2 p_2^3 p_4^2 p_5 p_6^3 p_3^4+23716 p_1 p_2^3 p_4^3 p_5^2 p_6^2 p_3^4+23716 p_1^2
 p_2^3 p_4^3 p_5 p_6^2 p_3^4) \textcolor{red}{z^{15}}+(72765 p_1 p_2^2 p_4^3 p_5 p_6^3 p_3^4+72765 p_1 p_2^3 p_4^2 p_5 p_6^3 p_3^4+32670 p_1 p_2^2 p_4^3
 p_5^2 p_6^2 p_3^4+108900 p_1 p_2^3 p_4^3 p_5 p_6^2 p_3^4+32670 p_1^2 p_2^3 p_4^2 p_5 p_6^2 p_3^4) \textcolor{red}{z^{14}}+(107800 p_1 p_2^2 p_4^2 p_5 p_6^3
 p_3^4+177870 p_1 p_2^2 p_4^3 p_5 p_6^2 p_3^4+177870 p_1 p_2^3 p_4^2 p_5 p_6^2 p_3^4+16940 p_1 p_2^2 p_4^3 p_5^2 p_6^2 p_3^3+16940 p_1^2 p_2^3 p_4^2
 p_5 p_6^2 p_3^3) \textcolor{red}{z^{13}}+(379456 p_1 p_2^2 p_4^2 p_5 p_6^2 p_3^4+45276 p_1 p_2^2 p_4^2 p_5 p_6^3 p_3^3+5082 p_1 p_2^2 p_4^2 p_5^2 p_6^2
 p_3^3+105875 p_1 p_2^2 p_4^3 p_5 p_6^2 p_3^3+105875 p_1 p_2^3 p_4^2 p_5 p_6^2 p_3^3+5082 p_1^2 p_2^2 p_4^2 p_5 p_6^2 p_3^3) \textcolor{red}{z^{12}}+705432 p_1
 p_2^2 p_3^3 p_4^2 p_5 p_6^2 \textcolor{red}{z^{11}}+(5082 p_1 p_2^2 p_4^2 p_6^2 p_3^3+5082 p_2^2 p_4^2 p_5 p_6^2 p_3^3+105875 p_1 p_2 p_4^2 p_5 p_6^2 p_3^3+105875
 p_1 p_2^2 p_4 p_5 p_6^2 p_3^3+45276 p_1 p_2^2 p_4^2 p_5 p_6 p_3^3+379456 p_1 p_2^2 p_4^2 p_5 p_6^2 p_3^2) \textcolor{red}{z^{10}}+(16940 p_1 p_2^2 p_4
 p_6^2 p_3^3+16940 p_2 p_4^2 p_5 p_6^2 p_3^3+177870 p_1 p_2 p_4^2 p_5 p_6^2 p_3^2+177870 p_1 p_2^2 p_4 p_5 p_6^2 p_3^2+107800 p_1 p_2^2 p_4^2 p_5
 p_6 p_3^2) \textcolor{red}{z^9}+(32670 p_1 p_2^2 p_4 p_6^2 p_3^2+32670 p_2 p_4^2 p_5 p_6^2 p_3^2+108900 p_1 p_2 p_4 p_5 p_6^2 p_3^2+72765 p_1 p_2 p_4^2
 p_5 p_6 p_3^2+72765 p_1 p_2^2 p_4 p_5 p_6 p_3^2) \textcolor{red}{z^8}+(23716 p_1 p_2 p_4 p_6^2 p_3^2+23716 p_2 p_4 p_5 p_6^2 p_3^2+16500 p_1 p_2^2 p_4
 p_6 p_3^2+16500 p_2 p_4^2 p_5 p_6 p_3^2+90112 p_1 p_2 p_4 p_5 p_6 p_3^2) \textcolor{red}{z^7}+(5929 p_2 p_4 p_6^2 p_3^2+26950 p_1 p_2 p_4 p_6 p_3^2+26950
 p_2 p_4 p_5 p_6 p_3^2+14784 p_1 p_2 p_4 p_5 p_6 p_3) \textcolor{red}{z^6}+(9702 p_2 p_4 p_6 p_3^2+8316 p_1 p_2 p_4 p_6 p_3+8316 p_2 p_4 p_5 p_6 p_3)
 \textcolor{red}{z^5}+(770 p_1 p_2 p_3 p_6+5775 p_2 p_3 p_4 p_6+770 p_3 p_4 p_5 p_6) \textcolor{red}{z^4}+(770 p_2 p_3 p_6+770 p_3 p_4 p_6) \textcolor{red}{z^3}+231 p_3 p_6 \textcolor{red}{z^2}+22
 p_6 \textcolor{red}{z}+1\)
 \end{itemize}
 

    \begin{center}
    {\bf Acknowledgments}
    \end{center}
  
   This research was funded by RUDN University, scientific project number FSSF-2023-0003.

 \end{document}